\documentclass[twoside,openright,12pt]{report}

\usepackage{hyperref} 

\usepackage{fancyhdr}
\usepackage{color}
\usepackage{latexsym}
\usepackage{amssymb}
\ifx\pdfoutput\undefined
\usepackage{graphicx}
\else
\usepackage[pdftex]{graphicx}
\usepackage{epstopdf}
\fi
\DeclareGraphicsExtensions{.pdf,.gif,.eps}
\usepackage{pifont}

\usepackage{subfig}
\usepackage{makecell}
\usepackage{array} 
\newcolumntype{P}[1]{>{\centering\arraybackslash}p{#1}}
\def\d{{\rm d}}
\usepackage{afterpage}
\usepackage[toc]{appendix}

\newcommand{\be}{\begin{eqnarray}}
\newcommand{\ee}{\end{eqnarray}}

\def\nue{{\nu_e}}

\def\numu{{\nu_{\mu}}}
\def\anumu{{\bar\nu_{\mu}}}
\def\nutau{{\nu_{\tau}}}

\def\lsim{\:\raisebox{-0.5ex}{$\stackrel{\textstyle<}{\sim}$}\:}
\def\gsim{\:\raisebox{-0.5ex}{$\stackrel{\textstyle>}{\sim}$}\:}
\newcommand{\ms}{\Delta m^2_{21}}
\newcommand{\ma}{\Delta m^2_{31}}
\newcommand{\meff}{\Delta m^2_{\rm eff}}

\newcommand{\stch}{\sin^2 2\theta_{13}}
\newcommand{\sa}{\sin^2 \theta_{23}}
\newcommand{\sta}{\sin^22 \theta_{23}}

\newcommand{\stcht}{\sin^2 2\theta_{13}{\mbox {(true)}}}
\newcommand{\sat}{\sin^2 \theta_{23}{\mbox {(true)}}}

\newcommand{\dcpt}{\delta_{\rm CP}{\mbox {(true)}}}
\newcommand{\dcp}{\delta_{\rm CP}}

\def\nn{\nonumber}

\def\gs{\mathrel{
   \rlap{\raise 0.511ex \hbox{$>$}}{\lower 0.511ex \hbox{$\sim$}}}}
\def\ls{\mathrel{
   \rlap{\raise 0.511ex \hbox{$<$}}{\lower 0.511ex \hbox{$\sim$}}}}
\newcommand{\bea}{\begin{equation} \begin{array}{c}}
\newcommand{\bead}{\begin{equation} \begin{array}{cccc}}
\newcommand{\eea}{ \end{array} \end{equation}}

\def\chisqical{\chi^2_{\rm ~ICAL}}
\def\chisqmh{\Delta\chi^2_{\rm ~ICAL-MH}}
\def\chisqpm{\Delta\chi^2_{\rm ~ICAL-PM}}
\def\chisqos{\Delta\chi^2_{\rm ~ICAL-OS}}

\def\be{\begin{equation}}
\def\ee{\end{equation}}
\def\bea{\begin{eqnarray}}
\def\eea{\end{eqnarray}}\def\nn{\nonumber}
\def\gsim{\ \rlap{\raise 2pt\hbox{$>$}}{\lower 2pt \hbox{$\sim$}}\ }
\def\lsim{\ \rlap{\raise 2pt\hbox{$<$}}{\lower 2pt \hbox{$\sim$}}\ }
\def\dslash{\kern-4pt \not{\hbox{\kern-2pt $\partial$}}}
\def\pslash{\not{\hbox{\kern-2pt p}}}

\newcommand{\nova}{NO$\nu$A\ }

\newcommand{\thb}{{\theta_b}}

\usepackage{chngcntr}

\begin{document}

\pagenumbering{roman}

\markboth{Physics Potential of the ICAL detector at INO}{The ICAL Collaboration}


\thispagestyle{empty} 

\setcounter{page}{1}
\begin{flushright}
{INO/ICAL/PHY/NOTE/2015-01} \\
{ArXiv:1505.07380 [physics.ins-det]} \\
{ \ } \\
{Pramana - J Phys (2017) 88 : 79}\\
{doi:10.1007/s12043-017-1373-4}
\end{flushright}        
   
\vfill

\begin{center}
{\LARGE \bf Physics Potential of the ICAL detector at} \\ [0.2cm] 
{\LARGE \bf the India-based Neutrino Observatory (INO)} \\ [1.5cm]

The ICAL Collaboration

\vfill

\newpage

[The ICAL Collaboration]

\vspace{0.2in}

Shakeel Ahmed, M. Sajjad Athar, Rashid Hasan, Mohammad Salim, S. K. Singh

{\it Aligarh Muslim University, Aligarh 202001, India }

\vspace{0.1in}

S. S. R. Inbanathan

{\it The American College, Madurai 625002, India}

\vspace{0.1in}

Venktesh Singh,  V. S. Subrahmanyam

{\it Banaras Hindu University, Varanasi 221005, India}

\vspace{0.1in}

Shiba Prasad Behera$^{\rm HB}$, Vinay B. Chandratre, Nitali Dash$^{\rm HB}$, 
Vivek M. Datar$^{\rm VD}$, \\
V. K. S. Kashyap$^{\rm HB}$, Ajit K. Mohanty, Lalit M. Pant

{\it Bhabha Atomic Research Centre, Trombay, Mumbai 400085, India}

\vspace{0.1in}

Animesh Chatterjee$^{\rm AC, HB}$, Sandhya Choubey, Raj Gandhi, 
Anushree Ghosh$^{\rm AG, HB}$, \\ Deepak Tiwari$^{\rm HB}$

{\it Harish Chandra Research Institute, Jhunsi, Allahabad 211019, India}

\vspace{0.1in}

Ali Ajmi$^{\rm HB}$, S. Uma Sankar

{\it Indian Institute of Technology Bombay, Powai, Mumbai 400076, India}

\vspace{0.1in}

Prafulla Behera, Aleena Chacko, Sadiq Jafer, James Libby, 
K. Raveendrababu$^{\rm HB}$, \\
K. R. Rebin

{\it Indian Institute of Technology Madras, Chennai 600036, India} 

\vspace{0.1in}

D. Indumathi, K. Meghna$^{\rm HB}$, S. M. Lakshmi$^{\rm HB}$, M. V. N. Murthy, 
Sumanta Pal$^{\rm SP,HB}$, \\
G. Rajasekaran$^{\rm GR}$, Nita Sinha

{\it Institute of Mathematical Sciences, Taramani, Chennai 600113, India}

\vspace{0.1in}

Sanjib Kumar Agarwalla, Amina Khatun$^{\rm HB}$

{\it Institute of Physics, Sachivalaya Marg, Bhubaneswar 751005, India}

\vspace{0.1in}

Poonam Mehta

{\it Jawaharlal Nehru University, New Delhi 110067, India}

\vspace{0.1in}

Vipin Bhatnagar, R. Kanishka, A. Kumar, J. S. Shahi, J. B. Singh
 
{\it Panjab University, Chandigarh 160014, India}

\vspace{0.1in}

Monojit Ghosh$^{\rm MG}$, Pomita Ghoshal$^{\rm PG}$, Srubabati Goswami, 
Chandan Gupta$^{\rm HB}$, \\ Sushant Raut$^{\rm SR}$

{\it Physical Research Laboratory, Navrangpura, Ahmedabad 380009, India}

\vspace{0.1in}

Sudeb Bhattacharya, Suvendu Bose, Ambar Ghosal, Abhik Jash$^{\rm HB}$, Kamalesh Kar, \\
Debasish Majumdar, Nayana Majumdar, Supratik Mukhopadhyay, Satyajit Saha 

{\it Saha Institute of Nuclear Physics, Bidhannagar, Kolkata 700064, India}

\vspace{0.1in}

B. S. Acharya, Sudeshna Banerjee, Kolahal Bhattacharya, 
Sudeshna Dasgupta$^{\rm SD,HB}$, \\
Moon Moon Devi$^{\rm MD,HB}$, Amol Dighe, Gobinda Majumder, 
Naba K. Mondal$^{\rm NM}$, \\ Asmita Redij$^{\rm AR}$, 
Deepak Samuel$^{\rm DS}$, 
B. Satyanarayana, Tarak Thakore$^{\rm TT}$ 

{\it Tata Institute of Fundamental Research, Colaba, Mumbai 400005, India}

\vspace{0.1in}

\newpage

C. D. Ravikumar, A. M. Vinodkumar

{\it University of Calicut, Kozhikode, Kerala 673635, India}

\vspace{0.1in}

Gautam Gangopadhyay, Amitava Raychaudhuri

{\it University of Calcutta, Kolkata 700009, India}

\vspace{0.1in}

Brajesh C. Choudhary, Ankit Gaur, Daljeet Kaur, Ashok Kumar,
Sanjeev Kumar, \\ Md. Naimuddin

{\it University of Delhi, New Delhi 110021, India}

\vspace{0.1in}

Waseem Bari, Manzoor A. Malik

{\it University of Kashmir, Hazratbal, Srinagar 190006, India}

\vspace{0.1in}

Jyotsna Singh

{\it University of Lucknow, Lucknow 226007, India }

\vspace{0.1in}

S. Krishnaveni, H. B. Ravikumar, C. Ranganathaiah

{\it University of Mysore, Mysuru, Karnataka 570005. India}

\vspace{0.1in}

Swapna Mahapatra

{\it Utkal University, Vani Vihar, Bhubaneswar, 751004, India}

\vspace{0.1in}

Saikat Biswas$^{\rm SB}$, Subhasis Chattopadhyay, Rajesh Ganai$^{\rm HB}$, 
Tapasi Ghosh$^{\rm TG}$, \\ Y. P. Viyogi$^{\rm YV}$

{\it Variable Energy Cyclotron Centre, Bidhannagar, Kolkata 700064, India}

\vspace{0.1in}

\end{center}

\vfill

\noindent \hrulefill

\noindent
$^{\rm NM}$ Spokesperson, nkm@tifr.res.in

\noindent
$^{\rm AC}$ Currently: University of Texas at Arlington, USA. 

\noindent
$^{\rm AG}$ Currently: Centro Brasileiro de Pesquisas F\'isicas, 
Rio de Janeiro, Brazil.

\noindent
$^{\rm AR}$ Currently: University of Bern, Switzerland.

\noindent
$^{\rm DS}$ Currently: Central University of Karnataka, Gulbarga, India.

\noindent
$^{\rm GR}$ Chennai Mathematical Institute, Chennai, India.

\noindent
$^{\rm HB}$ Homi Bhabha National Institute, Mumbai, India

\noindent
$^{\rm MD}$ Currently: Weizmann Institute of Science, Rehovot, Israel.

\noindent
$^{\rm MG}$ Currently: Tokyo Metropolitan University, Hachioji, Tokyo, Japan

\noindent
$^{\rm PG}$ Currently: LNM Institute of Information Technology, Jaipur, India.

\noindent
$^{\rm SR}$ Currently: 
Institute for Basic Science, Daejeon, Korea

\noindent
$^{\rm SB}$ Currently: National Institute of Science, Education and
Research, Bhubaneswar, India.

\noindent
$^{\rm SD}$ Currently: University of Bristol, UK.

\noindent
$^{\rm SP}$ Currently: Virginia Tech, Blacksburg, VA, USA 

\noindent
$^{\rm TG}$ Currently: 
Universidade Federal de Goias, Goiania, Brazil.

\noindent
$^{\rm TT}$ Currently: Louisiana State University, USA.

\noindent
$^{\rm VD}$ Currently: Tata Institute of Fundamental Research, Mumbai, India.

\noindent
$^{\rm YV}$ Raja Ramanna Fellow.

\newpage

\chapter*{Abstract}

The upcoming 50 kt magnetized iron calorimeter (ICAL) detector at the 
India-based Neutrino Observatory (INO) is designed to study the atmospheric
neutrinos and antineutrinos separately over a wide range of energies 
and path lengths. The primary focus of this experiment is to explore the 
Earth matter effects by observing the energy and zenith angle dependence 
of the atmospheric neutrinos in the multi-GeV range. This study will be 
crucial to address some of the outstanding issues in neutrino oscillation 
physics, including the fundamental issue of neutrino mass hierarchy. 
In this document, we present the physics potential of the detector
as obtained from realistic detector simulations. We describe the simulation 
framework, the neutrino interactions in the detector, and the expected 
response of the detector to particles traversing it. The ICAL detector can 
determine the energy and direction of the muons to a high precision, 
and in addition, its sensitivity to multi-GeV hadrons increases its 
physics reach substantially. 
Its charge identification capability, and hence its ability to 
distinguish neutrinos from antineutrinos, makes it an efficient detector
for determining the neutrino mass hierarchy. 
In this report, we outline the analyses carried out for the determination 
of neutrino 
mass hierarchy and precision measurements of atmospheric neutrino mixing 
parameters at ICAL, and give the expected physics reach of the detector
with 10 years of runtime.
We also explore the potential of ICAL for probing new physics 
scenarios like CPT violation and the presence of magnetic monopoles.


\newpage

\chapter*{Preface}

The past two decades in neutrino physics have been very eventful,
and have established this field as one of the flourishing areas of
high energy physics.
Starting from the confirmation of neutrino oscillations that resolved
the decades-old problems of the solar and atmospheric neutrinos,
we have now been able to show that neutrinos have nonzero masses, and
different flavors of neutrinos mix among themselves.
Our understanding of neutrino properties has increased
by leaps and bounds. Many experiments have been constructed and
envisaged to explore different facets of neutrinos, in particular
their masses and mixing. 

The Iron Calorimeter (ICAL) experiment at the India-based Neutrino 
Observatory (INO) \cite{ino-website} is one of the major detectors 
that is expected to see the light of the day soon. 
It will have unique features like the ability to 
distinguish muon neutrinos from antineutrinos at GeV energies, and 
measure the energies of hadrons in the same energy range. It is
therefore well suited for the identification of neutrino mass hierarchy, 
the measurement of neutrino mixing parameters, and many probes of 
new physics.
The site for the INO has been identified, and the construction is expected 
to start soon. In the meanwhile, the R\&D for the ICAL detector, including 
the design of its modules, the magnet coils, the active detector elements 
and the associated electronics, has been underway over the past decade. 
The efforts to understand the capabilities and physics potentials of the 
experiment through simulations are in progress at the same time. 

We present here the Status Report of our current understanding of the physics
reach of the ICAL, prepared by the Simulations and Physics Analysis
groups of the INO Collaboration. It describes the framework being used
for the simulations, the expected response of the detector to
particles traversing it, and the results we expect to obtain after
the 50 kt ICAL has been running for about a decade. The focus of
the physics analysis is on the identification of the mass hierarchy
and precision measurements of the atmospheric neutrino mixing parameters.
The feasibilities of searches for some new physics, in neutrino interactions 
as well as elsewhere, that can be detected at the ICAL, are also
under investigation. 

The first such report \cite{Athar:2006yb} had been published when the 
INO was being proposed, and the ICAL Collaboration was at its inception. 
Our understanding of the detector has now matured quite a bit, and more 
realistic results can now be obtained, which have been included in this 
report. The work on improving several aspects of the detector, the 
simulations, the reconstruction algorithms and the analysis techniques is in
progress and will remain so for the next few years. This Report is 
thus not the final word, but a work in progress that will be updated
at regular intervals. 

In addition to the ICAL detector, the INO facility is
designed to accommodate  experiments in other areas like neutrinoless
double beta decay, dark matter search, low-energy neutrino spectroscopy,
etc. Preliminary investigations and R\&D in this direction are in
progress. The special environment provided by the underground laboratory
may also be useful to conduct experiments in rock mechanics, geology,
biology etc. This Report focusses mainly on the ongoing physics 
and simulation related to the ICAL detector. The details of other experiments 
will be brought out separately.

The Government of India has recently (December 2014) given its approval for the
establishment of INO. This is a good opportunity to present the physics
capabilities of the ICAL experiment in a consolidated form.

\newpage


\setcounter{tocdepth}{3}
\tableofcontents
\afterpage{\null \thispagestyle{empty} \newpage}

\newpage

\pagenumbering{arabic}
\setcounter{page}{1}

\chapter*{Executive Summary}

\section*{The INO and the ICAL detector}
\addcontentsline{toc}{section}{The INO and the ICAL detector}

The India-based Neutrino Observatory (INO) is proposed to be built in
Bodi West Hills, in Theni district of Tamil Nadu in South India. 
The main detector proposed to be built at the INO is the magnetised Iron
CALorimeter (ICAL) with a mass of 50 kt. 
The major physics goal of ICAL is to study neutrino properties, through
the observation of atmospheric neutrinos that cover a wide range of
energies and path lengths.
A special emphasis will be on the determination of the neutrino mass hierarchy,
by observing the matter effects when they travel through the Earth. 
This would be facilitated through the ability of ICAL to distinguish neutrinos 
from antineutrinos. 

Table~\ref{tab:ical} gives the salient features of the ICAL detector. 
The active detector elements in ICAL will be the Resistive Plate
Chambers (RPCs). 
The detector is optimised to be sensitive primarily to the atmospheric 
muon neutrinos in the 1--15 GeV energy range. The structure of the detector, 
with its horizontal layers of iron interspersed with RPCs, allows it to 
have an almost complete coverage to the direction of incoming neutrinos, except
for those that produce almost horizontally traveling muons. This makes it
sensitive to a large range of path lengths $L$ for the neutrinos travelling
through the Earth, while the atmospheric neutrino flux provides a wide 
spectrum in the neutrino energy $E_\nu$. 

\begin{table}[h]
\centering
 \begin{tabular}{|l|l|}
 \hline
 \multicolumn{2}{|c|}{{\bf ICAL} \rule{0mm}{5mm}} \\
 \hline
 No. of modules & 3\\ 
 Module dimension & 16 m $\times$ 16 m $\times$ 14.5 m\\ 
 Detector dimension & 48 m $\times$ 16 m $\times$ 14.5 m\\ 
 No. of layers & 151\\ 
 Iron plate thickness & 5.6 cm \\ 
 Gap for RPC trays & 4.0 cm\\ 
 {Magnetic field} & {1.5 Tesla}  \\
 \hline
 \multicolumn{2}{|c|}{{\bf RPC} \rule{0mm}{5mm}} \\
 \hline
 RPC unit  dimension & 2 m $\times$ 2 m\\ 
 Readout strip width & 3 cm\\ 
 No. of RPC units/Layer/Module &  64\\ 
 {Total no. of RPC units}  & {$\sim$~30,000}  \\ 
 {No. of electronic readout channels}  & {3.9 $\times$ $10^6$} \\
 \hline
 \end{tabular}
\caption{Specifications of the ICAL detector.}
\label{tab:ical}
\end{table}

ICAL will be sensitive to both the energy and direction of the
muons that will be produced in charged-current (CC) interactions of the
atmospheric muon neutrinos (and antineutrinos) with the iron target in the
detector. In addition, the fast response time of the RPCs (of the order
of nanoseconds) will allow for a discrimination of the upward-going muon events
and downward-going ones. 
(Once the starting point of the track is identified, the initial hits 
in the track determine the initial muon direction accurately.)
This direction discrimination separates
the neutrinos with short path lengths from those with longer ones. 
Such a separation is crucial since the neutrino oscillation probability 
is strongly dependent on the path length $L$.

Moreover, since ICAL is expected to be magnetised to about 1.5 T in the plane 
of the iron 
plates, it will be able to discriminate between muons of different charges, 
and hence will be capable of differentiating events induced by muon neutrinos 
and muon antineutrinos. Through this sensitivity, one can probe the 
difference in matter effects in the propagation of neutrinos and 
antineutrinos that traverse the Earth before they reach the detector. 
This in turn will allow for a sensitivity to the neutrino mass hierarchy, 
which is the primary goal of the ICAL experiment. 

The magnetic field is also crucial for reconstructing the momentum of 
the muon tracks in the case of partially contained events. 
When the muon track is completely contained inside the detector, 
the length of the track can determine the energy of the muon reliably,
and the magnetic field plays a supplementary role of improving 
the momentum resolution.
However for the partially contained track events, the bending of the
track in the local magnetic field is crucial to reconstruct the muon
momentum in the energies of interest.
The good tracking ability and energy resolution of ICAL for 
muons makes it very well suited for the study of neutrino oscillation 
physics through the observation of atmospheric neutrinos.

In addition, ICAL is also sensitive
to the energy deposited by hadrons in the detector in the multi-GeV range, 
a unique property that enables a significant improvement in the physics reach 
of ICAL, as will be clear in this report. In the present configuration 
the sensitivity of ICAL to electrons is very limited; 
however, this is still under investigation.

Though the ICAL is yet to be built, its putative properties have been 
simulated using the CERN GEANT4 \cite{Agostinelli:2002hh} package. 
The details of
these simulations have been presented in Chapter~\ref{framework}. 
This report presents results on the response of ICAL to particles 
traversing through it in Chapter~\ref{response}. The resultant physics 
potential of the detector, obtained from these simulations, is given
in later chapters, where we focus on the identification of neutrino mass 
hierarchy, and the precise 
determinations of the atmospheric neutrino parameters: 
$|\Delta m_{\rm eff}^2|$ 
and $\sin^22\theta_{23}$, as well as the octant of $\theta_{23}$. 
In addition, we also discuss some novel and exotic physics possibilities 
that may be explored at ICAL.

\section*{The simulation framework}
\addcontentsline{toc}{section}{The simulation framework}

For the results presented in this report, the atmospheric neutrino events 
have been generated with the NUANCE \cite{Casper:2002sd} neutrino generator 
using the Honda 3d fluxes \cite{Honda:2011nf} for the Kamioka site in
Japan. The details of the fluxes have been presented in Chapter~\ref{fluxes}.
The Honda atmospheric neutrino fluxes at Theni, the INO site,
are expected to be finalised soon and will be used when available. A
preliminary comparison of the fluxes at the two sites is also presented
in Appendix~\ref{honda-ino-flux} of this report.
The number of muon track events are expected to be similar, within
statistical errors, for both fluxes, for energies more than 3
GeV. We therefore do not expect the reach of ICAL, especially for the
mass hierarchy, to change significantly with the use of the Theni fluxes.

A typical CC interaction of $\nu_\mu$ in the detector gives rise to
a charged muon that leaves a track, and single or multiple hadrons
that give rise to shower-like features. 
The simulations of the propagation of muons and hadrons in the detector 
have been used to determine the response of the detector to these 
particles. This leads to the determination of detection efficiencies, 
charge identification efficiencies, calibrations and resolutions of
energies and directions of the particles. The results of these simulations 
have been presented in Chapter~\ref{response}.

In order to perform the physics analysis, we generate a large number 
(typically, an exposure of 1000 years) of unoscillated events using NUANCE,
which are later scaled to a suitable exposure, and oscillations are included
using a reweighting algorithm. The typical values of oscillation parameters 
used are close to their best-fit values, and are given in 
Table~\ref{tab:centralval}.
\begin{table}[t]
\centering
\begin{tabular}{|c|c|c|c|c|c|} \hline
$\Delta m_{21}^2$ (eV$^2$) & $\Delta m_{\rm eff}^2$ (eV$^2$) & 
$\sin^2 \theta_{12}$ & $\sin^2\theta_{23}$ & $\sin^22\theta_{13}$ &
$\delta_{\rm CP}$ \\ \hline
$7.5 \times 10^{-5}$ & $2.4 \times 10^{-3}$ & 
0.3 & 0.5 & 0.1 & 0$^\circ$ \\ \hline
\end{tabular}
\caption{True values of the input oscillation parameters used in the
analyses, unless otherwise specified. 
For more details, including 3$\sigma$ limits on these
parameters, see Table~\ref{tab:best-fit}. }
\label{tab:centralval}
\end{table}
Here $ \Delta m^{2}_{\rm eff} \equiv \Delta m^{2}_{32}-(\cos^{2}\theta_{12}-\cos
\delta_{\rm CP}\sin\theta_{32}\sin2\theta_{12}\tan\theta_{23})\Delta m^{2}_{21}$ 
is the effective value of $\Delta m^2_{\rm atm}$ relevant for the two-neutrino
analysis of atmospheric neutrino oscillations 
\cite{Nunokawa:2005nx,deGouvea:2005hk}. The energies and directions 
of the relevant particles are then smeared according to the resolutions 
determined earlier. This approach thus simulates the average behaviour
of the measured quantities. {\em At the current stage of simulations, 
we also assume that the muon track and the hadron shower can be separated 
with full efficiency, and that the noise due to random hits near the signal
events in the short time interval of the event is negligible. While these
approximations are reasonable, they still need to be justified with actual
detector, possibly by collecting data with a prototype. In the meanwhile,
a complete simulation, which involves passing each of the generated events 
through a GEANT4 simulation of the ICAL detector, is in progress.}

The different analyses then determine the relevant physics results 
through a standard $\chi^2$ minimization procedure, with the systematic
errors included through the method of pulls, marginalizations over the
allowed ranges of parameters, and including information available from
other experiments using priors. The details of the analysis procedures
for obtaining the neutrino mixing parameters have been given in 
Chapter~\ref{analysis}, which presents the results using only the 
information on muon energy and angle, as well as the improvement due 
to the inclusion of information on 
hadron energies. Chapter~\ref{synergy} further includes combined analyses 
of the reach of ICAL with other current and near-future detectors such as 
T2K and NO$\nu$A, for the mass hierarchy and neutrino oscillation parameters. 
Chapter~\ref{exotic} discusses the reach of ICAL with respect to exotic 
physics possibilities such as the violation of CPT or Lorentz symmetries,
the detection of magnetic monopoles, etc..

We now list the highlights of the results compiled in this report. Many of 
these results have appeared elsewhere \cite{Chatterjee:2014vta,
Devi:2013wxa,Mohan:2014qua,Ghosh:2012px,Thakore:2013xqa,Kaur:2014rxa,
Devi:2014yaa,Chatterjee:2014oda,Dash:2015qha}, however
some have been updated with more recent information.

\section*{Detector response to propagating particles}
\addcontentsline{toc}{section}{Detector response to propagating particles}
\subsection*{Response to muons}
\addcontentsline{toc}{subsection}{Response to muons}

\begin{figure}[] 
\includegraphics[width=0.51\textwidth,height=0.38\textwidth]
{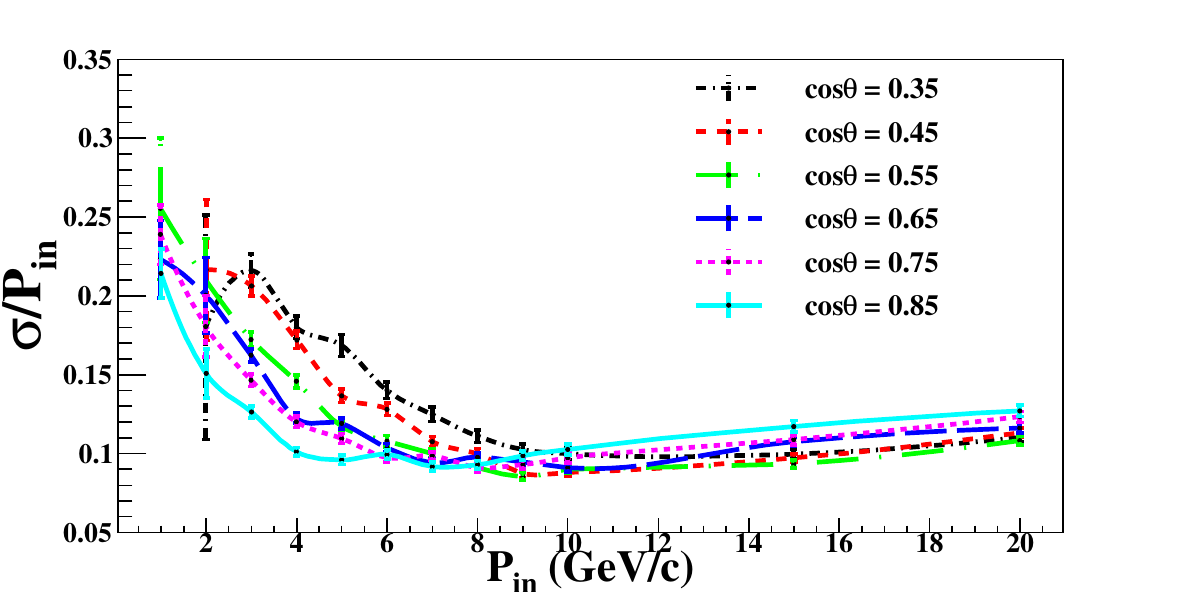}
\includegraphics[width=0.49\textwidth,height=0.35\textwidth]
{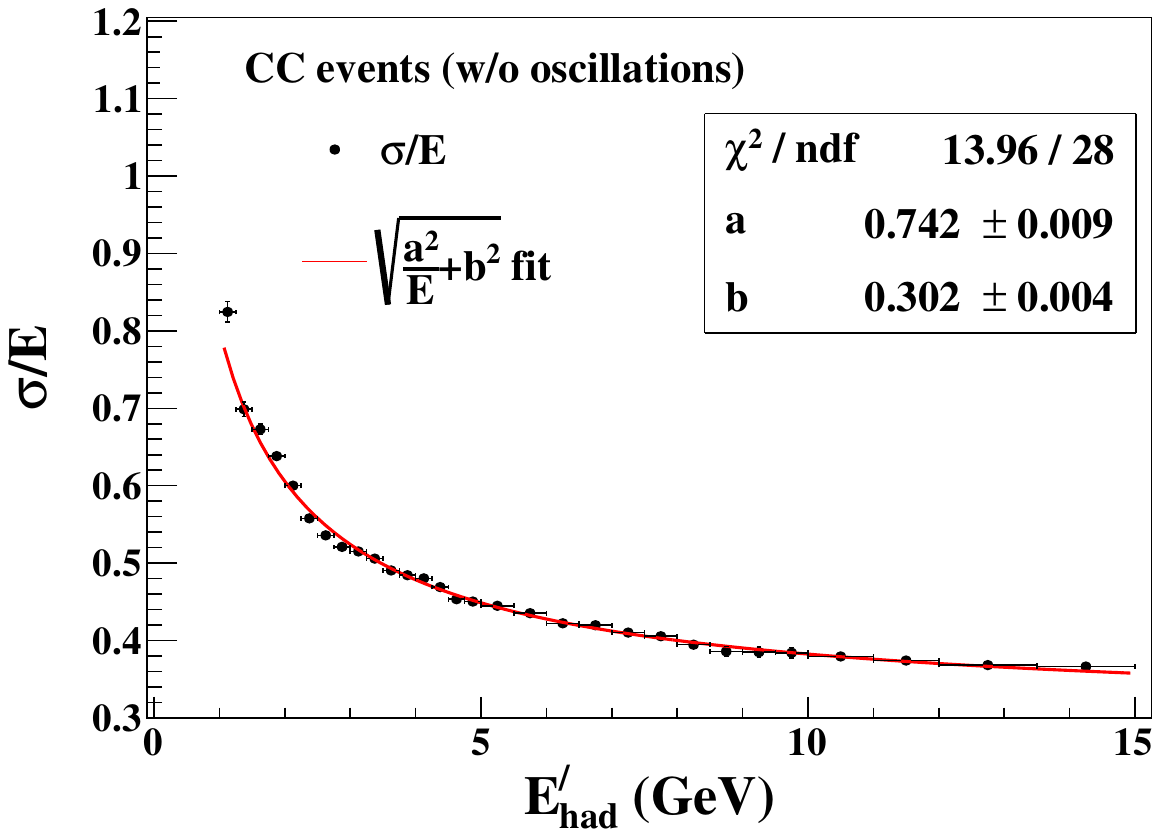} 
\caption{Left panel shows the momentum resolution of muons produced in 
the region $0<\phi <\pi/4$ (see Sec.~\ref{muon-resp}), as functions of the muon 
momentum in different zenith angle bins \cite{Chatterjee:2014vta}.
Right panel shows the energy resolution of hadrons (see Sec.~\ref{had-resp})
as functions of $E'_{\rm had}$, where events have been generated using NUANCE
in different $E'_{\rm had}$ bins. The bin widths are indicated by horizontal 
error bars \cite{Devi:2013wxa}.} 
\label{fig:muon-had-res} 
\end{figure}

The ICAL detector is optimised for the detection of muons propagating
in the detector, identification of their charges, and accurate determination 
of their energies and directions. The energies and directions of muons
are determined through a Kalman filter based algorithm. The reconstructed
energy (direction) for a given true muon energy (direction) is found to 
give a good fit to the Gaussian distribution, and hence the resolution is 
described in terms of the mean and standard deviation of a Gaussian 
distribution. The reconstruction efficiency for muons with energies
above 2 GeV is expected to be more than 80\%, while the charge of these
reconstructed muons is identified correctly on more than 95\% occasions.
The direction of these muons at the point of their production can be
determined to within about a degree. The muon energy resolution
depends on the part of the detector the muon is produced in, but is 
typically 25\% (12\%) at 1 GeV (20 GeV), as can be seen in the left
panel of Fig.~\ref{fig:muon-had-res} \cite{Chatterjee:2014vta}. 

\subsection*{Response to hadrons}
\addcontentsline{toc}{subsection}{Response to hadrons}

The detector response to hadrons is quantified in terms of the quantity 
$E'_{\rm had} \equiv E_\nu - E_\mu$ for the CC processes that produce a muon, 
which is calibrated against the number of hits in the detector.
The hit distribution for a given hadron energy is found to give a
good fit to the Vavilov distribution. Hence $E'_{\rm had}$ is 
calibrated against the mean of the corresponding Vavilov mean for
the number of hits, and the energy resolution is taken to be the 
corresponding value of $\sigma$. The energy resolution is shown 
in Fig.~\ref{fig:muon-had-res}. The complete description of Vavilov
distributions needs a total of four parameters, the details of which
may be found in Chapter~\ref{response}. The presence of different kinds of 
hadrons, which are hard to distinguish through the hit information,
is taken care of by the generation of events through NUANCE,
which is expected to produce the hadrons in the right proportions.
The energy resolution of hadrons is found to be about 85\% (36\%)
at 1 GeV (15 GeV) \cite{Devi:2013wxa}. The information on the shape of the 
hadron shower is not used for extracting hadron energy yet; the work on 
this front is still in progress.

\section*{Physics reach of ICAL}
\addcontentsline{toc}{section}{Physics potential of the ICAL}
\subsection*{Sensitivity to the Mass Hierarchy}
\addcontentsline{toc}{subsection}{Sensitivity to mass hierarchy}

In order to quantify the reach of ICAL with respect to the neutrino mass 
hierarchy, a specific hierarchy, normal or inverted, is chosen as the true
(input) hierarchy. The CC muon neutrino events are binned in the quantities 
chosen for the analysis, and a $\chi^2$ analysis is performed taking the 
systematic errors into account and marginalising over the $3\sigma$ 
ranges of the parameters $\vert \Delta m^2_{\rm eff} \vert$, 
$\sin^2\theta_{23}$ and $\sin^22\theta_{13}$. The significance of the result 
is then determined as the $\Delta \chi^2_{\rm ICAL-MH}$ with which the 
wrong hierarchy can be rejected (see Chapter~\ref{analysis}).

\begin{figure}[h!]
\centering
\includegraphics[width=0.49\textwidth,height=0.35\textwidth]
{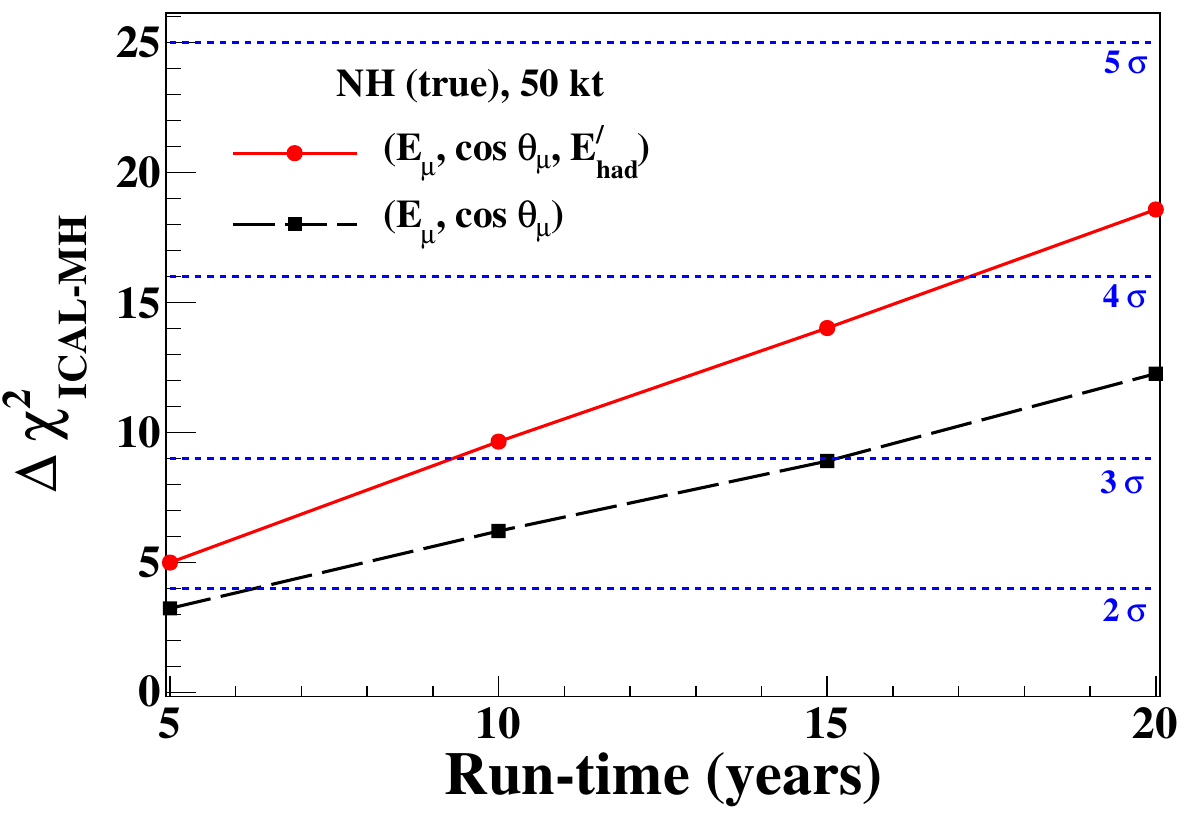}
\includegraphics[width=0.49\textwidth,height=0.35\textwidth]
{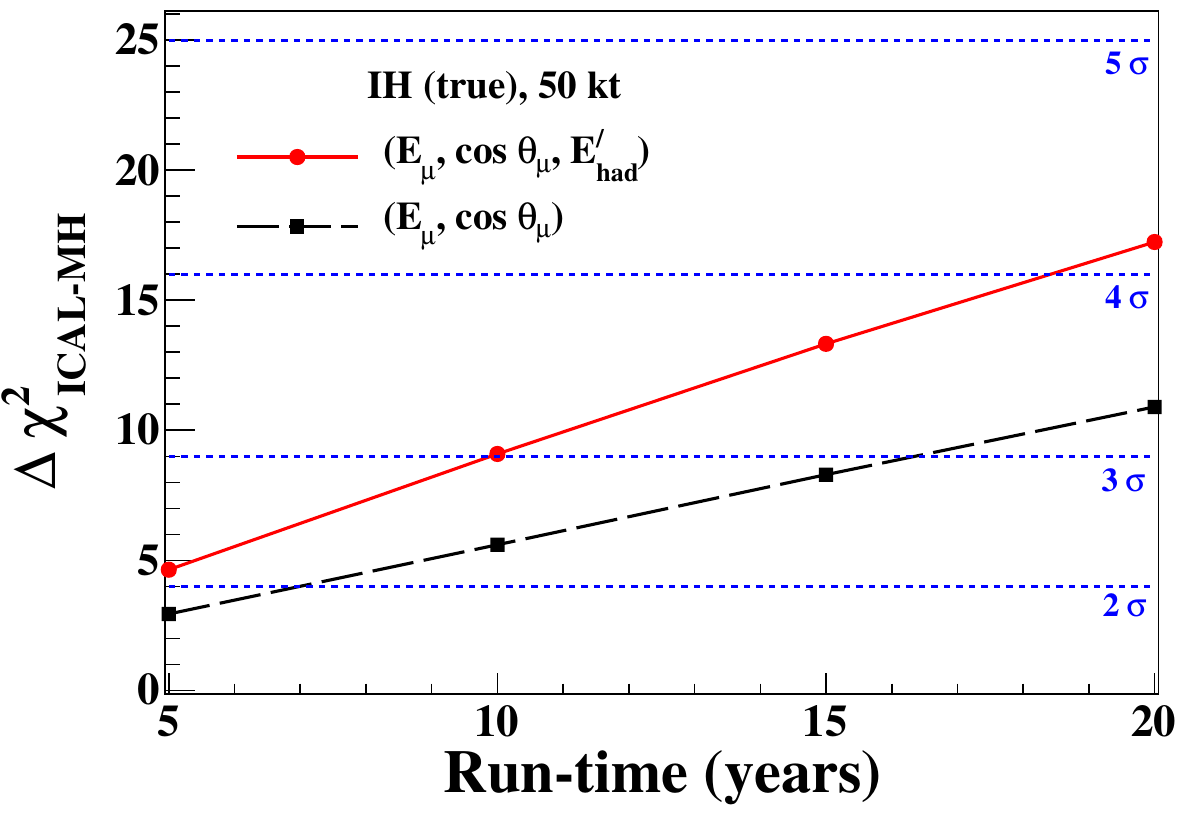}
\caption{The hierarchy sensitivity of ICAL with input normal (left)
and inverted (right) hierarchy including correlated hadron energy
information, with $\vert \Delta m^2_{\rm eff} \vert$,
$\sin^2\theta_{23}$ and $\sin^22\theta_{13}$ marginalised over their
$3\sigma$ ranges \cite{Devi:2014yaa}. 
Improvement with the inclusion of hadron energy 
is significant.}
\label{fig:mh-improve-3d}
\end{figure}

The analysis for mass hierarchy identification {\it using only the muon 
momentum information} \cite{Ghosh:2012px} yields 
$\Delta \chi^2_{\rm ICAL-MH} \approx 6.5$ 
with 10 years of exposure of the 50 kt ICAL, 
as can be seen from Fig.~\ref{fig:mh-improve-3d} (black
dashed curve), which also shows (red solid curve) that  
a considerable improvement in the physics reach is obtained 
if the correlated hadron energy information in each event is included
along with the muon energy and direction information; i.e. the binning
is performed in the 3-dimensional parameter space 
$(E_\mu, \cos\theta_\mu, E'_{\rm had})$. 
The same exposure now allows the 
identification of mass hierarchy with a significance of 
$\Delta\chi^2_{\rm ICAL-MH} \approx 9.5$ \cite{Devi:2014yaa} for
maximal mixing angle ($\sin^2\theta_{23}=0.5$) and $\sin^2 2\theta_{13}=0.1$. 
The significance depends on the actual value of $\theta_{23}$ and
$\theta_{13}$, and increases with the values of these mixing angles.
When $\sin^2\theta_{23}$ and $\sin^2 2\theta_{13}$ are varied in their
allowed $3\sigma$ ranges, the corresponding
signifiance varies in the range $\Delta \chi^2_{\rm ICAL-MH} \approx$ 7--12.

\subsection*{Precision Measurements of oscillation parameters}
\addcontentsline{toc}{subsection}{Precision measurements of oscillation 
parameters}

\begin{figure}[h!]
\centering
\includegraphics[width=0.6\textwidth,height=0.4\textwidth]
{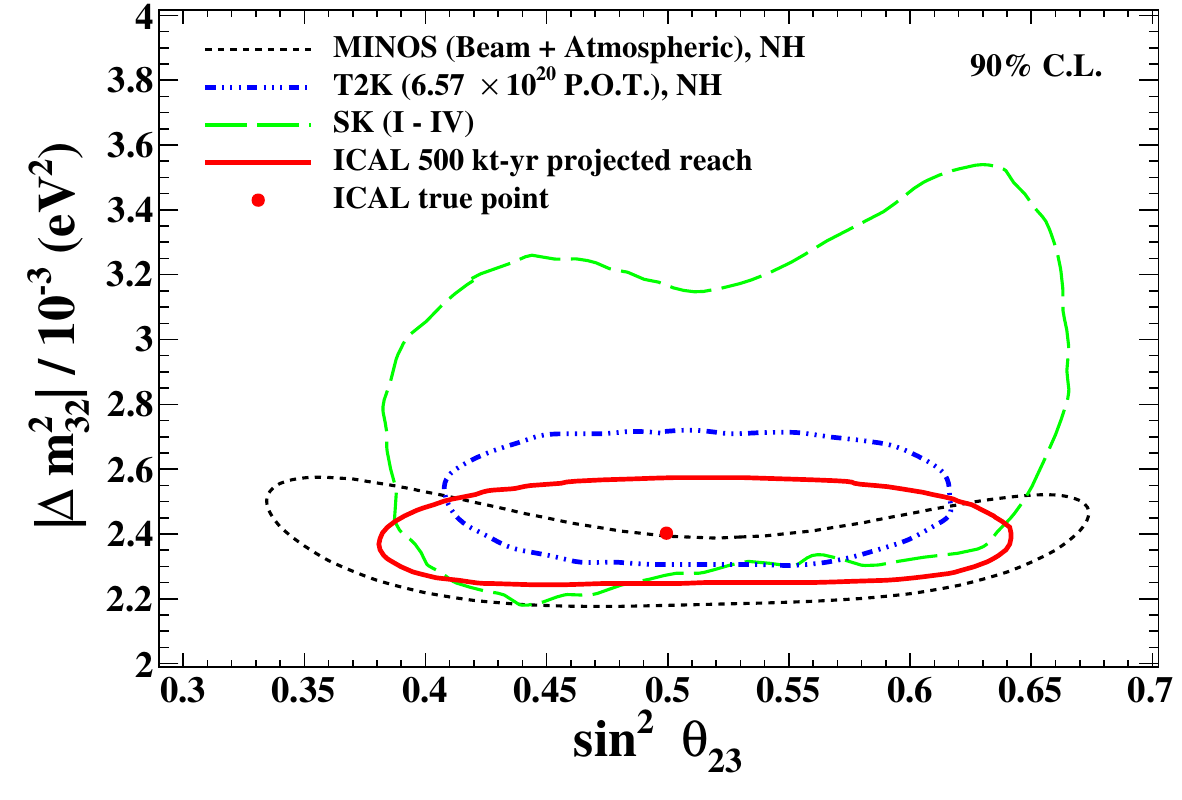}
\caption{The precision reach of ICAL in the 
$\sin^2\theta_{23}$--$\Delta m^2_{32}$ plane, in comparison with other 
current and planned experiments \cite{Devi:2014yaa}.
Information on hadron energy has been included.}
\label{fig:precision-cont}
\end{figure}

The precision on the measurements of the neutrino oscillation
parameters $\sin^2\theta_{23}$ and 
$|\Delta m^2_{32}|$ is quantified in terms of 
$\Delta \chi^2_{\rm ICAL-PM}(\lambda)$,
where $\lambda$ is the parameter under consideration. 
The precision on the measurement of $\theta_{23}$ is essentially a function 
of the total number of events, and is expected to be about 12-14\%,
whether one includes the hadron energy information or not. 
The precision on $|\Delta m^2_{32}|$, however, improves
significantly (from 5.4\% to 2.9\%) if the information on hadron
energy is included.

Figure~\ref{fig:precision-cont} shows the comparison of the 10-year
reach of 50 kt ICAL in the $\sin^22\theta_{23}$--$\Delta m^2_{32}$ plane,
with the current limits from other experiments.
It is expected that the $\Delta m^2_{32}$ precision of ICAL would be 
much better than the atmospheric neutrino experiments that use water 
Cherenkov detectors, due to its better energy measurement capabilities.
However the beam experiments will keep on accumulating more data, hence
the global role of ICAL for precision measurements of these parameters
will be not competitive, but complementary.

\subsection*{Sensitivity to the octant of $\theta_{23}$}
\addcontentsline{toc}{subsection}{Sensitivity to the octant of $\theta_{23}$}

While the best-fit value for $\sin^22\theta_{23}$ is close to maximal,
it is not fully established whether it deviates from maximality, and if
so, whether $\sin^2\theta_{23}$ is less than or greater than 0.5, that
is, whether it lies in the first or second octant. ICAL is sensitive to
the octant of $\theta_{23}$ through two kinds of effects: one is through
the depletion in atmospheric muon neutrinos (and antineutrinos) via the
survival probability $P_{\mu\mu}$ and the other is the contribution of
the atmospheric electron neutrinos to the observed CC muon events
through the oscillation probability $P_{e\mu}$. Both effects are proportional 
to $\sin^22\theta_{13}$, however act in opposite directions,
thereby reducing the effective sensitivity of ICAL to the $\theta_{23}$ octant. 
The reach of ICAL alone for determining the octant is therefore limited;
it can identify the octant to a $2\sigma$ significance with 500 kt-yr only if
$\sin^2 \theta_{23} < 0.37$ \cite{Devi:2014yaa}. 
The information from other experiments clearly
needs to be included in order to identify the octant.

\section*{Synergies with other experiments}
\addcontentsline{toc}{section}{Synergies with other experiments}

\subsection*{Neutrino Mass Hierarchy determination}
\addcontentsline{toc}{subsection}{Neutrino mass hierarchy determination}

\begin{figure}[ht]
\centering
\includegraphics[width=0.49\textwidth,height=0.35\textwidth]
{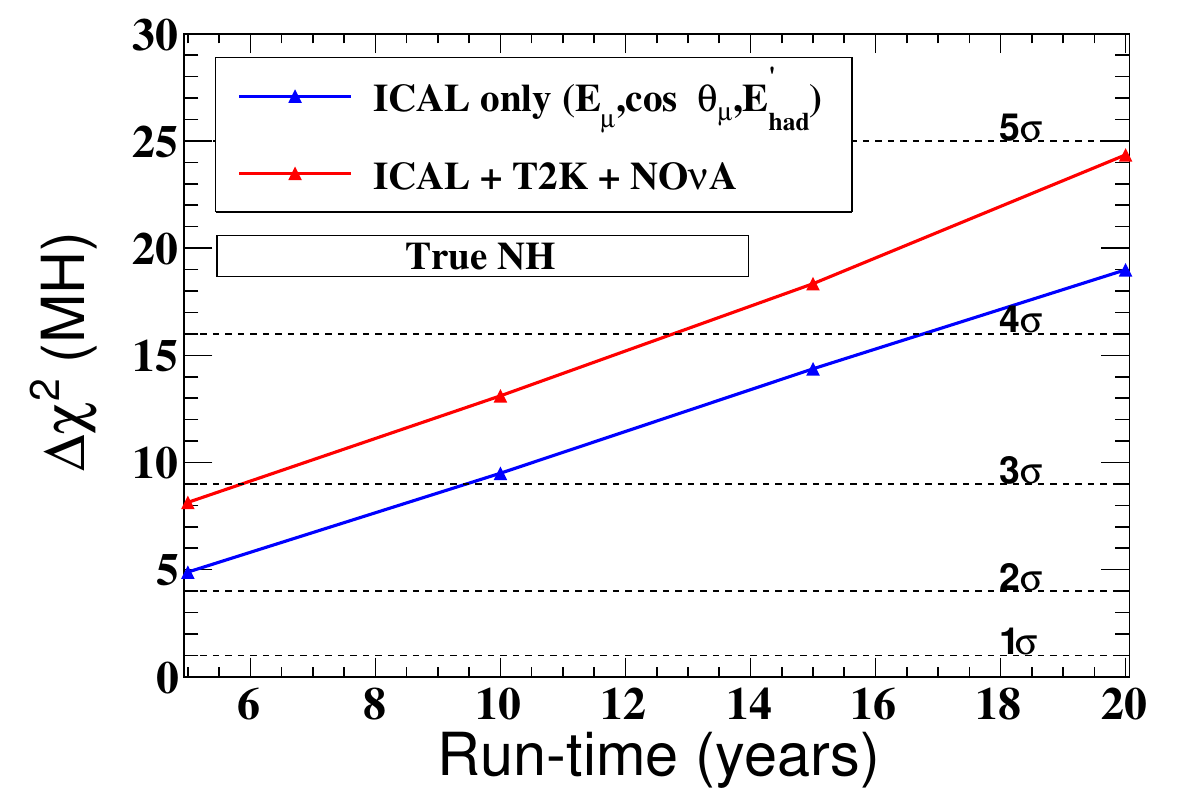}
\includegraphics[width=0.49\textwidth,height=0.35\textwidth]
{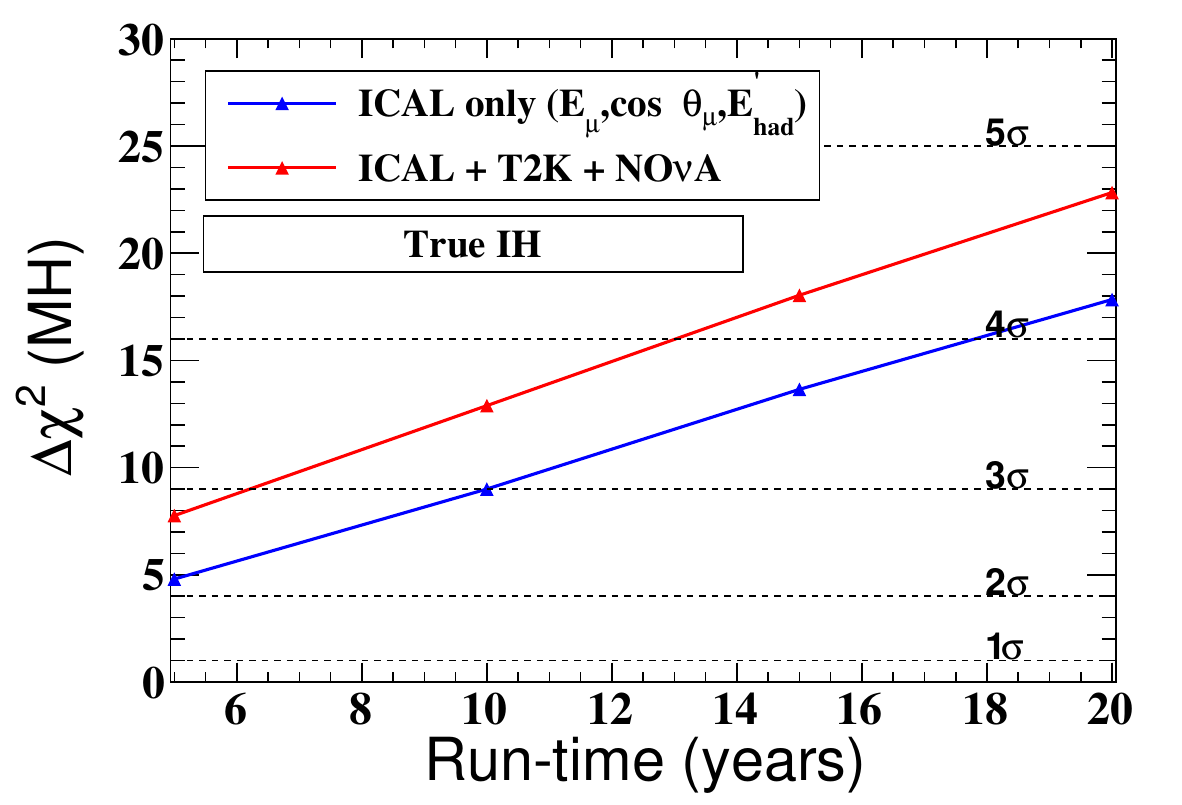}
\caption{Preliminary results on the hierarchy sensitivity with input 
normal (left)and inverted (right) hierarchy when ICAL data is combined with the
data from T2K (total luminosity of $8 \times 10^{21}$ protons on target 
in neutrino mode) and NO$\nu$A (3 years running in neutrino mode and 
3 years in antineutrino mode) \cite{synergy-new}.}
\label{fig:hier_combined}
\end{figure}

The ability of currently running long baseline experiments like T2K
and \nova to distinguish between the mass hierarchies depends crucially 
on the actual value of the CP-violating phase $\delta_{\rm CP}$.
For example, if $\delta_{\rm CP}$ is vanishing, this ability is
severely limited. However if one adds the data available from the
proposed run of these experiments, a preliminary estimation
suggests that even for vanishing $\delta_{\rm CP}$,
the mass hierarchy identification 
of $3\sigma$ may be achieved with a run-time as low as 6 years of
the 50 kt ICAL, for maximal mixing \cite{synergy-new}. This may be observed in
Fig.~\ref{fig:hier_combined}. Note that this improvement in the ICAL
sensitivity is not just due to the information provided by these 
experiments on the mass hierarchy, but also due to the 
improved constraints on $\Delta m^2_{32}$ and $\theta_{23}$.

\subsection*{Identifying mass hierarchy at all $\delta_{\rm CP}$ values}
\addcontentsline{toc}{subsection}{Identifying mass hierarchy at all 
$\delta_{\rm CP}$ values}

The large range of path length of the atmospheric neutrinos makes
ICAL insensitive to the CP phase $\delta_{\rm CP}$, as a result its reach
in distinguishing the hierarchy is also independent of the actual
value of $\delta_{\rm CP}$ \cite{Blennow:2012gj}. 
On the other hand the sensitivity of
fixed-baseline experiments such as T2K and \nova is extremely limited if
$0<\delta_{\rm CP} < \pi$ and the true hierarchy is normal. 
However adding of the ICAL information ensures that
the hierarchy can be identified even in these unfavoured $\delta_{\rm CP}$
regions \cite{Ghosh:2012px}. 
Of course, in the $\delta_{\rm CP}$ regions favourable to 
the long baseline experiments, the ICAL data can only enhance the
power of discriminating between the two hierarchies.

\subsection*{Determination of the CP phase}
\addcontentsline{toc}{subsection}{Determination of the CP phase}

Though ICAL itself is rather insensitive to $\delta_{\rm CP}$, data from
ICAL can still improve the determination of $\delta_{\rm CP}$ itself, by
providing input on mass hierarchy. This is especially crucial in the range 
$0 \le \delta_{\rm CP} \le \pi$, precisely where the ICAL data would also 
improve the 
hierarchy discrimination of NO$\nu$A and other experiments
\cite{Ghosh:2013yon}.

\section*{Other physics possibilities with ICAL}
\addcontentsline{toc}{section}{Other physics possibilities with ICAL}

ICAL is a versatile detector, and hence could be employed to test for
a multitude of new physics scenarios. For example, the violation of 
CPT or Lorentz symmetry in the neutrino sector 
\cite{Chatterjee:2014oda} can be probed to a great
precision, owing to its excellent energy measurement capability.
The passage of magnetic monopoles through the detector may be looked
for by simply looking for slowly moving, undeflecting tracks
\cite{Dash:2015qha}.
Dark matter annihilation inside the Sun may be constrained by comparing
the flux from the sun with the flux from other directions. Many such
scenarios are under investigation currently. 

\section*{Concluding remarks}
\addcontentsline{toc}{section}{Concluding remarks}

A strong and viable physics program is ready for ICAL at INO.
The simulations based on the incorporation of the ICAL geometry in
GEANT4 suggests that the detector will have excellent abilities for
detection, charge identification, energy measurement and direction
determination for charged muons of GeV energies. The magnetic
field enables separation of $\mu^-$ from $\mu^+$, equivalently that of
$\nu_\mu$ from $\bar\nu_\mu$, thus increasing the sensitivity to the
difference in matter effects on neutrino and antineutrino oscillations.
It will also be sensitive to hadrons, an ability that will increase its 
physics reach significantly and will offer advantages over other atmospheric
neutrino detectors. Apart from its main aim of identifying the neutrino
mass hierarchy, ICAL can also help in precision measurements of 
other neutrino mixing parameters, and can probe exotic physics issues
even beyond neutrinos.

\newpage


\counterwithin{figure}{chapter}
\counterwithin{equation}{chapter}
\counterwithin{table}{chapter}

\chapter{Introduction}
\label{intro}

\begin{flushright}
{\it
The earth is just a silly ball \\
To them, through which they simply pass \\
Like dustmaids through a drafty hall \\
-John Updike
}
\end{flushright}

Many important developments have taken place in neutrino physics 
and neutrino astronomy in recent years. The discovery of neutrino 
oscillations and consequent inference about the non-vanishing mass of 
the neutrinos, from the study of neutrinos from the Sun and cosmic rays, 
have had far-reaching consequences for particle physics, astroparticle 
physics and nuclear physics.
The observation of neutrinos from natural sources as well as those
produced at reactors and accelerators have given us the first confirmed
signals of physics beyond the Standard Model of particle physics.
They have also enabled us access to the energy production mechanisms 
inside stars and other astrophysical phenomena.

Experimental observations of neutrino interactions began in the mid 
1950s at Savannah river reactor by Reines and Cowan \cite{Cowan:1992xc}
followed by experiments deep in the mines of Kolar Gold Fields (KGF) 
in India \cite{Achar:1965ova} and in South Africa \cite{Reines:1965qk}. 
The pioneering solar neutrino experiments of Davis and collaborators 
in the USA \cite{Davis:1968cp,Cleveland:1998nv}, the water Cherenkov 
detector Kamiokande 
\cite{Hirata:1991ub} and its successor the gigantic Super-Kamiokande 
(SK) \cite{Fukuda:1998fd,Fukuda:1998mi}, the gallium detectors
SAGE \cite{Abdurashitov:1999zd} in Russia and Gallex \cite{Anselmann:1992um},
GNO \cite{Kirsten:1999cx} at the Laboratorio National di Gran Sasso 
(LNGS) in Italy, the heavy-water detector at the Sudbury Neutrino 
Observatory (SNO) in Canada \cite{Ahmad:2001an,Ahmad:2002jz}, 
the KamLAND \cite{Eguchi:2002dm} and K2K \cite{Ahn:2002up} experiments 
in Japan, etc. have together contributed in a very 
fundamental way to our knowledge of neutrino properties and interactions. 
The observation of solar neutrinos has given a direct experimental proof 
that the Sun and the stars are powered by 
thermonuclear fusion reactions that emit neutrinos. The recent results from 
reactor neutrino experiments, beginning with Double CHOOZ 
\cite{Abe:2011fz} in France and 
culminating in the results from Reno \cite{Ahn:2012nd} in Korea and 
Daya Bay \cite{An:2012eh} in China, 
and from accelerator experiments like MINOS \cite{Michael:2006rx}, 
T2K \cite{Abe:2011sj}, and \nova\ \cite{Adamson:2016tbq,Adamson:2016xxw}
have further revealed properties of neutrinos that not only serve as windows 
to physics beyond the Standard Model of particle physics, but also 
provide possibilities of understanding the matter-antimatter
asymmetry in the universe through the violation of the charge-parity (CP)
symmetry in the lepton sector. 

Impelled by these discoveries and their implications for the future of 
particle physics and astrophysics, plans are underway worldwide for 
new neutrino detectors to study such open issues as the hierarchy of 
neutrino masses, the masses themselves, the extent of CP violation
in the lepton sector, the Majorana or Dirac nature of neutrinos, etc..
This involves R\&D efforts for producing intense beams of neutrinos
at GeV energies, suitable detectors to detect them at long baseline 
distances, and sensitive neutrinoless double beta decay experiments. 
A complementary approach to these is the use of atmospheric 
neutrinos, whose fluxes are more uncertain than beam neutrinos, but
which provide a wider range of energies, and more importantly, a wider
range of baselines.

The India-based Neutrino Observatory (INO) is one such proposal aiming 
to address some of the challenges in understanding the nature of neutrinos,
using atmospheric neutrinos as the source. 
The unique feature of ICAL, the main detector in INO, will be the
ability to distinguish neutrinos from antineutrinos, which enables
a clearer distinction between the matter effects on neutrinos and
antineutrinos travelling through the Earth, leading to the identification
of neutrino mass hierarchy.
In this Chapter, we shall introduce the INO laboratory, the ICAL detector,
and describe the role of such a detector in the global context of neutrino
physics experiments.

\section{The ICAL detector at the INO facility}

\subsection{Neutrino experiments in India:  past and present}

Underground science in India has a long history. The deep underground 
laboratory at Kolar Gold Fields (KGF), where Indian scientists conducted 
many front ranking experiments in the field of cosmic rays and neutrinos, 
was a pioneering effort. The KGF mines are situated at about 870m above sea 
level near the city of Bangalore in South India. It has a flat topography 
around the area surrounding the mines.  The mines have extended network of 
tunnels underground which permitted experiments up to a depth of 3000 m 
below the surface. Initially attempts were made to find the depth variation 
of muon fluxes starting from the surface up to the deepest reaches. 
The absence of any count around a  depth of 8400 hg/cm$^2$ lead to the 
conclusion that the atmospheric muon intensity is attenuated to such a 
level where one could search for very weak processes like the interactions 
of high energy neutrinos. This was in the beginning of the sixties when very 
little was known about the interaction of neutrinos at high energies 
($>$ a few GeV) from accelerators, and that too with only muon neutrino beams.
Nothing was known about the electron neutrino or antineutrino interactions.

Thus began the neutrino experiments in KGF in the early sixties, conducted by a 
collaboration consisting of groups from Durham University (UK), Osaka City 
University (Japan) and TIFR in India. The techniques used were perfected 
during the years of muon experiments and involved a basic trigger with 
scintillation counters and Neon Flash Tubes (NFT) for tracking detectors 
initially. Seven such detectors were placed in a long tunnel at a depth of 
2.3 km, in the Heathcote shaft of Champion reef mines, in three batches over 
a period of two year starting from the end of 1964 \cite{Narasimham:2004zz}.

The first ever atmospheric neutrino event was recorded underground was 
in early 1965 \cite{Achar:1965ova}. Two well defined tacks emerging from 
the rock in an upward direction indicated unambigiously a clear inelastic 
neutrino event. 
Later this collaboration put together the first experiment that searched 
for proton decay. Nature did not oblige and experiments are still 
looking for proton decay. The KGF laboratory operated for nearly four 
decades, almost till the end of 1980's, collecting data on atmospheric 
muon and neutrino interactions at various depths, starting from about 300 
metres all the way down to 2700 metres. In the process they also detected 
some anomalous events which could not be attributed to neutrinos at 
depths around 2000 metres \cite{Krishnaswamy:1975zu,Krishnaswamy:1975qe,
krishnaswamy2}. Such events have neither 
been proved wrong nor have they been confirmed by other experiments.

The INO project, the discussions about which were formally held first 
in the Workshop on High Energy Physics Phenomenology (WHEPP-VI) in Chennai
\cite{Murthy:2000ta} is an ambitious proposal to recapture this pioneering 
spirit and do experiments in neutrino physics at the cutting edge. 
The immediate goal of INO project is the creation of an underground 
laboratory which will house a large magnetised iron calorimeter (ICAL) 
detector to study the properties of naturally produced neutrinos in the 
earth's atmosphere.  Apart from experiments involving neutrinos, in the 
long term the laboratory is envisaged to develop into a full-fledged 
underground laboratory for studies in Physics, Biology and Geology as 
well. The INO is the first basic science laboratory planned on
such a large scale in India.

\subsection{Location and layout of the INO}

\begin{figure} 
\includegraphics[width=0.45\textwidth,height=0.35\textwidth]
{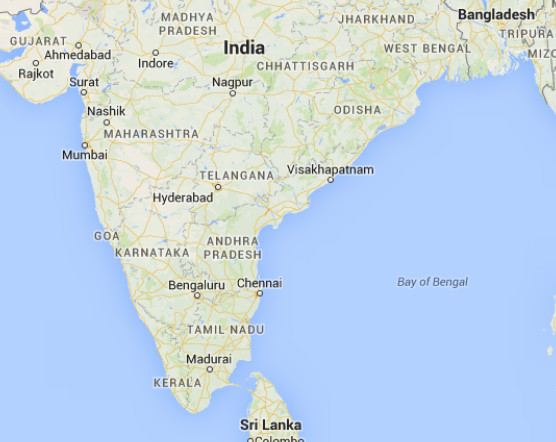} \hfill
\includegraphics[width=0.45\textwidth,height=0.35\textwidth]
{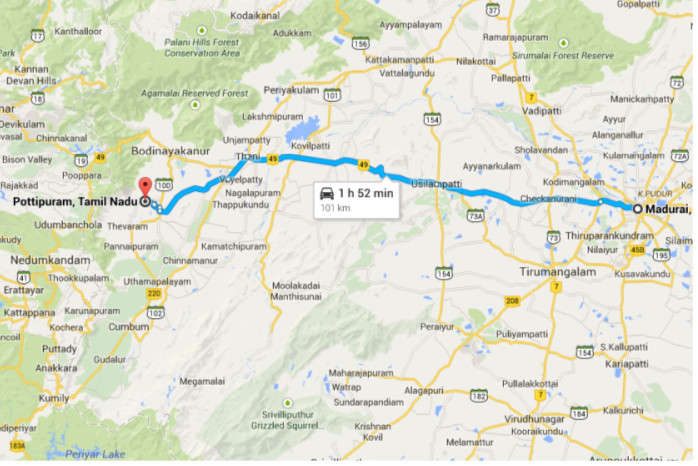} \\
\vspace{0.2cm} \\
\includegraphics[width=0.45\textwidth,height=0.45\textwidth]
{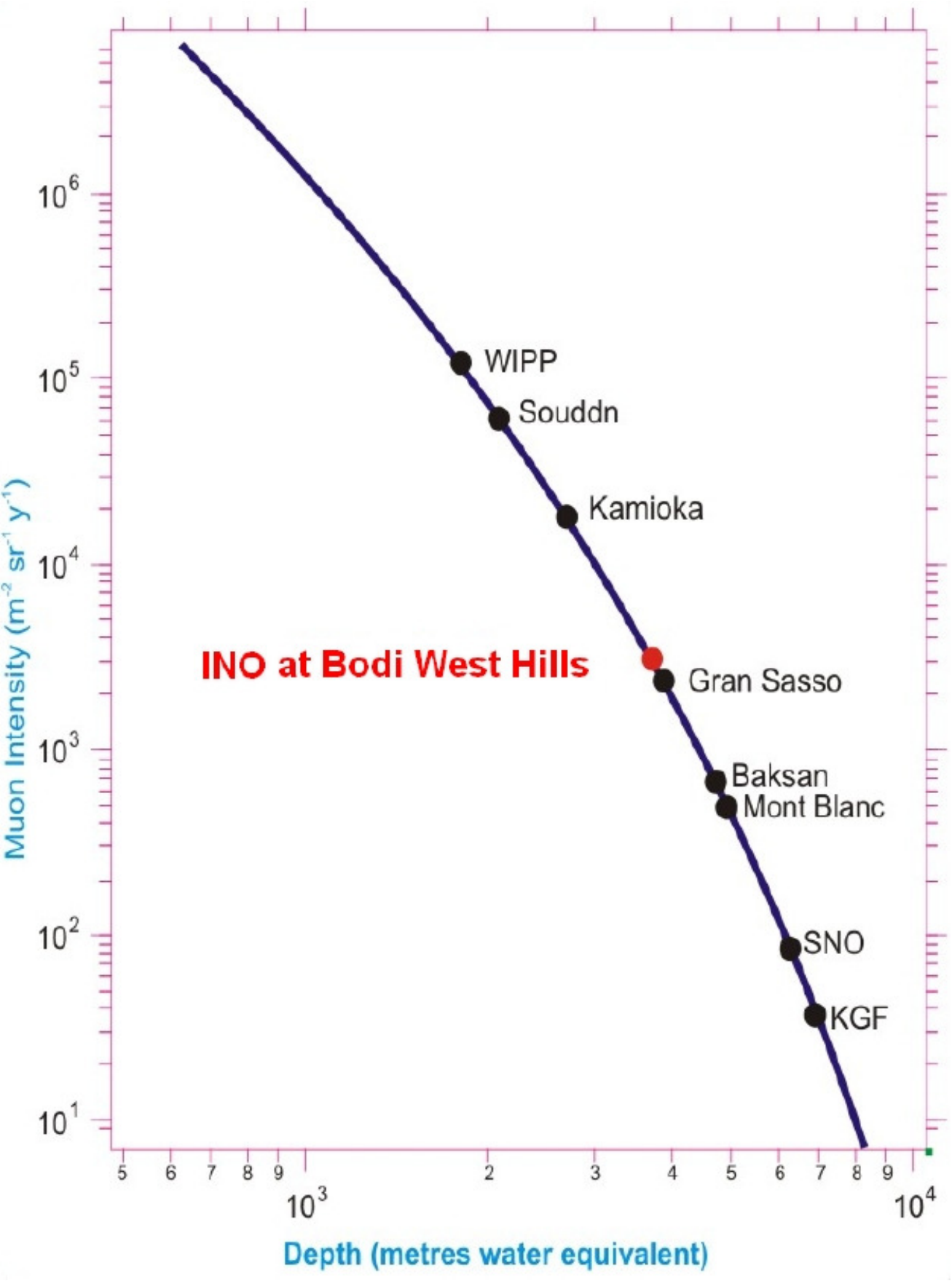} \hfill
\includegraphics[width=0.45\textwidth,height=0.45\textwidth]
{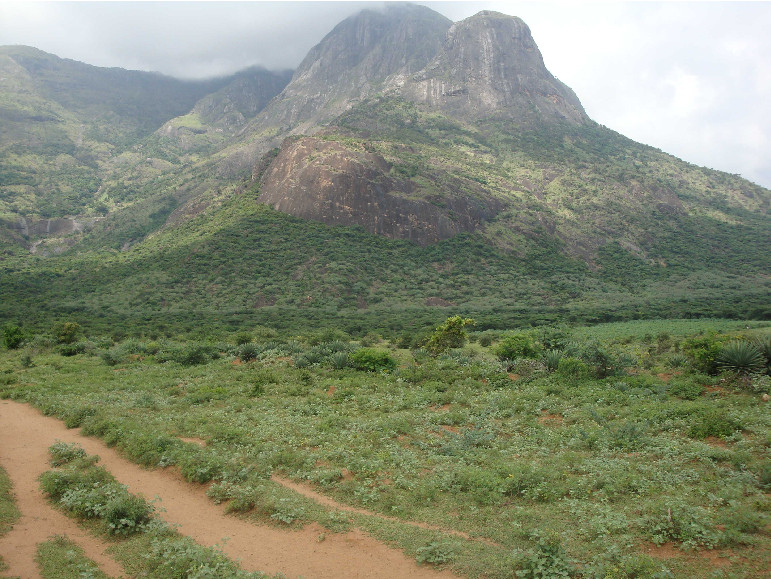} 
\caption{The location of the INO site and the nearby major landmarks. 
The IICHEP is located about 120 km east of the INO site, 
in the city of Madurai, as shown in the top right panel. 
The photo in the bottom right panel shows the view of the hill under which 
the cavern will be located. The terrain is totally flat with minimal 
undergrowth as seen in the picture (Photo: M V N Murthy). The photo is 
taken before the start of any construction.
The bottom left panel shows the suppression in intensity of atmospheric
muon flux at various underground sites, compared to the INO cavern
\cite{Athar:2006yb}.} 
\label{ino-site}
\end{figure}

The INO will be located at the Bodi West Hills (BWH), near Pottipuram village, 
in the Theni district of Tamil Nadu, India. The site has been chosen both for 
geotechnical reasons as well as from environmental considerations. It is
near the historic city of Madurai, as shown in Fig.~\ref{ino-site}. Madurai
is about 120 km from the INO site, and will also be the location for the
Inter-Institutional Centre for High Energy Physics (IICHEP), where activities
related to the INO will be carried out. The figure shows the location 
and also the features of the local terrain.
The construction of the laboratory below the Bodi Hills involves building a 
horizontal tunnel, approximately 1900 m long, to reach the laboratory 
that is located under a mountain peak. One large and three small 
laboratory caverns are to be built with a rock burden of 1000 m or more all 
around (with a vertical overburden of $\sim$1300 m) to house the experiments.  
The reduction of cosmic ray background at this site is almost the same as 
at the Gran Sasso laboratory, as can be seen from the bottom right panel,
which shows the cosmic ray muon flux as a function of depth, with the 
locations of other major laboratories.

In addition to the main Iron CALorimeter (ICAL) detector whose prime goal
is the determination of the neutrino mass hierarchy,  the laboratory is
designed to accommodate  experiments in other areas like neutrinoless
double beta decay, dark matter search, low-energy neutrino spectroscopy,
etc. Preliminary investigations and R\&D in this direction are in
progress. The special environment provided by the underground laboratory
may also be useful to conduct experiments in rock mechanics, geology,
biology etc. 

\begin{figure}
\includegraphics[width=0.98\textwidth]{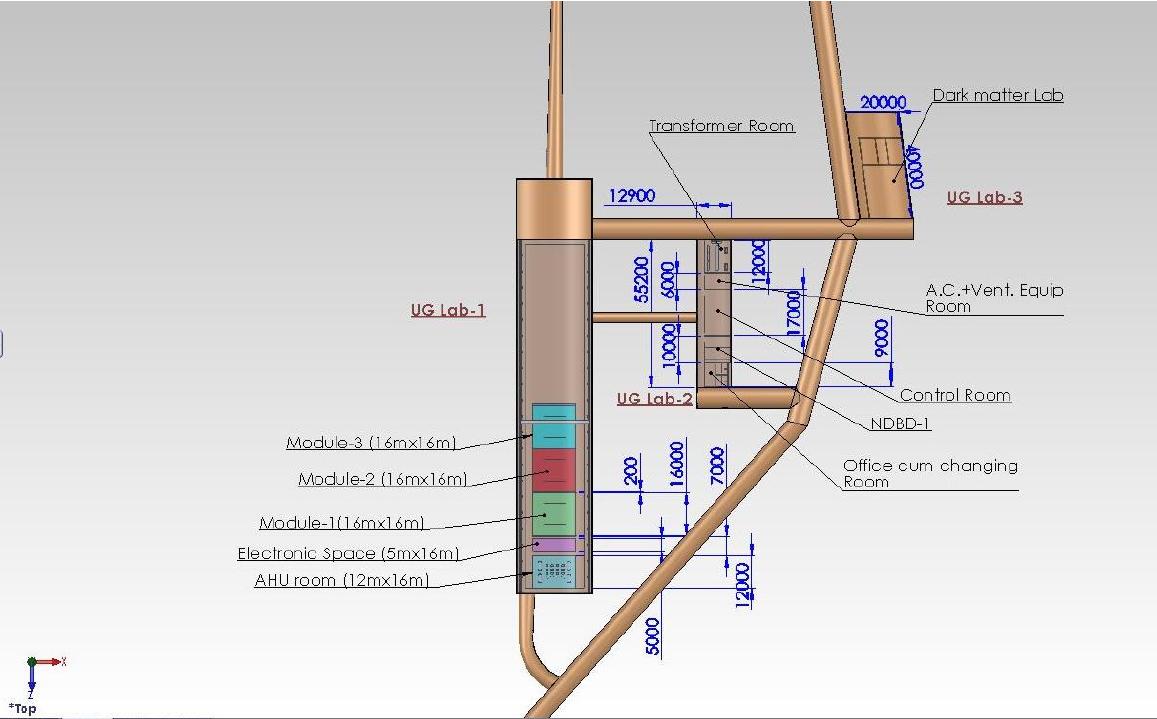} 
\caption{Underground caverns layout showing the footprints of proposed
experiments and other components}
\label{uglayout}
\end{figure}

The present configuration of the laboratory caverns is shown in
Fig.~\ref{uglayout}. The largest cavern that will house the main iron
calorimeter detector (ICAL) is 132~m (L) $ \times$  26~m (W) $ \times$
32.5~m (H). This cavern, called ``UG-Lab 1'', is designed to accommodate 
a 50 kt ICAL (planned) and a second possible ICAL-II neutrino detector
of equal size. Each ICAL consists of three modules with dimensions of
16 m (L) $\times$ 16 m (W) $\times$ 14.5 m (H), so that the total
footprint of both the detectors would be 96 m (L) $\times$ 16 m (W). 
Fig.~\ref{uglayout} shows three modules of the 50 kt ICAL.

\subsection{The ICAL detector}
\label{sec:detector}
\begin{figure}[h!]
\vspace{1.0cm}
\centering
\includegraphics[width=0.8\textwidth]{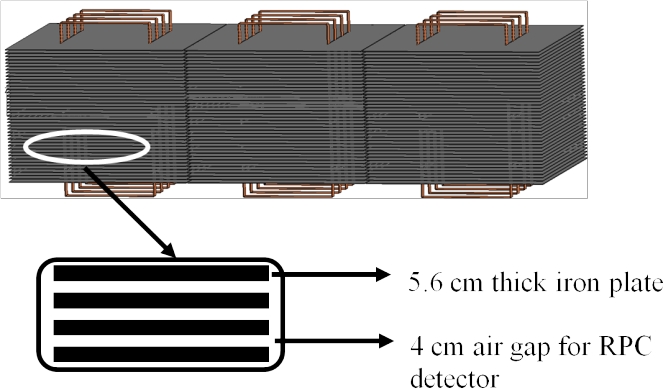}
\caption{Schematic view of the 50 kt ICAL detctor}
\label{fig:ICAL-detector}
\end{figure}

The ICAL detector is similar in concept to the earlier proposed 
Monolith \cite{TabarellideFatis:2001wy,monolith} detector at Gran Sasso.
The layout of the proposed ICAL detector is shown in 
Fig.~\ref{fig:ICAL-detector}.
The detector will have a modular structure of total lateral size 
$48~\hbox{m} \times 16~\hbox{m}$, subdivided into three modules of area 
$16 \hbox{ m} \times 16 \hbox{ m}$. It will consist of a stack of 151
horizontal layers of 5.6 cm thick magnetized iron plates interleaved
with 4 cm gaps to house the active detector layers, making it 14.5 m high.
Iron spacers acting as supports will
be located every 2 m along both X and Y directions; the 2 m wide roads
along the transverse ($Y)$ direction will enable the insertion and
periodic removal of RPCs, when required.
 
The active detector elements, the resistive plate chambers (RPCs) made up of 
a pair of 3 mm thick glass plates of area $2 \hbox{ m} \times 2 \hbox{ m}$ 
separated by 2 mm spacers, will be inserted in the gaps between the
iron layers. These 
will be operated at a high voltage of about 10 kV in avalanche mode. 
A high energy charged particle, passing through the RPCs, will leave
signals that will be read by orthogonal X and Y pickup strips, about
3 cm wide, one on each side of an RPC. Detailed R \& D has shown that the
RPCs have an efficiency of around 90--95\% with a time resolution of
about a nanosecond.
This will allow the determination of the X and Y 
coordinates of the track of the charged particles passing through the RPC. 
The layer number of the RPC will provide the Z coordinate. The 
observed RPC time 
resolution of $\sim$ 1 ns will enable the distinction between upward-going 
particles and downward-going particles. From the hit pattern observed in 
the RPCs, the energies as well as directions of the charged particles 
produced in the neutrino interactions can be reconstructed. 

Each module will have two vertical slots cut into it to enable
current-carrying copper coils to be wound around as shown in 
Fig.~\ref{fig:ICAL-detector}.
Simulation studies \cite{Behera:2014zca} have shown that the iron
plates can be magnetized to a field strength of about 1.5 T, with fields
greater than 1 T over at least 85\% of the volume of the detector.
The bending of 
charged particles in this magnetic field will enable the identification of 
their charge. In particular, the sign of the charge of the muon produced by 
neutrino interactions inside the detector will help in identifying and 
studying the $\nu_\mu$ and $\bar\nu_\mu$ induced events separately. The 
magnetic field will also help the measurement of the momentum of the final 
state particles, especially the muons.

With about 14000 iron plates of 2 m $\times$ 4 m area and 5.6 cm thickness,
30000 RPCs of 2 m $\times$ 2 m area, 4,000,000 electronic readout channels, 
and a magnetic field of 1.5 T, the ICAL is going to be the largest 
electromagnet in the world, and is expected to play a pivoting role in
our understanding of neutrino properties.


\section{Role of ICAL in neutrino mixing and beyond}

In this section, we briefly discuss our present understanding of  
neutrino oscillation parameters and identify the fundamental issues in  
the neutrino sector that can be addressed by the ICAL detector.

\subsection{Current status of neutrino mixing parameters} 
\label{current_status} 
 
The neutrino flavour states $\vert \nu_\alpha \rangle$ (where $\alpha = e,  
\mu, \tau$) are linear superpositions of the neutrino mass eigenstates  
$\vert \nu_i\rangle$ (with $i=1,2,3$), with masses $m_i$ : 
\begin{equation} 
\vert \nu_\alpha \rangle = \sum_i U_{\alpha i} \vert \nu_i\rangle. 
\end{equation} 
Here $U$ is the $3 \times 3$ unitary mixing matrix. A physically  
motivated form of the mixing matrix that is conventionally used is  
\cite{Pontecorvo:1957cp,Maki:1962mu,Agashe:2014kda} 
\begin{eqnarray} 
U = \left( 
 \begin{array}{ccc} 
 c_{12}c_{13} & s_{12}c_{13} & s_{13}e^{-i\delta}  \\ 
 -c_{23}s_{12} - s_{23}s_{13}c_{12}e^{i\delta} & c_{23}c_{12} - 
 s_{23}s_{13}s_{12}e^{i\delta}&  s_{23}c_{13} \\ 
 s_{23}s_{12} - c_{23}s_{13}c_{12}e^{i\delta}& -s_{23}c_{12} - 
 c_{23}s_{13}s_{12}e^{i\delta} & c_{23}c_{13} \end{array} \right) \; , 
\label{eq:mns} 
\nonumber 
\end{eqnarray} 
where $c_{ij}=\cos\theta_{ij}$, $s_{ij}=\sin\theta_{ij}$, and $\delta$ denotes 
the CP violating (Dirac) phase, also called $\delta_{\rm CP}$. Note that the
Majorana phases are not included in the above parameterization, since they
do not play a role in neutrino oscillation experiments.

The probability of an initial neutrino $\nu_\alpha$ of flavour $\alpha$ 
and energy $E$ being detected as a neutrino $\nu_\beta$ of the same energy 
but with flavour ${\beta}$ after travelling a distance $L$ in vacuum is 
\begin{eqnarray} 
P_{\alpha\beta} & = & \delta_{\alpha \beta} - 4~ \sum_{i>j}~ 
\hbox{Re}[U_{\alpha i} U^*_{\beta i} U^*_{\alpha j} U_{\beta j}] 
\sin^{2}(\Delta_{ij}) \nonumber \\
& & \phantom{\delta_{\alpha \beta}}
+ 2~ \sum_{i>j}~ 
\hbox{Im} [U_{\alpha i} U^*_{\beta i} U^*_{\alpha j} U_{\beta j}] 
\sin(2\Delta_{ij}) \; ,
\label{eq:pab} 
\end{eqnarray} 
where 
$\Delta_{ij} = 1.27~ \Delta m^2_{ij}$(eV$^2$)$\times L$(km)$/E$(GeV), 
with  $\Delta m^2_{ij} = {m_i}^2 - {m_j}^2$
the mass squared differences between the $i$th and $j$th neutrino mass
eigenstates.
Oscillation measurements are not sensitive to the individual neutrino
masses, but only to their mass-squared differences.
Note that the above expression is valid only for propagation through vacuum. 
In matter, the probabilities are drastically modified. The relevant 
expressions may be found in Appendix~\ref{app:prob}.

The neutrino flavour conversion probabilities can be expressed in terms
of the two mass squared differences, the three mixing angles, and the single 
CP-violating phase. Also of crucial importance is the 
{\it mass ordering}, i.e., the sign of $\Delta m^2_{32}$ 
(the same as the sign of $\Delta m^2_{31}$).
While we know that $\Delta m^2_{21}$ is positive so as to  
accommodate the observed energy dependence of the electron neutrino  
survival probability in solar neutrino experiments, at present  
$\Delta m^2_{32}$ is allowed to be either positive or negative. Hence, it  
is possible to have two patterns of neutrino masses: 
$m_3 > m_2 > m_1$, called normal ordering, where $\Delta m^2_{32}$ is positive, and  
$m_2 > m_1 > m_3$, called inverted ordering, where $\Delta m^2_{32}$ is negative.
Determining the sign of $\Delta m^2_{32}$ is one  
of the prime goals of the ICAL experiment. 
{\it Note that, though the ``mass ordering'' is perhaps the more appropriate
term to use in this context, the more commonly used term in literature
is ``mass hierarchy''. In this report, therefore, we will use the
notation ``normal hierarchy'' (NH) to denote normal ordering, and
``inverted hierarchy'' (IH) to denote inverted ordering. The word 
``hierarchy'' used in this context 
has no connection with the absolute values of neutrino masses.}

Table~\ref{tab:best-fit} summarises the current status of neutrino  
oscillation parameters~\cite{Gonzalez-Garcia:2014bfa,NuFIT} based 
on the world neutrino data that was available after the NOW 2014 conference.  
The numbers given in Table~\ref{tab:best-fit} are obtained by keeping  
the reactor fluxes free in the fit and also including the short-baseline  
reactor data with $L \le 100$ m~\cite{Gonzalez-Garcia:2014bfa,NuFIT}. 
 
\begin{table}[t] 
\begin{center} 
\begin{tabular}{|l|c|c|c|} \hline
Parameter & Best-fit values & $3\sigma$ ranges & Relative  \\
& & &  {$1\sigma$ precision} \\
\hline 
$\Delta m^2_{21}$ (eV$^2$) & $7.5 \times 10^{-5}$ & $[7.0, \, 8.1] \times 
10^{-5}$ & $2.4\%$ \cr  
\hline 
$\Delta m^2_{31}$ (eV$^2$) & $2.46 \times 10^{-3}$ (NH) &  
$[2.32, \, 2.61] \times 10^{-3}$ (NH) &  $2.0\%$  \cr 
\hline 
$\Delta m^2_{32}$ (eV$^2$) & $ - 2.45 \times 10^{-3}$ (IH) &  
$- [2.59, \, 2.31] \times 10^{-3}$ (IH) &  
$1.9\%$ \cr 
\hline 
$\sin^2\theta_{12}$ &  0.3 & [0.27, \, 0.34]  &  
$4.4\%$ \cr 
\hline 
$\sin^2\theta_{23}$ & 0.45 (NH), 0.58 (IH) & [0.38, \, 0.64] &  
$8.7\%$ \cr  
\hline 
$\sin^2\theta_{13}$ & 0.022 & [0.018, \,0.025] &  
$5.3\%$ \cr  
\hline 
$\delta_{\mathrm{CP}} (^\circ)$ & $ 306$ & [0, \, $360$] & -- \cr 
\hline 
\end{tabular} 
\caption{The values of neutrino oscillation parameters used for the
analyses in this paper~\cite{Gonzalez-Garcia:2014bfa}. 
The second column shows the central values of the oscillation parameters. 
The third column depicts the $3\sigma$ ranges of the parameters 
with the relative $1\sigma$ errors being listed in the last column. 
{\em Note that the parameter $\Delta m^2_{31}$ ($\Delta m^2_{32}$) 
is used while performing the fit with normal (inverted) hierarchy.}
The current best-fit values and allowed ranges of these parameters
may be found in \cite{Capozzi:2016rtj,Gonzalez-Garcia:2015qrr,NuFIT}. 
\label{tab:best-fit} 
} 
\end{center} 
\end{table} 
 
Table~\ref{tab:best-fit} also provides the relative $1\sigma$ 
precision\footnote{Here the $1\sigma$ precision is defined as 
1/6th of the $\pm 3\sigma$ variations around the best-fit value.}
on the measurements of these quantities at this stage. 
The global fit suggests the best-fit value of $\Delta m^2_{21} = 7.5 \times  
10^{-5}~\mbox{eV}^2$ with a relative $1\sigma$ precision of  
2.4\%. In a three-flavor framework, we have the best-fit values of 
$\Delta m^2_{31} = 2.42 \times 10^{-3}~\mbox{eV}^2$ for NH and 
$\Delta m^2_{32} = - 2.41 \times 10^{-3}~\mbox{eV}^2$ for IH. 
A relative $1\sigma$ precision of 2.5\% has been achieved for the 
atmospheric mass-squared splitting. 
 
As far as the mixing angles are concerned, $\theta_{12}$ is now pretty 
well measured with a best-fit value of $\sin^2\theta_{12} = 0.3$ and 
a relative $1\sigma$ precision of 4\% has been achieved for the solar
mixing angle. Our understanding of the atmospheric mixing angle  
$\theta_{23}$ has also  improved a lot in recent years. Combined  
analysis of all the neutrino oscillation data available so far disfavours 
the maximal mixing solution for $\theta_{23}$ at $\sim 1.5\sigma$ 
confidence level~\cite{GonzalezGarcia:2012sz,Gonzalez-Garcia:2014bfa,
NuFIT,Capozzi:2013csa,Forero:2014bxa}.
This result is mostly governed by the  MINOS accelerator data in $\nu_\mu$ and 
$\bar{\nu}_\mu$ disappearance modes~\cite{Adamson:2013whj}. 
The dominant term in 
$\nu_\mu$ survival channel mainly depends on the value of $\sin^2 2\theta_{23}$. 
Now, if $\sin^2 2\theta_{23}$ turns out to be different from 1 as suggested by  
the recent oscillation data, then it gives two solutions for 
$\sin^2\theta_{23}$: 
one whose value is less than half, known as the lower octant (LO) solution, 
and the other whose value is greater than half, known as the higher octant 
(HO) solution. This creates  the problem of octant degeneracy of 
$\theta_{23}$~\cite{Fogli:1996pv}.  
At present, the best-fit value of $\sin^2\theta_{23}$ in LO (HO) is 0.45  
(0.58) assuming NH (IH). The relative $1\sigma$ precision on $\sin^2\theta_{23}$ 
is around 8.7\% assuming maximal mixing as the central value. 
Further improvement 
in the knowledge of $\theta_{23}$ and settling the issue of its octant 
(if it turns out to be non-maximal) are also important issues that can be 
addressed by observing atmospheric neutrinos.
 
For many years, we only had an upper bound on the value of the 1-3 mixing 
angle \cite{Apollonio:1997xe,Narayan:1997mk,Apollonio:1999ae,Piepke:2002ju}. 
A nonzero value for this angle has been discovered rather 
recently \cite{Abe:2011fz,Ahn:2012nd,An:2012eh,Abe:2013hdq,An:2013zwz}, 
with a moderately large best-fit value of $\sin^2\theta_{13}= 0.022$, 
which is mostly driven by the high-statistics data provided by the 
ongoing Daya Bay reactor experiment~\cite{An:2012eh,An:2013zwz}. 
It is quite remarkable that already we have achieved a relative 
$1\sigma$ precision of 5.3\% on $\sin^2\theta_{13}$. 
On the other hand, the whole range of $\delta_{\rm CP}$ is still allowed at 
the $3\sigma$ level.


\subsection{Unravelling three-neutrino mixing with ICAL}
\label{iron-cal}

As has been discussed earlier, the main advantage of a magnetised iron
calorimeter is its ability to distinguish $\mu^+$ from $\mu^-$, and
hence to study $\nu_\mu$ and $\bar\nu_\mu$ separately. This allows 
a cleaner measurement of the difference in the matter effects 
experienced by neutrinos and antineutrinos. However, this
difference depends on the value of $\theta_{13}$.
The recent measurement of a
moderately large value of $\theta_{13}$ therefore boosts the capability 
of ICAL for observing these matter effects, and hence
its reach in addressing the key issues related to the neutrino masses 
and mixing. In this section, we shall highlight the role that an iron 
calorimeter like ICAL will have in the context of global efforts to measure
neutrino mixing parameters.

The moderately large $\theta_{13}$ value has opened the door to the  
fundamental measurements of (i) the neutrino mass ordering, 
(ii) the deviation of 2-3 mixing angle from its maximal value and
hence the correct octant of $\theta_{23}$, and (iii) the CP phase 
$\delta_{\rm CP}$ and to look for CP violation in the lepton sector, 
for several experiments which would have had limited capability 
to address these questions had this parameter been significantly 
smaller. Central to all these measurements are effects which differ 
between neutrinos and antineutrinos. These could either be matter 
related effects which enhance or suppress the oscillation probabilities 
(relevant for i and ii above), or those induced by a non-zero value of 
$\delta_{\rm CP}$ (relevant for iii). The ICAL will be sensitive to the matter 
effects, but will have almost no sensitivity to the actual value of 
$\delta_{\rm CP}$. 

Prior to  summarizing  the role of ICAL, it is useful to mention several other 
experiments which are underway or will come online in the next couple of
decades with the aim of making the three important measurements mentioned above.
The T2K experiment has observed electron neutrino appearance in a muon
neutrino beam \cite{Abe:2013hdq}, thus showing a clear evidence of neurino 
oscillations.
The accelerator based long-baseline beam experiments NOvA 
\cite{Ayres:2002ws,Ayres:2004js,Ayres:2007tu} has already started taking
data that will be sensitive to mass hierarchy, and the first results
have been presented \cite{Adamson:2016xxw}.
The IceCube DeepCore experiment has also recently \cite{Aartsen:2014yll} 
published their results on atmosphetic neutrino oscillations.
Future large atmospheric neutrino detectors on the cards are 
Hyper-Kamiokande (HK) \cite{Abe:2011ts}, 
Precision IceCube Next-Generation Upgrade (PINGU) \cite{Aartsen:2014oha} 
and Oscillation Research with Cosmics in the Abyss (ORCA) 
\cite{Adrian-Martinez:2016fdl}.
The Deep Underground Neutrino Experiment (DUNE) \cite{Acciarri:2015uup},
a combined initiative of the earlier Long Baseline Neutrino Experiment
(LBNE) \cite{Akiri:2011dv,Adams:2013qkq} and Long Baseline Neutrino 
Oscillation (LBNO) \cite{Agarwalla:2011hh,Stahl:2012exa,Agarwalla:2013kaa}
collaborations, is also slated to aim at the mass hierarchy identification.
Additionally, the medium-baseline reactor oscillation experiments 
\cite{Li:2013zyd}, JUNO \cite{An:2015jdp} and RENO-50 \cite{Kim:2014rfa} 
aim to determine the hierarchy by performing a very precise, 
high statistics measurement of the neutrino energy spectrum.
The CP phase $\delta_{\rm CP}$ can be measured (in principle) 
by accelerator experiments
like T2K \cite{Abe:2014tzr,Agarwalla:2012bv}, NOvA 
\cite{Huber:2009cw,Agarwalla:2012bv,Machado:2013kya,Ghosh:2014dba} 
T2HK \cite{Abe:2011ts}, and DUNE \cite{Acciarri:2015uup}.
These experiments, if they run in both the neutrino and the antineutrino mode, 
would additionally 
be sensitive to the octant of $\theta_{23}$ 
\cite{Agarwalla:2013ju,Agarwalla:2013hma}, and so would 
the large-mass atmospheric experiments like ICAL \cite{Devi:2014yaa} 
and Hyper-K \cite{Kearns:NNN2013}.

The ICAL detector at the INO cavern will provide an excellent opportunity
to study the atmospheric neutrinos and antineutrinos separately with high 
detection efficiency and good enough energy and angular resolutions in the 
multi-GeV range in the presence of the Earth's matter effect. 
There is no doubt that the rich data set which would be available from 
the proposed ICAL atmospheric neutrino experiment will be extremely useful 
to validate the three flavor picture of the neutrino oscillation
taking into account the Earth's large matter effect in the multi-GeV range.
The first aim of the ICAL detector would be to observe the oscillation pattern
over at least one full period, in order to make a precise measurement of 
the atmospheric oscillation parameters. 
The ICAL detector performs quite well in a wide range of $L/E$ 
and can confirm the evidence of the sinusoidal flavor transition probability 
of neutrino oscillation already observed by the Super-Kamiokande detector 
by observing the dips and peaks in the event rate versus $L/E$ 
\cite{Ishitsuka:2004un}, 
as well as by the IceCube DeepCore \cite{Aartsen:2014yll}. 
In the case of Super-Kamiokande, the sub-GeV events have played an important 
role to perform this $L/E$ analysis, while for IceCube the very high energy
events ($E \gtrsim 10$ GeV) have contributed significantly. 
The ICAL detector is sensitive mainly to the energy range 1--10 GeV, which
fills the gap between the other two large Cherenkov detetors.
In its initial phase, the ICAL experiment will also provide an independent
measurement of $\theta_{13}$ by exploring the Earth's matter effect 
using the atmospheric neutrinos. This will certainly complement the ongoing 
efforts of the reactor and the accelerator experiments to learn about the 
smallest 
lepton mixing angle $\theta_{13}$. 

The relevant neutrino oscillation probabilities that the ICAL will be sensitive
to are $P_{\mu\mu}$, $P_{\bar\mu \bar\mu}$, $P_{e\mu}$, and $P_{\bar{e} \bar\mu}$, 
especially the former two. These probabilities have a rich structure for 
neutrinos and
antineutrinos at GeV energies, traveling through the Earth for a distance
of several thousands of km (for a detailed description, 
see Appendix~\ref{app:prob}). 
The matter effects on these neutrinos and antineutrinos lead to significant 
differences between these oscillation probabilities, 
which may be probed by a detector 
like ICAL that can distinguish neutrinos from antineutrinos. 
This feature of ICAL would be 
instrumental in its ability to distinguish between the two possible 
mass hierarchies.

Detailed simulations of the ICAL detector performance, as discussed
in the following chapters, show that ICAL would be an excellent tracker 
for muons. The energy and direction of a muon would also be reconstructed rather
accurately, with the muon direction resolution of better than a degree
at high energies. Furthermore, the capability of ICAL to study 
the properties of the final state hadrons in multi-GeV neutrino 
interactions would be one of its unique features. This would allow
the reconstruction 
of the neutrino energy in every event, albeit with large error. (Note 
that the extraction of hadronic information at multi-GeV energies in 
currently running or upcoming water or ice based atmospheric neutrino 
detectors is quite challenging; the efficiency of reconstruction of 
multi-ring events is rather small in such detectors.)
As a result, the ICAL would have a significant stand-alone sensitivity
to the mass hierarchy, which, when combined with data from experiments 
like NOvA and T2K, would significantly enhance the overall sensitivity 
to this important quantity. 

Although the ICAL would not be sensitive to the value of $\delta_{\rm CP}$, this
very feature would make it an important supporting experiment for others
that are sensitive to $\delta_{\rm CP}$, in a unique manner. 
Note that in experiments where event rates are sensitive simultaneously 
to both matter and CP phase effects, disentangling one from the other 
restricts the sensitivities to individual and unambiguous measurements of 
each of the three quantities (i), (ii) and (iii) mentioned above. 
The virtue of ICAL here would lie in its ability to offer a data-set that is 
free of entanglements between matter enhancements and dynamical CP violating 
effects due to a non-zero $\delta_{CP}$. Thus, when used in combination with 
other experiments, the ICAL measurements will facilitate the lifting of 
degeneracies which may be present otherwise.
In particular, the ICAL data, when combined with that from \nova\ and T2K, 
would make a significant difference to their discovery potential of 
CP violation \cite{Ghosh:2013yon}.

\subsection{Addressing new physics issues with ICAL}

The full role of an iron calorimeter in the global scenario of neutrino 
physics is rich and complex. In addition to what is described here, it 
can add to our knowledge on very high energy muons 
\cite{Gandhi:2005at,Panda:2007fa} on 
hitherto undiscovered long range forces \cite{Joshipura:2003jh}, 
on CPT violation \cite{Chatterjee:2014oda,Datta:2003dg}
and on non-standard interactions \cite{Chatterjee:2014gxa}, 
among other issues. The future of 
neutrino physics is, in our opinion, crucially dependent on the 
synergistic combination of experiments with differing 
capabilities and strengths. A large iron calorimeter brings in unique 
muon charge identification capabilities and an event sample independent 
of the CP phase. Both these aspects will play an important role in our 
concerted global effort to understand the mysteries of neutrino physics 
and consequently understand physics beyond the Standard Model.

Though the ICAL has been designed mainly with neutrino physics in view, 
it is expected that many non-neutrino issues may find relevance with 
this detector. 
For example, a few decades ago, both in the cosmic ray neutrino experiments
\cite{Krishnaswamy:1975zu,Krishnaswamy:1975qe}
and later in the proton decay experiment \cite{krishnaswamy2}
at Kolar Gold
Fields (KGF) in south India, some unusual events were seen. These 
so-called Kolar events were multi-track events with some unusual
features which could not be explained away by any known processes of
muons or neutrinos.
Recently it has been speculated that such events may be caused by the 
decay of unstable cold dark matter particles with mass in the range of 
10 GeV with a life time approximately equal to the age of the universe 
\cite{Murthy:2013uca}. Such an interpretation may be easily tested with ICAL at 
INO, even without further modifications \cite{Dash:2014sza}.
Signals of dark matter annihilation inside the Sun can also be detected
at ICAL. The possible observation of GUT monopoles is another such issue
that can be addressed at ICAL with its current setup.

\chapter{Atmospheric Neutrino Fluxes}
\label{fluxes}

\begin{flushright}
{\it
Everything is in a state of flux, 
Even the status quo.\\
- Robert Byrne}
\end{flushright}

\section{Introduction to Atmospheric Neutrinos} 
\label{atm-neutrinos} 
 
Atmospheric neutrinos are produced in the cosmic ray interactions with  
the nuclei of air molecules in the atmosphere. The first report of
cosmic ray induced atmospheric neutrinos was from the deep  
underground laboratories at Kolar Gold Field (KGF) in India by TIFR,  
Osaka University and Durham University \cite{Achar:1965ova}, and immediately  
afterwards by Reines et al. \cite{Reines:1965qk} in an experiment conducted in  
South African mines in 1965. Atmospheric neutrinos have been studied  
since then in several other underground laboratories, and important  
discoveries such as the evidence for neutrino oscillations 
\cite{Fukuda:1998mi},  
have been made. We will briefly review the atmospheric neutrinos  
in this section. 
 
\begin{figure}[t] 
\centering\includegraphics[width=0.6\textwidth]{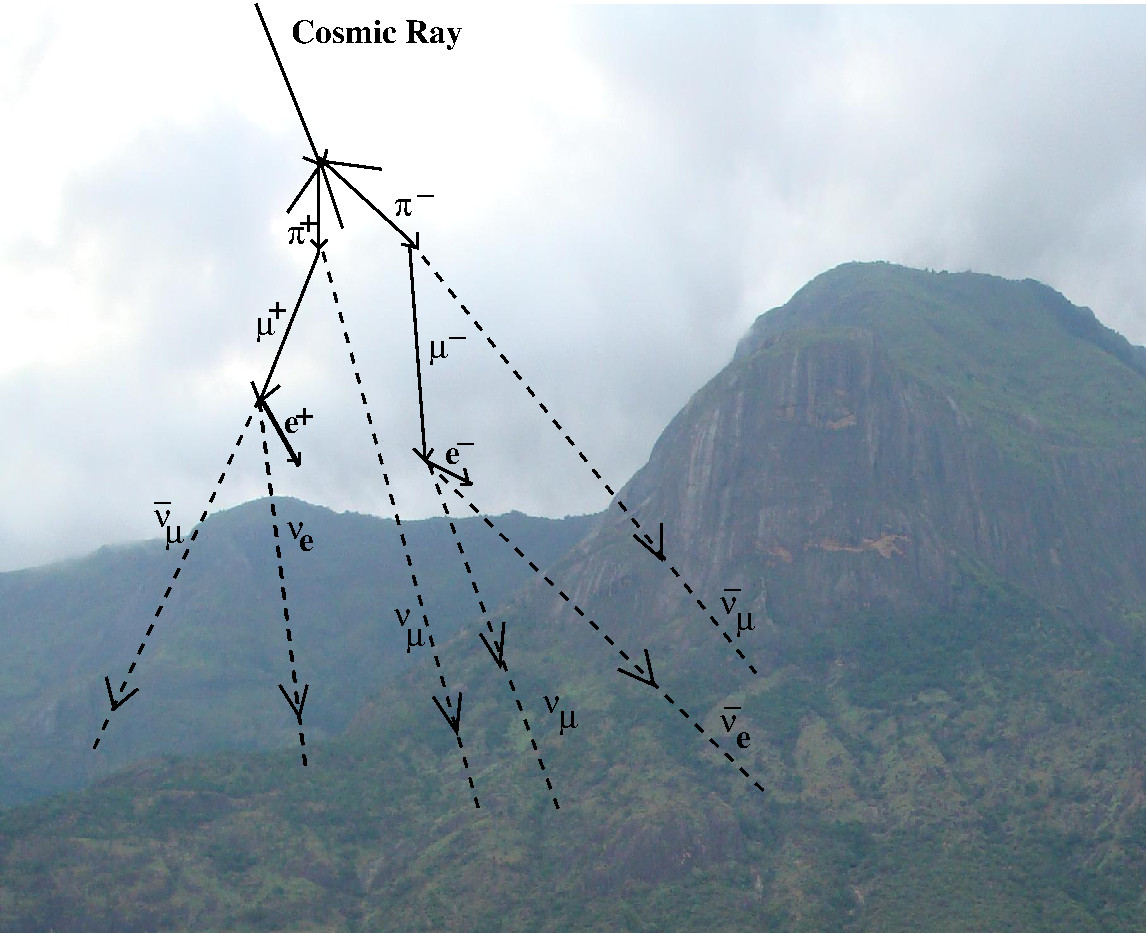} 
\caption{A schematic illustration of the production of neutrinos 
due to cosmic rays.} 
\label{cosmic} 
\end{figure}

Primary cosmic rays are high energy particles impinging on the Earth  
from galactic and extragalactic sources. Their origins are still  
clouded in mystery. In the GeV energy range, the cosmic ray particles are  
made up of mainly protons and about 9\% helium nuclei, with a small  
fraction of heavy nuclei. Although the energy spectrum of cosmic rays
extends to very high energies, beyond even $10^{10}$ GeV, it falls rapidly 
as energy increases. When cosmic rays enter the atmosphere, interactions 
with the nuclei in air molecules produce secondary particles. 
These secondary particles are mainly pions with a small admixture of kaons. 
These mesons decay mainly to muons and their  
associated neutrinos following the decay chain 
\begin{eqnarray} 
\pi^{\pm}\rightarrow \mu^{\pm}+\nu_\mu(\bar\nu_\mu)  \; , \nonumber\\ 
\mu^{\pm}\rightarrow e^{\pm}+\bar\nu_\mu(\nu_\mu)+\nu_e(\bar\nu_e) \; . 
\label{atm-reactions} 
\end{eqnarray}  
Kaons also decay in a similar manner producing the two neutrino flavours, 
but their contribution to the atmospheric neutrino flux is small compared 
to the pions for neutrinos of a few GeV. We call the neutrinos produced in 
this manner as atmospheric neutrinos. It may be noted that only the 
$\nu_e$ and $\nu_\mu$  neutrinos, along with 
their antiparticles, are produced in the atmosphere. The flux of $\nu_\tau$ 
requires the production of mesons with heavy quarks, as a result their flux 
is extremely small and we do not consider these neutrinos here. 
A schematic illustration 
of this cascading neutrino production is shown in Fig.~\ref{cosmic}. 

\begin{figure}[] 
\includegraphics[width=0.49\textwidth,height=0.35\textwidth]
{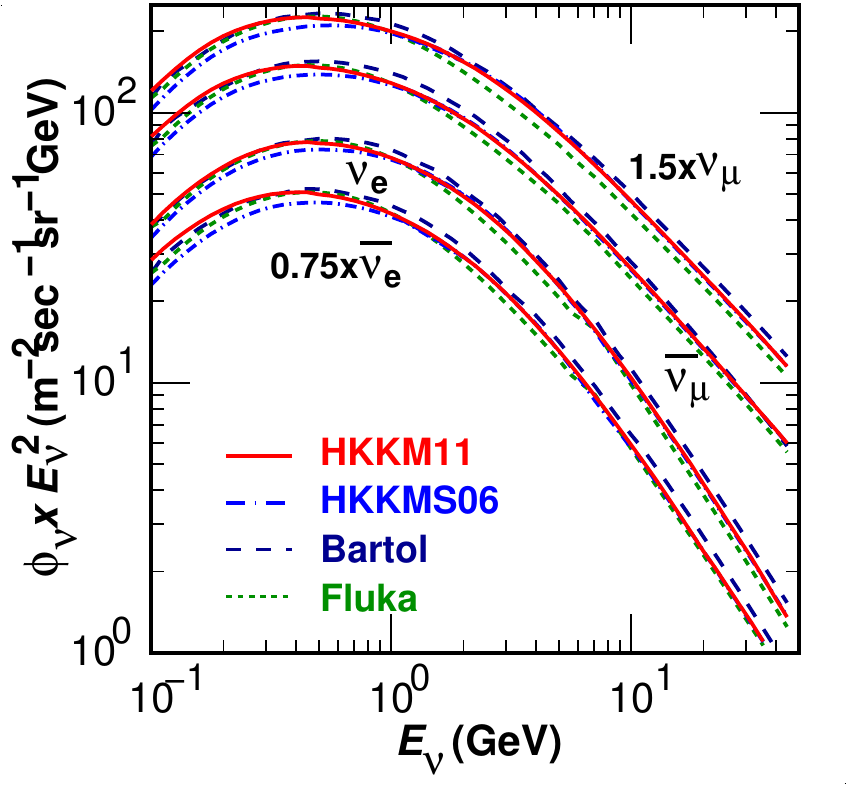} 
\includegraphics[width=0.49\textwidth,height=0.35\textwidth]
{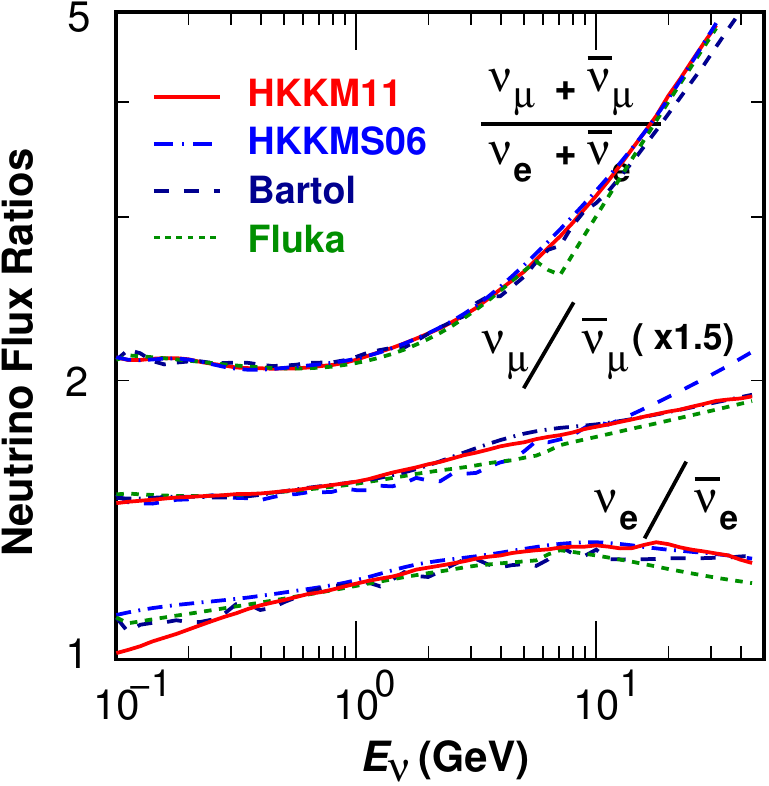} 
\caption{The direction integrated neutrino fluxes in various models
are shown on the left panel. The ratios of fluxes of different neutrino
species as functions of energy are shown on the right panel.  
Figures are reproduced from Honda et al. \cite{Honda:2011nf}, based on the  
analysis of cosmic ray neutrino fluxes from \cite{Honda:2006qj}, 
\cite{Barr:2004br} and \cite{Battistoni:2002ew}.}
\label{rnu} 
\end{figure}

From Eq.~\ref{atm-reactions} it is clear that the ratio  
\begin{equation}
R=\frac{\Phi(\nu_\mu)+\Phi(\bar\nu_\mu)}{\Phi(\nu_e)+\Phi(\bar\nu_e)}
\approx 2 \; ,
\end{equation} 
where $\Phi$ denotes the flux of neutrinos. The ratio is only approximate, 
since at high energies muon may not decay before reaching the surface of the 
Earth. It, however, remains greater than 2, which may be observed
from Fig.~\ref{rnu}. The figure 
displays the direction-integrated neutrino fluxes in various models,
as well as the ratios of fluxes of different kinds of neutrinos.

An important property of the atmospheric neutrino flux is that it is  
symmetric about a given direction on the surface of the Earth, that is 
\begin{equation} 
\Phi_\nu(E,\cos\theta)=\Phi_\nu(E,-\cos\theta) \; , 
\end{equation}  
where $\theta$ is the zenith angle. This result is robust above 3 GeV,  
though at lower energies the geomagnetic effects result in deviations 
from  this equality. Therefore, at higher energies, any asymmetry in the
fluxes of the upward-going and downward-going neutrinos can be attributed
to the flavour changes during propagation. Even at lower energies, 
large deviations from the above equality are not expected, except from
neutrino oscillations. This up-down asymmetry is thus the basis of
atmospheric neutrino analysis, and it was effectively used by the
SuperKamiokande collaboration to establish the first confirmed signal of 
neutrino oscillations \cite{Fukuda:1998mi}. Of course detailed analyses 
need the 
calculations of atmospheric neutrino fluxes as functions of energies, 
zenith angles as well as azimuthal angles.  

The atmospheric neutrinos not only provide neutrinos in two distinct 
flavours, but also over a whole range of energies from hundreds of MeV 
to TeV and beyond. Yet another advantage over the conventional accelerator 
neutrino beams is the fact that the atmospheric  neutrinos traverse 
widely different distances in different directions: from $\sim$ 10 km on the
way downwards to more than 12000 km on the way upwards through the
centre of the earth. They also traverse matter densities varying from very 
small (essentially air) to almost 13 g/cc when passing through the 
earth's core. 
 
These facts make the analysis of atmospheric neutrinos not only  
interesting but also unique. The ICAL is expected to exploit the
advantages of the freely available atmospheric neutrino flux,
not only to explore the neutrino oscillation parameters 
but also determine the hierarchy of neutrino masses, and perhaps 
probe new physics.

\section{Calculations of atmospheric neutrino fluxes}
\label{flux-calc}

The neutrino oscillation studies with atmospheric neutrinos can be put  
on a firm foundation provided the atmospheric neutrino flux estimates and  
their interaction cross sections are known as precisely as possible. The  
main steps in the determination of atmospheric neutrino fluxes are:  
 
\begin{itemize} 
 
\item The energy spectrum of primary cosmic rays: The flux of primary  
cosmic rays decreases approximately as $E^{-2.7}$ in the 10 GeV to TeV  
region. Consequently the flux of neutrinos decreases rapidly in the high  
energy region. The flux of cosmic rays outside the atmosphere is  
isotropic and constant in time. These are well measured experimentally 
up to tens of GeV. The primary spectrum of cosmic ray protons can be 
fitted to a form  
\begin{equation} 
\Phi(E)=K[E+b\exp(-c\sqrt{E})]^{-\alpha}, 
\end{equation} 
where $\alpha=2.74,~K=14900,~b=2.15,~c=0.21$~\cite{Gaisser:2002jj}.
 
\item The energy spectrum of secondary muons: The interactions of  
primary cosmic rays with the air nuclei produce in pions and kaons,
which in turn yield muons. An important input needed for this calculation
is the hadronic cross sections. These are well measured  
in accelerator experiments from low energy up to hundreds of GeV. Beyond the  
range of accelerator energies these cross sections are model-dependent.  
Hence the composition of the secondary cosmic rays and their energy spectrum 
is well  known up to TeV energies. The muons are produced by the decay of 
these mesons. 
 
\item The energy spectrum of neutrinos: This needs modelling the altitude  
dependence of interactions in the atmosphere, the geomagnetic effect on the 
flux of cosmic rays and secondaries, and the longitudinal dependence of 
extensive air showers. 
 
\end{itemize} 
 
Uncertainties in each one of the above steps limit the precision in the 
determination of neutrino fluxes on the surface. Typically these  
introduce an uncertainty of the order of 15-20\% in the overall normalization.
 
\begin{figure}[] 
\centering\includegraphics[width=0.98\textwidth,height=0.6\textwidth]
{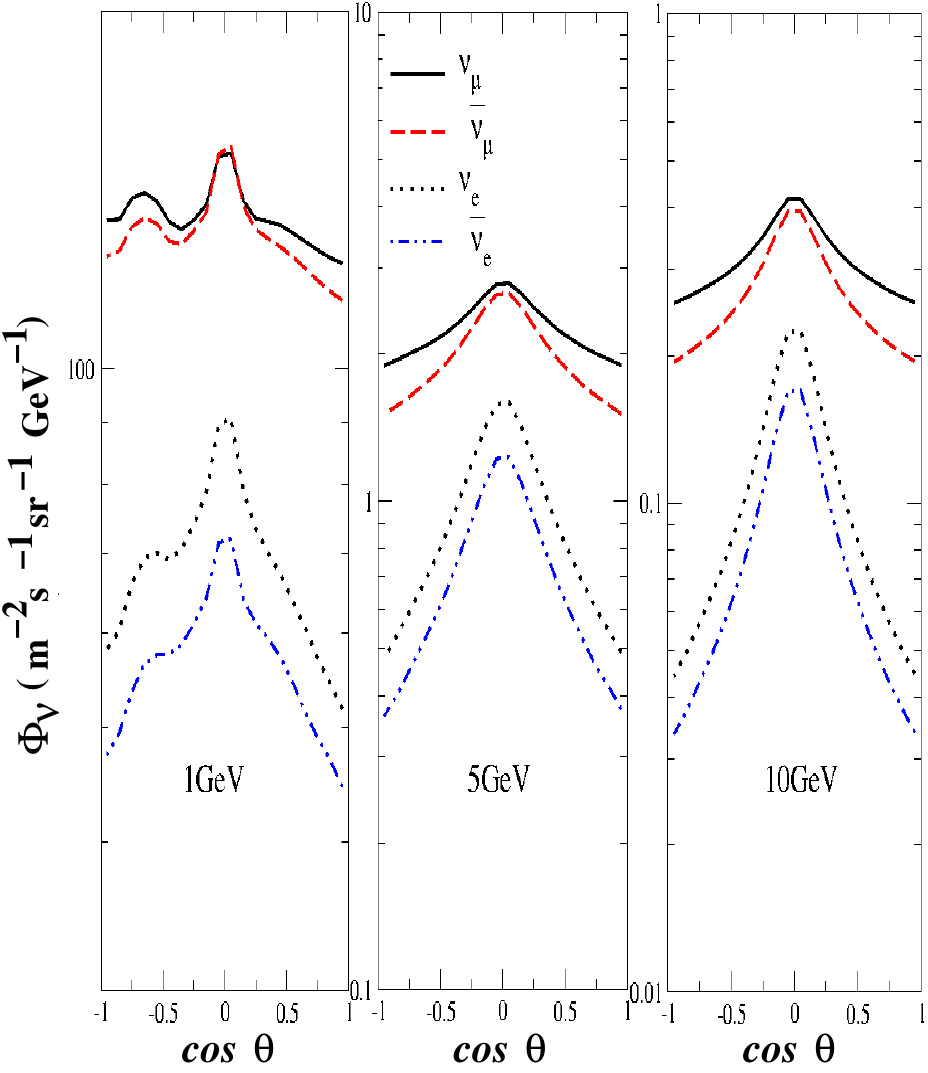} 
\caption{Zenith angle dependence of the neutrino flux 
averaged over azimuthal angles are shown for three different energies. 
This has been calculated for the Kamioka location. 
Figures are taken from Honda et al. \cite{sajjad-new}.
Note that the flux scales are different for different energies.
}
\label{nuflux} 
\end{figure} 

Some typical features of the zenith angle distributions of atmospheric 
neutrino fluxes may be seen in Fig.~\ref{nuflux}, which show the fluxes
(averaged over azimuthal angles) for three neutrino energies, calculated
for Kamioka, the location of the Superkamiokande detector.
The figure shows that the flux is typically maximum near $\cos\theta=0$,
i.e. for horizontal neutrinos, where the muons have had the maximum
proper time to decay. Also, the ratio of muon to electron neutrino flux
is observed to increase at higher energies and at more vertical
(down-going or up-going) neutrinos, where muons have less proper time to decay,
so the second reaction in Eq.~\ref{atm-reactions} is less efficient.
It can also be seen that at $E > 3$ GeV, the fluxes are essentially 
symmetric in zenith angle. However at lower energies, there is some
asymmetry, arising mainly from the bending of muons in the geomagnetic 
field. 

Recently Honda et al. \cite{Athar:2012it}  have calculated the atmospheric  
neutrino spectrum at the INO location (Theni). 
It is observed that the total flux at INO is slightly smaller than that 
at Kamioka at low energies ($E \lesssim 3$ GeV), but the difference
becomes small with the increase in neutrino energy. 
Also, at low energies ($E \sim 1$ GeV), the up-down asymmetries are larger
at the INO site. These asymmetries decrease with the increase in neutrino 
energy. The detailed characteristics
of these fluxes have been given in Appendix~\ref{honda-ino-flux}. 
The analyses presented in 
this report use fluxes at the Kamioka location. We plan to use
the recently compiled fluxes \cite{sajjad-new,Athar:2012it} for the 
Theni site in our future analysis.

\chapter{The ICAL Simulation Framework}
\label{framework}

\begin{flushright}
{\it
A good simulation $\cdots$ gives us \\
a sense of mastery over experience. \\
Heinz R. Pagels 
}
\end{flushright}


The broad simulation framework for the ICAL, starting with event generation,
is indicated schematically in Table~\ref{simfw}. The events in the
detector are generated using the NUANCE Monte Carlo generator
\cite{Casper:2002sd}. This uses the atmospheric neutrino
fluxes as described in Chapter~\ref{fluxes} along with various possible
neutrino-nucleus interaction cross-sections to generate the vertex and
the energy-momentum of all final states in each event; these are then
propagated through the virtual ICAL detector 
using the GEANT4 simulation tool. The GEANT4 simulates the propagation
of particles through the detector, including the effects of the iron, 
the RPCs, and the magnetic field. The information in the events is 
then digitised in the form of $(X,Y,Z)$ coordinates of the hits in the
RPCs and the timing corresponding to each of these ``hits". 
This is the information available for the event reconstruction algorithms, 
which attempt, from the hit pattern, to separate the muons tracks from
the showers generated by the hadrons, and reconstruct the energies and
directions of these particles. The process is described in detail below.

\begin{table}[tp]
\begin{tabular}[t]{ccP{0.27\textwidth}cP{0.27\textwidth}} \hline
  & & & & \\
{\color{blue}{\sf NUANCE}} & &
{\large \bf Neutrino Event Generation}

{\color{blue}$\nu_{\ell} + N \to \ell + X$~.}

Generates particles that result from a random interaction of a neutrino
with matter using theoretical models for both neutrino fluxes and
cross-sections.
&   & 
{\large \bf Output:}

(i) Reaction Channel

(ii) Vertex and time information

(iii) Energy and momentum of all final state particles
\\ \hline 
\\ [-5.5ex]
 & {$\left \Downarrow \rule{0mm}{6mm} \right.$} &
 & \\
{\color{blue}{\sf GEANT}} & &
{\large \bf Event Simulation}

{\color{blue}{$\ell + X$ through simulated ICAL}}

Simulates propagation of particles through the ICAL detector with RPCs
and magnetic field.
&   & 
{\large \bf Output:}

(i) $x, y, z, t$ of the particles as they propagate through detector

(ii) Energy deposited

(iii) Momentum information
\\ \hline 
\\ [-5.5ex]
 & {$\left \Downarrow \rule{0mm}{6mm} \right.$} &
 & \\

{\color{blue}{\sf DIGITISATION}} & &
{\large \bf Event Digitisation}

{\color{blue}{$(X, Y, Z, T)$ of final states on including noise and detector
efficieny}}

Add detector efficiency and noise to the hits.
&  & 
{\large \bf Output:}

(i) Digitised output of the previous stage

\\ \hline 
\\ [-5.5ex]
 & {$\left \Downarrow \rule{0mm}{6mm} \right.$} &
 & \\

{\color{blue}{\sf ANALYSIS}} & &
{\large \bf Event Reconstruction}

{\color{blue}{$(E, \vec{p})$ of $\ell$, $X$ (total hadrons)}}

Fit the muon tracks using Kalman filter techniques to reconstruct muon
energy and momentum; use hits in hadron shower to reconstruct hadron
information.
&  & 
{\large \bf Output:}

(i) Energy and momentum of muons and hadrons, for use in physics
analyses.

\\ \hline 
\end{tabular}
\caption{The simulation frame-work as implemented in the ICAL simulation
package.}
\label{simfw}
\end{table}

\section{Neutrino interactions and event generation}
\label{nuance}

Neutrino and antineutrino interactions in the ICAL detector are 
modelled using NUANCE neutrino generator version 3.5 \cite{Casper:2002sd}. 
Some preliminary studies and comparisons have also been initiated using the 
GENIE neutrino generator \cite{Andreopoulos:2009rq}, 
but are not a part of this Report. 
The interactions modelled in NUANCE include 
(i) quasi-elastic scattering (QE) for both charged and neutral 
current neutrino interactions with nucleons, which dominate below
neutrino energies of 1 GeV,
(ii) resonant processes (RES) with baryon resonance production mainly
from neutrinos with energy between 1 and 2 GeV,
(iii) deep inelastic scattering (DIS) processes with considerable
momentum (squared) transfer from the neutrino to the target, with
the nuclei breaking up into hadrons, which is the dominant contribution
in the multi-GeV region,
(iv) coherent nuclear processes on nuclei, and
(v) neutrino-electron elastic scattering and inverse muon decay.
These are the main neutrino interaction processes of relevance for our
simulation frame work, with the contribution of the last two being the
least in the few GeV energy region of interest. A simple ICAL geometry
has been described within NUANCE, including mainly the iron and glass
components of the detector, as most of the interactions will occur in
these two media. NUANCE identifies these bound nucleons (with known
Fermi energies) differently from free nucleons and also applies final
state nuclear corrections.

\begin{figure}[h]
\includegraphics[width=0.49\textwidth,height=0.35\textwidth]
{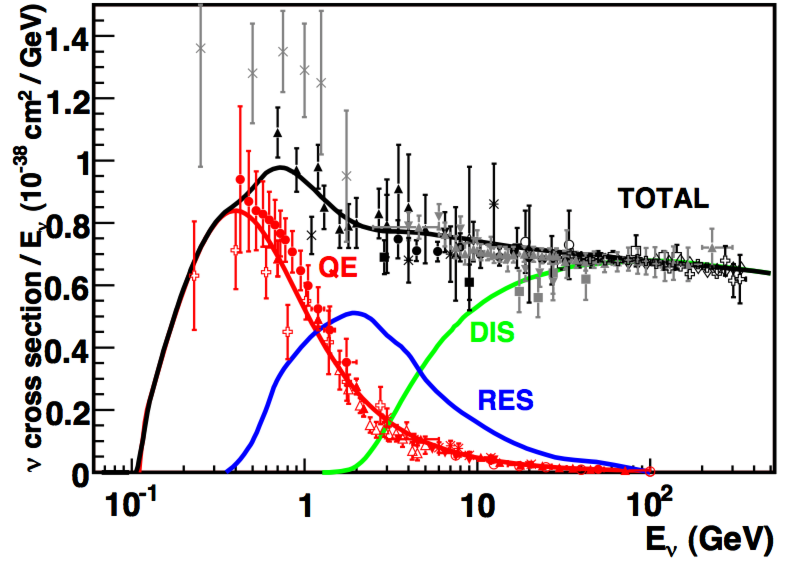} 
\includegraphics[width=0.49\textwidth,height=0.35\textwidth]
{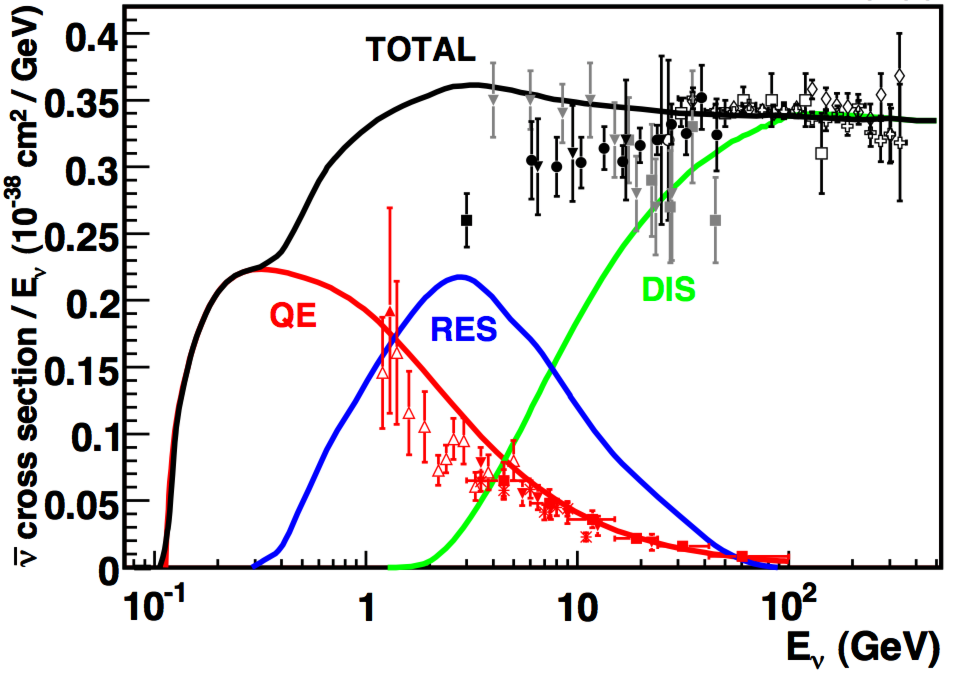}
\caption{Muon neutrino (left) and muon antineutrino (right) total
charged current cross sections in (cm$^2$/GeV), obtained from NUANCE,
are shown (smooth lines) as a function of incident neutrino energy,
$E_\nu$, in comparison with the existing measurements of these cross
sections along with their errors~\cite{FormaggioZeller:2012}. Note that
the $y$-axis scale of the two panels is different.}
\label{crossec}
\end{figure}

The NUANCE generator calculates event rates by integrating different
cross sections weighted by the fluxes for all charged current (CC) and
neutral current (NC) channels at each neutrino energy and angle. Some
typical total cross sections for different CC processes are illustrated
in Fig.~\ref{crossec}. Based on the interaction channel, there can
be 10-40$\%$ uncertainty in cross sections in the intermediate energy
ranges \cite{Hewett:2012ns}.

As mentioned in Chapter~\ref{fluxes}, for the present we have used the
Honda fluxes \cite{Honda:2011nf} generated at the location of Kamioka. This
will be changed soon to that at the actual location of INO.  We choose
to generate only unoscillated neutrino events using NUANCE for the
simulations, even though there is a provision for generating oscillated
events in it. The oscillations are applied externally, separately in
each analysis.



\section{Simulation of the ICAL Detector}

We now describe the ICAL detector geometry within the GEANT4
\cite{Agostinelli:2002hh} simulation framework. This includes the geometry
itself, and the magnetic field map and the RPC characteristics that are
inputs to the simulation.

\subsection{The Detector Geometry}

The simulations have been performed for the 50 kt ICAL detector, which has a
modular structure with the full detector consisting of three modules, each
of size 16~m (length) $\times$ 16~m (width) $\times$ 14.5~m (height), with
a gap of 20~cm between the modules. 
The ICAL coordinates are defined as follows.
The direction along which the modules are placed is labelled as the
$x$-direction with the remaining horizontal transverse direction being
labelled $y$. At present, $x$ is also considered to coincide with
South, since the final orientation of the INO cavern is not yet decided.
The $z$-axis points vertically upwards so that the polar angle equals the
zenith angle $\theta$ while the zero of the azimuthal angle $\phi$ points
South. The origin is taken to be the centre of the second module. Each
module comprises 151 horizontal layers of 5.6~cm thick iron plates.
The area of each module is 16~m $\times$ 16~m, while the area of each 
iron plate is 2~m $\times$ 4~m. There is a vertical gap of 4~cm between two
layers. The iron sheets are supported every 2~m in both the $x$ and
$y$ directions, by steel support structures. The basic RPC units have
dimensions of 1.84~m $\times$ 1.84~m $\times$ 2.5~cm, and are placed in
a grid format within the air gaps, with a 16~cm horizontal gap between
them in both $x$ and $y$ directions to accommodate the support structures.

Vertical slots at $x=x_0 \pm 4$ m (where $x_0$ is the central $x$
value of each module) extending up to $y=\pm 4$ m and cutting through
all layers are provided to accommodate the four copper coils that wind
around the iron plates, providing a magnetic field in the $x$-$y$ plane,
as shown in Fig.~\ref{fig:magfield}. The detector excluding the coils
weighs about 52 kt, with 98\% of this weight coming from iron where
the neutrino interactions are dominantly expected to occur, and less
than 2\% from the glass of the RPCs. In the central region of each
module, typical values of the field strength are about 1.5 T in the
$y$-direction, as obtained from simulations using MAGNET6.26 software
\cite{magnetcode}. 

\subsection{The Magnetic Field}

\begin{figure}[t]
\begin{center}\includegraphics[width=0.65\textwidth]
{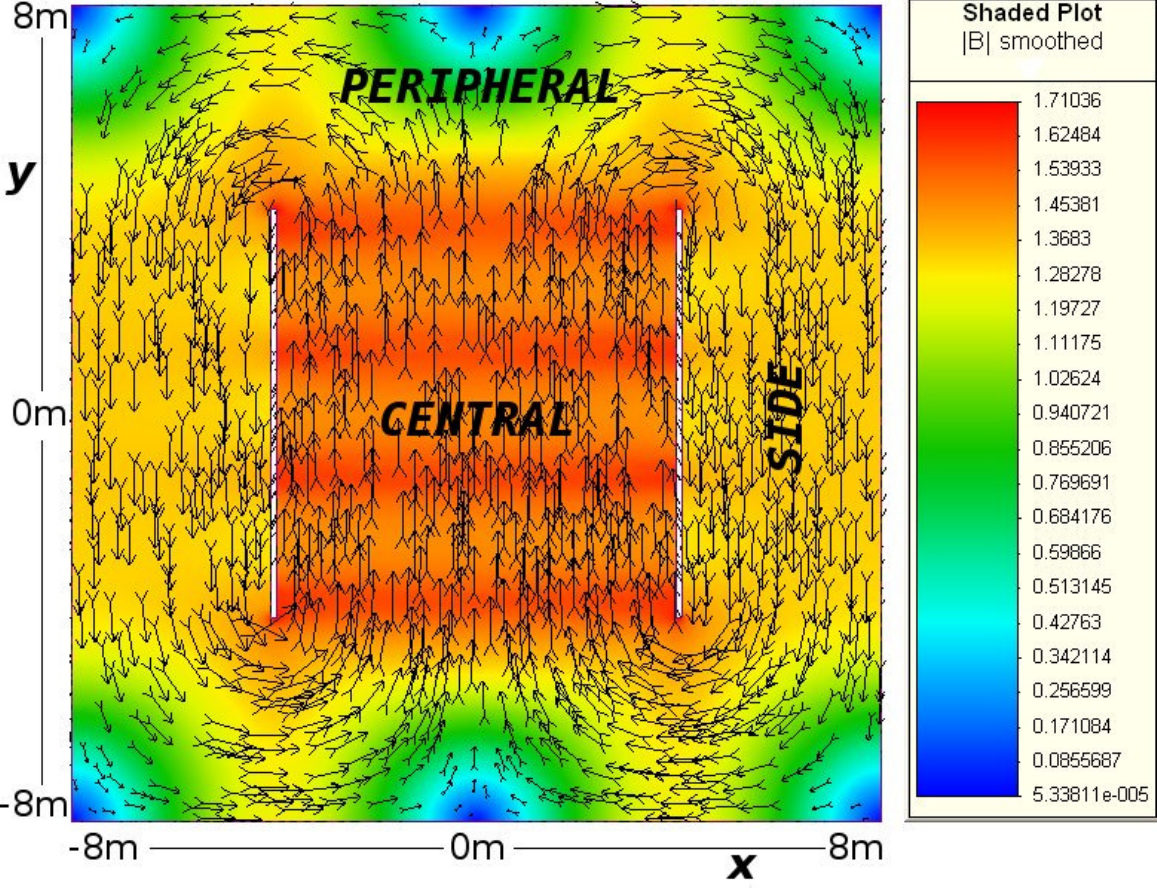}
\end{center}
\caption{Magnetic field map in the central plate of the central module
($z=0$), as generated by the MAGNET6 software. The length and direction
of the arrows indicate the magnitude and direction of the field; the
magnitude (in T) is also shown according to the colour-coding indicated
on the right.}
\label{fig:magfield}
\end{figure}

Fig.~\ref{fig:magfield} depicts the magnetic field lines in the central
iron plate near the centre of the central module. The arrows denote 
the direction of magnetic field lines while the length of the arrows (and the
shading) indicates the magnitude of the field. The maximum magnitude of
the magneic field is about 1.5 T. Notice that the field
direction reverses on the two sides of the coil slots (beyond
$x_0\pm$4~m) in the $x$-direction. In between the coil slots (an
8~m $\times$ 8~m square area in the $x$-$y$ plane) the field is
maximum and nearly uniform in both magnitude (to about 10\%) and
direction; we refer to this as the {\it central region}. Near the
edges in the $x$ direction (outside the coil slots) the field is
also fairly uniform, but in the opposite direction; this is called
the {\it side region}. Near the edges in the $y$ direction, i.e., in
the regions $4$~m $\le \vert y \vert \le 8$~m, both the direction
and magnitude of the magnetic field vary considerably; this region is
labelled as the {\it peripheral region}.

In our simulations, the field has been assumed to be uniform over the 
entire thickness of an iron plate at every $(x,y)$ position, and has 
been generated in the centre of the iron plate, viz., at $z=0$. 
In the 4~cm air gap between the iron plates, the field is taken to be 
zero since it falls off to several hundred gauss in these regions,
compared to more than 1 T inside the iron plates. 
The magnetic field is also taken to be zero in the (non-magnetic) steel 
support structures. These support structures, along with the coil slot, 
form the bulk of the dead spaces of the detector.

The side and peripheral regions are beset by edge effects as well as
by non-uniform and lower magnetic field. We confine ourselves, in the
present study, to tracks generated only in the central region ($-4$
m $\le x-x_0 \le 4$ m and $-4$ m $\le y \le 4$ m), although the particle
may subsequently travel outside this region or even exit the detector.

\subsection{The Resistive Plate Chambers}

In order to appreciate the hit pattern in the simulated detector
it is necessary to describe the active detector elements, the RPCs.
These are glass chambers made by sealing two 3 mm thick glass sheets
with a high DC voltage across them, with a uniform gap of 2 mm using 
plastic edges and spacers through which a mixture of R134A ($\sim$ 95\%), 
isobutane, and trace amounts of SF$_6$ gas continually flows. In brief, the
working principle of an RPC is the ionisation of the gaseous medium when
a charged particle passes through it. The combination of gases keeps the
signal localised and the location is used to determine the trajectory of
the charged particle in the detector. 
For more details, see Ref.~\cite{rpc_char}.

A 150 micron thick copper sheet is the component most relevant to the
simulation and track reconstruction as it inductively picks up the
signal when a charged particle traverses the chamber. This copper sheet
is pasted on the {\it inside} of a 5 mm thick foam (used for structural
strength and electrical insulation) placed both above and below the glass
chamber. It is pasted on the side of the foam facing the glass and is
insulated from the glass by a few sheets of mylar. This layer is scored
through with grooves to form strips of width 1.96 cm in such a way that
the strips above and below are transverse to each other, that is, in the
$x$ and $y$ directions\footnote{Note that the strip width in the current
ICAL design is 3 cm; however this is subject to change.}. These pick-up
strips thus provide the $x$ and $y$ location of the charged particle
as it traverses the RPC while the RPC layer number provides the $z$
information. A timing resolution of about 1.0 ns is assumed as also an
efficiency of 95\%, consistent with the observations of RPCs that have
been built as a part of the R\&D for ICAL \cite{rpc_char}.



\section{Event Simulation and Digitisation}

Muons and hadrons, generated in neutrino interactions with the
detector material, pass through dense detector material and an
inhomogeneous magnetic field. Simulation of such particles through the
detector geometry is performed by a package based on the 
GEANT4~\cite{Agostinelli:2002hh}
toolkit. Here the ICAL geometry is written to a machine readable GDML file
--- which includes the RPC detector components, the support structure,
and the gas composition as described above ---
that can be read off by other associated packages, like the event
reconstruction package.
The pickup strips are considered as a continuous material for GEANT4
simulations, however for signal digitization the strips are considered
independently.

When a charged particle, for example, a muon, passes through an RPC, it
gives a signal which is assigned $x$ or $y$ values from the respective
pick-up strip information, a $z$-value from the layer information, and a
time stamp $t$. 
The minimum energy deposited in the RPC gap which will produce
an electron-ion pair, and hence give a hit, is taken to be 30 eV, with
an average efficiency of 95\%.
The global coordinates of the signals are then translated
through digitisation into information of the $X^{th}$ $x$ strip and the
$Y^{th}$ $y$ strip at the $Z^{th}$ plane. These digitized signals along
with the time stamp form what are called ``hits'' in the detector as
this is precisely the information that would be available when the
actual ICAL detector begins to take data.

The spatial resolution in the horizontal plane is of the order of cm (due
to the strip width) while that in the $z$ direction is of the order of mm
(due to the gas gap between the glass plates in the RPCs). The effect
of cross-talk, i.e., the probability of either or both adjacent strips
giving signals in the detector, is also incorporated, using the results
of the on-going studies of RPCs \cite{rpc_char}. Finally, since the $X$
and $Y$ strip information are independent, all possible pairs of
nearby $X$ and $Y$ hits in a plane are combined to form a {\em cluster}.

\begin{figure}[htp]
\begin{center}
\includegraphics[width=0.7\textwidth]{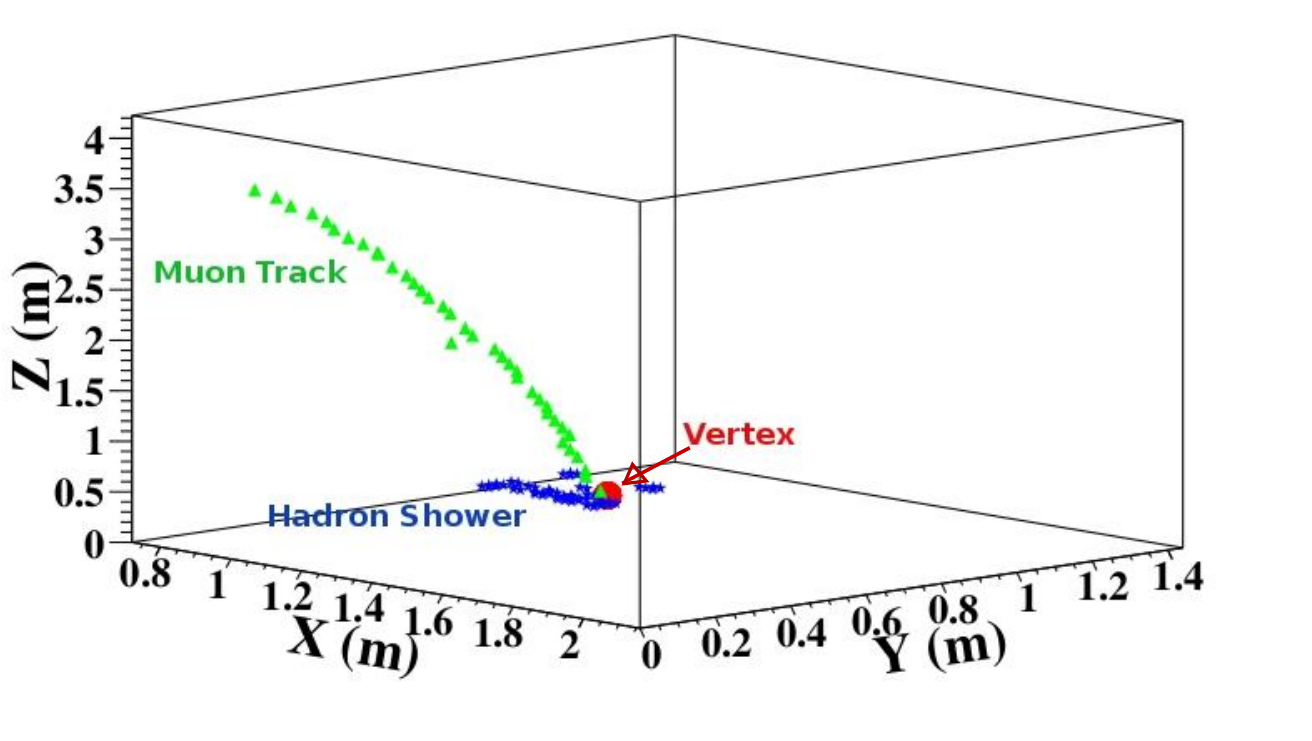}
\end{center}
\caption{Sample track of a neutrino event with a muon track and hadron
shower in the ICAL detector, where $z=0$ indicates the central
layer of the detector.}
\label{fig:samplevicetrack}
\end{figure}

A typical neutrino CC interaction giving rise to an event with a muon track 
and associated hadron shower is shown in Fig.~\ref{fig:samplevicetrack}.
It can be seen that the muon track is clean with typically one or two
hits per layer, whereas the hadron hits form a diffused shower.

\section{Event Reconstruction}

The reconstruction of individual hadrons in the hadronic showers is
not possible since the response of the detector to different hadrons is
rather similar. Only an averaged information on the energy and direction
of the hadrons is in principle possible; furthermore, hadrons, due to
the different nature of their interactions, propagate relatively short
distances in the detector. The response of ICAL to hadrons as determined
by simulations studies is described in Sec.~\ref{had-resp}.

Muons, on the other hand, being minimum ionizing particles, leave long
clean tracks, and hence the ICAL detector is most sensitive to them.
The muon momentum can be determined from the curvature of its trajectory
as it propagates in the magnetized detector, and also by measuring its
path length. The nanosecond time resolution of the RPCs also allows the
distinction between up-going and down-going muons. The muon momentum
reconstruction is achieved by using a track finder package, followed
by a track fitting algorithm that reconstructs both the momentum and
charge of the muon, using the information on the local magnetic field.

{\it The track finder} uses clusters, i.e., the combinations of all
possible pairs of nearby $X$ and $Y$ hits in a $Z$-plane, as its basic
elements. A set of clusters generated in three successive layers is
called a {\em tracklet}. The track finder algorithm uses a simple
curve fitting algorithm to find possible tracklets by finding clusters
in three adjacent planes. It includes the possibility of no hit (due
to inefficiency) in a given plane, in which case the next adjoining
planes are considered. Typically, charged current muon
neutrino interactions in ICAL have a single long track due to the muons
and a shower from the hadrons near the vertex. Since typically muons
leave only about one or two hits per layer they traverse ($\sim 1.6$
on average) as opposed to hadrons that leave several hits per layer,
the hadron showers are separated by using criteria on the average number
of hits per layer in a given event.

Ends of overlapping tracklets are matched to form longer tracks, and the 
longest possible track is found \cite{marshall2008study} by iterations 
of this process. The track finder package thus forms muon tracks as an
array of three dimensional clusters. In the rare cases when there are two
or more tracks, the longest track is identified as the
muon track. The direction (up/down) of the track is calculated from the
timing information which is averaged over the $X$ and $Y$ timing values
in a plane. For muon tracks which have at least 5 hits in the event,
the clusters in a layer are averaged to yield a single hit per layer
with $X$, $Y$ and timing information; the coordinates of the hits in the
track are sent to the track fitter
for further analysis. (This translates to a minimum momentum of about
0.4 GeV/c for a nearly vertical muon, below which no track is fitted.)

{\it The track fitter}, a Kalman-filter based algorithm, is used to fit
the tracks based on the bending of the tracks in the magnetic field. Every
track is identified by a starting vector $X_0 = (X, Y, \d X/\d Z, \d
Y/\d Z, q/p)$ which contains the position of the earliest hit $(X,Y,Z)$
as recorded by the finder, with the charge-weighted inverse momentum $q/p$
taken to be zero. Since the tracks are virtually straight in the starting
section, the initial track direction (the slopes $\d X/\d Z, \d Y/\d Z$)
is calculated from the first two layers. This initial state vector is
then extrapolated to the next layer using a standard
Kalman-filter based algorithm, which calculates the Kalman gain matrix
using the information on the local magnetic field and the geometry, the
composition of the matter through which the particle propagates, and the
observed cluster position in that later.

In the existing code, the state prediction is based upon the 
Kalman filter algorithm and 
the corresponding error propagation is performed by a
propagator matrix~\cite{marshall2008study}. The state extrapolation
takes into account process noise due to multiple scattering as described
in \cite{Wolin:1992ti} and energy loss in matter, mostly iron,
according to the Bethe formula \cite{Bethe}. A new improved set of
formulae for the propagation of the state and errors 
\cite{Bhattacharya:2014tha},
optimised for atmospheric neutrinos with large energy and range,
have also been developed, and are being used in the Kalman filter.
The extrapolated point is compared with the actual location of a hit in
that layer, if any, and the process is iterated.

The process of iteration also obtains the best fit to the track. The track 
is then extrapolated backwards to another half-layer of iron (since the interaction 
is most likely to have taken place in the iron) to determine the vertex of 
the interaction and the best fit value of the momentum at the vertex is 
returned as the reconstructed momentum (both in magnitude and direction). 
Only fits for which the quality of fit is better than $\chi^2/\hbox{ndf}
<10$ are used in the analysis.

While $q/p$ determines the magnitude of the momentum at the vertex,
the direction is reconstructed using $\d X/\d Z$ and $\d Y/\d Z$,
which yield the zenith and the azimuthal angles, i.e., $\theta$ and $\phi$. 
The results on the quality of reconstruction are presented in the next chapter.

\chapter{ICAL Response to Muons and Hadrons}
\label{response}

\begin{flushright}
{\em ``Muons are clean because they leave a trail, \\
Hadrons are dirty because they shower.'' \\
-- M. V. N. Murthy}
\end{flushright}

In this chapter, we discuss the simulations response of ICAL to
the final state particles produced in neutrino-nucleus interactions
as discussed in the previous chapter. Being minimum ionising particles,
muons typically register clean long tracks with just about one hit per
RPC layer in the detector while hadrons produce a shower with
multiple hits per layer, due to the very different nature of their
interactions. Multiple scattering further affects the quality of the
track.

First we discuss the detector response with respect to single particles
(muons or hadrons) with fixed energies. In order to simulate the neutrino
events fully, we then use the particles generated in atmospheric
neutrino events using the NUANCE \cite{Casper:2002sd} 
event generator, for calibration.
For the case of single muons, we study the response of the detector
to the energy/momentum, direction and charge of muons propagated with
fixed energy/momentum and direction ($\theta, \phi$) from the central
region of the detector (described in Chapter \ref{framework}). Next we
propagate the hadrons, mainly single pions, also with fixed energy and
direction, through the central region of the detector and determine the
energy response of the detector with respect to hadrons. Amongst the
particles generated via NUANCE, the muons tracks can be separated while
the hits from all the hadrons in the event are treated as just one shower.

\section{Response of ICAL to Muons}
\label{muon-resp}

In this Section, we present the results of the reconstruction of the
charge, energy and direction of muons \cite{Chatterjee:2014vta}. 
For this study, we confine ourselves to tracks generated only in the 
central region of the ICAL detector, i.e. $-4~{\rm m} \le x \le 4~{\rm m}$, 
$-4~{\rm m} \le y \le 4~{\rm m}$ and $-4~{\rm m} \le z \le 4~{\rm m}$,
with the origin taken to be the center of
the central detector module. The particle may subsequently travel outside this 
region or even exit the detector: both fully contained and partially 
contained events are analyzed together. At low energies, the tracks are 
fully contained while particles start to leave the detector region for 
$P_{\rm in} \gtrsim 6$ GeV/c, depending on the location of the vertex
and the direction of the paticles.

\begin{figure}[ht]
\centering
\includegraphics[width=0.49\textwidth,height=0.35\textwidth]
{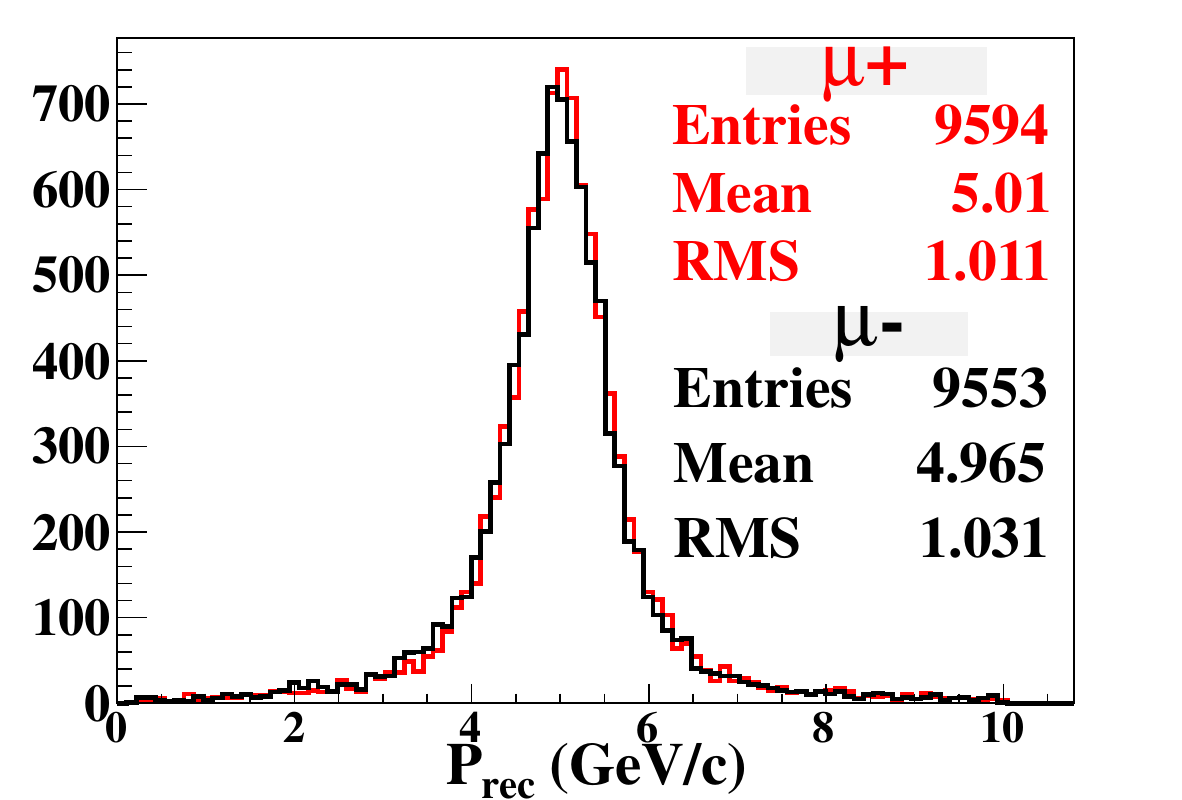}
\includegraphics[width=0.49\textwidth,height=0.35\textwidth]
{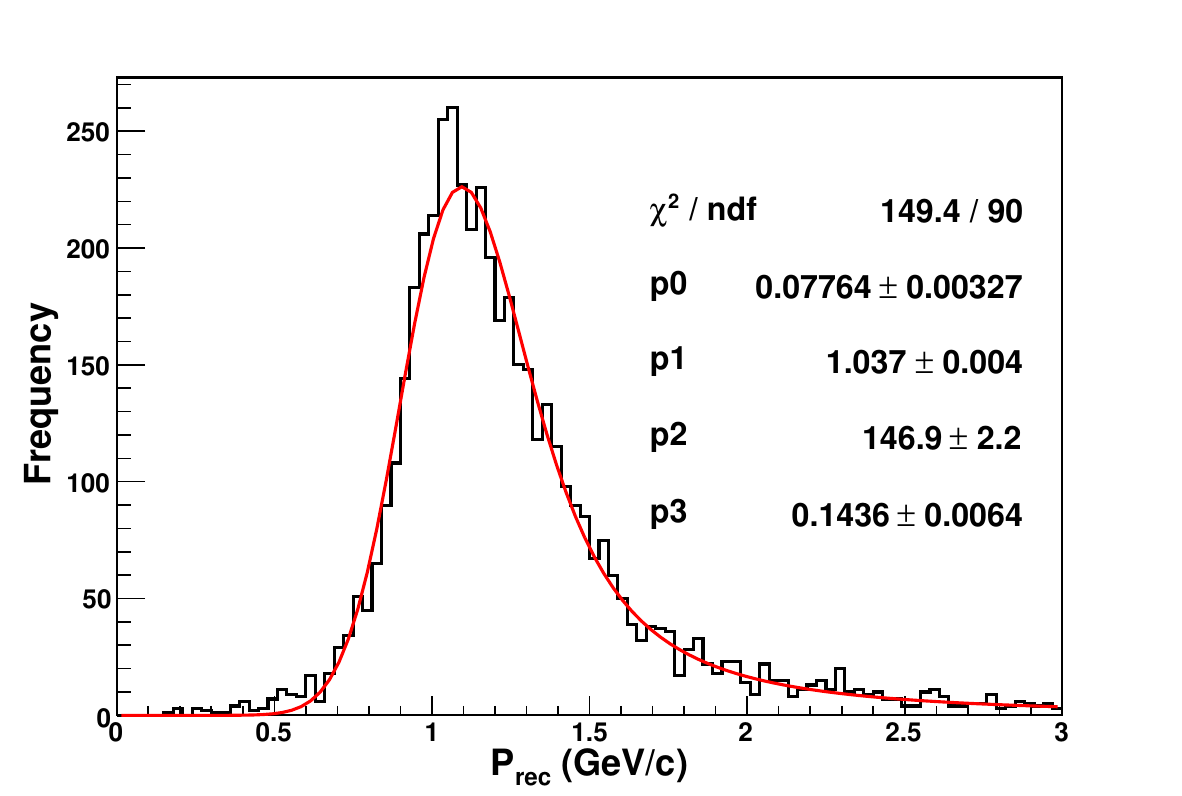}
\caption{The left panel shows the reconstructed momentum distributions 
for $(P_{\rm in}, \cos\theta) = (5\hbox{ GeV/c}, 0.65)$ smeared over the 
central volume of the detector for $\mu^-$ and $\mu^+$ particles
\cite{Chatterjee:2014vta}. 
The right panel shows the same distribution, but for 
$(P_{\rm in}, \cos\theta) = (1\hbox{ GeV/c}, 0.65)$, for $\mu^-$,
fitted with the Landau-convoluted-with-Gaussian distribution.}
\label{fig:mu+-}
\end{figure}

The $\mu^{-}$ and $\mu^{+}$ are analysed separately for fixed values
of the muon momentum $P_{\rm in}$ and direction: while $\cos\theta$
is kept fixed for a set of typically 10000 muons, the azimuthal
angle $\phi$ is smeared over all possible values from $-\pi$--$\pi$. The
distribution of reconstructed muon momentum for the particular choice
$(P_{\rm in}, \cos\theta)=(5\hbox{ GeV/c}, 0.65)$ is shown in the left
panel of Fig.~\ref{fig:mu+-}. Since the results are almost identical,
as can be seen from the figure, only the results for $\mu^{-}$ are
presented in the further analysis.

The mean and the rms width $\sigma$ are
determined by fitting the reconstructed momentum distribution; the
momentum resolution is defined as $R \equiv \sigma/P_{\rm in}$.
Apart from the intrinsic uncertainties due to particle interactions
and multiple scattering effects, the distribution---especially its
width---is also sensitive to the presence of detector support structures,
gaps for magnetic field coils, etc., that have been described in the
previous chapter.

In addition, the reconstructed momentum distribution of low energy
muons has a clear asymmetric tail, as can be seen in the right panel
of Fig.~\ref{fig:mu+-} for muons with $(P_{\rm in}, \cos\theta) =
(1\hbox{ GeV/c}, 0.65)$.  It can be seen that the distribution at low
energies is significantly broader, and there is also a shift in the
mean. It is therefore fitted with a convolution of Landau and Gaussian
distributions. For muons with $P_{\rm in} > 2$ GeV/c, the distributions are
fitted with purely Gaussian distributions. In the case of Landau-Gaussian
fits, the width is defined as $\sigma \equiv \hbox{FWHM}/2.35$, in order
to make a consistent and meaningful comparison with the Gaussian fits at
higher energies, where the square root of the variance, or the rms width,
equals FWHM/2.35. Before we present results on the muon resolution, we
first discuss the impact on the resolutions of the muon angle and
location within the detector.

\subsection{Momentum resolution in different azimuthal regions}

The number of hits in the detector by a muon with a fixed energy will
clearly depend on the zenith angle, since muons traversing at different
angles travel different distances through each iron plate. As a result,
the momentum resolution would depend on the zenith angle. However,
it also has a significant dependence on the azimuthal angle for two
different reasons. One is that the magnetic field explicitly breaks
the local azimuthal symmetry of the detector geometry. There is an
additional effect due to the coil gaps that are located at $x=x_0 \pm 4$
m where $x_0$ is the centre of each module. The second reason is that
the support structures are also not azimuthally symmetric; moreover,
the length of track ``lost" within these dead spaces is also a function
of the location from where the muon was propagated and the zenith angle
at which it traverses these spaces. The cumulative dependence on the
azimuthal angle $\phi$ is a complex consequence of all these dependences
and impacts low momentum and large zenith angle muons more than higher
energy, small angle ones.

\begin{figure}[t]
\centering
\includegraphics[width=0.55\textwidth]{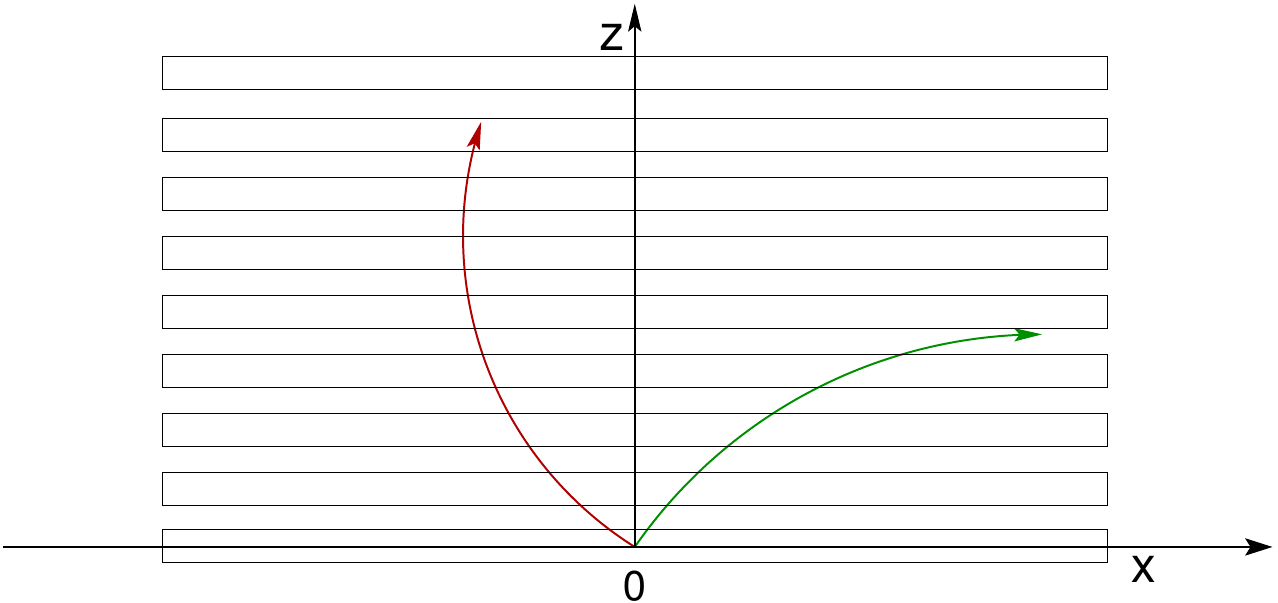}
\caption{Schematic showing muon tracks (for $\mu^-$) in the $x$-$z$ plane
for the same values of $(P_{\rm in}, \cos\theta)$ but with $\vert \phi
\vert < \pi/2$ and $> \pi/2$ (momentum component in the $x$ direction
positive and negative, respectively). The different bending causes the
muon to traverse different number of layers in the two cases.}
\label{fig:bend}
\end{figure}

\begin{figure}[h!]
\includegraphics[width=0.49\textwidth,height=0.35\textwidth]
{./response/muon-p-resol-1.pdf}
\includegraphics[width=0.49\textwidth,height=0.35\textwidth]
{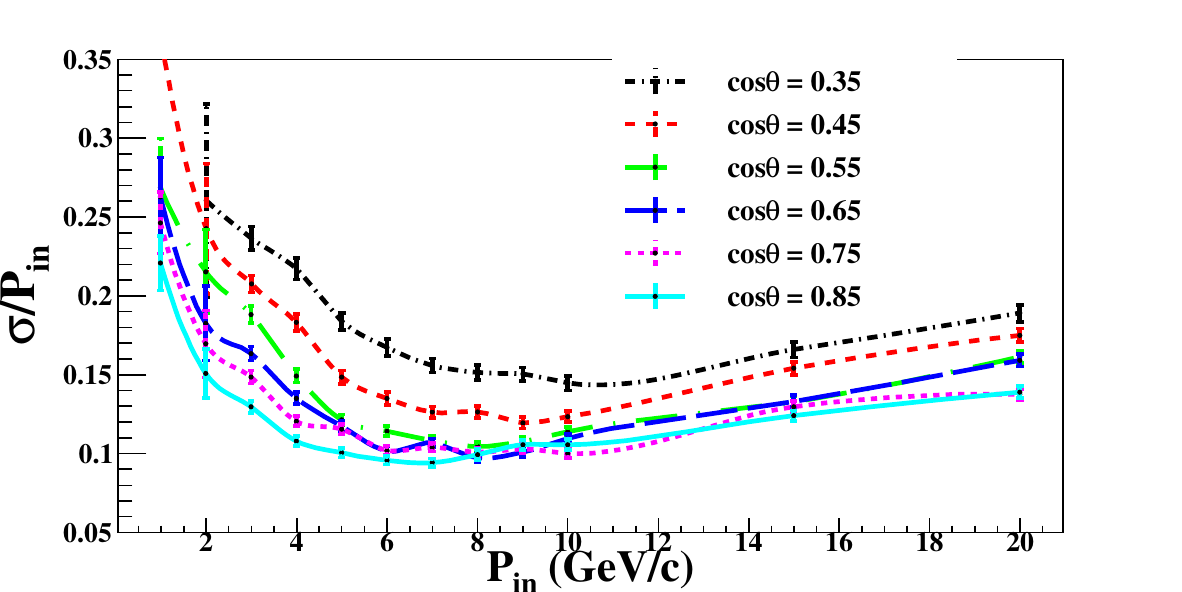} \\
. \hspace{0.1\textwidth} (a) $|\phi| \leq \pi/4$  \hspace{0.3\textwidth}
(b) $\pi/4 < |\phi| \leq \pi/2$  \\ 
{ } \\
\includegraphics[width=0.49\textwidth,height=0.35\textwidth]
{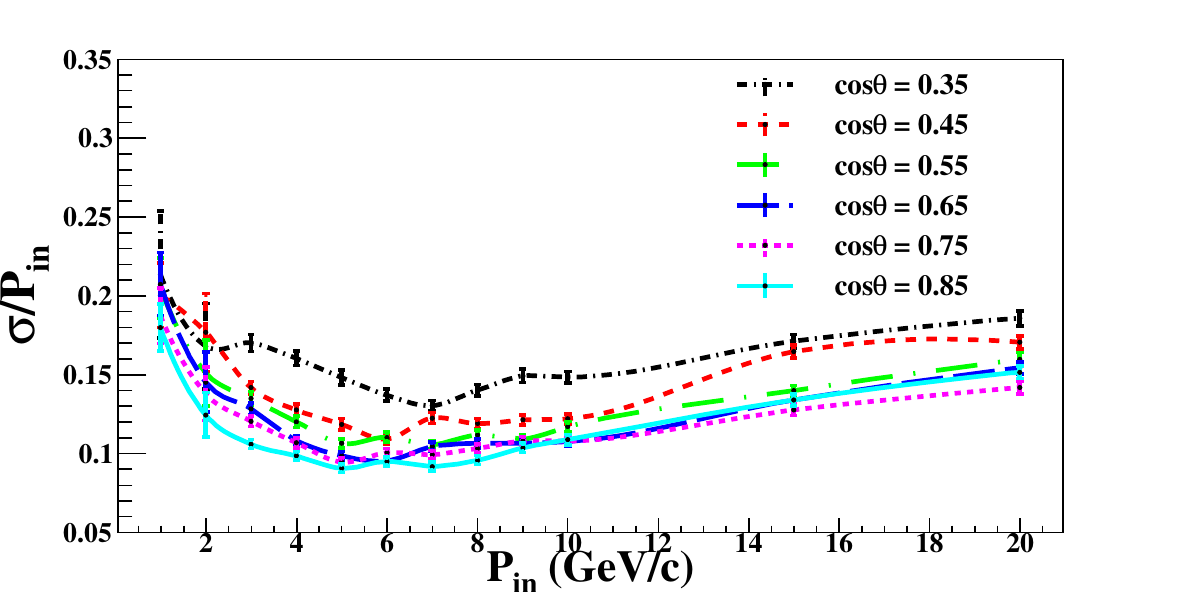}
\includegraphics[width=0.49\textwidth,height=0.35\textwidth]
{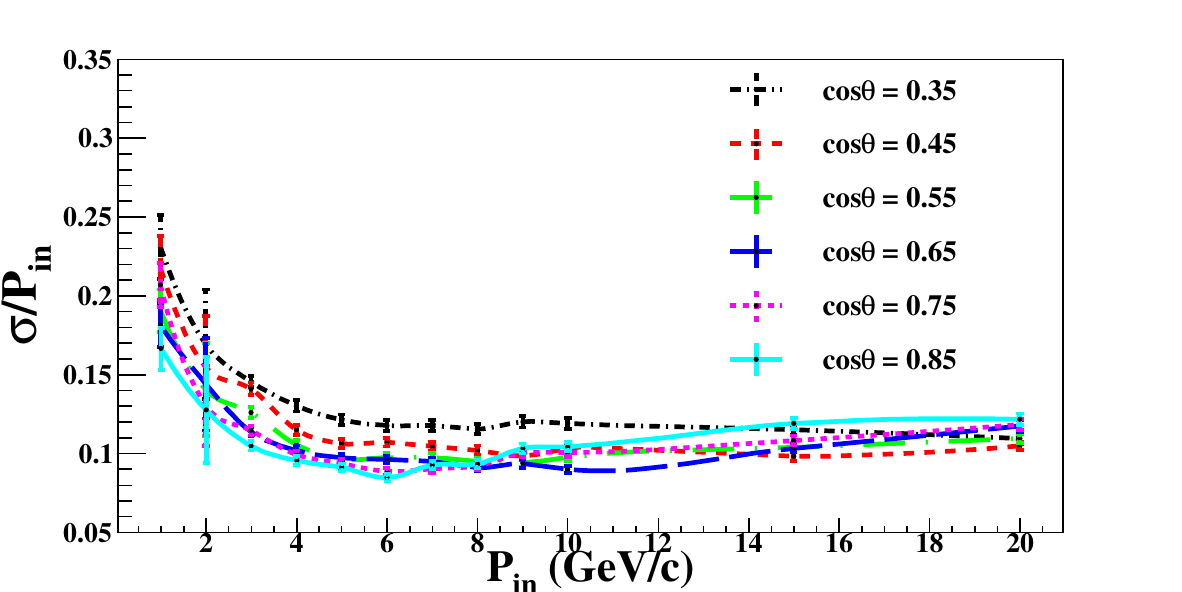} \\
. \hspace{0.1\textwidth} (c) $\pi/2< |\phi| \leq 3\pi/4$ \hspace{0.3\textwidth}
(d) $3\pi/4 < |\phi| \leq \pi$  \\ 
\caption{Muon resolution as a function of input momentum and
$\cos\theta$, in bins of $\phi$ as described in the text
\cite{Chatterjee:2014vta}.
 }
\label{fig:sv_E_ct_phi}
\end{figure}

For instance, a muon initially directed along the $y$-axis experiences
less bending since the momentum component in the plane of the iron plates
(henceforth referred to as in-plane momentum) is parallel to the magnetic
field. Furthermore, upward-going muons that are in the negative (positive)
$x$ direction experience a force in the positive (negative) $z$ direction
(the opposite is be true for $\mu^+$) and so muons injected with $\vert
\phi \vert > \pi/2$  traverse more layers than those with the same energy
and zenith angle but with $\vert \phi \vert < \pi/2$ and hence are better
reconstructed. This is illustrated in the schematic in Fig.~\ref{fig:bend}
which shows two muons ($\mu^-$) injected at the origin with the same
momentum magnitude and zenith angle, 
one with positive momentum component in the $x$ direction,
$P_x > 0$ and the other with negative $x$ momentum component. The muon
with $P_x > 0$ (initially directed in the positive $x$ direction) bends
differently than the one with $P_x < 0$ (along negative $x$ direction)
and hence they traverse different number of layers, while having roughly
the same path length. Hence, muons with different $\phi$ elicit different
detector response. Because of these effects, the momentum resolution
is best studied in different azimuthal angle bins.
We separate our muon sample into four regions/bins: bin I with $\vert
\phi \vert \le \pi/4$, bin II with $\pi/4 < \vert \phi \vert \le \pi/2$,
bin III with $\pi/2 < \vert \phi \vert \le 3\pi/4$, and bin IV with
$3\pi/4 < \vert \phi \vert \le \pi$.
The resolutions of $P_{\rm in}$ in the above $\phi$-regions, for six
values of the zenith angle,  are shown in Fig.~\ref{fig:sv_E_ct_phi}.


It may be noticed that, at lower energies the resolution improves as 
$\cos\theta$ increases, as expected, but at higher energies the behaviour 
is rather complicated. At higher energies, the muons injected at larger 
zenith angles in $\phi$ regions I and IV, have better resolutions than 
their counterparts at more vertical
angles (for instance, $\cos\theta = 0.45$ versus $\cos\theta = 0.85$)
because larger portions of the tracks, being more slanted, are still
contained within the detector. In contrast in $\phi$ regions II and III,
muons with larger zenith angles have worse resolution than those at
smaller zenith angles because the former exits the detector from the $y$
direction and are partially contained. In general, at angles larger than 
about $75^\circ$ ($\cos\theta = 0.25$), the resolution is relatively poor 
since there are several times fewer hits than at more vertical angles. 

Finally, note that the simulations studies of the physics reach of
ICAL discussed in the later chapters use the {\em azimuth-averaged}
resolutions for muons. This is because the main focus of ICAL is the
study of neutrino oscillations using atmospheric neutrino fluxes that
are azimuthally symmetric for $E_\nu \geq 2$ GeV. While studying the
neutrinos from fixed astrophysical sources, for example, the azimuthal
dependence of the detector needs to be taken into account.

In the rest of this section, we present simulations studies of the {\em
azimuth-averaged} response of ICAL to the muon direction and its charge,
since these are not very sensitive to the azimuth. We also present the
overall reconstruction efficiency for muons; this determines the overall
reconstruction efficiency of the entire neutrino event as there may or
may not be an associated hadron shower in the event.

\subsection{Zenith Angle Resolution}

\begin{figure}[h]
\centering
\subfloat[$\cos\theta = 0.25$; $\theta=1.318$]{\label{fig:12a}
\includegraphics[width=0.48\textwidth]{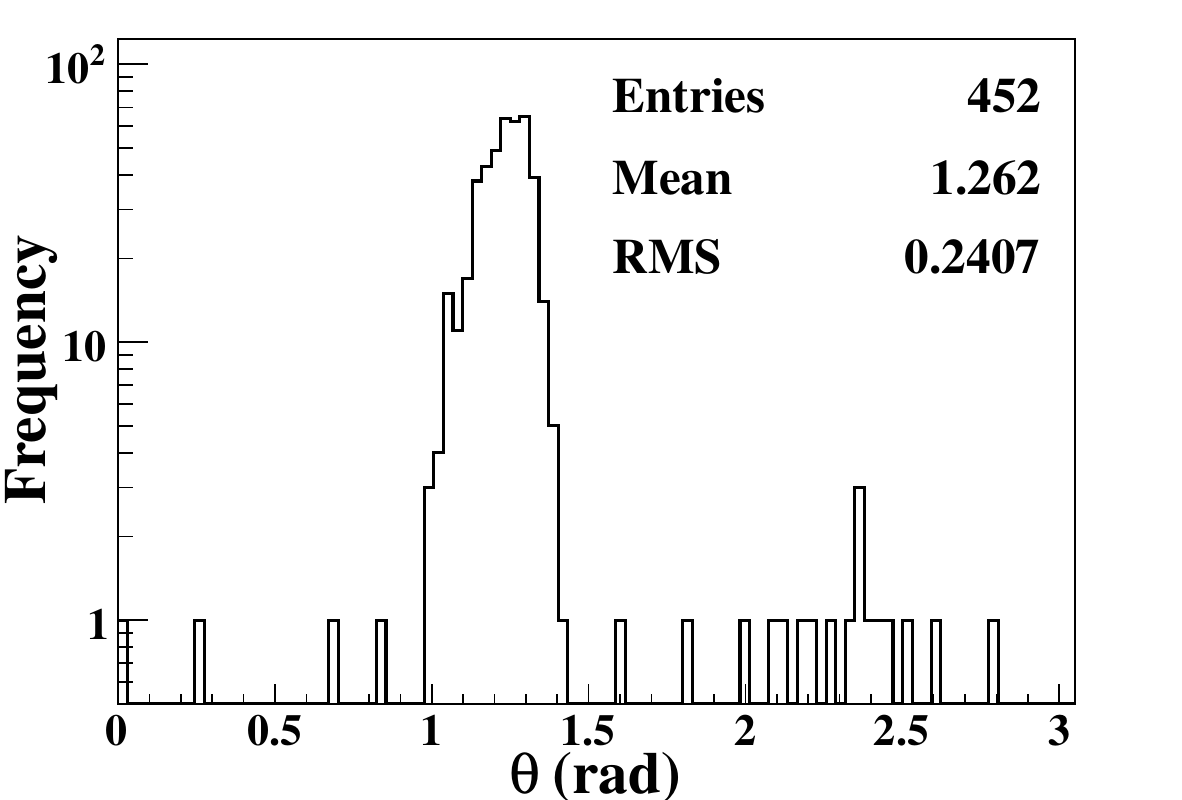}}
\subfloat[$\cos\theta = 0.85$; $\theta = 0.555$]{\label{fig:12b}
\includegraphics[width=0.48\textwidth]{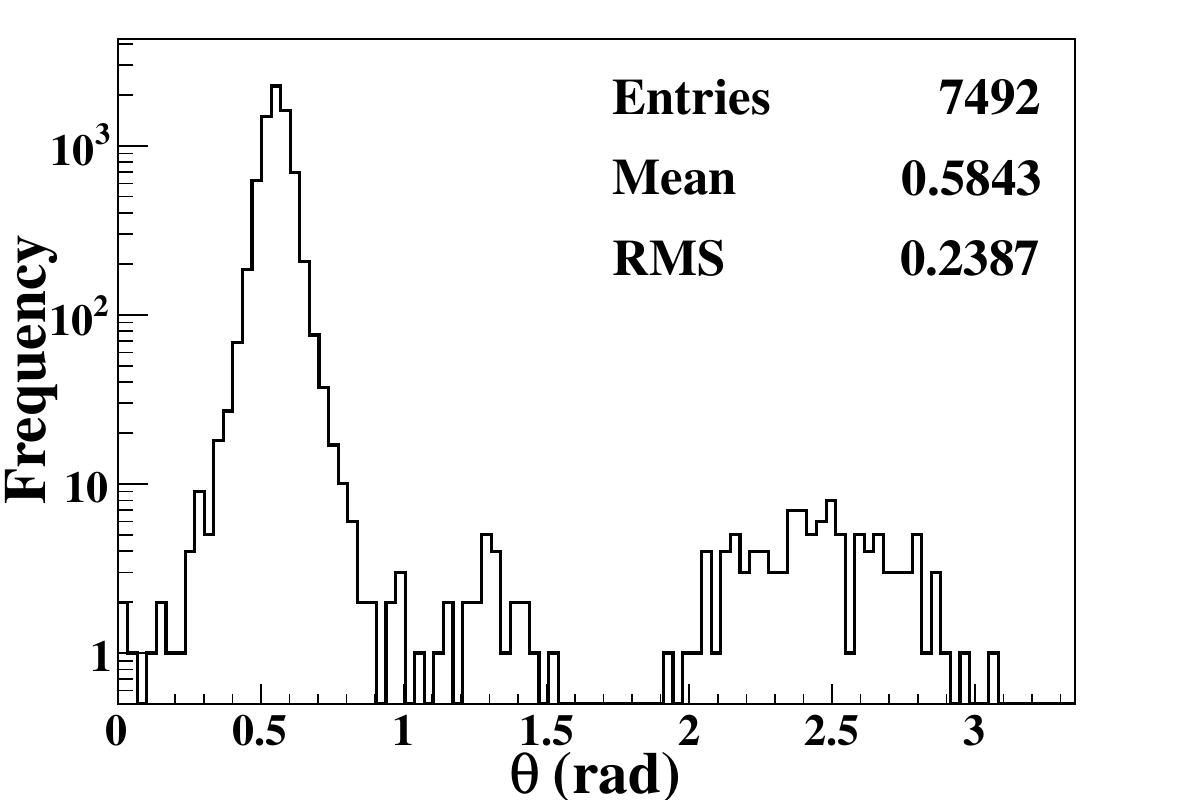}}
\caption{Reconstructed theta distribution for $P_{\rm in} = 1$~GeV/c at two
values of $\cos\theta$ \cite{Chatterjee:2014vta}. 
The time resolution of the RPCs has been taken to be 1 ns. Note that
the fraction of muons reconstructed in the wrong hemispehere decreases
sharply with energy.}
\label{fig:theta}
\end{figure}

The events that are successfully reconstructed for their momenta
are analysed for their zenith angle resolution. The reconstructed
zenith angle distribution for muons with $P_{\rm in} = 1$ GeV/c, at
zenith angles $\cos\theta = 0.25$ and $\cos\theta = 0.85$, are shown in
Fig.~\ref{fig:theta}, 
where the time resolution of the RPCs is taken to be $1$ ns.
(Muons with $\cos\theta=1$ are up-going). It can
be seen that a few events are reconstructed in the opposite (downward)
direction i.e., with zenith angle ($\theta_{rec} \sim \pi - \theta$). 
For muons with $P_{\rm in} = 1$ GeV/c at large (small) angles
with $\cos\theta = 0.25 (0.85)$, this fraction is about 4.3 (1.5) \%.
As energy increases, the fraction of events reconstructed in the wrong
direction drastically comes down. For example, this fraction reduces
to 0.3\% for muons with $P_{\rm in} = 2$ GeV/c at $\cos\theta = 0.25$.
Comparison of the goodness of fits to a track, assuming it to be 
in upward and downward direction, reduces this uncertainty further.
The analysis incorporating this technique is in progress.

The muons that contribute to mass hierarchy identification have energies 
greater than about 4 GeV  and the time interval between the first and the 
last hit of such muons is more than 5 ns, so that the up-going vs. 
down-going muons 
would be easily identified. The time resolution of the detector 
therefore is not expected to affect the mass hierarchy determination.

\begin{figure}[t]
\begin{center}
\includegraphics[width=0.49\textwidth,height=0.35\textwidth]
{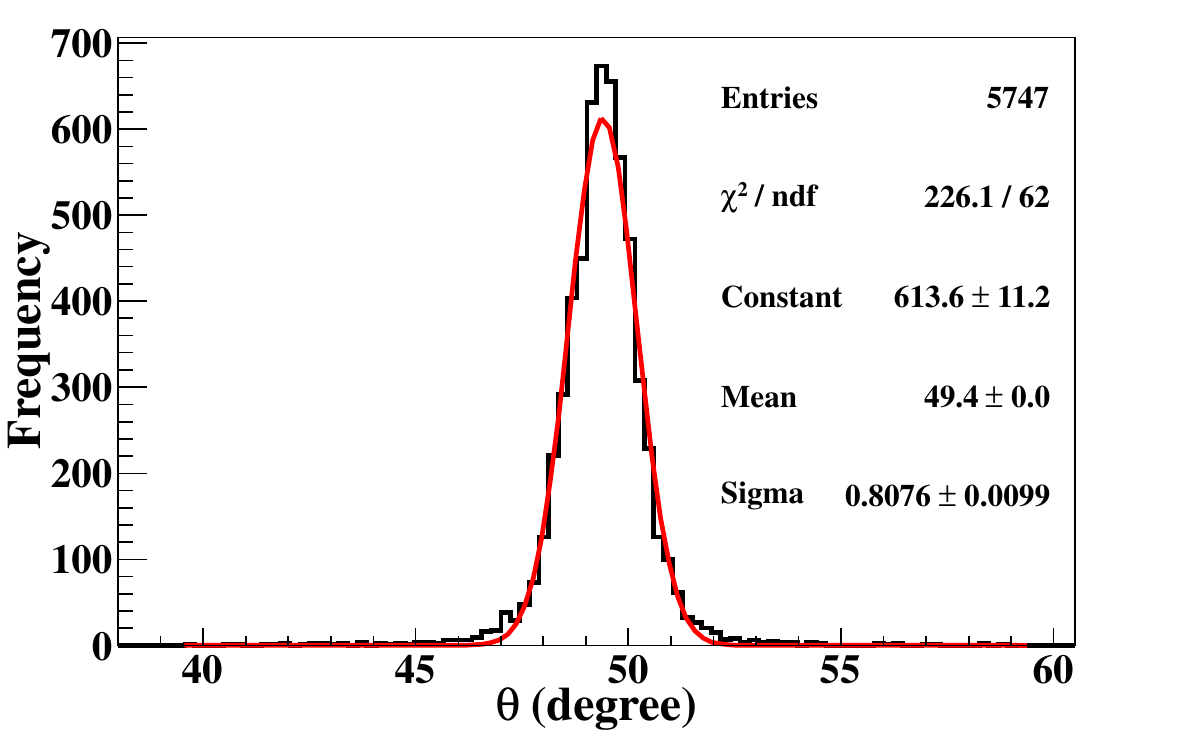}
\includegraphics[width=0.35\textwidth,height=0.49\textwidth,angle=90]
{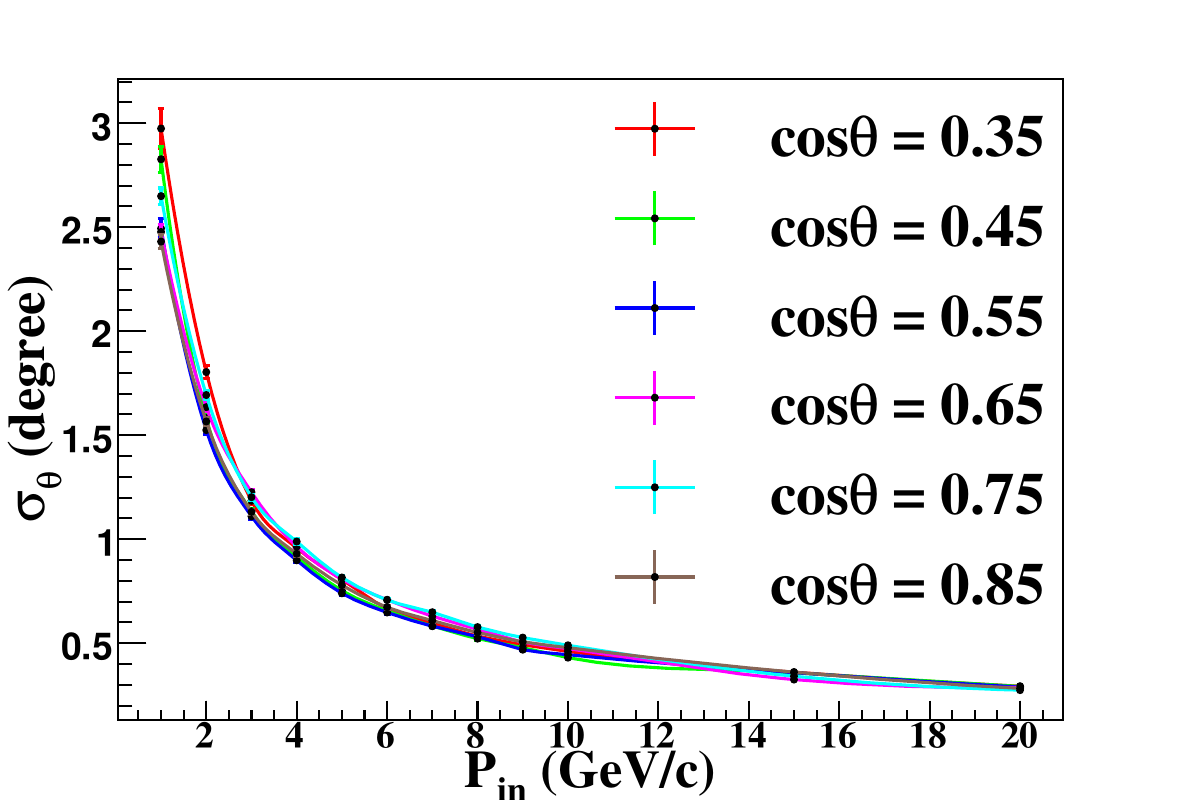}
\end{center}
\caption{The left panel shows the reconstructed $\theta$ distribution 
for input $(P_{\rm in}, \cos\theta) = (5\hbox{ GeV/c}, 0.65)$.
The right panel shows the $\theta$ resolution as a function of input 
momentum.}
\label{fig:muon-ang-resol}
\end{figure}

The events distribution as a function of the reconstructed zenith angle
is shown in the left panel of Fig.~\ref{fig:muon-ang-resol} for a sample
$(P_{\rm in}, \cos\theta) = (5\hbox{ GeV/c}, 0.65)$. It is seen that
the distribution is very narrow, indicating a good angular resolution
for muons.  The right panel of Fig.~\ref{fig:muon-ang-resol} shows the
$\theta$ resolution as a function of input momentum for different zenith
angles.  

As noted earlier, due to multiple scattering and the smaller number of
layers with hits, the momentum resolution is worse at lower energies.
This is also true for the zenith angle, whose resolution improves with
energy.  For a given energy, the resolution is worse for larger zenith
angles since again the number of layers with hits decreases. Even so, it
is seen that the angular resolution for $\cos\theta \ge 0.25$ (i.e. $\theta
\lsim 75^\circ$) is better than $1^\circ$ for muon momenta greater
than 4 GeV/c.

\subsection{Reconstruction efficiency}

\begin{figure}[htp]
\begin{center}
\includegraphics[width=0.35\textwidth,height=0.49\textwidth,angle=90]
{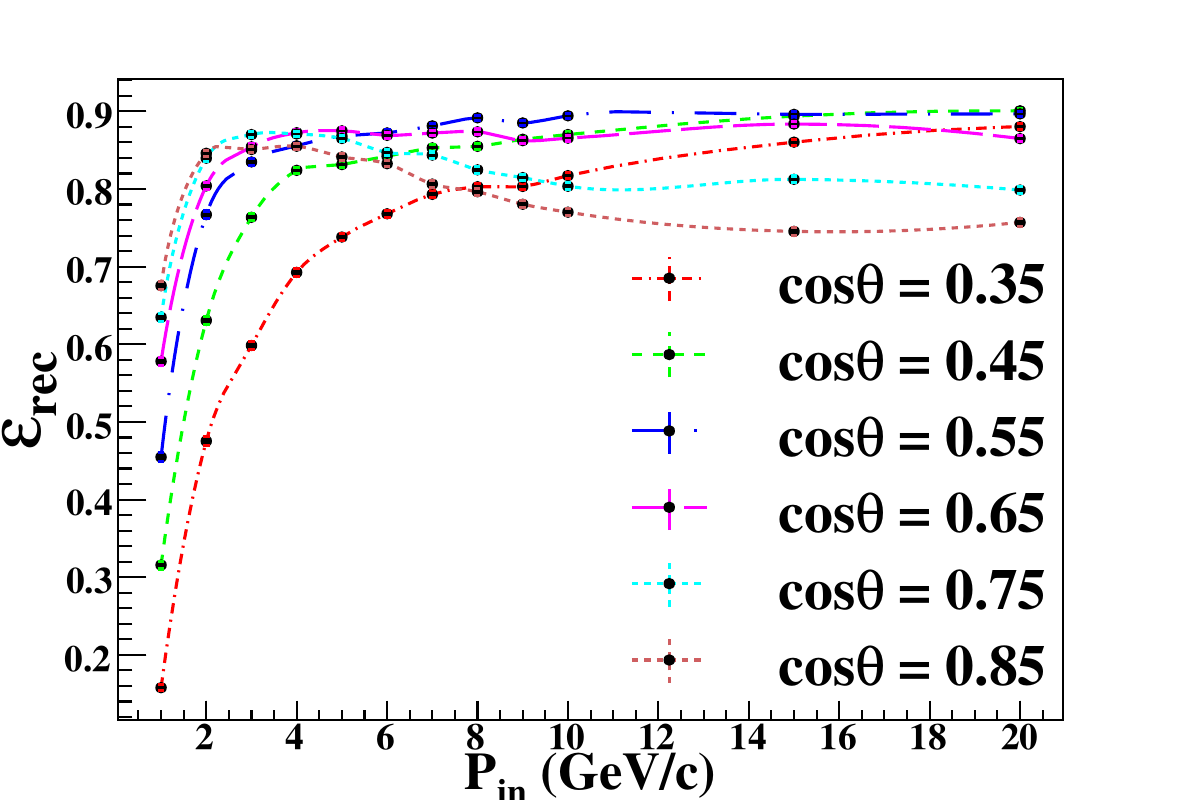}
\includegraphics[width=0.35\textwidth,height=0.49\textwidth,angle=90]
{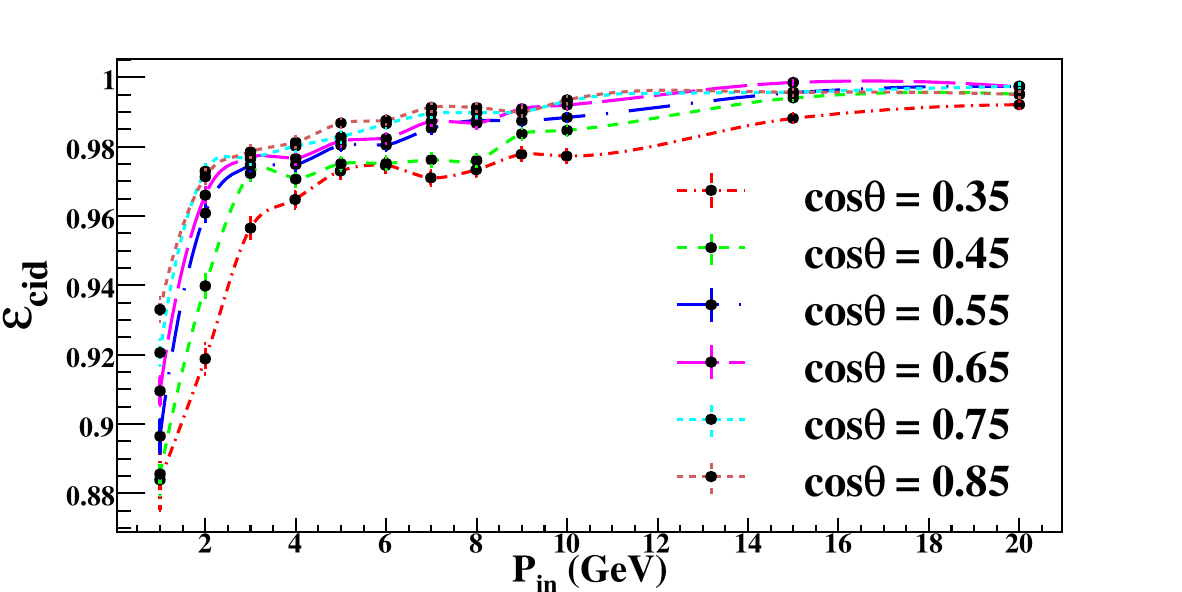}
\end{center}
\caption{The left (right) panel shows the reconstruction efficiency 
(the relative charge identification efficiency) as a function 
of the input momentum for different $\cos\theta$ values
\cite{Chatterjee:2014vta}. }
\label{fig:muon-eff}
\end{figure}

The reconstruction efficiency for muons is defined as the ratio of the number 
of reconstructed events $n_{\rm rec}$ (irrespective of charge) to the total
number of events, $N_{total}$ (typically 10000). We have
\begin{eqnarray}
\epsilon_{\rm rec} & = & \frac{N_{\rm rec}} {N_{\rm total}}~, \\ \nonumber
\delta \epsilon_{\rm rec} & = & \sqrt{\epsilon_{\rm rec}(1-\epsilon_{\rm
rec})/N_{\rm total}}~.
\end{eqnarray} 
The left panel of Fig.~\ref{fig:muon-eff} shows the muon reconstruction
efficiency as a function of input momentum for different $\cos\theta$ bins. 

When the input momentum increases, the reconstruction efficiency also 
increases for all angles,
since the number of hits increases as the particle crosses more number
of layers. At larger angles, the reconstruction efficiency for small
energies is smaller compared to vertical angles since the number of hits
for reconstructing tracks is less. But as the input energy increases, 
above almost 4 GeV
since the particle crosses more number of layers, the efficiency of
reconstructing momentum also increases and becomes comparable with
vertical angles. At higher energies the reconstruction efficiency becomes
almost constant. The drop in efficiency at high energies for vertical
muons is due to the track being partially contained, their
smaller bending in the magnetic field, as well as the
impact of the detector support structure. it is expected that this may
improve as the track recognition algorithms are refined and better tuned.

The fraction of muon-less charged current events / neutral current events 
that get misidentified as charged current muon events is $\sim 2-3\%$, 
as long as the energy of the reconstructed muon is $\gsim 1$ GeV. 
This fraction may be further reduced with a proper choice of cuts, and for
high energy muons that are relevant for mass hierarchy determination,
this is expected to be negligible. Work in this direction is in progress.

\subsection{Relative charge identification efficiency}

The charge of the particle is determined from the direction of curvature 
of the track in
the magnetic field. Relative charge identification efficiency is defined
as the ratio of number of events with correct charge identification,
$n_{\rm cid}$, to the total number of reconstructed events:
\begin{eqnarray}
\epsilon_{\rm cid} & = & \frac{N_{\rm cid}} {N_{\rm rec}}~,\\ \nonumber
\delta \epsilon_{\rm cid} & = & \sqrt{\epsilon_{\rm cid}(1-\epsilon_{\rm cid})/N_{\rm rec}}~.
\end{eqnarray}                            
Figure~\ref{fig:muon-eff} shows the relative charge identification efficiency as
a function of input momentum for different $\cos\theta$ bins. As seen
earlier, there is a very small contribution to the set with the wrongly
identified charge from the events where the track direction is wrongly
identified ($ \theta \to \pi - \theta$); such events will also reconstruct
with the wrong charge as there is a one-to-one correspondence between
the up-down identification and the muon charge.

When a low energy muon propagates in the detector it undergoes multiple 
scattering. So the number of layers with hits is small, and the 
reconstruction of charge goes wrong, which results in poor charge 
identification efficiency as can be see from Fig.~\ref{fig:muon-eff}. 
As the energy increases, the length of the track also increases, due to 
which the charge identification efficiency also improves. Beyond a few 
GeV/c, the charge identification efficiency becomes roughly constant, 
about 98\%.

\section{Response of ICAL to hadrons}
\label{had-resp}
 
An important feature of ICAL is its sensitivity to hadrons over a wide
energy range. This allows the reconstruction of the energy of the 
incoming muon neutrino in a charged-current event by combining the 
energies of the muon and the hadrons. It also enables the detection
of neutral-current events, charged-current DIS events generated
by $\nu_e$ interactions, and charged-current $\nu_\tau$ events where
the $\tau$ decays hadronically. The information contained in all 
these events adds crucially to our knowledge of neutrino oscillations.
The charged-current events is a direct measure of the oscillation
probabilities among the three active neutrino species. 
On the other hand, the neutral-current events are not affected by 
active neutrino oscillations, and hence help in flux normalization,
as well as in the search for oscillations to sterile neutrinos.
It is therefore important to characterize the response of the ICAL 
to hadrons.

The hadrons generated from the interactions of atmospheric neutrinos 
consist mainly of neutral and charged pions, which together account
for about 85\% of the events. The rest of the events consist of
kaons and nucleons, including the recoil nucleons that cannot be 
distinguished from the remaining hadronic final state. The neutral pions 
decay immediately giving rise to two photons, while the charged pions 
propagate and develop into a cascade due to strong interactions.  
For the neutrino-nucleon interaction $\nu_{\mu} N \rightarrow \mu X$, 
the incident neutrino energy is given by 
\begin{equation} 
\label{simple_equation} 
E_{\nu} = E_{\mu} + E_{hadrons} - E_N \; , 
\end{equation} 
where $E_N$ is the energy of the initial nucleon which is taken 
to be at rest,  neglecting its small Fermi momentum. 
The visible hadron energy depends on factors like the shower 
energy fluctuation, leakage of energy, and invisible energy loss mechanisms, 
which in turn affect the energy resolution of hadrons . 
We choose to quantify the hadron response of the detector in terms of
the quantity \cite{Devi:2013wxa}
\begin{equation} 
\label{simple_equation1} 
E_{\rm had}^{\prime} = E_{\nu} -  E_{\mu}\; . 
\end{equation}  

As the first step in understanding the ICAL response to hadrons, 
single charged hadrons of fixed energies are generated via Monte Carlo and 
propagated through the detector to compare its response to them.
The response to charged pions is then studied in more detail, and 
the pion energy is calibrated against the number of hits. 
Next, the multiple hadrons produced through atmospheric neutrino 
interactions are generated using NUANCE \cite{Casper:2002sd}, 
and the quantity $E'_{\rm had}$  
is used to calibrate the detector response. 
This should take care of the right combination of the contributions
of different hadrons to the hits, on an average. It would of course
be dominated by neutral and charged pions, and hence we expect it to
be similar to the response to fixed-energy pions.

\subsection{Energy response to fixed-energy hadrons} 
\label{sec:E_res_fix} 

In an RPC, the X- and Y-strip information on a hit is {\em independently}
obtained from the top and bottom pick-up strips, as described in
Sec.~\ref{sec:detector} and then combined to give the $(X,Y)$ coordinates of
the hit. For a muon, the $(X,Y)$ positions of the hits in a given layer
can easily be identified since a muon usually leaves only one or two hits
per layer. However a hadron shower consists of multiple hits per layer,
and combining all possible X and Y strip hits leads to overcounting,
resulting in what are termed as ``ghost hits''.  To avoid the ghost hit
counts, the variables ``x-hits'' and ``y-hits''---the number of hits in
the X and Y strips of the RPC, respectively---can be used. We choose to
perform the energy calibration with the variable ``orig-hits'', which
is the maximum of x-hits or y-hits.

\begin{figure}[h]  
\includegraphics[width=0.49\textwidth,height=0.35\textwidth]
{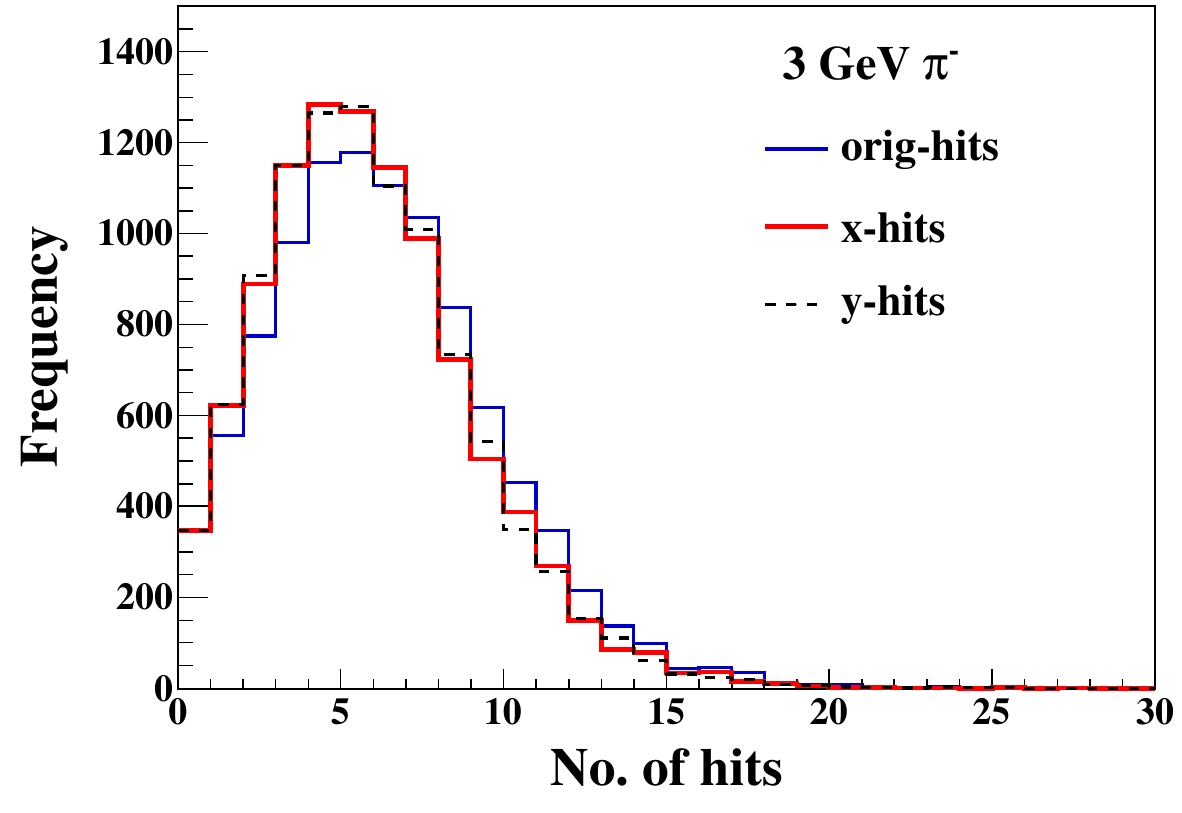}  
\includegraphics[width=0.49\textwidth,height=0.35\textwidth]
{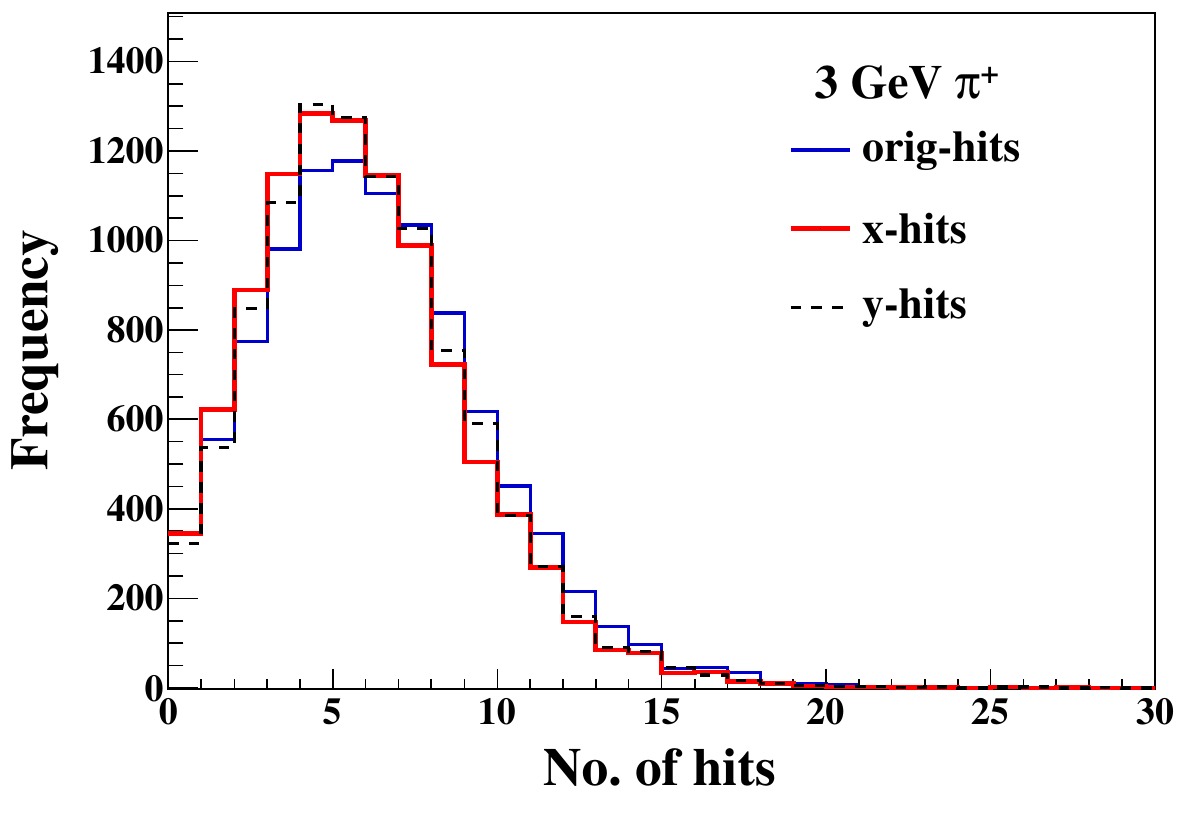} 
\caption{The comparison of the distributions of x-hits, y-hits and  
orig-hits for $\pi^{-}$ (left) and $\pi^{+}$ (right) of energy 3 GeV
\cite{Devi:2013wxa}.}  
\label{xyorcomp}  
\end{figure}

Figure~\ref{xyorcomp} shows the comparison of these three types of
hit variables for $\pi^{\pm}$ of energy 3 GeV. Clearly, there is no
significant difference among these variables, however orig-hits has been
used as the unbiased parameter. It is also observed that the detector
response to the positively and negatively charged pions is identical,
so we shall not differentiate between them henceforth.

Fixed-energy single pion events in the energy range of 1 to 15~GeV 
were generated using the particle gun in GEANT4. The total number of events 
generated for each input energy value is 10000 in this section, unless
specified otherwise. Each event is randomly generated to have vertices over a volume 
2 m $ \times $ 2 m $\times$ 2 m in the central region of the ICAL detector. 
As in the earlier section, the reference frame chosen has the origin 
at the centre of the detector, the $z$-axis pointing vertically up, and 
the $x$-$y$ plane along the horizontal plates, with the three modules
lined up along the $x$ axis. 
The hadron direction is uniformly smeared over the zenith angle 
$0 \le \theta \le \pi$ and azimuth of $ 0 \le \phi \le 2\pi$. 
This serves to smear out any angle-dependent bias in the energy resolution 
of the detector by virtue of its geometry which makes it the least (most) 
sensitive to particles propagating in the horizontal (vertical) direction.

\begin{figure}[h!]  
\includegraphics[width=0.49\textwidth,height=0.35\textwidth]
{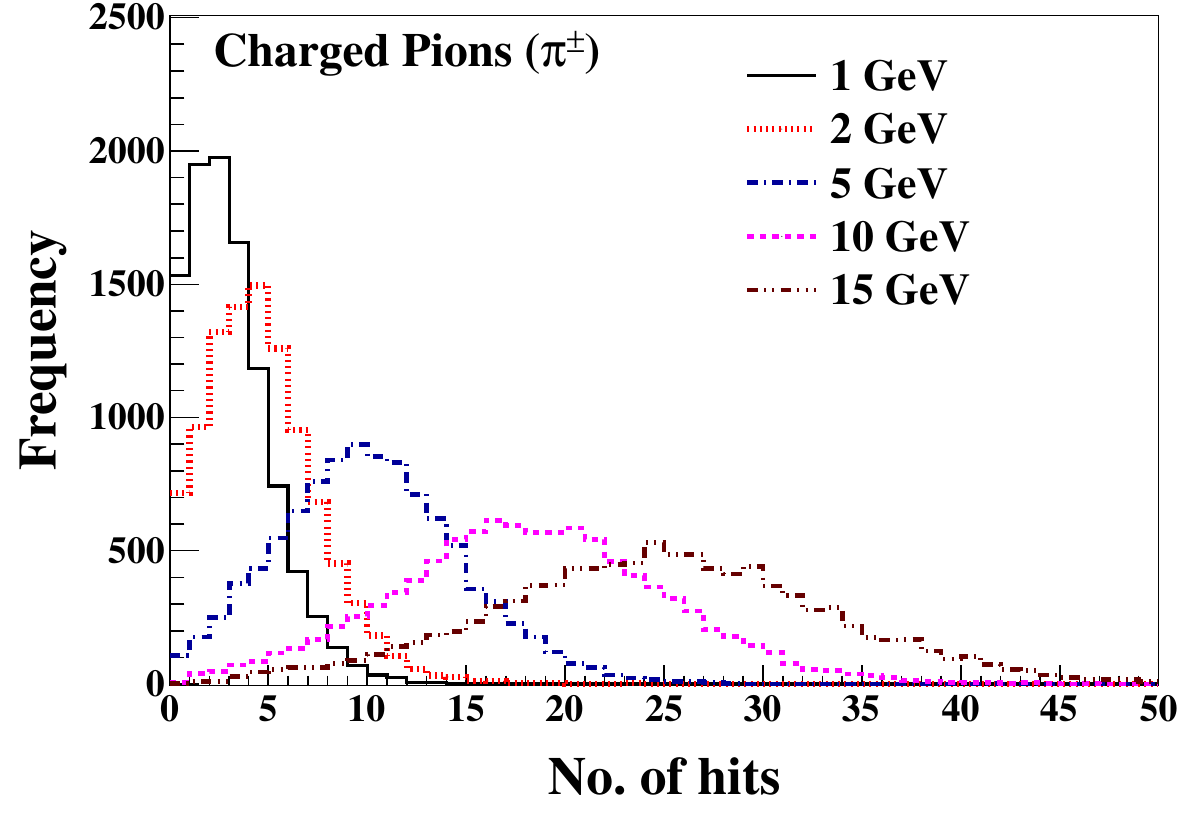}  
\includegraphics[width=0.49\textwidth,height=0.35\textwidth]
{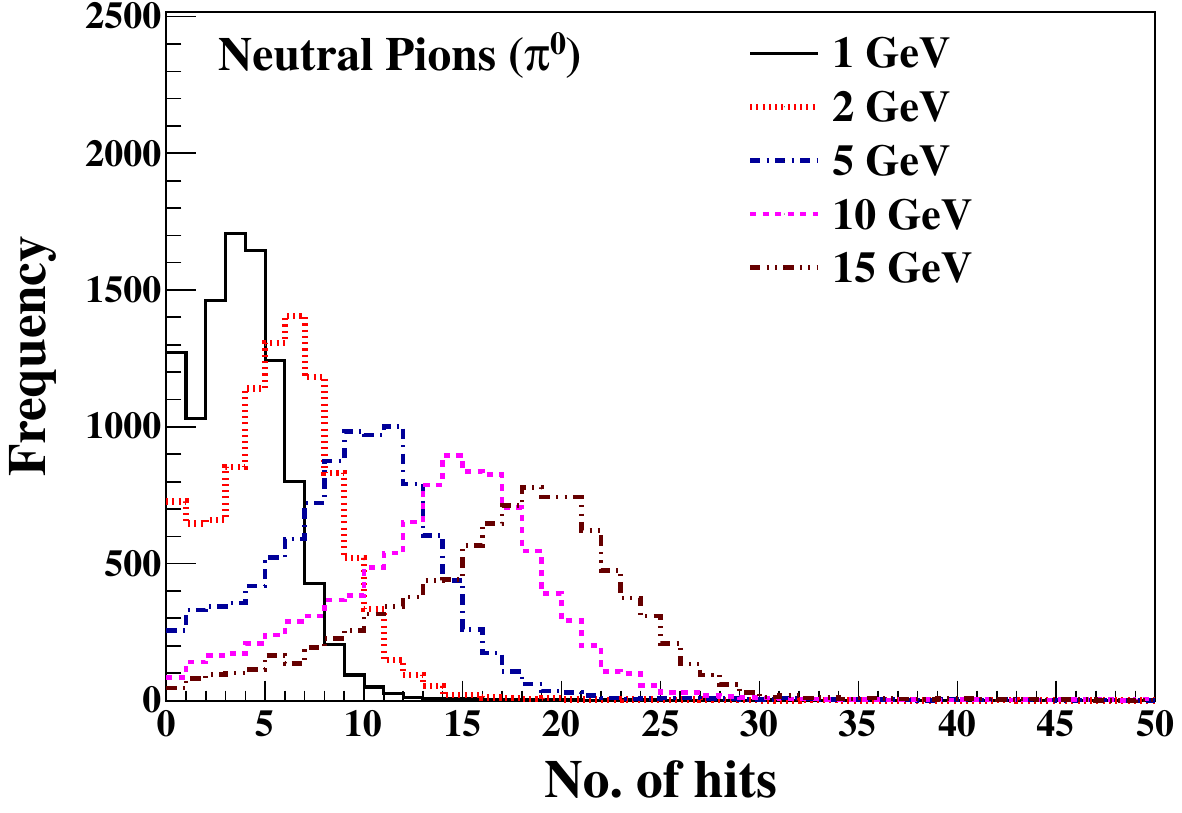}  \\
\includegraphics[width=0.49\textwidth,height=0.35\textwidth]
{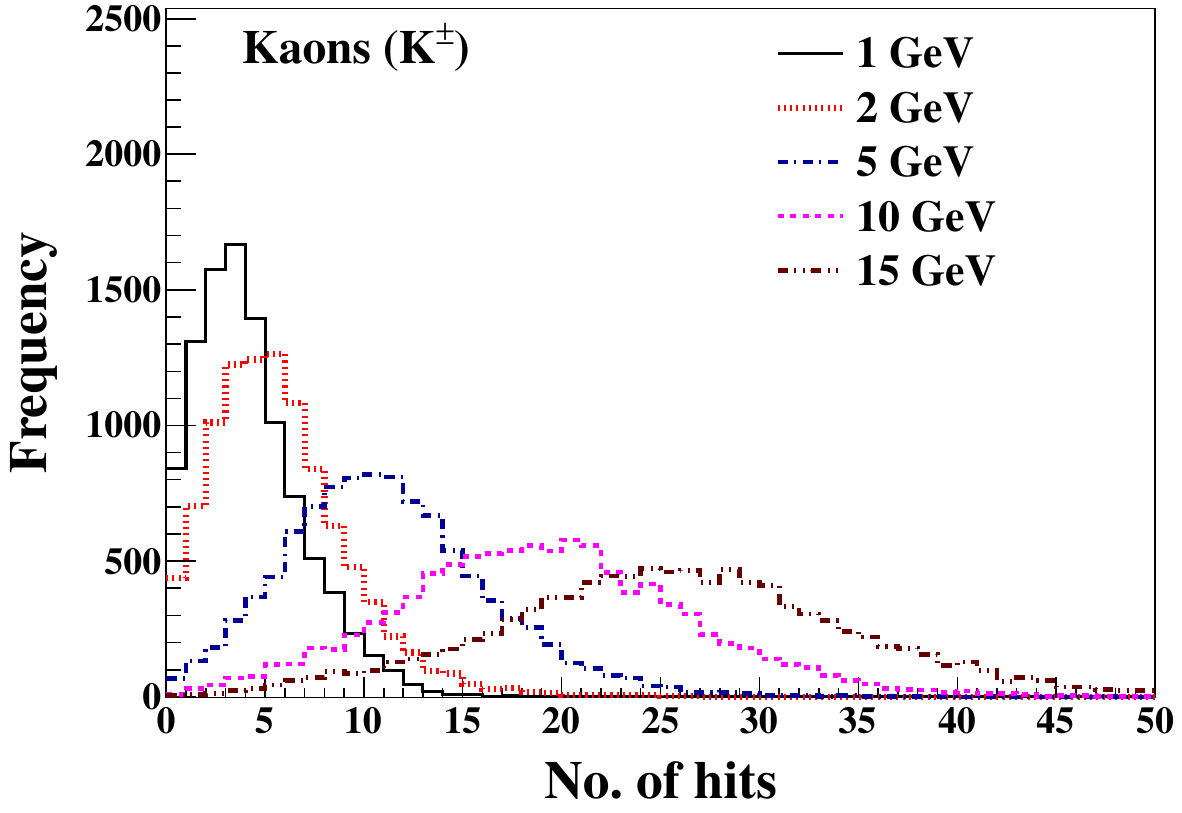}  
\includegraphics[width=0.49\textwidth,height=0.35\textwidth]
{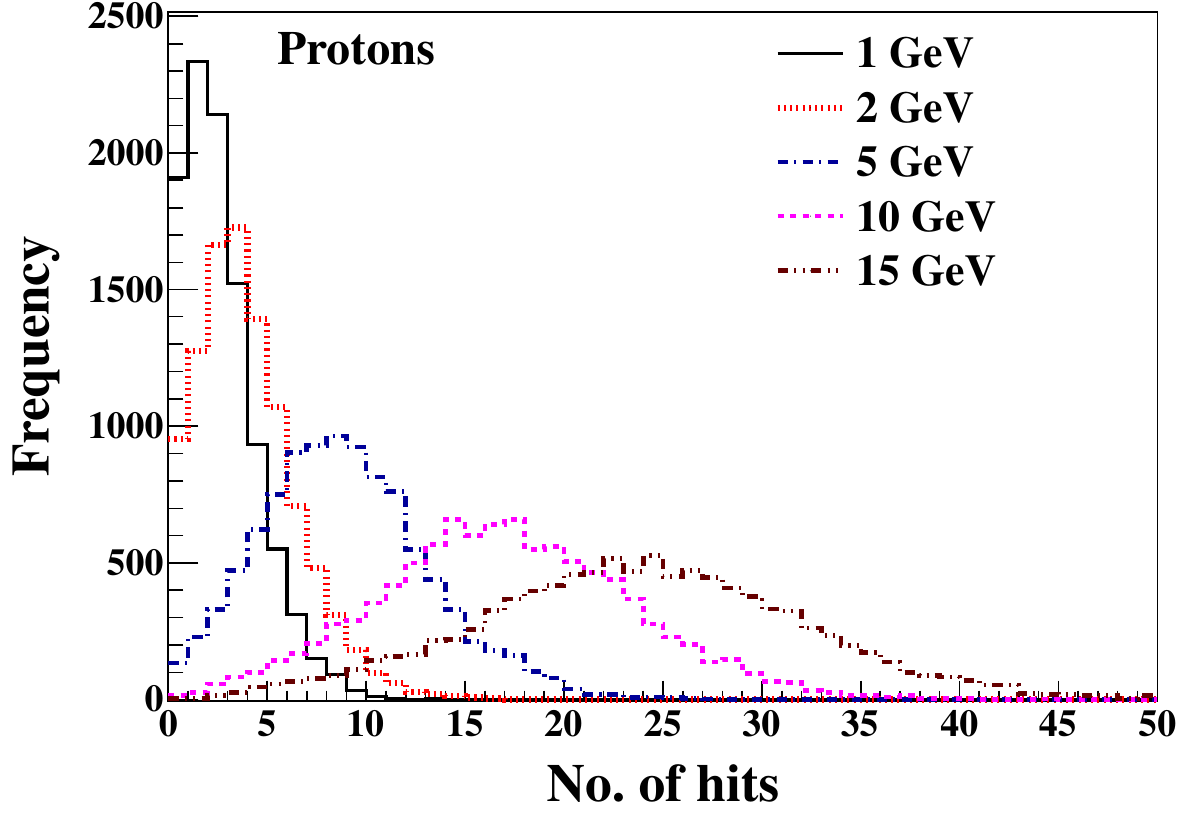}  
\caption{The direction-averaged hit distributions at various energies 
for $\pi^{\pm}$, $\pi^{0}$, $K^{\pm}$ and protons propagated from 
vertices smeared  over the chosen detector volume \cite{Devi:2013wxa}.} 
\label{histocomp}  
\end{figure}  

Figure~\ref{histocomp} shows the hit distributions in the detector for 
pions, kaons, and protons at various energies in the range of 1 to 15 GeV. 
It is observed that the hit patterns are similar for all these hadrons, 
though the peak positions and spreads are somewhat dependent 
on the particle ID. Hence the detector cannot distinguish the specific 
hadron that has generated the shower. The large variation in the number 
of hits for the same incident particle energy is mainly a result of
different strong interaction processes for different hadrons (for
$\pi^0$ the interactions are electromagnetic since it decays immediately
to a $\gamma \gamma$ pair), and partly an effect of angular smearing.

\subsubsection{The $e/h$ ratio}
\label{sec:ebyh}

The NUANCE \cite{Casper:2002sd} simulation suggests that the fraction 
of the different types 
of hadrons produced in the detector is $\pi^+: \pi^-: \pi^0::0.38:0.25:0.34$, 
with the remaining 3\% contribution coming mainly from kaons
\cite{Mohan:2014qua}. While the response
of the detector to $\pi^+$ and $\pi^-$ is almost identical as seen earlier, 
its response to the electromagnetic part of the hadron shower that originates
from $\pi^0$ is different. This may be quantified in terms of the $e/h$
ratio, i.e. the ratio of the electron response to the charged-pion
response. This ratio would help us characterise the effect of neutral hadrons
on the energy resolution.

\begin{figure}[h]
\centering
\includegraphics[width=0.49\textwidth,height=0.35\textwidth]
{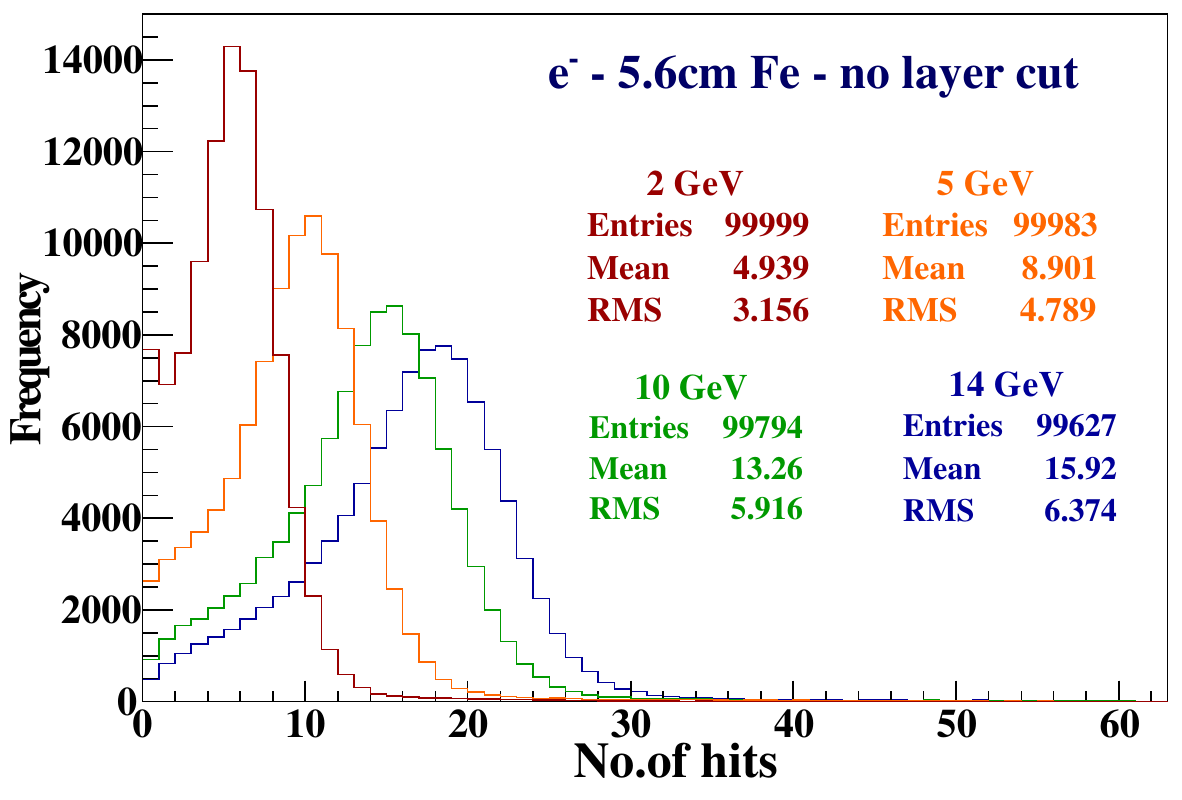}
\includegraphics[width=0.49\textwidth,height=0.35\textwidth]
{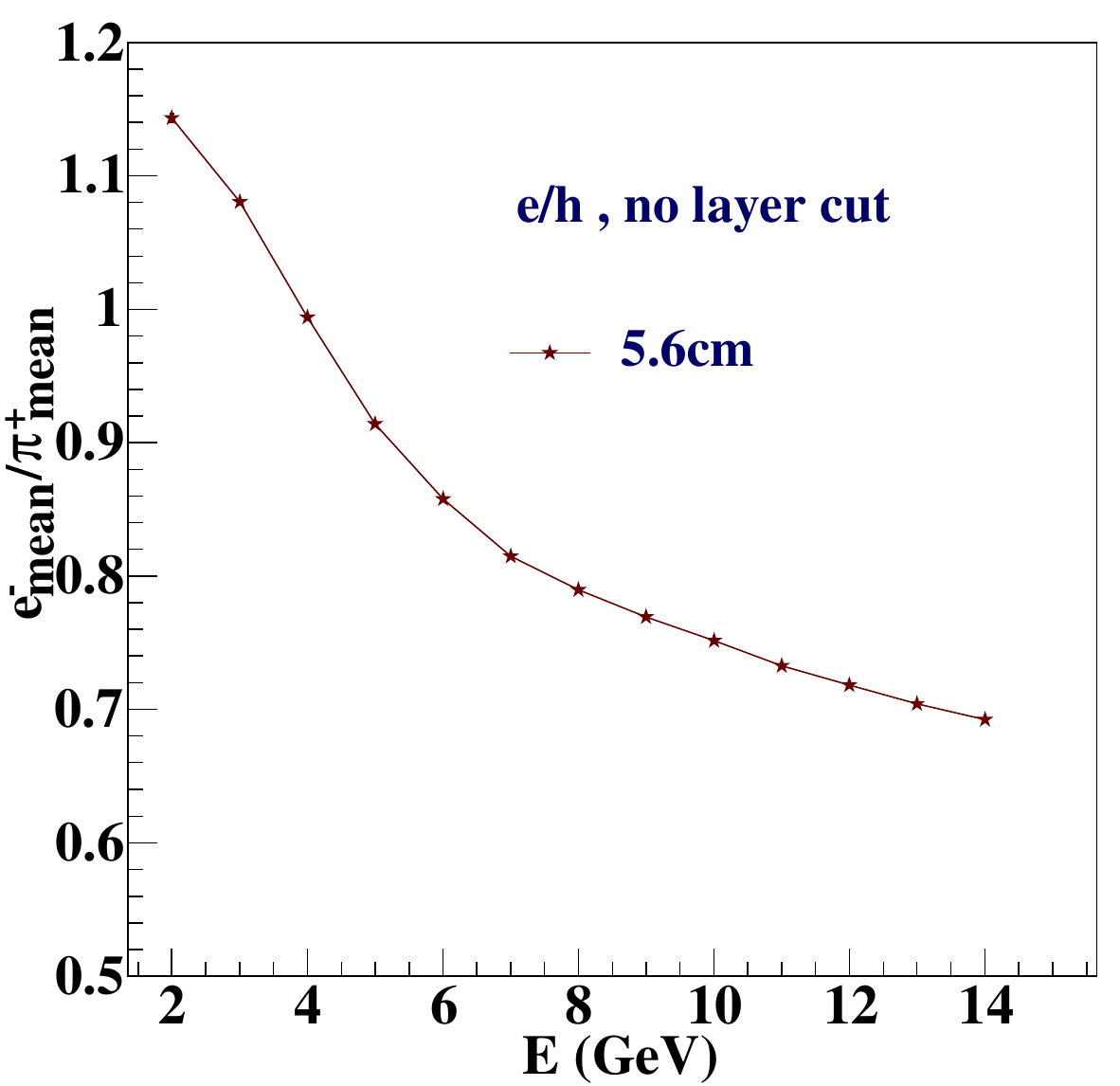}
\caption{The left panel shows the hit distributions of fixed energy single 
electrons at 2, 5, 10 and 14 GeV, averaged over all directions. 
The right panels shows the variation of the $e/h$ ratio with the particle 
energies \cite{Mohan:2014qua}. 
}
\label{e-hit-histos}
\end{figure}

In order to study this ratio, we generated 100,000 electron events 
at fixed energies in the energy range 2--15 GeV, propagating in arbitrary 
directions (with $\theta$ smeared from $0-\pi$ and $\phi$ from $0-2\pi$) 
from vertices within a volume of 2 m $\times$ 2 m $\times$ 2 m in the central region 
of the ICAL detector. The hit distributions averaged
over all directions for $2, 5, 10$ and 14 GeV electrons is shown in
Fig.~\ref{e-hit-histos}. This may be compared with the hit distributions
shown in Fig.~\ref{histocomp}. The response is almost the same as that
for $\pi^0$, with narrower high-energy tails than those for charged pions.

The $e/h$ ratio is obtained as
\begin{equation}
e/h = e^-_{mean}/{\pi^{+}_{mean}},
\end{equation}
where $e^-_{mean}$ is the arithmetic mean of the electron hit distribution
and ${\pi^{+}_{mean}}$ is the arithmetic mean of the hit distribution
for $\pi^{+}$, for a given fixed energy of the two particles. 
If $e/h$ = 1, then the detector is said to be compensating.
The variation of the $e/h$ ratio with incident energy is shown in 
the right panel of Fig.~\ref{e-hit-histos}.

It can be seen that the value of $e/h$ decreases with energy.  However,
it should be noted that there is no direct measurement of the energy
deposited in ICAL.  Here the energy of a shower is simply {\it calibrated}
to the number of hits, and electrons which travel smaller distances in a
high Z material like iron have lower number of hits compared to charged
pions. At lower energies the electron as well as pion shower hits are 
concentrated around a small region. 
The mean of the electron hit distribution is
roughly the same or slightly larger than that of the $\pi^{+}$
hit distribution. With the increase in energy, the charged pions
travel more distance and hence give more hits (as they traverse more
layers) since the hadronic interaction length is much more than the
electromagnetic interaction length at higher energies and hence the ratio
of hits in the two cases drops with energy. 

In a neutrino interaction where all types of hadrons can be produced
(although the dominant hadrons in the jet are pions), the response of
ICAL to hadrons produced in the interaction depends on the relative
fractions of charged and neutral pions. Using the relative fractions 
$\pi^+: \pi^-: \pi^0::0.38:0.25:0.34$ as mentioned above, the average 
response of hadrons obtained from the charged current muon
neutrino interaction can be expressed as:
\begin{eqnarray}
 R_{had} & = & \left[(1-F_0) \times h + F_0 \times e \right], 
 \nonumber \\
 & = & h \left[(1-F_0) + F_0 \times \frac{e}{h} \right],
 \label{eqn-e-by-pi}
\end{eqnarray}
where $e$ is the electron response, $h$ the charged hadron response and
$F_0$ is the neutral pion fraction in the sample.

The atmospheric neutrino events of interest in ICAL are dominated by
low energy events with hadrons typically having energies $E < 10$
GeV for which the average value of $e/h$ is $e/h\approx0.9$. Using
$F_0~=~0.34$ in Eq.~(\ref{eqn-e-by-pi}), we get the average hadron response
for NUANCE-generated events to be $R_{had} \approx 0.97 h$ which is not very
different from $h$. For this reason, the analysis of response with
multiple hadrons in NUANCE-generated events sample is not
expected to be very different from that of the single pions sample.
However we shall confirm this by first focusing on the detector response 
to fixed-energy charged pions in Sec.~\ref{sec:charge_pions}, 
and then moving on to a more general admixture 
of different hadrons in Sec.~\ref{sec:E_res_nuance}.

\subsection{Analysis of the charged pion hit pattern} 
\label{sec:charge_pions} 

\begin{figure}[h!] 
\includegraphics[width=0.49\textwidth,height=0.35\textwidth]
{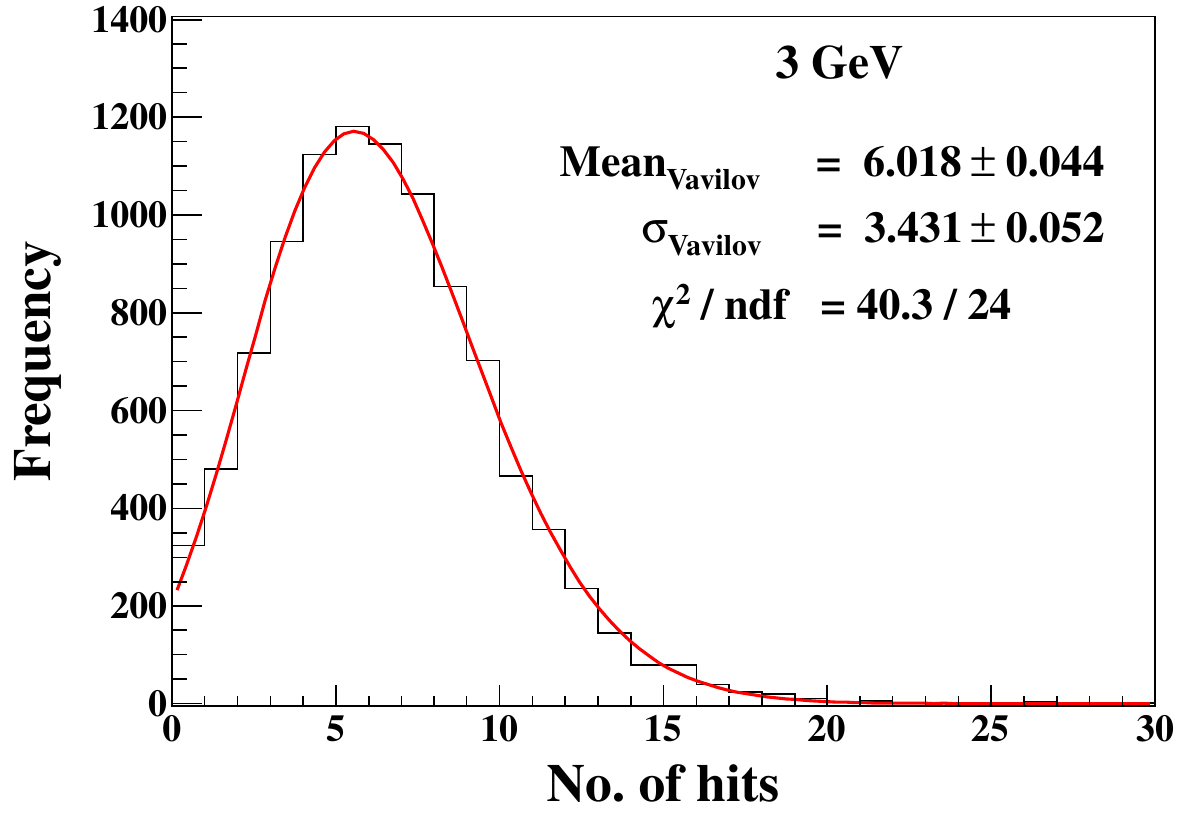}  
\includegraphics[width=0.49\textwidth,height=0.35\textwidth]
{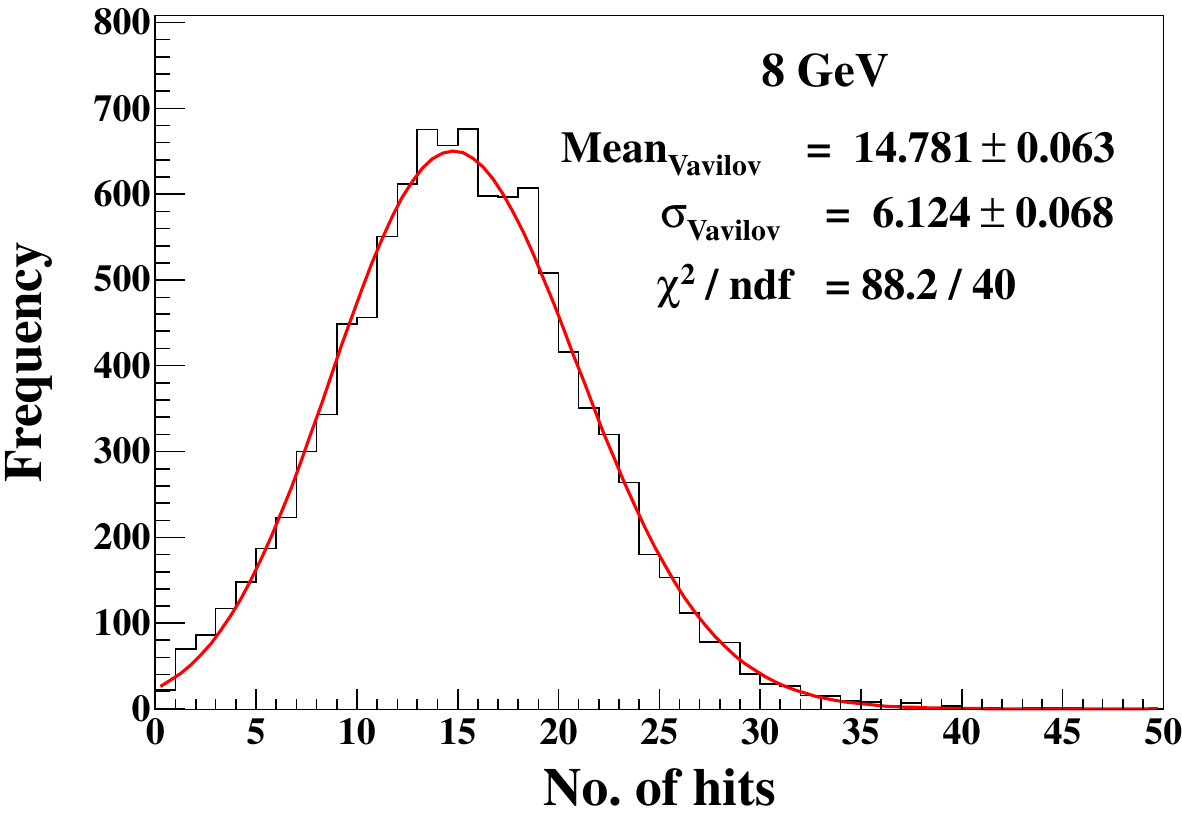} 
\caption{The hit distributions at 3 GeV (left) and 8 
GeV (right), for pions propagating in the detector, starting from randomized 
vertices over a volume of 2 m$~\times~$2 m$~\times~$2 m in the detector. 
The red curve denotes a fit to the Vavilov distribution \cite{Devi:2013wxa}.} 
\label{pionhisto} 
\end{figure}

The hit distributions for charged pions, at sample values of $E=3,8$ GeV, 
are shown in Fig.~\ref{pionhisto}. The distributions are asymmetric 
with long tails, with a mean of about two hits per GeV. 
In addition, at low energies several events yield zero hits in the detector.

A search for a good fitting function for the distribution was made, 
and it was found that the Vavilov distribution function gives a good fit
for all energies, as is illustrated in Fig.~\ref{pionhisto}. 
This distribution (see Appendix \ref{sec:appA}) is described by 
the four parameters $\rm{P}_0$, $\rm{P}_1$, $\rm{P}_2$ and $\rm{P}_3$,
which are energy-dependent \cite{Devi:2013wxa}.
The Vavilov distribution reduces to a Gaussian 
distribution for $\rm{P}_0~\geq~10$, which happens for $E > 6$ GeV.
However at lower energies, it is necessary to use the full Vavilov 
distribution.

\begin{figure}[] 
\includegraphics[width=0.49\textwidth,height=0.35\textwidth]
{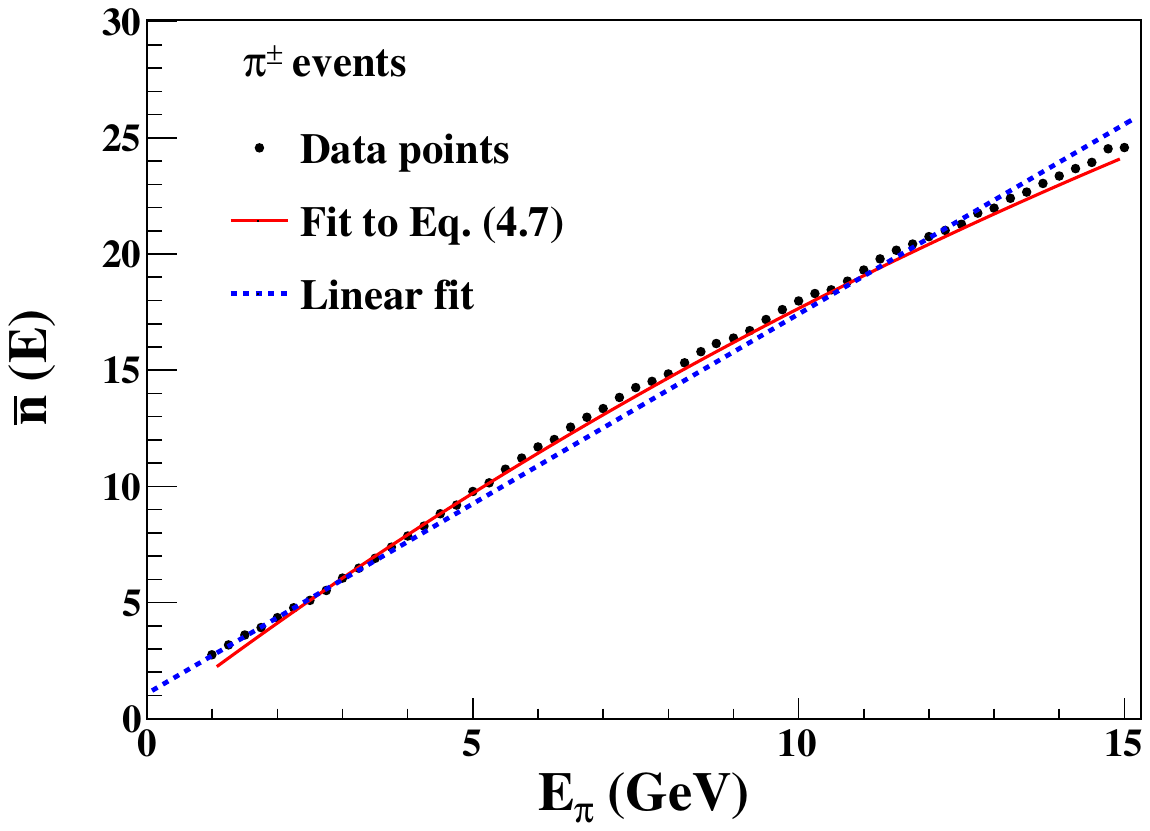}  
\includegraphics[width=0.49\textwidth,height=0.35\textwidth]
{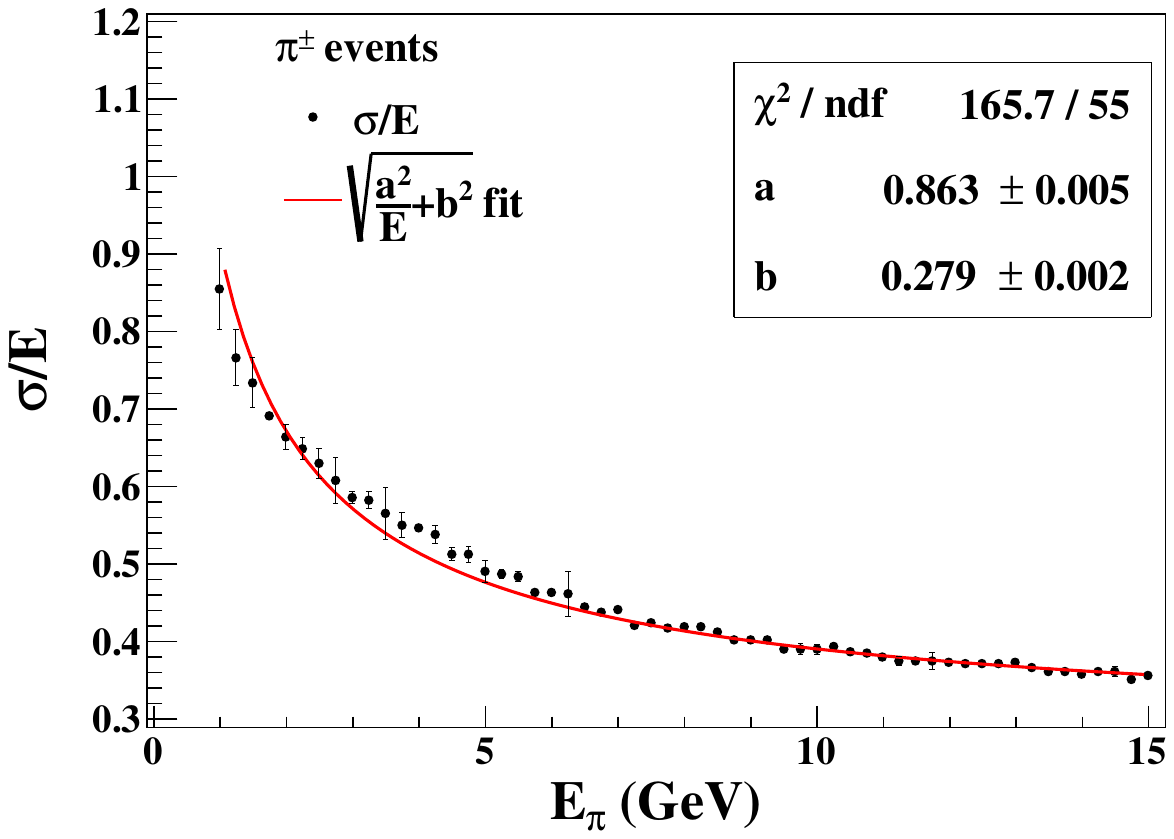} 
\caption{The mean hit distribution (left) and the energy resolution (right) 
for fixed-energy charged pion events, as a function of pion energy. 
The right panel also shows a fit to Eq.~(\ref{eq5})
\cite{Devi:2013wxa}.} 
\label{meancomp} 
\end{figure} 

The mean $\bar{n}(E)$ of the number of hits from the Vavilov fit 
at different energies is shown in the left panel of Fig.~\ref{meancomp}. 
It increases with increasing pion energy, and saturates at higher energies. 
It may be approximated by 
\begin{equation} 
\bar{n}(E)=n_0[1-\exp(-E/E_0)] \; , 
\label{eq2} 
\end{equation} 
where $n_0$ and $E_0$ are constants. This fit has to be interpreted 
with some care, since $n_0$ and $E_0$ are sensitive to the energy 
ranges of the fit. The value of $E_0$ is found to be $\sim$ 30 GeV
when a fit to the energy range 1--15 GeV is performed.
Since the energies of interest for atmospheric neutrinos 
are much less than $E_0$, Eq.~(\ref{eq2}) may be used in its approximate 
linear form $\bar{n}(E)=n_0 E/E_0$. 
A fit to this linear form is also shown in Fig.~\ref{meancomp}. 
 
Since in the linear regime ($E \ll E_0$) one has
\begin{equation} 
\frac{\bar{n}(E)}{n_0}=\frac{E}{E_0} \; , 
\label{eq3} 
\end{equation} 
The energy resolution may be written as 
\begin{equation} 
\frac{\sigma}{E} 
 = \frac{\Delta {n}(E)}{\bar{n}(E)} \; , 
\label{eq4} 
\end{equation} 
where $(\Delta n)^2$ is the variance of the distribution.
The notation $\sigma/E$ will be used 
for energy resolution throughout, and Eq.~(\ref{eq4}) 
will be taken to be valid for the rest of the analysis. 
 
The energy resolution of pions may be parameterized by  
\begin{equation} 
\frac{\sigma}{E} = \sqrt{\frac{a^2}{E} + b^2}~, 
\label{eq5} 
\end{equation} 
where $a$ and $b$ are constants.  
The energy resolutions for charged pions as functions of the pion energy
are shown in the right panel of
Fig.~\ref{meancomp}. 
The parameters $a$ and $b$ extracted by a fit to Eq.~(\ref{eq5}) 
over the pion energy range 1--15 GeV are also shown.
The values of $a$ and $b$ depend on the iron plate thickness;
this dependence has been studied in detail in Appendix~\ref{app:thickness}.

\subsubsection{Dependence of the energy resolution on hadron direction}

\begin{figure}[h!] 
\centering
\includegraphics[width=0.6\textwidth,height=0.4\textwidth]
{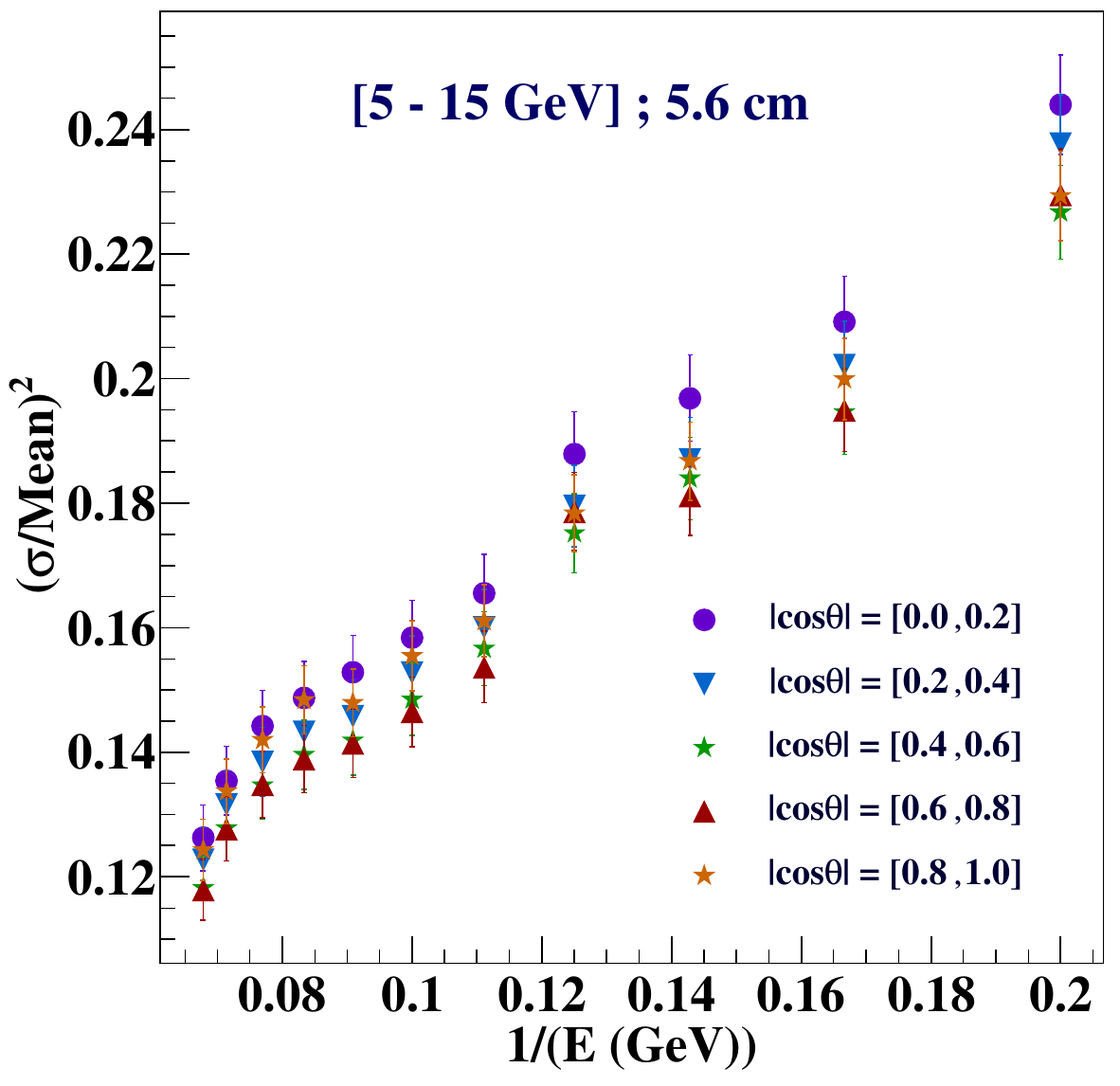} 
\caption{The dependence of the pion energy resolution on the
zenith angle \cite{Mohan:2014qua}.} 
\label{had-res-angle} 
\end{figure}

Since the number of layers traveresed by a particle would depend on the 
direction of the particle, it is expected that the energy calibration
and energy resolution for hadrons will depend on the direction of 
the hadron. To check this dependence, we simulate pions of fixed energies 
in the detector, which travel in different directions. The directions are
binned into 5 zenith angle bins, and the distributions of number of hits
is recorded. The ratio of the rms width of the distribution and its mean
is used as a measure of the energy resolution \cite{Mohan:2014qua}. 
Figure~\ref{had-res-angle}
shows the zenith angle dependence of the hadron energy resolution.

Since there is only a mild dependence on the hadron direction, and the
direction of hadron itself cannot be determined yet with a good
confidence, we continue to use the direction-averaged results in 
the future analyses in this Report.

\subsection{Response to hadrons produced by atmospheric neutrinos} 
\label{sec:E_res_nuance} 

Atmospheric neutrino interactions in the detector may contain no
hadrons (for quasi-elastic scattering events), one hadron or
multiple ones (in resonance scattering and DIS events). While the former
events dominate for $E_\nu \sim 1$ GeV, at higher energies the DIS
events dominate. In this section we focus on the charged-current
$\nu_\mu$ interactions in the detector that produce hadrons in addition
to the charged muons.

We assume here that the $\nu_\mu$ CC events can be clearly separated 
from the NC as well as $\nu_e$ CC events, and that the muon and hadron
shower may be identified separately. 
(In our procedure, we determine the number of hadron hits by 
taking away the true muon hits, as in the Monte-Carlo simulation,
from the total hits in the event.)
Preliminary studies show that 
this is a reasonable assumption for $E_\nu \gtrsim 1$ GeV. At lower 
energies where the number of hits is small, the misidentification of a
muon hit as a hadron one, or vice versa, can significantly affect the
hadron energy calibration. The analysis of this effect is in
progress.

The atmospheric neutrino ($\nu_\mu$) and antineutrino 
(${\overline{\nu}}_{\mu}$) events in ICAL have been simulated 
using the neutrino event 
generator NUANCE (v3.5) \cite{Casper:2002sd}. 
The hadrons produced in these interactions are primarily pions, but 
there are some events with kaons (about 3\%) and a
small fraction of other hadrons as well. As discussed earlier, it is 
not possible to identify the hadrons individually in ICAL. However since the 
hit distribution of various hadrons are similar to each other 
(see Fig. \ref{histocomp}), and the NUANCE generator is expected to 
produce a correct mixture of different hadrons at all energies, it is 
sufficient to determine the hadron energy resolution at ICAL through 
an effective averaging of NUANCE events, without having to identify 
the hadrons separately. 
 
\begin{figure}[h!] 
\includegraphics[width=0.46\textwidth,height=0.33\textwidth]
{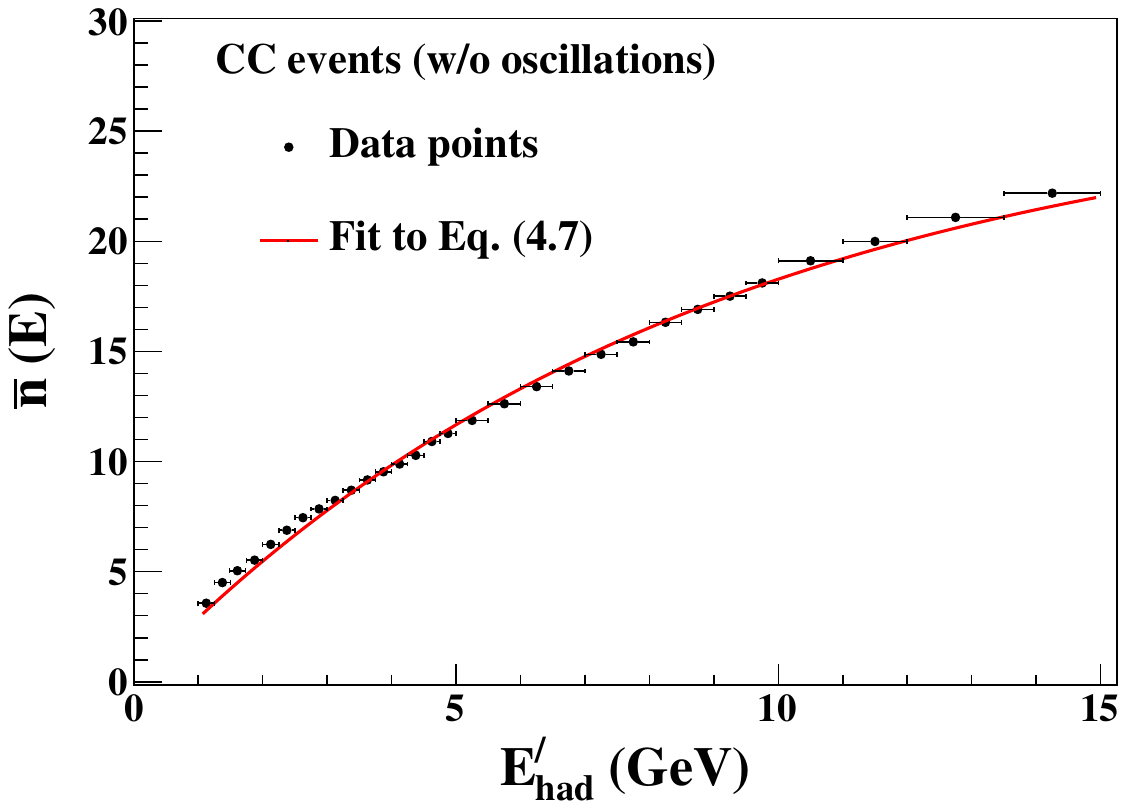} 
\includegraphics[width=0.46\textwidth,height=0.33\textwidth]
{./response/nuance_had_res_new.pdf} 
\caption{The mean hit distribution (left) and the energy resolution (right) 
for hadron events generated by NUANCE, as a function of 
$\rm{E}_{\rm{had}}^{\prime}$ . The right panel 
also shows a fit to Eq.~(\ref{eq5}). 
The bin widths are indicated by horizontal error bars \cite{Devi:2013wxa}.
} 
\label{histomean_vav_nuance} 
\end{figure}  

A total of 1000 kt-years of ``data'' events (equivalent to 20 
years of exposure with the 50 kt ICAL module) were generated with NUANCE. 
The events were further binned into the various 
$E^{\prime}_{\rm had}$ energy bins and the hit 
distributions (averaged over all angles) in these bins are fitted to 
the Vavilov distribution function. 
The mean values ($\textnormal{Mean}_{\textnormal{Vavilov}}$) of  
these distributions as a function of $E_{\rm{had}}^{\prime}$ 
are shown in the left panel of Fig.~\ref{histomean_vav_nuance}. 
As expected, these are similar to the mean values obtained earlier 
with fixed energy pions. Since the mean hits grow approximately linearly 
with energy, the same linearized approximation used in 
section~\ref{sec:charge_pions} can be used to obtain the energy resolution 
$\sigma/E$ = $\Delta n/\bar{n}$. 
The energy resolution as a function of $E_{\rm{had}}^{\prime}$ 
is shown in Fig.~\ref{histomean_vav_nuance}. 
The energy resolution ranges from 85\% (at 1 GeV) to 36\% (at 15 GeV).

The effective energy response obtained from the 
NUANCE-generated data is an average over the mixture of many hadrons 
that contribute to hadron shower at all energies. The fractional weights 
of different kinds of hadrons produced in neutrino interactions may, in 
principle, depend upon neutrino oscillations. In addition, the relative 
weights of events with different energy that contribute in a single energy 
bin changes because neutrino oscillations are energy dependent.  
In order to check this, events with oscillations using  
the best-fit values of standard oscillation parameters  
(mixing angles and mass-squared differences) \cite{oscparam} 
were also generated. The resolutions obtained without and with 
oscillations are very close to each other. Thus, the hadron energy 
resolution can be taken to be insensitive to oscillations.

\subsection{Hadron energy calibration} 
\label{sec:E_calib} 

\begin{figure}[h!] 
\centering 
\includegraphics[width=8cm]{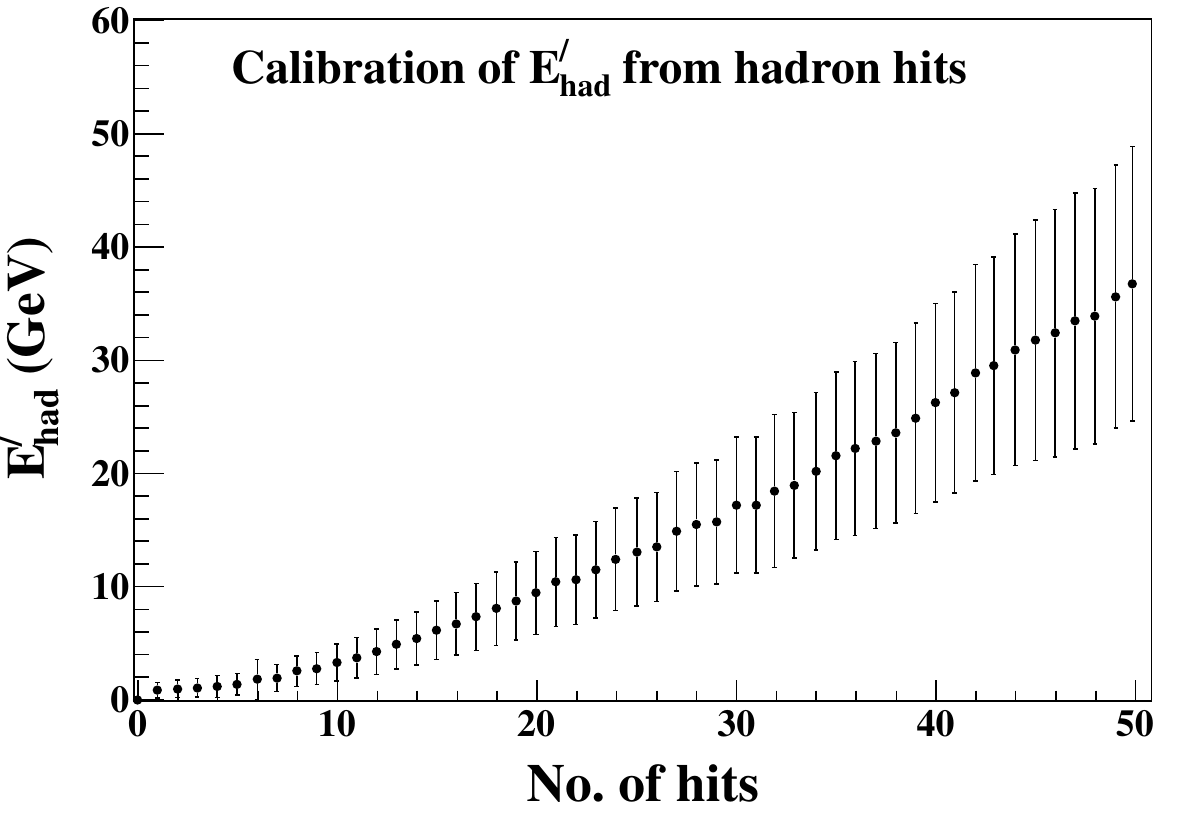} 
\caption{Calibration plot for $\textnormal{E}_{\textnormal{had}}^{\prime}$, 
where mean and $\sigma$ from the Vavilov fits are represented by the 
black filled circles and error bars, respectively
\cite{Devi:2013wxa}.}  
\label{calib_nuance} 
\end{figure}

To calibrate the hadron energy $E'_{\rm had}$ against the hit multiplicity,
hadrons from the simulated NUANCE \cite{Casper:2002sd} ``data'' were
divided into bins of different hit multiplicities $n$.
Even here, a good fit was obtained
for the Vavilov distribution function at all values of $n$.
We show the mean, $\rm{Mean}_{\rm{Vavilov}}$, and 
the standard deviation, $\sigma_{\rm{Vavilov}}$, obtained
from the fit in the calibration plot in Fig.~\ref{calib_nuance}.

For charged-current $\nu_\mu$ events, 
the energy of the incident neutrino can be reconstructed through
\begin{equation}
E_\nu = E_\mu + E'_{\rm had} \; ,
\end{equation}
where $E_\mu$ is reconstructed from the Kalman filter algorithm 
and $E'_{\rm had}$
is calibrated against the number of hadronic hits.
The neutrino energy resolution will in principle depend on the
energy and direction of the muon as well as that of the hadron shower.
The poor energy resolution of hadrons also makes the energy
resolution of neutrinos rather poor, and loses the advantage of an
accurate muon energy measurement. Therefore, reconstructing
neutrino energy is not expected to be the most efficient method
for extracting information from the ICAL analysis. Indeed, we expect to
use the muon and hadron information separately, as will be seen
in Sec.~\ref{3d-analysis}.

\subsection{Salient features of the detector response}
\label{salient-response} 
 
The ICAL detector is mainly sensitive to muons produced in the charged-current 
interactions of atmospheric $\nu_\mu$ or $\overline{\nu}_{\mu}$. We have studied
the response of the detector to muons \cite{Chatterjee:2014vta} 
generated in the central 
volume of a module of the ICAL detector where the magnetic field is uniform. 
The momentum, charge and direction of the muons are determined from the 
curvature of the track in the magnetic field using Kalman filter algorithm. 
The response of the detector to muons in the energy range 1--20 GeV with 
$\cos\theta > 0.25$ is studied in the different azimuthal $\phi$ regions. 
The momentum resolution, reconstruction efficiency, charge identification 
efficiency and direction resolution are calculated. The momentum resolution 
is about 20\% (10\%) for energies of 2 GeV (10 GeV), while the reconstruction 
efficiency is about 80\% for $E_\mu > 2$ GeV. The relative charge identification
efficiency is found to be 98\% for almost all energies above the threshold. 
The direction resolution is found to be better than a degree
for all angles for energies greater than about 4 GeV.

The hadron events of interest in the ICAL detector primarily contain
charged pions. The hit pattern of pions and kaons in the detector is 
similar; hence it is not possible to separate different hadrons in the 
detector. Similarly, neutrino-nucleus interactions produce events with 
multiple hadrons in the final state (generated by the NUANCE neutrino 
generator), whose energies cannot be reconstructed individually.  However, 
the {\em total energy} deposited in hadrons can be determined by a 
calibration against the hit multiplicity of hadrons in the detector
\cite{Devi:2013wxa}. 

The hit patterns in single and multiple hadron events are roughly similar, 
and may be described faithfully by a Vavilov distribution.
Analyses, first with fixed-energy pions, and later with a mixture of 
hadrons from atmospheric $\nu_{\mu}$ interaction events, show that 
a hadron energy resolution in the range 85\% (at 1 GeV) -- 36\% (at 15 GeV) 
is obtainable.
The parameters of the Vavilov fit presented here as a function of hadron energy
can be used for simulating the hadron energy response of the detector, in order
to perform physics analyses that need the hadron energy resolution of ICAL. 
We have also presented the calibration for the energy of the hadron shower as a 
function of the hit multiplicity. This analysis will be improved upon by
incorporating edge effects and noise in a later study, after data from 
the prototype detector is available. 

The reconstruction of hadrons allows us to reconstruct the total visible
energy in NC events. Combined with the information on the muon energy
and direction in the CC events, it will also allow one to reconstruct the 
total neutrino energy in the CC events. As we shall see in 
Chapter~\ref{analysis}, the correlated information in muon and hadron
in a CC event will also help to enhance the capabilities of the ICAL. 
The ICAL will be one of the largest neutrino detectors sensitive to the 
final state muons as well as hadrons in neutrino interactions at multi-GeV
energies, and this advantage needs to be fully exploited. 

Note that the calibration of the hadron response presented in this 
section has been determined by Monte-Carlo simulations. To confirm its
validity, we have compared in Appendix~\ref{app:thickness} the results of 
our simulations with the hadron response at MINOS and the baby MONOLITH 
detector at appropriate plate thicknesses. This appendix also studies 
the dependence of hadron response at ICAL as a function of 
the iron plate thickness.

Some muon events at ICAL will also arise from the $\nu_\tau$'s arising
due to oscillations of $\nu_\mu$. These $\nu_\tau$'s may produce $\tau$'s
through charged-current interactions, which would further decay to
muons within the detector. These events will then contaminate the direct 
muon signal \cite{Indumathi:2009hg}.
The number of such events (indirect muon events 
through tau production) is however heavily suppressed, first due to the 
mass of the $\tau$ that implies a large threshold energy for the neutrino,
then due to the small branching fraction of $\tau \to \mu \nu_\tau \bar\nu_\mu$, 
and finally due to the three-body kinematics of $\tau$ decay that reduces 
the energy of the resultant muon even further. This results in only 
about 150 such indirect muon events in 5 years, as compared to a few 
thousands of direct muon events. Hence at this level of analysis, we have
neglected these events. The $\nu_e$ charged current events which 
may be mistakenly reconstructed as charged-current muon events have
also been neglected at this stage. These are in the process of being 
included in a more sophisticated analysis.

\chapter{Neutrino Oscillation Physics at ICAL}
\label{analysis}

\begin{flushright}
{\it
The pendulum of mind oscillates between sense and nonsense, \\
not between right and wrong. \\
-- Carl Gustav Jung
}
\end{flushright}

In this chapter we will present the physics capabilities of ICAL for 
the mixing parameters within the three-generation flavor oscillation paradigm.
We shall restrict ourselves to the charged-current events produced in 
the ICAL from $\nu_\mu$ and $\bar\nu_\mu$ interactions, which produce 
$\mu^-$ and $\mu^+$, respectively. We shall start by describing our analysis 
method in Sec.~\ref{muon-events}, and then 
proceed to present the results showing the physics reach of this experiment for 
various quantities of interest. We shall focus on the identification of
the neutrino mass hierarchy, as well as on the precision measurements 
of $|\Delta m^2_{32}|$ and $\theta_{23}$.

The results will be presented using three different analyses. 
First in Sec.~\ref{sec:emu_thmu},
we use only the information on the measured muon energy and muon 
direction ($E_\mu, \cos\theta_\mu$), both of which should be rather precisely 
measured in this detector, 
as described in detail in Chapter~\ref{response}. 
Note that the results for the muon reconstruction used in these physics 
analyses have been obtained with an averaging over the azimuthal angle, 
and with the vertex taken to be in the central (8m x 8m x 10m) region of 
each module of the detector. These muons may propagate out of this 
region into the peripheral regions, and even exit the detector.
The latter ``partially contained'' events roughly form about 12\% of our 
sample, and we have not analyzed them separately.

We next show 
the improvement expected in the precision measurement of the atmospheric 
mass squared difference and the mixing angle if we use the information on 
the hadron energy in addition, to reconstruct the neutrino energy in each
event. In this analysis, first in Sec.~\ref{sec:enu_thmu} we analyse the 
data in terms of the reconstructed neutrino energy and the measured muon angle
($E_\nu, \cos\theta_\mu$). 
However the reconstruction of neutrino energy involves the addition of the
rather coarsely known hadron energy information to the measured muon energy, 
which results in a dilution of the muon energy information, which is more
accurately known due to the good tracking capabilities of the ICAL. 
In order to retain the benefits of the accurately measured muon energy,
we separately use the information on the measured muon energy, muon 
direction and the hadron energy ($E_\mu, \cos\theta_\mu, E'_{\rm had}$) 
corresponding to each atmospheric neutrino event at the ICAL detector. 
The results of this final analysis, which leads to the best physics
reach for ICAL at this stage, are presented in Sec.~\ref{3d-analysis}.

Note that the detector characteristics used for the analyses presented in this 
Chapter have been determined in the central region of the central module 
of the ICAL detector, as mentioned in the previous Chapter. 
When the three modules are placed adjacent to each other
along the $x$-axis, similar detector response is seen in the
extended central region that includes the central region of each
module as well as the ``side'' regions that are sandwiched between
two central regions. This comprises the region 
$-20 {\rm m} \le x \le 20 {\rm m}$, 
$-4 {\rm m} \le y \le 4 {\rm m}$, and the entire $z$ region, that
is, about 42\% of ICAL. As expected, the muon response is worse in
the peripheral regions of the detector \cite{Kanishka:2015qsa}.
Studies show that the reconstruction efficiencies drop by about
10\% while the charge identification efficiency drops from 98\% 
in the central to
about 96\% in the peripheral region for few-GeV muons. Further, the
momentum resolution worsens from $\sigma/P \sim$ 10\% to about
12--15\% while the direction resolution remains the same. Hence this
would worsen the physics results that we would obtain, although not
drastically. Note that the hadron resolutions are not altered on
inclusion of the entire volume of ICAL, mainly since it is independent
of the magnetic field. We do not comment further on this in this paper,
and present all results using the central region resolutions
described earlier.

\section{Charged-current $\nu_\mu$  events in ICAL}
\label{muon-events}

We focus on the charged-current events from the atmospheric
$\nu_\mu$ interactions, that produce muons in the ICAL. 
We shall start by dividing them into bins of energy and momenta, taking
into account the efficiencies and resolutions obtained in 
Chapter~\ref{response}. 
As its output, the generator provides the 4-momentum ($p^\mu$) of the initial, 
intermediate and the final state particles for each event. To reduce the Monte 
Carlo fluctuations in the events obtained, we generate an event 
sample corresponding to 1000 years of running of ICAL and and scale it down 
to the desired exposure for the $\chi^2$ analysis. 
The ICAL sensitivities presented here can then be interpreted as median
sensitivities (in the frequentist sense), as described in 
\cite{Blennow:2013oma}.
Using 1000 years of data takes us closer to the ideal ``Asimov'' 
data set \cite{Cowan:2010js} that has no statistical fluctuations.

In the oscillated event sample,  
the total number of $\mu^-$ events come from the combination of the 
$\nu_\mu \rightarrow \nu_\mu$ 
and the $\nu_e \rightarrow \nu_\mu$ channels as 
\begin{equation} \label{eq_event_rate}
 \frac{d^2N}{dE_\nu \: d(\cos \theta_\nu)} = 
 N_T \, N_D \, \sigma_{\nu_\mu} \, 
 \left[P_{\mu\mu}  \frac{d^2\Phi_{\nu_\mu}}{dE_\nu \: d(\cos \theta_\nu)}
 + P_{e\mu} \frac{d^2\Phi_{\nu_e}}{dE_\nu \: d(\cos \theta_\nu)}
 \right] \,,
\end{equation}
where $N_D$ is the number of targets and $N_T$ is the exposure time of the
detector. Here $\Phi_{\nu_\mu}$ and $\Phi_{\nu_e}$ are the fluxes of $\nu_\mu$ and
$\nu_e$, respectively, and $P_{\alpha\beta}$ is the $\nu_\alpha
\rightarrow \nu_\beta$ oscillation probability. The first term in
Eq.~(\ref{eq_event_rate}) corresponds to the number of $\mu^-$ events
from $\nu_\mu$ that have survived oscillations, while the second term
corresponds to the oscillated $\nu_e$ flux into $\nu_\mu$. 

The oscillation probabilities $P_{\mu\mu}$ and $P_{e\mu}$ are calculated
numerically for any given set of oscillation parameters for each event,
corresponding to the neutrino energy and zenith angle associated with it.
Since it takes a long time to run the NUANCE code to generate such a large 
event sample, generating events for each set of possible oscillation 
parameters is practically impossible. Therefore, we run the event generator 
only once for no oscillations and thereafter incorporate the oscillations
using the ``reweighting'' algorithm, which works as follows.

In order to implement the effects of oscillation on a $\nu_\mu$, 
a random number $R$ between 0 and 1 is generated. 
If $R < P_{\mu e}$, the event is classified as a $\nue$ event. 
If $ R >  (P_{\mu e} + P_{\mu\mu})$, then we classify the 
event as a $\nutau$ event. 
If $P_{\mu e} \leq R \leq (P_{\mu e} + P_{\mu\mu})$, then 
it means that this event has come from an atmospheric $\numu$ which has 
survived as a $\numu$ and is hence selected as a muon neutrino event.
The effects of oscillation on the $\nu_e$ events are implemented similarly,
where the muon events are a result of oscillated $\nu_e$ events with
a probability $P_{e\mu}$.
The net number of muon events are obtained by adding the ``survived"
and the ``oscillated" $\nu_\mu$ events, as shown in Eq.~(\ref{eq_event_rate}). 

\begin{figure}[h]
\centering
\includegraphics[width=0.7\textwidth]{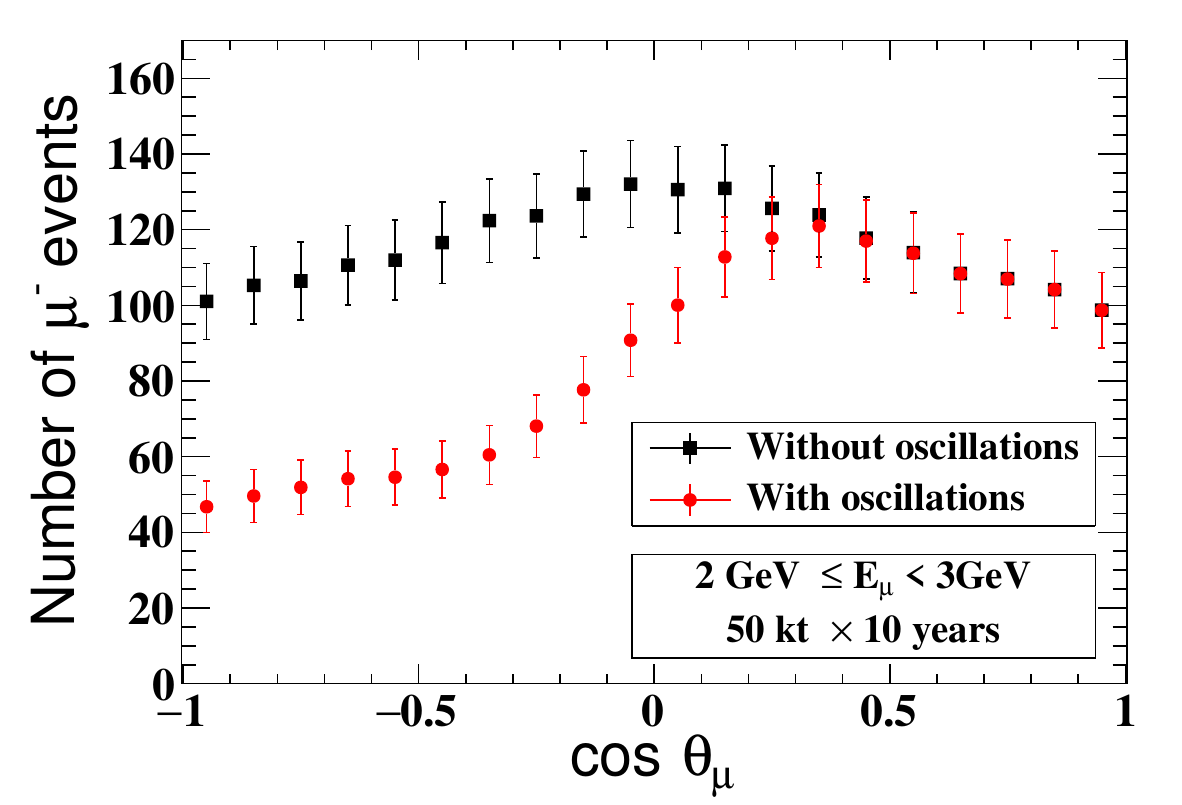}
\caption{
Zenith angle distribution of $\mu^-$ events for the bin 2 GeV $\leq E_\mu<$ 
3 GeV, without and with flavor oscillations. The detector efficiencies have 
not been included here. The error bars shown are statistical \cite{Ghosh:2012px}
.}
\label{fig:fig_MC_Osc}
\end{figure}

The $\mu^+$ events
in the detector are generated using a similar procedure. This final data
sample is then binned in energy and zenith angle bins.
Figure~\ref{fig:fig_MC_Osc} shows the zenith angle distribution of $\mu^-$
events in the muon energy bin $E_\mu=2-3$ GeV, before and after invoking
oscillations. We use the oscillation  parameters described in 
Table~\ref{tab:best-fit} and take the exposure to be 50 kt $\times$ 
10 years.

These are the events in an ideal detector. To proceed further, we apply
the muon reconstruction efficiencies and resolutions (in both energy
and direction) obtained in the previous chapter.  The reconstruction
efficiency ($\epsilon_{R-}$) and the charge identification (CID) efficiencies
($\epsilon_{C-}$ for $\mu^-$ and $\epsilon_{C+}$ for $\mu^+$ event sample)
are applied as follows:
\begin{equation}
N_{\mu^-}^C = \epsilon_{C-} ~ N_{\mu^-} + 
(1 - \epsilon_{C+}) ~ N_{\mu^+} \,,
\label{eq:evcid}
\end{equation}
with
\begin{equation}
N_{\mu-} = \epsilon_{R-} ~ N_{\mu-}^{\rm true} \,,
\label{eq:evrecon}
\end{equation}
where $N_{\mu-}^{\rm true}$ is the number of $\mu^-$ events in a given 
($E_\mu$, $\cos \theta_\mu$) bin. 
All the quantities appearing in Eq.~(\ref{eq:evcid}) are functions 
of $E_\mu$ and $\cos \theta_\mu$, and are determined bin-wise.
The same procedure is applied for determining the $\mu^+$ events.

\begin{figure}[h]
\centering
\includegraphics[width=0.7\textwidth]{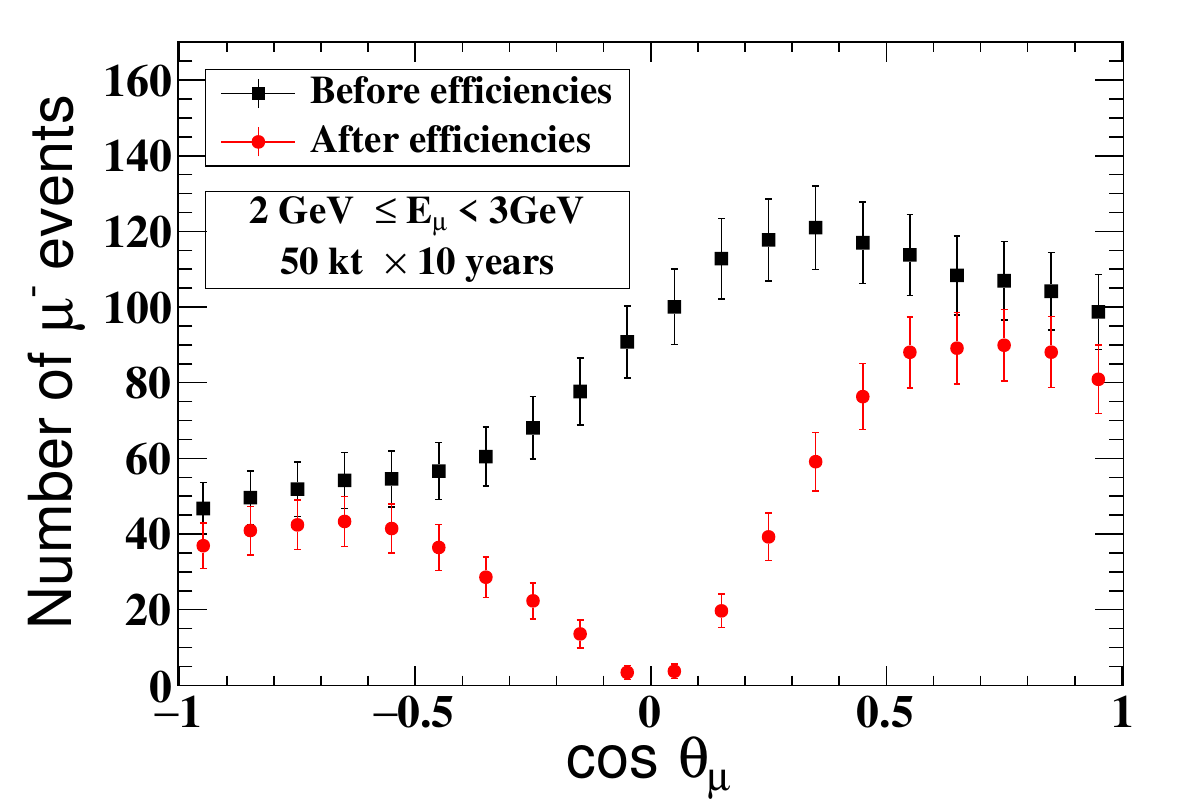}
\caption{Zenith angle distribution of oscillated $\mu^-$ events for the 
bin 2 GeV $\leq E_\mu<$ 3 GeV, after taking into account detector 
efficiencies. The error bars shown are statistical \cite{Ghosh:2012px}.}
\label{fig:fig_event_distribution_with_Eff}
\end{figure}

Figure~\ref{fig:fig_event_distribution_with_Eff} shows the
zenith angle distribution of events obtained before and after
applying the reconstruction and CID efficiencies. Compared to Fig.
\ref{fig:fig_MC_Osc}, one can notice that the number of events fall
sharply for the almost horizontal ($\cos \theta_\mu \approx 0$) bins.
This is because the reconstruction efficiency for muons falls as we
go to more horizontal bins since the iron slabs and RPCs in ICAL are 
stacked horizontally. As a result there are hardly any events for bins
with $-0.2 \leq \cos \theta_\mu < 0.2$.

\begin{figure}[h]
\centering
\includegraphics[width=0.7\textwidth]{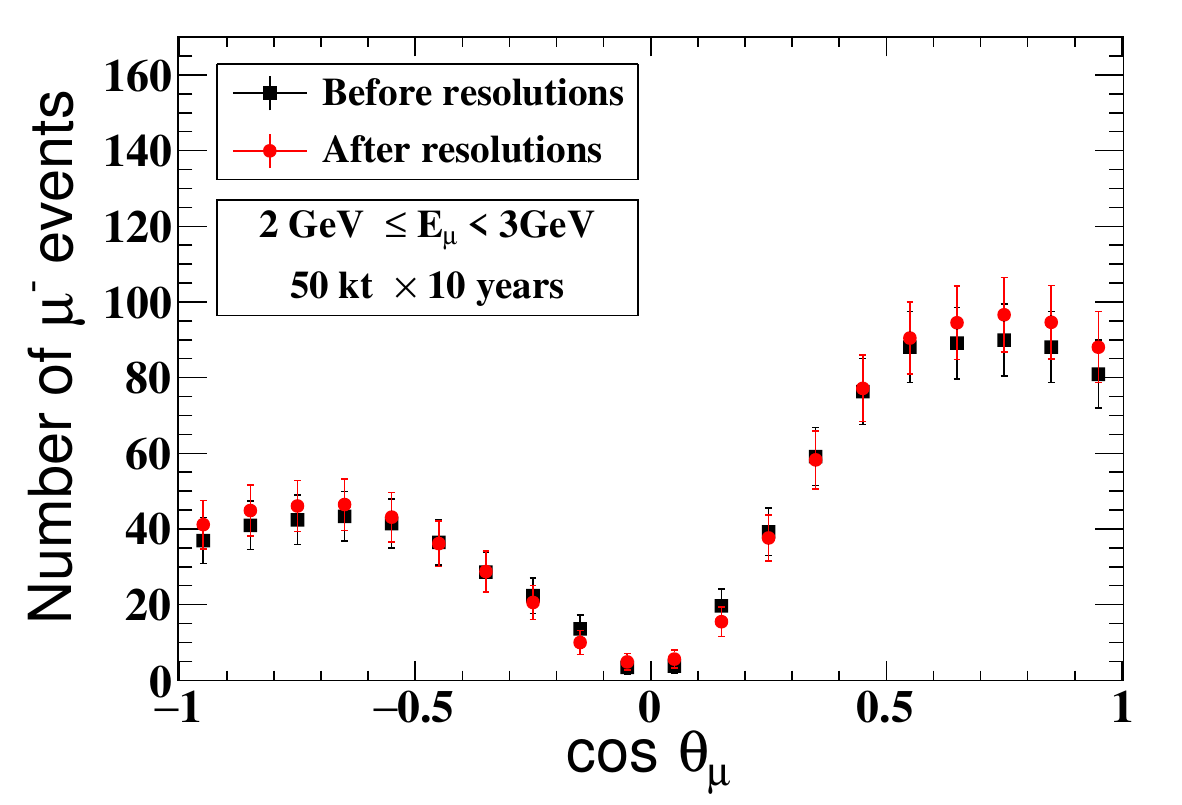}
\caption{Zenith angle distribution of $\mu^-$ events for the bin 2 GeV 
$\leq E_\mu<$ 3 GeV before and after including energy and zenith angle 
resolution function. Here $E$ and $\theta$ are the measured energy and 
measured zenith angle, respectively. The error bars shown are statistical
\cite{Ghosh:2012px}.
}
 \label{fig:fig_event_distribution_with_Res}
\end{figure}

Finally, the muon resolutions $\sigma_E$ and $\sigma_{\cos \theta}$ are 
applied as follows:
\begin{equation}\label{eq_det_res}
(N_{\mu^-}^D)_{ij} (E, \cos\theta) = \sum_k 
\sum_l N_{\mu^-}^C(E_\mu^k,\cos \theta_\mu^l) \,\,K_i^k(E_\mu^k) \,\,
M_j^l(\cos \theta_\mu^l) \,,
\end{equation}
where $(N_{\mu^-}^D)_{ij}$ denotes the number of muon events in the 
$i^{th}$ $E$-bin and the $j^{th}$ $\cos \theta$-bin after applying the 
energy and angle resolutions. Here $E$ and $\cos \theta$ are the measured 
muon energy and zenith angle. The summation is over the true energy bin 
$k$ and true zenith angle bin $l$, with $E_\mu^k$ and $\cos \theta_{\mu}^l$ 
being the central values of the $k^{\rm th}$ true muon energy and $l^{\rm th}$ 
true muon zenith angle bin. The quantities $K_i^k$ and $M_j^l$ are the 
integrals of the detector resolution functions over the bins of $E$ and 
$\cos \theta$, the measured energy and direction of the muon, respectively. 
These are evaluated as
\begin{equation}\label{eq_det_int_K}
 K_i^k (E_\mu^k) = \int_{E_{L_i}}^{E_{H_i}} dE \frac{1}{\sqrt{2\pi} \sigma_{E_\mu^k}} 
\exp \left( {- \frac{(E_\mu^k - E)^2}{2 \sigma_{E_\mu^k}^2} } \right) \,,
\end{equation}
and
\begin{equation}\label{eq_det_int_M}
M_j^l (\cos \theta_\mu^l) = 
\int_{\cos \theta_{L_j}}^{\cos \theta_{H_j}} d\cos  \theta\frac{1}{\sqrt{2\pi} 
\sigma_{\cos\theta_\mu^l}} 
\exp \left (  - \frac{(\cos \theta_\mu^l - \cos \theta)^2}
{2 \sigma_{\cos\theta_\mu^l}^2} \right ) \,,
\end{equation} 
where $\sigma_{E_\mu^k}$ and $\sigma_{\cos\theta_\mu^l}$ are the energy 
and zenith angle resolutions, respectively, in these bins, as obtained
in Chapter~\ref{response}. 
We perform the integrations between the lower and upper boundaries of the 
measured energy ($E_{L_i}$ and $E_{H_i}$) and the measured zenith angle 
($\cos \theta_{L_j}$ and $\cos \theta_{H_j}$). For the extreme $\cos \theta$ 
bins, the bins are taken to be ($-\infty$, -0.9) and [0.9, +$\infty$) while 
integrating, and the events are assigned to the bins [-1, -0.9] and [0.9, 1], 
respectively. This ensures that no event is lost to the unphysical region 
and the total number of events does not change after applying the angular 
resolution. For $E_\mu^k <1$ GeV, the integrand in Eq.~(\ref{eq_det_int_K}) is
replaced with the Landau distribution function.
Figure~\ref{fig:fig_event_distribution_with_Res} shows the zenith angle 
distribution of $\mu^-$ events before and after folding in the resolution 
functions. The angular dependence seems to get only slightly diluted.
This is due to the good angular resolution of the detector.

Table~\ref{tab_number_events} shows the total number of muon events with 
the measured energy range 0.8--10.8~GeV at various stages of the analysis for 
an exposure of 50 kt $\times$ 10 years. Note the sharp fall in statistics 
due to the reconstruction efficiencies. The reconstruction efficiencies are 
particularly poor for the near-horizontal bins where the reconstruction of 
the muon tracks is very hard. The small increase in the number of events 
after applying the energy resolution function is due to the spillover of 
events from the low-energy part of the spectrum to measured energies greater 
than 0.8 GeV. The spillover to the energy bins with $E_\mu>$10.8 GeV is 
comparatively small. The zenith angle resolution leaves the number of muon 
events nearly unchanged. 

\begin{table}[htp]
\begin{center}
\begin{tabular}{|c|c|c|c|c|} \hline
   & $\mu^-$ & $\mu^+$ \\
  \hline  
 Unoscillated & 14311 & 5723 \\
  \hline
  Oscillated & 10531 & 4188 \\
  \hline
  After Applying Reconstruction and CID Efficiencies & 4941 & 2136 \\
  \hline
  After Applying ($E$, $\cos \theta$) Resolutions & 5270 & 2278 \\
  \hline
  \end{tabular}
  \end{center}
\caption{Number of muon events produced in CC $\nu_\mu$ interactions
at various stages of the analysis for an exposure of 50 kt $\times$
10 years in the energy range 0.8--10.8 GeV.}
\label{tab_number_events}
\end{table}

\section{Analysis with $E_\mu$ and $\cos\theta_\mu$}
\label{sec:emu_thmu}

In this analysis, we use only the information in muon energy and muon
direction. These two quantities can be measured with a good precision, 
much better than the precision on hadron energy. Therefore we ignore the 
latter for now. We shall include the hadron information in our analysis in 
the next two sections.

We generate the data at the benchmark true values for oscillation parameters 
given in Table \ref{tab:best-fit}. We define a $\chi^2$ for the ICAL data as
\begin{eqnarray}
\chi^2(\mu^-) & = & {\min_{\{\xi_k\}}} \displaystyle\sum_{i,j}\left[
2\left(N_{ij}^{\rm theory}(\mu^-)-N_{ij}^{\rm data}(\mu^-)\right) \right. \nonumber \\ 
& & \phantom{ {\min_{\{\xi_k\}}} \displaystyle\sum_{i,j}\left[ \right.} \left.
+2N_{ij}^{\rm data}(\mu^-)ln\left(\frac{N_{ij}^{\rm data}(\mu^-)}{N_{ij}^{\rm theory}(\mu^-)}\right)
\right] + \displaystyle\sum\limits_{k} \xi_k^{2} \,,
\label{eq:chisqino}
\end{eqnarray}
where 
\begin{equation}
N_{ij}^{\rm theory}(\mu^-)=N_{ij}^{0 ~{\rm theory}}(\mu^-)
\left(1+\displaystyle\sum\limits_{k=1}^l \pi_{ij}^{k}{\xi_k}\right) \,.
\label{eq:evth}
\end{equation}
Here we use the linearized approximation while using the method of pulls.
We have assumed a Poissonian distribution for the errors in this definition 
of $\chi^2$. The reason is that the number of events falls sharply with 
energy due to the falling flux (cf. Fig.~\ref{nuflux})
and for small exposure times these bins could have very few events per bin. 
Since ICAL will have separate data in $\mu^-$ and $\mu^+$, we 
calculate this $\chi^2(\mu^-)$ and $\chi^2(\mu^+)$ separately 
for the $\mu^-$ sample and the $\mu^+$ 
sample respectively and then add the two to get the total $\chi^2$ as
\begin{equation}
\chi^2 = \chi^2(\mu^-) + \chi^2(\mu^+) \,.
\label{chisq-sum}
\end{equation}
In the above equations,  $N_{ij}^{\rm data}(\mu^-)$ and $N_{ij}^{\rm data}(\mu^+)$ 
are the observed number of $\mu^-$ and $\mu^+$ events respectively in the 
$i^{th}$ energy and $j^{th}$ angle bin and 
$N_{ij}^{0~{\rm theory}}(\mu^-)$ and $N_{ij}^{0~{\rm theory}}(\mu^+)$ are the 
corresponding theoretically predicted event spectrums. 
This predicted event spectrum could shift due to the systematic  uncertainties,
which is taken care of by the method of pulls 
\cite{Fogli:2002au,Huber:2002mx}. 
The shifted spectrum $N_{ij}^{\rm theory}$ is given by Eq. (\ref{eq:evth}), where
$\pi^k_{ij}$ is the $k^{th}$ systematic uncertainty in the 
${ij}^{th}$ bin and $\xi_k$ is the pull variable corresponding to the 
uncertainty $\pi^k$. The $\chi^2 $ is minimized over the full set of pull 
variables $\{\xi_k\}$. In our analysis we have considered the muon energy 
range 0.8 GeV to 10.8 GeV with 10 bins of bin size 1 GeV. 
The zenith angle range in $\cos \theta$ is taken from $-1$ to $+1$,
with 80 bins of bin size 0.025. 
Note that the zenith angle resolution of ICAL is $\sim 0.01$ in 
$\cos\theta_\mu$ over the entire parameter range of interest, the
number of zenith angle bins is limited to ensure enough number of
events in individual bins. 

The index $k$ in Eqs. (\ref{eq:chisqino}) and (\ref{eq:evth}) runs from 
1 to $l$, where $l$ is the total number of systematic uncertainties. 
We have included the following five systematic uncertainties in our analysis
\cite{Gandhi:2007td}: 
\begin{itemize} 
\item an overall flux normalization error of 20\%,
\item an overall cross-section normalization error of 10\%,
\item a 5\% uncertainty on the zenith angle dependence of the fluxes,
\item an overall 5\% energy-independent systematic uncertainty, and
\item an energy dependent ``tilt factor", incorporated according to the 
following prescription. The event spectrum is calculated with the predicted 
atmospheric neutrino fluxes
and then with the flux spectrum shifted according to
\begin{equation}
\Phi_\delta (E) = \Phi_0 (E)\bigg (\frac{E}{E_0}\bigg)^\delta \simeq 
\Phi_0 (E) \bigg ( 1 + \delta \, \ln \frac{E}{E_0} \bigg ) \,,
\label{eq:chiprior}
\end{equation}
where $E_0=2$ GeV and $\delta$ is the $1\sigma$ systematic 
error which we have taken as 5\% \cite{GonzalezGarcia:2004wg}. The difference 
between the predicted events rates for the two cases is then included in the 
statistical analysis. 
\end{itemize}

\subsection{Mass hierarchy sensitivity}
\label{mh-2d}

\begin{figure}[htp]
\centering
\includegraphics[width=0.49\textwidth,height=0.35\textwidth]
{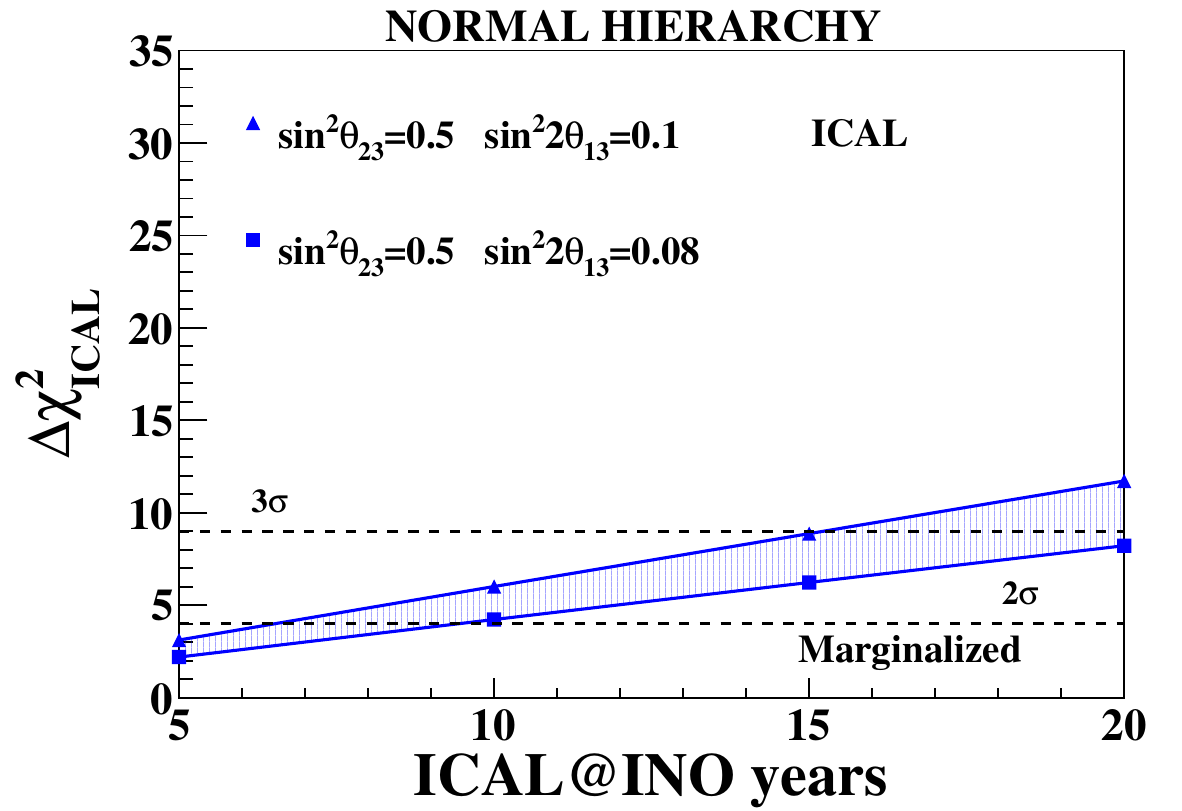}
\includegraphics[width=0.49\textwidth,height=0.35\textwidth]
{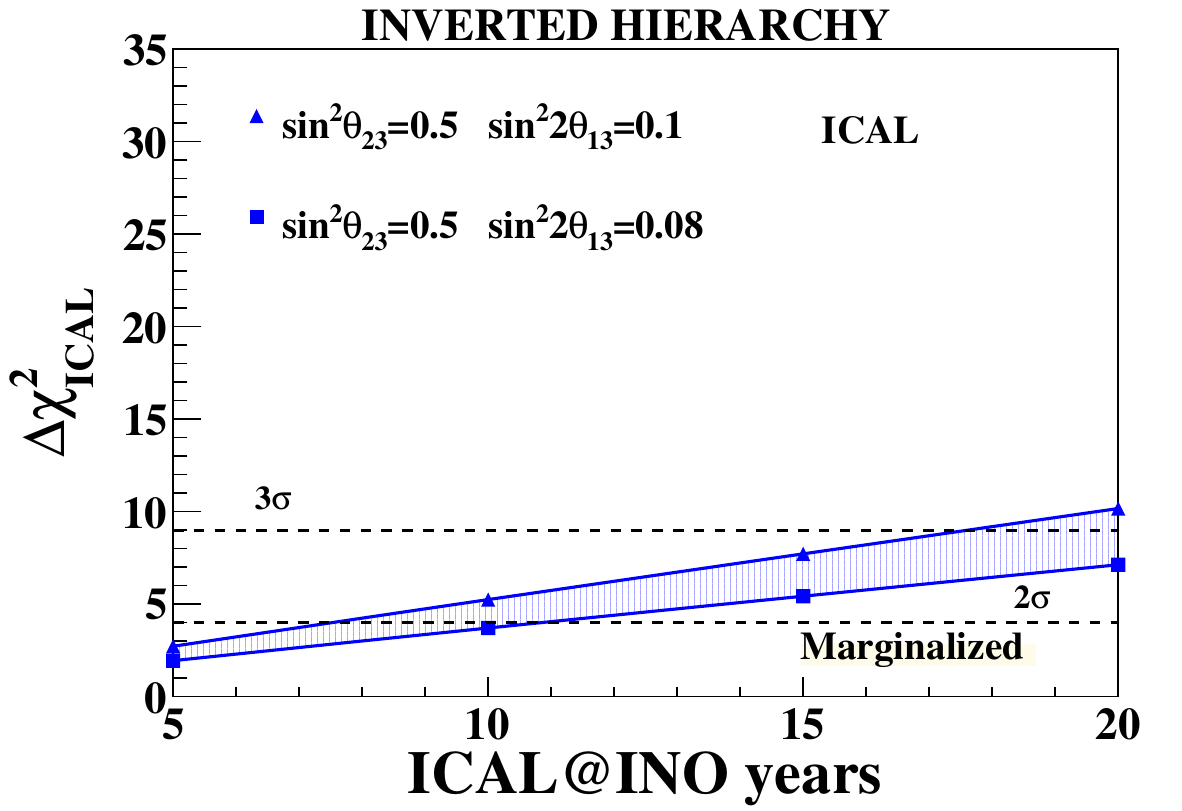}
\caption{Left panel shows the $\Delta \chi^2 $ for the wrong hierarchy 
when normal hierarchy is taken to be true, while the right panel shows 
the corresponding reach when inverted hierarchy is taken as true. 
The bands correspond to $\stcht$ in the range between 0.08 and  0.1 
as shown in the legend box, while $\sat=0.5$ for all cases.
We take only ICAL {\it muon} data into the analysis and marginalize over 
oscillation 
parameters $|\meff|$, $\sa$ and $\stch$, which are allowed to vary freely 
within their $3\sigma$ ranges given in Table \ref{tab:best-fit}}
\cite{Ghosh:2012px}. 
\label{fig:hiermarg}
\end{figure}

Figure~\ref{fig:hiermarg} shows the discovery potential of ICAL
alone for the neutrino mass hierarchy, as a function of the number of
years of data taking of the 50 kt ICAL. The data is generated for the
values of the oscillation parameters given in Table \ref{tab:best-fit}
and for $\sat=0.5$ for a definite hierachy. This simulated data 
is then fitted with the wrong mass hierarchy to check the statistical 
significance with which this wrong hierarchy can be disfavoured.

The bands in Fig.~\ref{fig:hiermarg} correspond 
to $\stcht$ in the range between 0.08 and  0.1.
The left-hand panel is
for true normal hierarchy while the right-hand panel is
for true inverted hierarchy.  In Fig.~\ref{fig:hiermarg} the plots
show the sensitivity reach of ICAL when $\chi^2$ is marginalized
over oscillation parameters $|\meff|$, $\sa$ and $\stch$, meaning
these oscillation parameters are allowed to vary freely in the fit
within the ranges shown in Table \ref{tab:best-fit}, and the minimum
of the $\chi^2$ taken. The CP phase $\delta_{\rm CP}$ does not
significantly impact the ICAL mass hierarchy sensitivity. 
(This will be discussed in some detail later.)
Therefore, we keep $\delta_{\rm CP}$ fixed at 0 in the fit.  The parameters 
$\ms$ and $\sin^2\theta_{12}$ also do not affect $\chi^2 $ and hence are 
kept fixed at their
true values given in Table \ref{tab:best-fit}.  From the figure we see
that for full marginalization within the current $3\sigma$ allowed
range for $|\meff|$, $\sa$ and $\stch$, the sensitivity reach of ICAL
with 10 (5) years data would be only about $2.2\sigma$ ($1.6\sigma$)
for $\sat=0.5$ and $\stcht=0.1$, for true normal hierarchy. The impact
for the inverted hierarchy case is seen to be marginally worse.

We shall further explore the effects of priors, systematic 
uncertainties, and the true value of $\theta_{23}$ in some detail. 
The lessons learnt from this study will be applied directly to
later analyses.

\subsubsection{Impact of priors}
\label{priors}

All values of the oscillation parameters are not allowed
with equal C.L. by the current data.  Moreover, all oscillation
parameters are expected to be measured with much better precision
by the ongoing and upcoming neutrino experiments.  In fact, by the
time ICAL is operational,  all of the current accelerator-based and
reactor experiments would have completed their scheduled runs and hence
we expect that by then significant improvements in the allowed ranges
of the oscillation parameters would have been made.  In particular,
we expect improvement in the values of $\stch$, $|\meff|$ and $\sta$.
One could incorporate this information into the analysis by including 
``priors" on these parameters, through
\be
\chi^2_{\rm ICAL} = \chi^2 + \sum \chi^2_{prior} ~ ,  \quad 
\chi^2_{prior} (p)= \frac{(p_0 - p)^2}{\sigma_0^2}~,
\ee
where $\chi^2 = \chi^2_+ + \chi^2_-$ as in Eq.~(\ref{chisq-sum}),
$p$ is the parameter on which a prior is included and $p_0$ and
$\sigma_0$ are its best fit and $1\sigma$ error, respectively. 
For our analysis, we take the $1\sigma$ error on  
$\stch$ to be 0.1 (a bit conservative, given the current measurements), 
and take $|\meff|$ and $\sta$ to be determined
with an accuracy of 2\% and 0.65\%, respectively.

\begin{figure}[htp]
\centering
\includegraphics[width=0.49\textwidth,height=0.35\textwidth]
{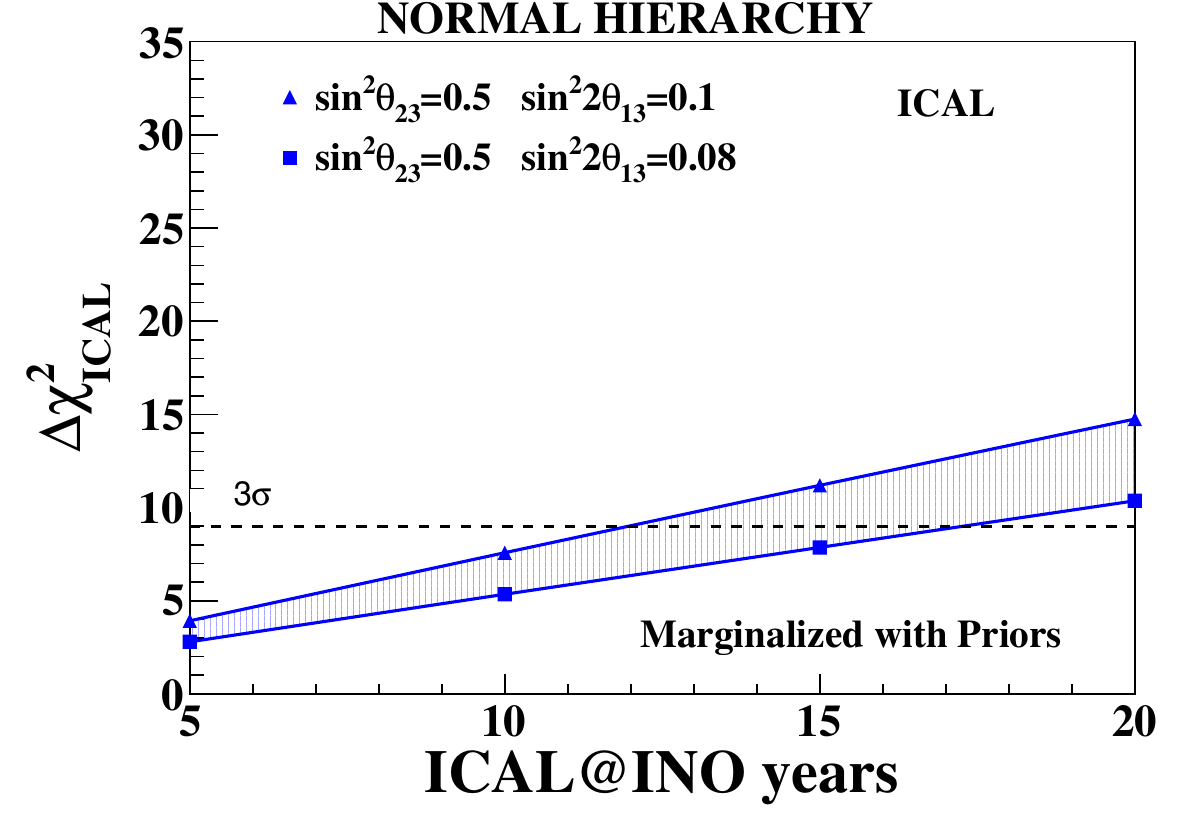}
\includegraphics[width=0.49\textwidth,height=0.35\textwidth]
{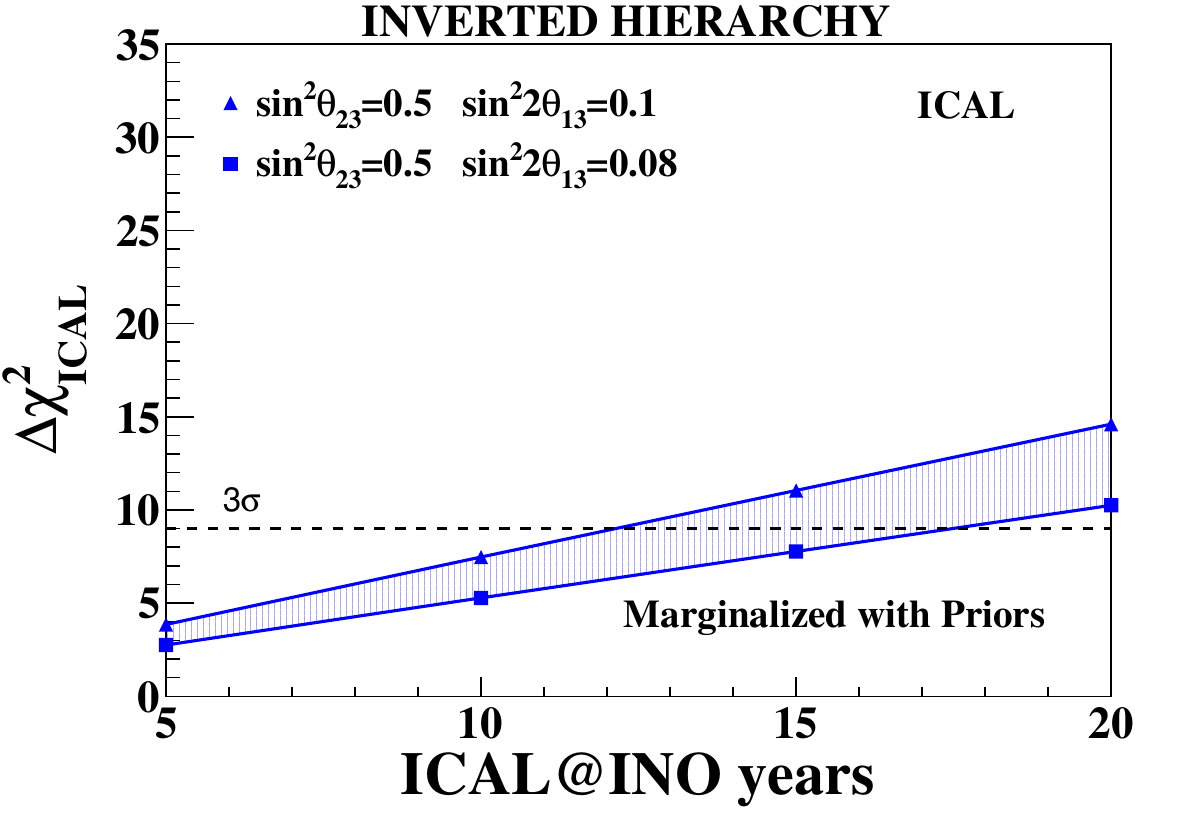}
\caption{Same as Fig.~\ref{fig:hiermarg} but here we impose
priors while marginalizing over $|\meff|$, $\sa$ and 
$\stch$, as discussed in the text \cite{Ghosh:2012px}.  
}
\label{fig:hierprior}
\end{figure} 

The sensitivity reach of ICAL with projected priors on $|\meff|$,
$\sta$ and $\stch$ keeping other parameters fixed is shown in
Fig.~\ref{fig:hierprior}.  We can note from these plots that with 5
years of ICAL data alone, we will have a $1.8\sigma$ ($1.8\sigma$) signal
for the wrong hierarchy if normal (inverted) hierarchy is true. After 10
years of ICAL data, this will improve to $2.5\sigma$ ($2.5\sigma$)
signal for the wrong hierarchy if normal (inverted) hierarchy is true.
The sensitivity obviously increases with the true value of $\stcht$. The
$\Delta \chi^2$ is seen to increase almost linearly with exposure. This
is not hard to understand as the hierarchy sensitivity comes from
the difference in the number of events between normal and inverted
hierarchies due to earth matter effects. Since this is a small difference,
the relevant statistics in this measurement is small. As a result the
mass hierarchy analysis is statistics dominated and 
the $\Delta \chi^2$ increases linearly with exposure.

\subsubsection{Impact of systematic uncertainties}
\label{sec:syst}

\begin{figure}[htp]
\centering
\includegraphics[width=0.46\textwidth,height=0.33\textwidth]
{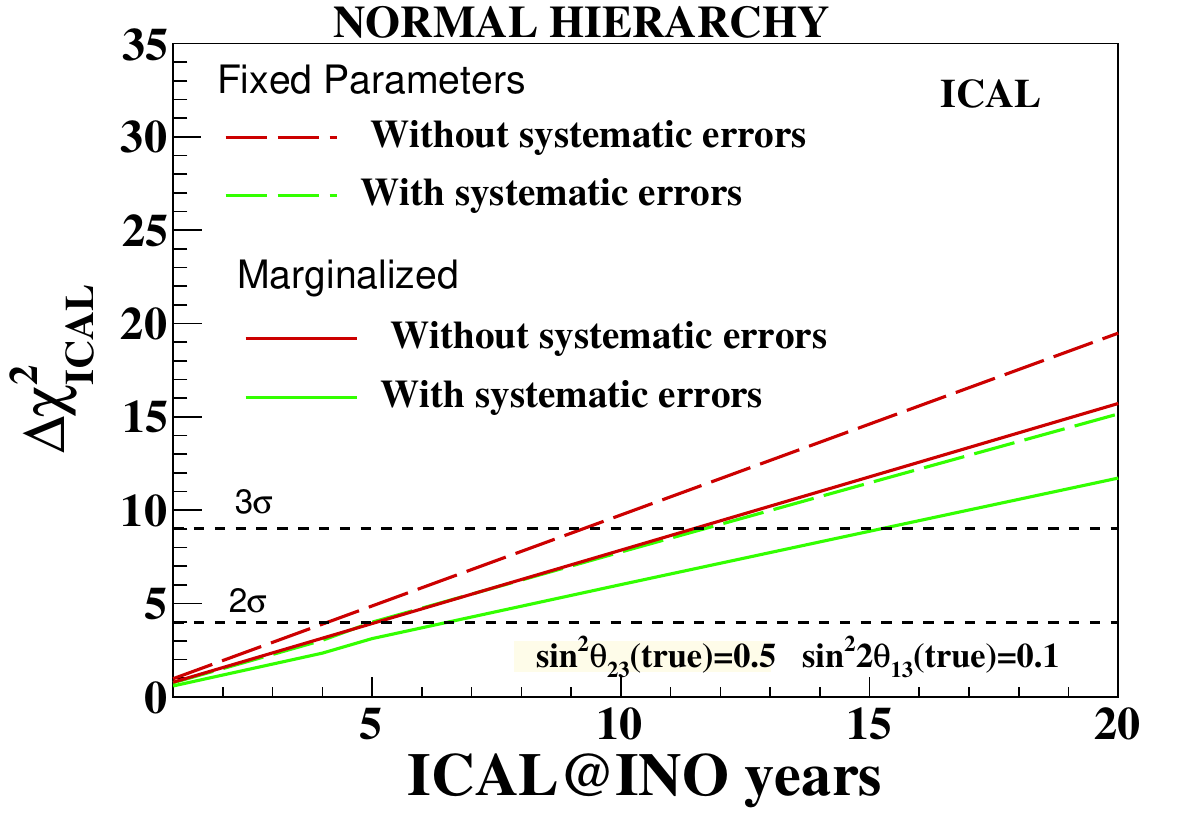}
\includegraphics[width=0.46\textwidth,height=0.33\textwidth]
{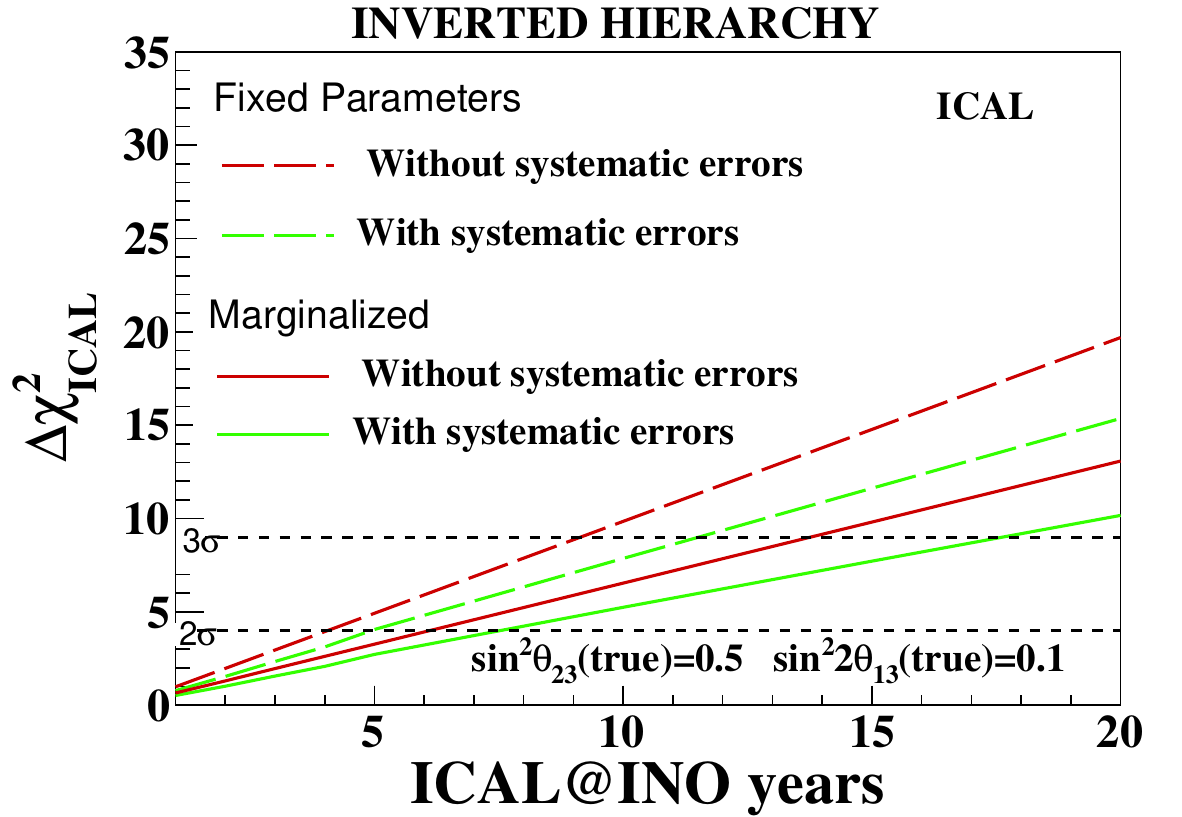}
\caption{The impact of systematic uncertainties on mass hierarchy sensitivity. 
The red lines are obtained without taking systematic uncertainties in the 
ICAL analysis, while the green lines are obtained when systematic 
uncertainties are included. Long-dashed lines are for fixed parameters in 
theory as in data, while solid lines are obtained by marginalizing over 
$|\meff|$, $\sa$ and $\stch$ \cite{Ghosh:2012px}.}
\label{fig:hiersyst}
\end{figure}

The systematic uncertainties, mainly due to the uncertainties in
the atmospheric neutrino fluxes, have already been included in the
above analysis, through the method of pulls 
\cite{Fogli:2002au,Huber:2002mx} as described in 
Sec.~\ref{mh-2d}. An analysis of the extent of the impact of these
uncertainties will give us an idea of how much the reduction in
these uncertainties will help. 
In Fig.~\ref{fig:hiersyst} we show the mass hierarchy sensitivity with
and without systematic uncertainties in the ICAL analysis. The $\Delta
\chi^2$ is shown as a function of the number of years of exposure of the 
experiment. The data was generated at the benchmark oscillation point.

The effect of taking systematic uncertainties is to reduce the statistical
significance of the analysis. We have checked that of the five systematic
uncertainties, the uncertainty on overall normalization of the fluxes
and the cross-section normalization uncertainty have minimal impact on
the final results.  The reason for that can be understood from the fact
that the atmospheric neutrinos come from all zenith angles and over a
wide range of energies. The overall normalization uncertainty is the
same for all bins, while the mass hierarchy dependent earth matter
effects, are important only in certain zenith angle bin and certain
range of energies. Therefore, the effect of the overall normalization
errors get cancelled between different bins.  On the other hand, the
tilt error could be used to modify the energy spectrum of the muons in
the fit and the zenith angle error allows changes to the zenith angle
distribution. Therefore, these errors do not cancel between the different
bins and can dilute the significance of the data. In particular, we
have checked that the effect of the zenith angle dependent systematic
error on the atmospheric neutrino fluxes has the maximum effect on the
lowering of the $\Delta \chi^2$ for the mass hierarchy sensitivity.

\subsubsection{Impact of the true value of $\sa$}
\label{sec:sa}

\begin{figure}[h]
\centering
\includegraphics[width=0.46\textwidth,height=0.33\textwidth]
{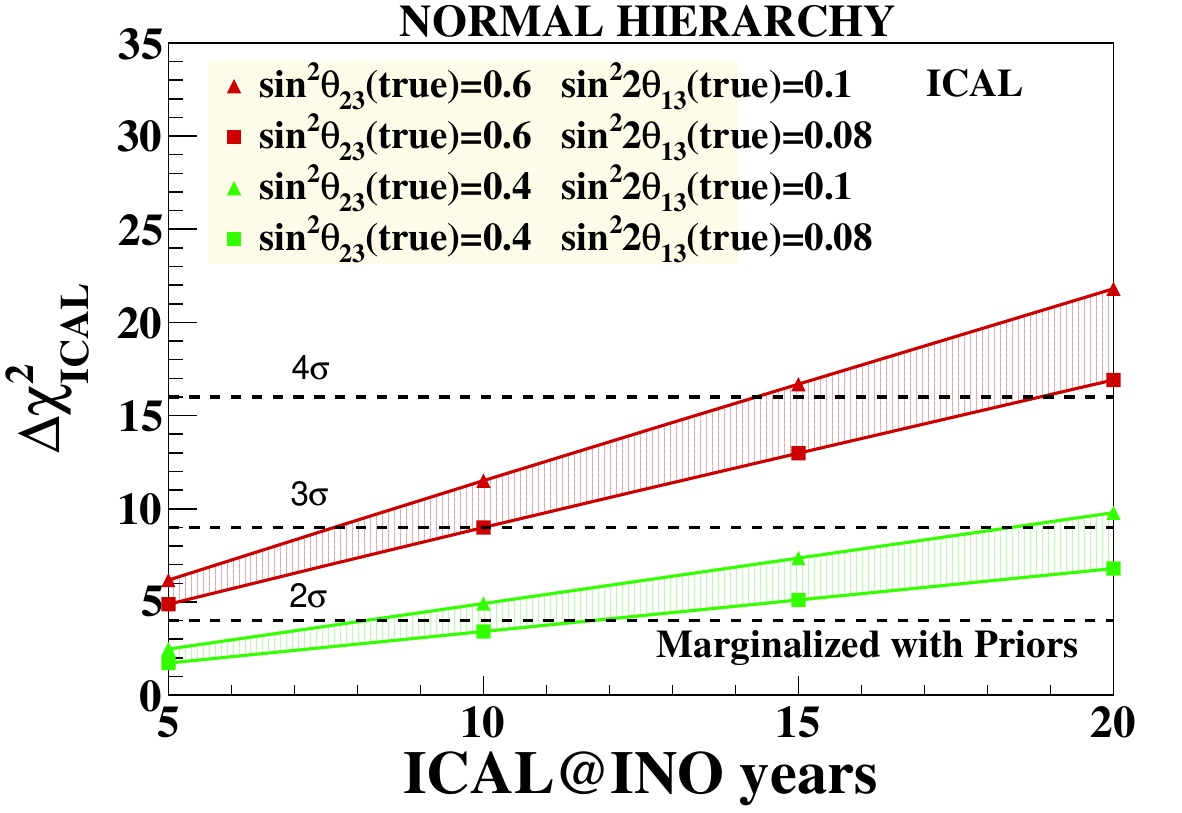}
\includegraphics[width=0.46\textwidth,height=0.33\textwidth]
{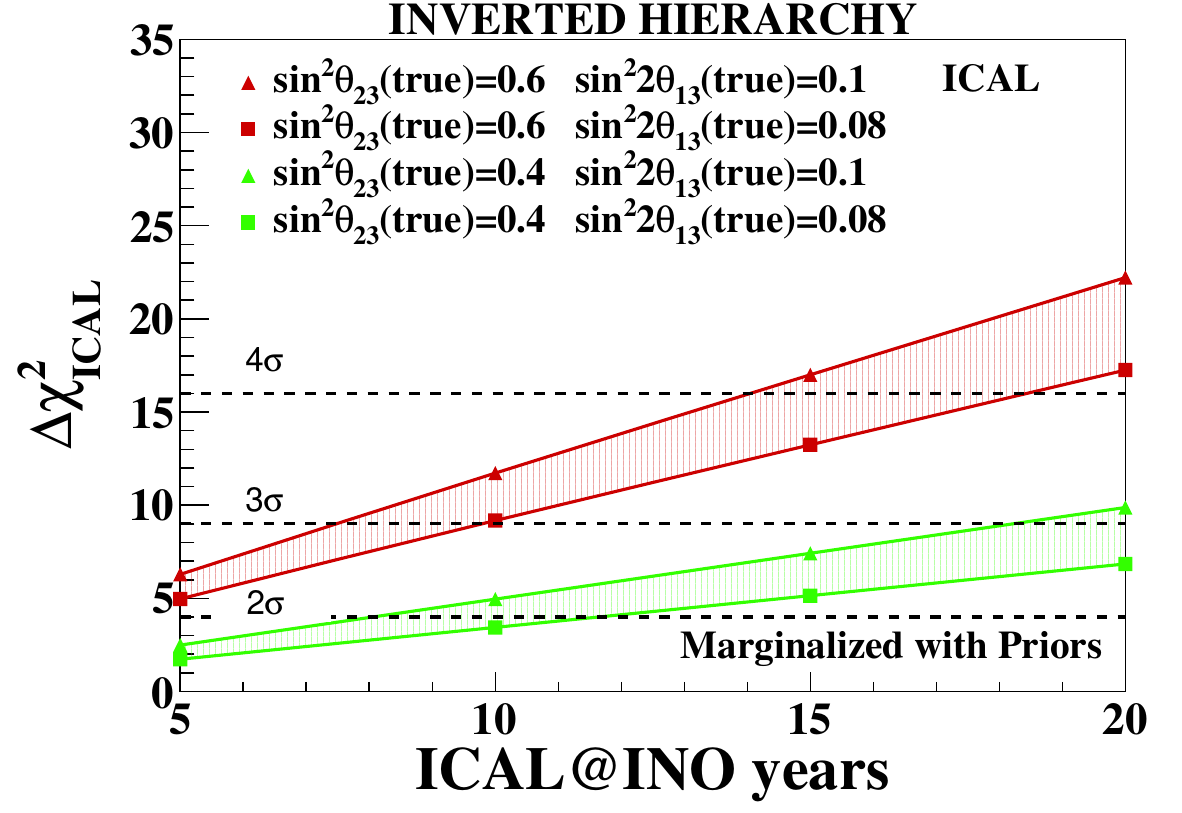}
\caption{Same as Fig.~\ref{fig:hierprior} but for $\sat=0.4$ 
(green band) and $\sat=0.6$ (red band).
The width of each of the bands is mapped by increasing the value of 
$\stcht$ from 0.08 to 0.1.}
\label{fig:hiersaprior}
\end{figure}

The mass hierarchy sensitivity of the ICAL will depend strongly on
the actual value of $\theta_{23}$. The amount of earth matter effects 
increases with increase in both $\theta_{13}$ and $\theta_{23}$.
In the previous plots, we have seen the mass
hierarchy sensitivity for different allowed values of $\stcht$, while
$\sat$ was fixed at maximal mixing. In Fig.~\ref{fig:hiersaprior} we
show the sensitivity to the neutrino mass hierarchy as a function of
number of years of running of ICAL for different values of $\stcht$ as
well as $\sat$.  As seen in the previous subsection,
the $\Delta \chi^2$ for the wrong mass hierarchy increases with $\stcht$
for a given value of $\sat$ and ICAL exposure. A comparison of the $\Delta
\chi^2$ for different values of $\sat$ reveals that the $\Delta \chi^2$
also increases with $\sat$.

\subsubsection{(In)sensitivity to the CP violating phase}
\label{insensitivity}

\begin{figure}
\centering
\includegraphics[width=0.49\textwidth,height=0.35\textwidth]
{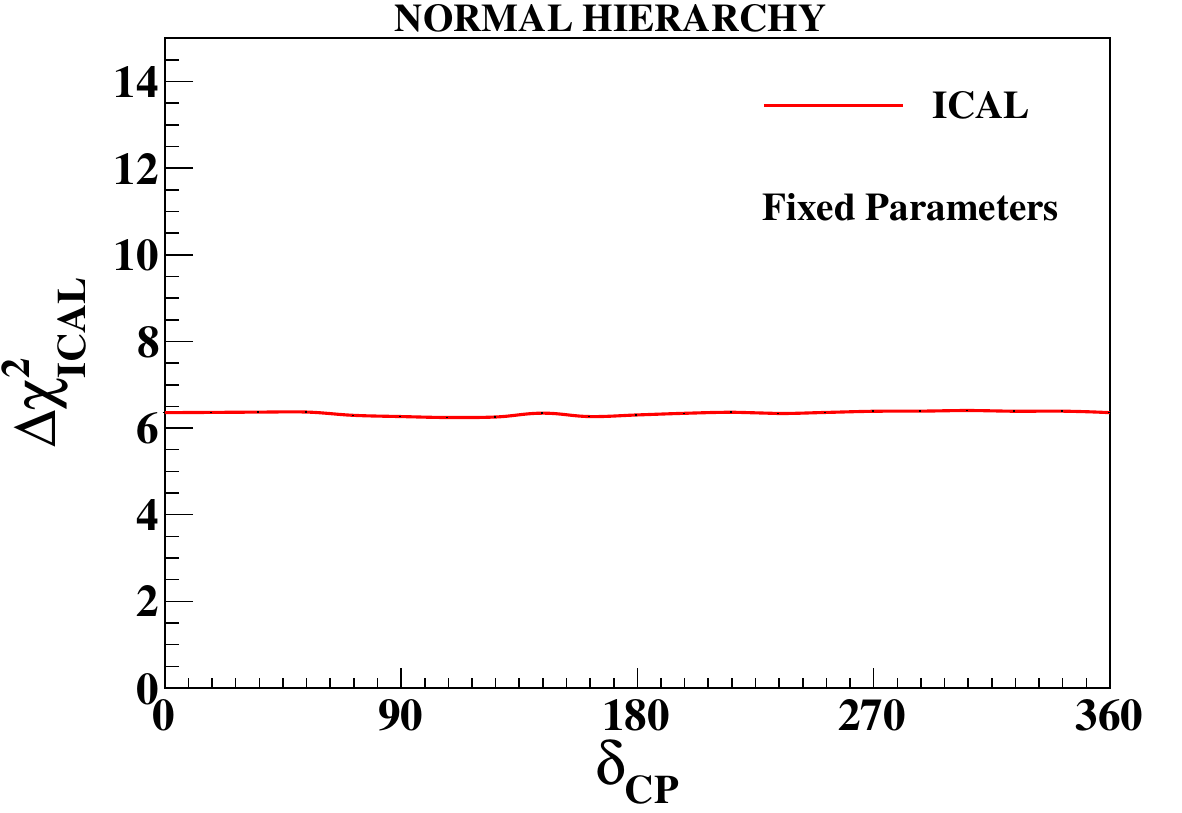}
\caption{The hierarchy sensitivity at different assumed values of
$\delta_{\rm CP}$ with 500 kt-yr of ICAL data, when the data is generated
with the actual value $\delta_{\rm CP}=0$. 
\label{fig:cp-insensitive}}
\end{figure}

In the analyses above, the true value for the CP-violating phase
$\delta_{\rm CP}$ has been taken to be $0^\circ$, and this parameter has
not been marginalised over. The reason for this is that atmospheric
neutrino data is insensitive to this phase 
\cite{Blennow:2012gj,Gandhi:2007td}. In order to illustrate this,
we show Fig.~\ref{fig:cp-insensitive} where the data with 500 kt-yr exposure
is generated for $\delta_{\rm CP}=0$, and the fit is tried for all 
$\delta_{\rm CP}$ values. It can be observed that the hierarchy 
sensitivity is not affected by what value of $\delta_{\rm CP}$
we choose to fit the data with. We have checked that the same results
hold for any actual $\delta_{\rm CP}$ value, that is the hierarchy 
sensitivity of ICAL is independent of the actual value of $\delta_{\rm CP}$.

The reason behind the insensitivity of ICAL to the actual value of $\dcp$ 
lies in the fact that the muon neutrinos at ICAL come dominantly from the 
original unoscillated muon neutrinos, while the muon neutrinos coming from 
the oscillated electron neutrinos are a smaller fraction, partly due to
the smallness of the electron neutrino flux, but mostly due to the small
value of the conversion probability $P_{e\mu}$ owing to the smallness of
$\theta_{13}$ and $\Delta m^2_{21}/\Delta m^2_{31}$ (see Appendix~\ref{app:prob}). 
The survival probability $P_{\mu\mu}$ therefore controls the oscillations 
detected at ICAL. In $P_{\mu\mu}$, the CP-violating phase $\dcp$ appears as 
a subdominant term that oscillates as $\cos\Delta$, where the oscillation 
phase $\Delta \equiv \Delta m^2_{31}L/(4E)$ (see Appendix~\ref{app:prob}).
If the distance $L$ travelled by the neutrino is uncertain by $\delta L$,
this phase becomes uncertain by $\delta \Delta \approx \Delta (\delta L/L)$.
For a few GeV neutrinos travelling distances of 
a few thousands of km through the earth matter, $\Delta \sim 10$. Hence only 
a 10\% uncertainty in $L$ can wipe out the information about the phase 
$\Delta$, and hence about the dependence on $\dcp$. It is indeed difficult 
to determine the
direction of incoming neutrino, and hence the value of $L$, for an 
atmospheric neutrino. Another added factor is the uncertainty in energy,
which also contributes to the uncertainty in $\Delta$ as $\delta \Delta
\approx \Delta (\delta E/E)$ in a similar way.

Note that this insensitivity of ICAL to the $\dcp$-dependent oscillating 
term is already a part of our analysis, so the good capability of ICAL
to distinguish between the two hierarchies is {\em in spite of} this
disadvantage. This may be contrasted with the hierarchy sentitivities of 
fixed baseline experiments like T2K or \nova, whose sensitivities to
the mass hierarchy depend crucially on the actual $\dcp$ value. 
This issue will be discussed later in Chapter~\ref{synergy}.

\subsection{Precision measurement of $|\Delta m_{32}^2|$ and 
$\sin^2\theta_{23}$}
\label{pm-2d}

\begin{figure}[htp]
 \centering
\subfloat[]{
\includegraphics[width=0.49\textwidth,height=0.35\textwidth]
{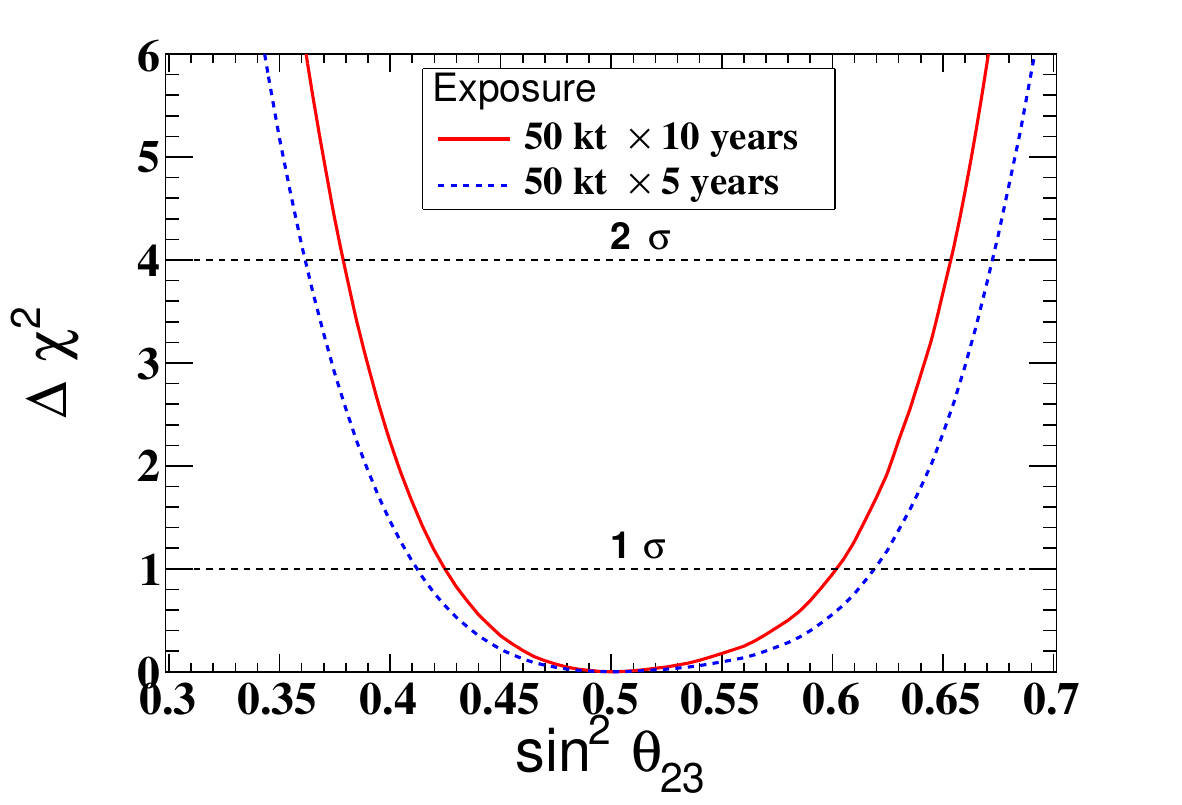}
\label{fig:fig_chi2_function_th23}
} 
\subfloat[]{
\includegraphics[width=0.49\textwidth,height=0.35\textwidth]
{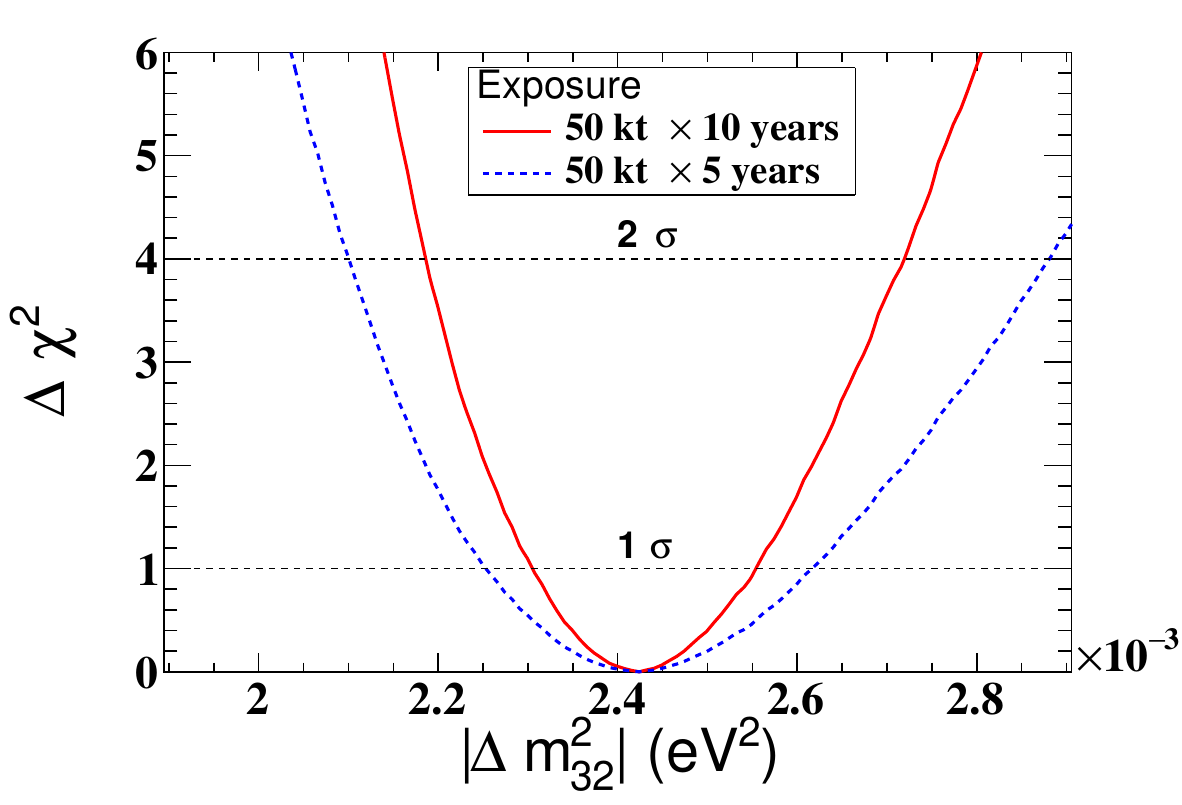}
\label{fig:fig_chi2_function_dms32}
}
\caption{The panel (a) shows the $\chi^2$ as a function of $\sin^2 \theta_{23}$ 
for $|\Delta m^2_{32}|$ = 2.4 $\times$ $10^{-3}$ $\rm eV^2$ and 
$\sin^2 \theta_{23}(\rm true)=0.5$. 
The panel (b) shows the $\chi^2$ as a function of $|\Delta m^2_{32}|$ for  
$\sin^2 \theta_{23}$ = 0.5 and $|\Delta m^2_{32}|(\rm true)$ = 2.4 
$\times$ $10^{-3}$ $\rm eV^2$. Only muon information has been used
\cite{Thakore:2013xqa}.}
\label{fig:fig_chi2_function}
\end{figure}

The reach of ICAL for the parameters $\sin^2 \theta_{23}$ and 
$|\Delta m^2_{32}|$ separately is shown in Fig.~\ref{fig:fig_chi2_function}
in terms of the $\Delta\chi^2$ compared to the best-fit value to the 
simulated data. The precision on these parameters may be quantified by
\begin{equation}
{\rm precision} = \frac{p_{max}-p_{min}}{p_{max}+p_{min}} \,,
\end{equation}
where $p_{max}$ and $p_{min}$ are the largest and smallest value of
the concerned oscillation parameter, determined at the given C.L. from
the atmospheric neutrino measurements at ICAL, for a given exposure.
We find that after 5 years of data taking, ICAL would
be able to measure $\sin^2\theta_{23}$ to a precision of $20\%$ and
$|\Delta m^2_{32}|$ to 7.4\% at 1$\sigma$. With 10 years exposure,
these numbers are expected to improve to 17\% and 5.1\% for 
$\sin^2 \theta_{23}$ and
$|\Delta m^2_{32}|$, respectively. The precision on  $\sin^2 \theta_{23}$
is mainly governed by the muon reconstruction efficiency and is expected
to improve with it. It will also improve as the systematic uncertainties 
are reduced. If the flux normalization error were to come down from 20\%
to 10\%, the precision on $\sin^2 \theta_{23}$ would improve to 14\%
for 10 years of exposure. Reducing the zenith angle error from 5\%
to 1\% would also improve this precision to $\sim$ 14\%. On the other
hand, the precision on $|\Delta m^2_{32}|$ is governed by the ability
of the detector to determine the value of $L/E$ for individual events
accurately. This depends on the energy- and $\cos \theta$- resolution
of the detector.

A few more detailed observations may be made from the $\chi^2$ plots in 
Fig.~\ref{fig:fig_chi2_function}. From Fig.~\ref{fig:fig_chi2_function_th23} 
one can notice that the precision on $\theta_{23}$ when it is in the first 
octant ($\sin^2 \theta_{23} < 0.5$) is slightly better than 
when it is in the second octant ($\sin^2 \theta_{23} > 0.5$), 
even though the muon neutrino survival probability depends 
on $\sin^2 2\theta_{23}$ at the leading order. 
This asymmetry about $\sin^2 \theta_{23}=0.5$ stems mainly
from the full three-flavor analysis that we have performed in this study. 
In particular, we have checked that the non-zero value of $\theta_{13}$ 
is responsible for the asymmetry observed in this figure.
On the other hand, $\chi^2$ asymmetry about the true value of 
$|\Delta m^2_{32}|$ observed in Fig.~\ref{fig:fig_chi2_function_dms32} is 
an effect that is present even with a two-flavor analysis. 

\begin{figure}[t]
\centering
\subfloat[Precision reach for 5 years run]{
\includegraphics[width=0.49\textwidth,height=0.35\textwidth]
{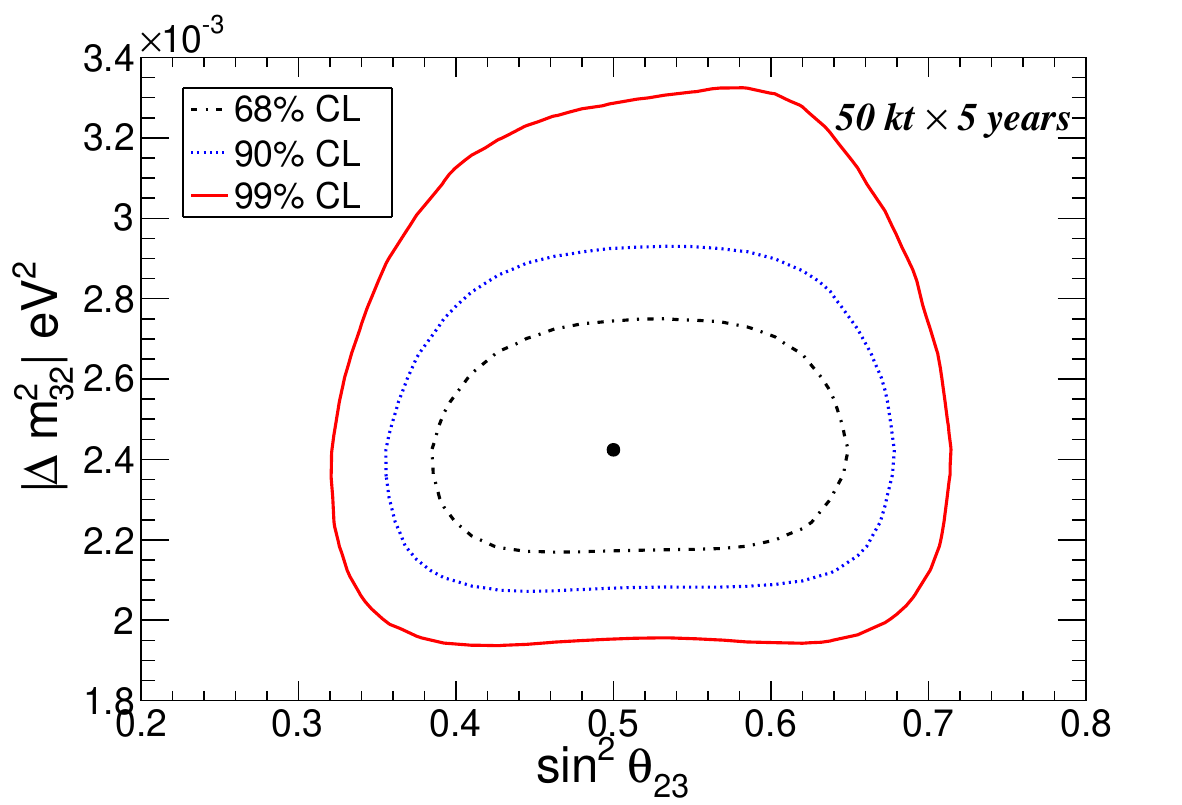}
\label{fig:5yrs}
 }
  \subfloat[Precision reach for 10 years run]{
\includegraphics[width=0.49\textwidth,height=0.35\textwidth]
{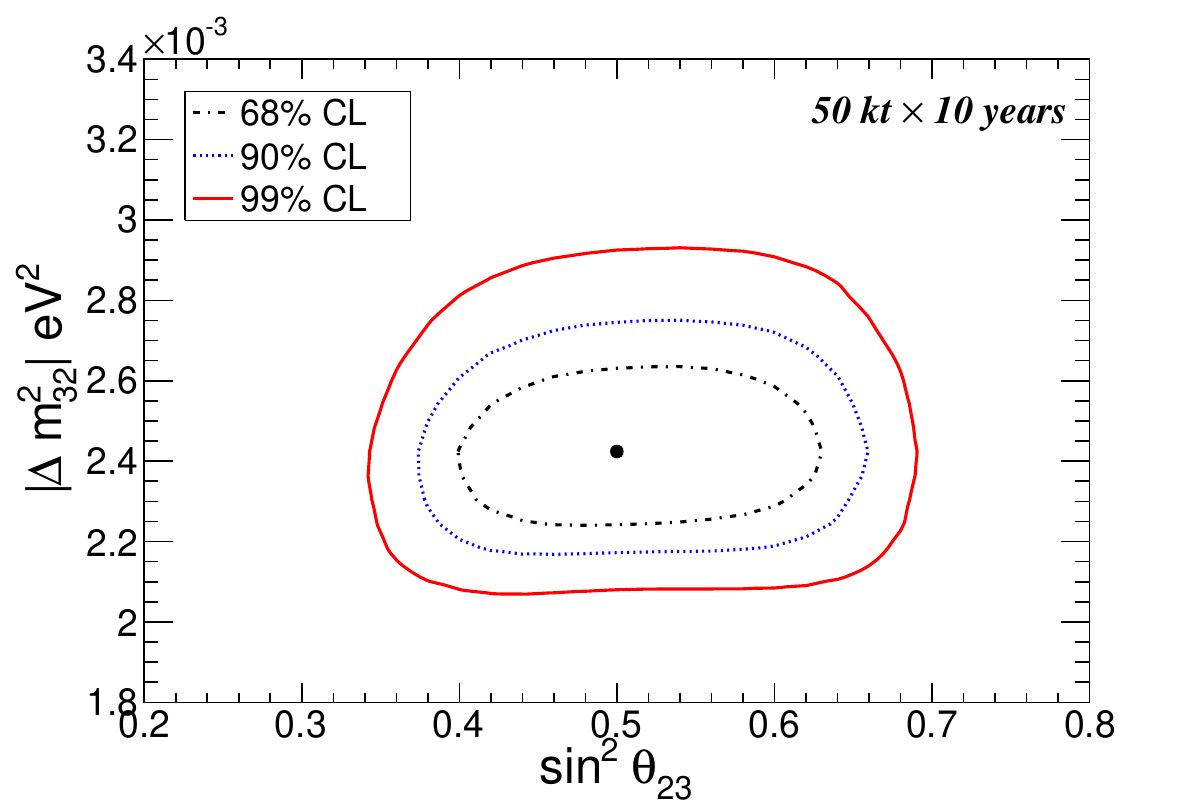}
\label{fig:10yrs}
 }
 \caption{The precision reach expected at ICAL in the 
$\sin^2\theta_{23}-|\Delta m^2_{32}|$ plane at various confidence levels, 
using only muon information. 
The black(broken), blue(dotted) and red(solid) lines show 68\%, 90\% and 
99\% C.L contours.
The true values of $\sin^2\theta_{23}$ and $|\Delta m^2_{32}|$ used for 
generating data are shown by 
the black dots. The true values of other parameters used are given in 
Table \ref{tab:best-fit}. Panel 
(a) is for five-year running of the 50 kt detector while (b) is for ten 
years exposure \cite{Thakore:2013xqa}.}
 \label{fig:fig_CL_ssq2th23_1d0}
\end{figure}

The precisions obtainable at ICAL for $\sin^2 \theta_{23}$ and 
$|\Delta m^2_{32}|$ are expected to be correlated. We therefore present the
correlated reach of ICAL for these parameters in Figs.~\ref{fig:5yrs}
and \ref{fig:10yrs}. As noted
above, our three-neutrino analysis should be sensitive to the octant
of $\theta_{23}$. Therefore we choose to present our results in terms
of $\sin^2 \theta_{23}$ instead of $\sin^2 2\theta_{23}$. Though the
constant-$\chi^2$ contours still look rather symmetric about $\sin^2
\theta_{23}=0.5$, that is mainly due to the true value of $\sin^2
\theta_{23}$ being taken to be 0.5. The values of  $\sin^2 \theta_{23}$
away from 0.5 would make the contours asymmetric and would give rise to
some sensitivity to the octant of $\theta_{23}$, as we shall see later.

\subsection{Sensitivity to the octant of $\theta_{23}$}
\label{octant-2d}

\begin{figure}[ht!] 
\centering
\subfloat[$\sin^2 2\theta_{23}$ = 0.9, first octant ($\sin^2\theta_{23}=0.342$)]
{\includegraphics[width=0.49\textwidth,height=0.33\textwidth]
{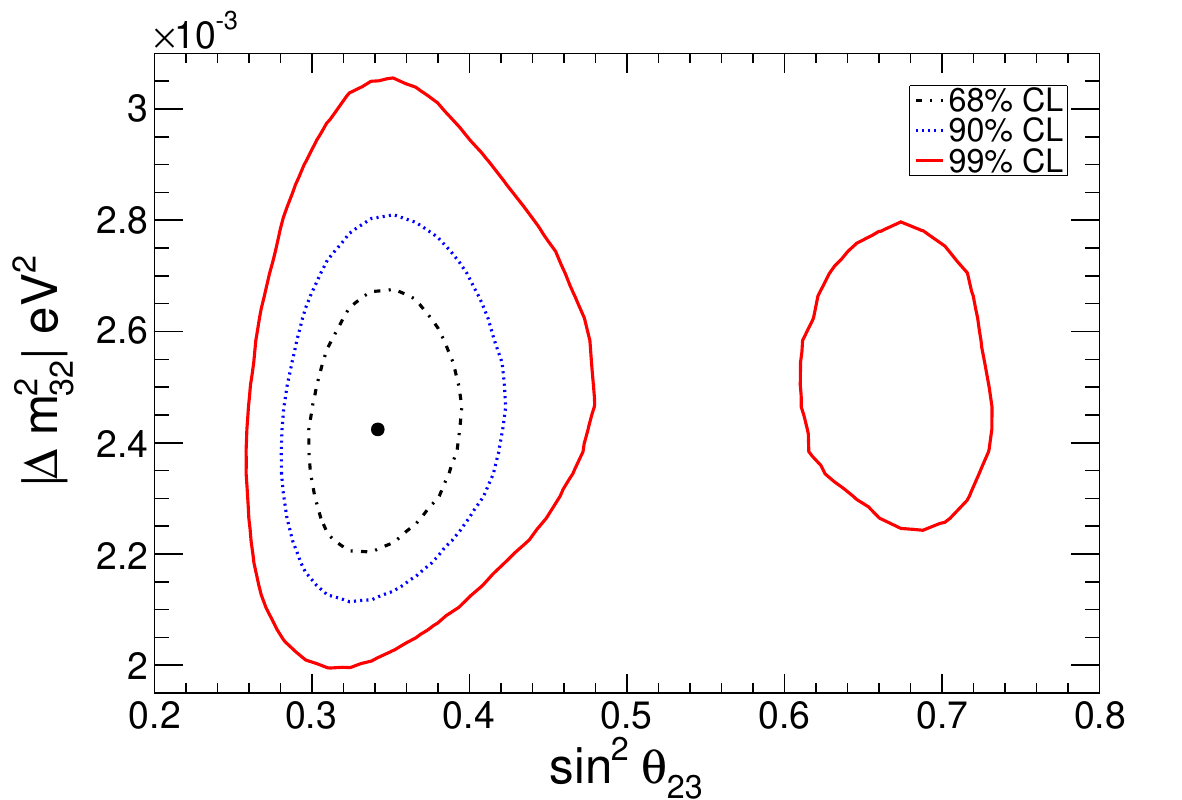}} 
\subfloat[$\sin^2 2\theta_{23}$ = 0.9, second octant ($\sin^2 \theta_{23}=0.658$)]
{\includegraphics[width=0.49\textwidth,height=0.33\textwidth]
{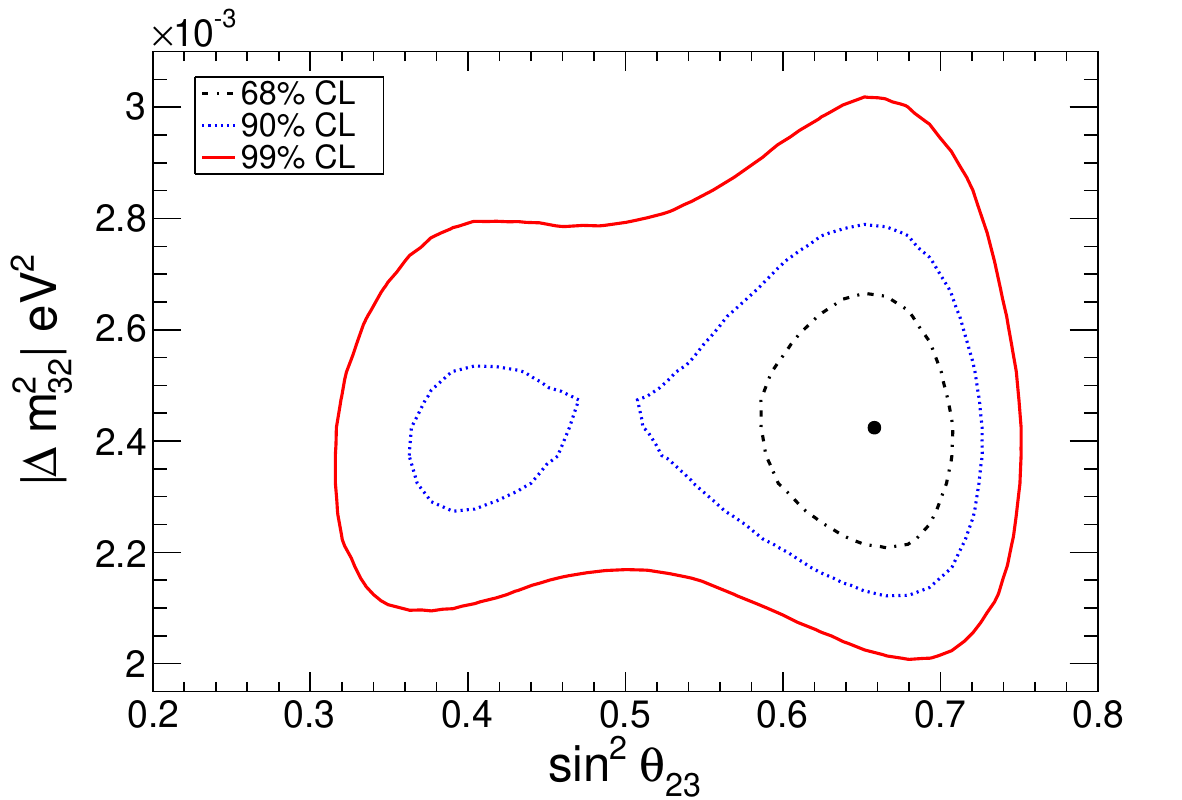}} \\
\subfloat[$\sin^2 2\theta_{23}$ =0.95, first octant ($\sin^2 \theta_{23}=0.388$)]
{\includegraphics[width=0.49\textwidth,height=0.33\textwidth]
{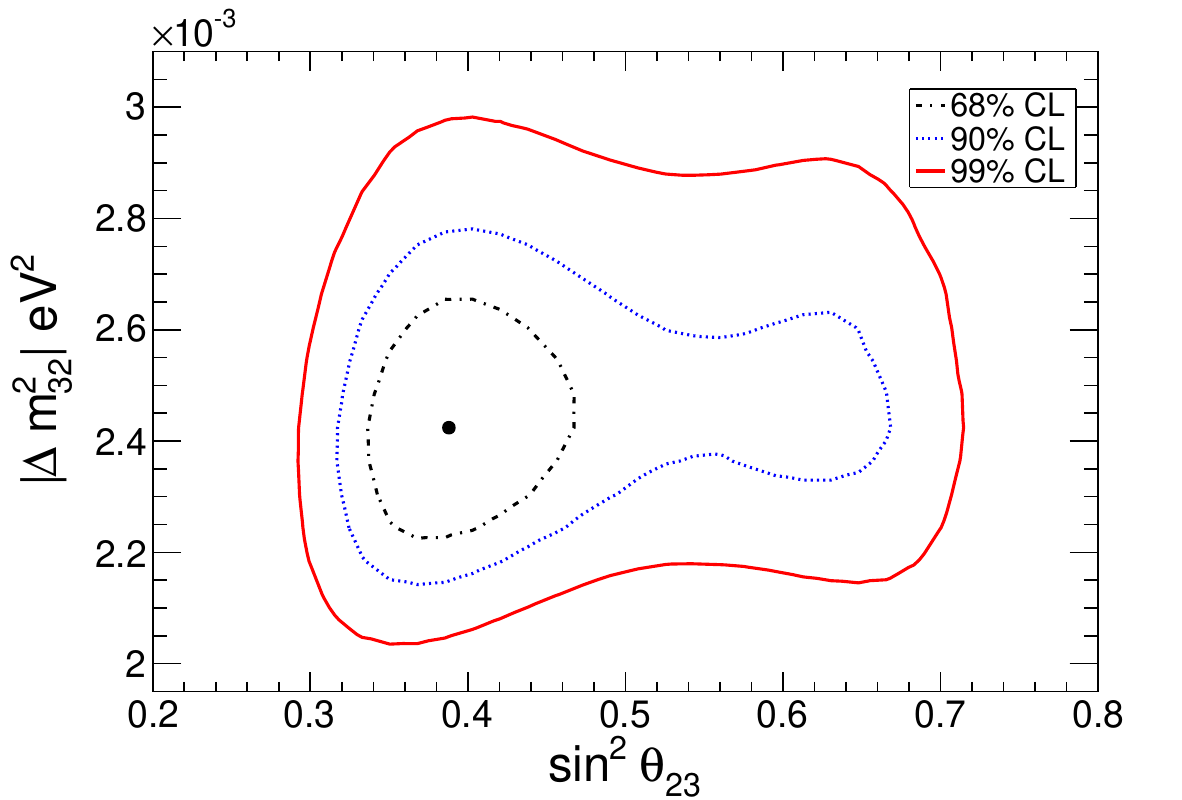}}
\subfloat[$\sin^2 2\theta_{23}$=0.95, second octant ($\sin^2  \theta_{23}=0.612$)]
{\includegraphics[width=0.49\textwidth,height=0.33\textwidth]
{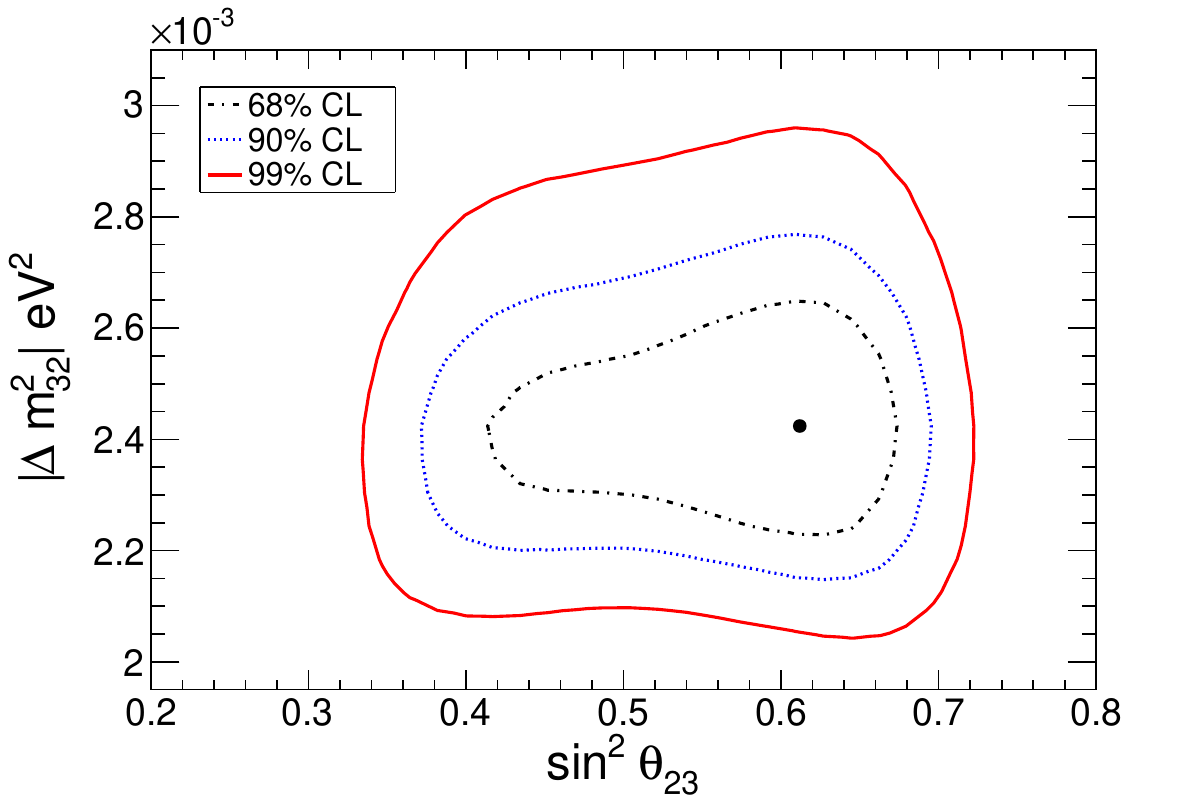}
\label{fig:fig_CL_ssqth23_0d95_2} } 
\caption{The projected reach in
the $\sin^2\theta_{23}-|\Delta m^2_{32}|$ plane for four different
non-maximal choices of $\theta_{23}$. The black(broken), blue(dotted)
and red(solid) lines show 68\%, 90\% and 99\% C.L. contours for 10 years
of 50 kt ICAL run, using only muon information. 
Note that we use normal hierarchy, and assume that
it is already known \cite{Thakore:2013xqa}.}
 \label{fig:fig_CL_octant_sensitivity}
\end{figure}

Earth matter effects in atmospheric neutrinos can be used to distinguish 
maximal from non-maximal $\theta_{23}$ mixing and can lead to the determination 
of the correct $\theta_{23}$ octant 
\cite{Choubey:2005zy,Barger:2012fx,Indumathi:2006gr}.  
We show in Fig.~\ref{fig:fig_CL_octant_sensitivity} the potential of 10 years 
of ICAL run for distinguishing a non-maximal value of $\theta_{23}$ from 
maximal mixing in the case where $\sin^2 2\theta_{23}=0.90$ 
($\sin^2 \theta_{23}$ = 0.342, 0.658) and $\sin^2 2\theta_{23}$ = 0.95 
($\sin^2 \theta_{23}$ = 0.388, 0.612). 
The figure shows that, if the value of $\theta_{23}$ is near the current 
3$\sigma$ bound and in the first octant, then it may be possible to exclude 
maximal mixing to 99\% C.L. with this 2-parameter analysis. If $\theta_{23}$ 
is in the second octant, or if $\sin^2 2\theta_{23}$ is larger than 0.9, 
the exclusion of the maximal mixing becomes a much harder task.

Figure~\ref{fig:fig_CL_octant_sensitivity} can also be used to quantify
the reach of ICAL for determining the correct octant of $\theta_{23}$,
if the value of $\sin^2 2\theta_{23}$ is known. This can be seen by
comparing the $\chi^2$ value corresponding to the true value of $\sin^2
\theta_{23}$, but in the wrong octant, with that corresponding to the true
value of $\sin^2 \theta_{23}$. We find that, for $\sin^2 2\theta_{23}
= 0.9$, i.e. just at the allowed 3$\sigma$ bound, the octant can be
identified at $>$95\% C.L.  with 10 years of ICAL run if $\theta_{23}$
is in the first octant. However if $\theta_{23}$ is in the second octant,
the identification of the octant would be much harder: $\theta_{23}$
in the wrong octant can be disfavored only to about 85\% C.L.. The
situation is worse if $\sin^2 2\theta_{23}$ is closer to unity.

The precision on $|\Delta m^2_{32}|$ will keep improving with ongoing
and future long baseline experiments. The inclusion of the information
may improve the chance of ICAL being able to identify deviation of
$\theta_{23}$ from maximal mixing and its octant to some extent.

Note that all the results in this
section (Sec.~\ref{mh-2d}) use only the information on muon energy
and direction. The addition of hadron information is expected to
improve the physics reach of ICAL in all aspects, and this will be
explored in the next sections.

\section{Analysis with $E_\nu$ and $\cos\theta_\mu$}
\label{sec:enu_thmu}

The addition of the value of hadron energy $E'_{\rm had}$, calibrated
with respect to the number of hadronic hits as described in
Chapter~\ref{response}, to the muon energy $E_\mu$ reconstructs
the incoming neutrino energy in a charged-current interaction. Since
it is the neutrino energy that appears in the neutrino oscillation
probabilities, it is conceivable that direct access to neutrino
energy will improve the reach of the ICAL to the oscillation
parameters.  

The analyses in this section and the next that use the hadron energy
information assume that the hits created by a muon and hadron can 
be separated with an 100\% efficiency by the ICAL particle reconstruction 
algorithm. Then, whenever a muon is reconstructed, the corresponding hadron 
hits can always be considered to be a hadron shower, so that the energy
$E'_{\rm had}$ can be reconstructed. This implies that the neutrino event 
reconstruction efficiency is the same as the muon reconstruction effciency. 
(Note that the calibration of $E'_{\rm had}$ against the number of hadron 
shower hits also allows for the possibility of no hits observed in the 
hadron shower.)  Finally, the background hits 
coming from other sources such as the neutral-current events, charged-current
$\nu_e$ events, cosmic muons, noise, have not been taken into account so far. 
The systematics due to these effects will have to be taken care of in future, 
as the understanding of the ICAL detector improves.

The neutrino energy $E_\nu$ is reconstructed
as the sum of the reconstructed muon energy $E_\mu$ 
(Sec.~\ref{muon-resp}) and the calibrated hadron energy $E'_{\rm had}$
(Sec.~\ref{had-resp}). The muon energy resolutions and zenith angle
resolutions have been implemented by smearing the true muon energy and 
direction of each $\mu^{+}$ and $\mu^{-}$ event, as discussed earlier. 
Energy of hadron events has also been smeared separately, following the 
discussion in Sec~\ref{had-resp}. The reconstructed neutrino energy is 
then taken as the sum of reconstructed muon energy and hadron energy,
with the uncertainties calculated separately for each event.
The events are then binned in $(E_\nu, \cos\theta_\mu)$ bins:
15 equal $E_\nu$ bins in the range $[0.8,5.8]$ GeV, 5 equal $E_\nu$
bins in the range $[5.8,10.8]$ GeV, and 20 equal
$\cos\theta_\mu$ bins in the range $[-1.0,1.0]$.
The same analysis as in Sec.~\ref{sec:emu_thmu} is then performed, 
marginalizing over the $3\sigma$ allowed ranges of $|\Delta m^2_{32}|$,
$\sin^2\theta_{23}$, and $\theta_{13}$, with a 10\% prior
on $\theta_{13}$. The systematic errors have been taken into account using
the method of pulls as before. 

\begin{figure}[t]
\centering
\includegraphics[width=0.49\textwidth,height=0.35\textwidth]
{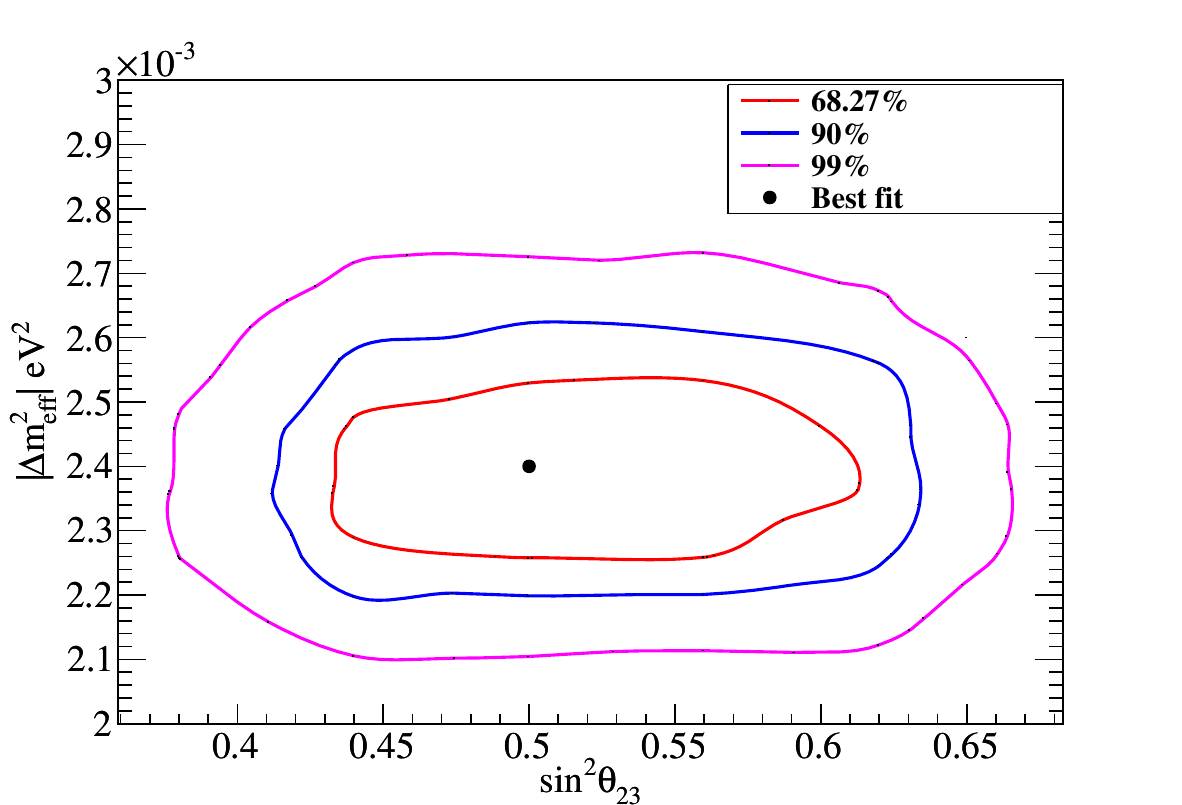}
\caption{The reach of the ICAL for precision measurements of 
$|\Delta m^2_{\rm eff}|$ and $\sin^2 \theta_{23}$, using information on
$E_\nu$ and $\cos\theta_\mu$, for an exposure of 500 kt-yr. The contours
with 68$\%$, 90$\%$ and 99$\%$ confidence level are shown
\cite{Kaur:2014rxa}.
\label{2d-enu_thmu}}
\end{figure}

Figure~\ref{2d-enu_thmu} shows the sensitivity of ICAL to the
atmospheric mixing parameters $|\Delta m^2_{\rm eff}|$ and $\sin^2 \theta_{23}$,
where 
$ \Delta m^{2}_{\rm eff} \equiv \Delta m^{2}_{32}-(\cos^{2}\theta_{12}-\cos
\delta_{\rm CP}\sin\theta_{32}\sin2\theta_{12}\tan\theta_{23})\Delta m^{2}_{21}$ 
is the  effective value of $\Delta m^2_{\rm atm}$ relevant for the two-neutrino
analysis of atmospheric neutrino oscillations 
\cite{Nunokawa:2005nx,deGouvea:2005hk}. 
It is seen that with 10 years of data and the ($E_\nu,\cos\theta_\mu$)
analysis technique, the 50 kt ICAL can measure the magnitude of the 
atmospheric neutrino mass squared difference to 4\% 
and $\sin^2\theta_{23}$ to 13\%, at $1\sigma$. A comparison 
of these numbers against those obtained using the 
($E_\mu,\cos\theta_\mu$) analysis shows that that using the reconstructed 
neutrino energy improves the precision on $\sin^2\theta_{23}$ 
and $|\Delta m^2_{32}|$ by about 20\%.
However the addition of the coarsely measured hadron energy to
the accurately measured muon energy results in some loss of
information, which results in some degradation in the performance
for mass hierarchy identification.

\section{Analysis using correlated Information on $E_\mu$, $\cos\theta_\mu$, 
and $E^{\prime}_{\rm had}$}
\label{3d-analysis}

The preceeding analyses have been performed by using only the 
information on muon direction, and the energy of muons or neutrinos.
However in each CC event at ICAL, the information on $E_\mu$, 
$\cos\theta_\mu$  and $E^{\prime}_{had} \equiv E_\nu - E_\mu $ 
is available independently. Thus, the inelasticity 
$y \equiv E'_{\rm had}/E_\nu$ is an additional measurable quantity 
in each event. This advantage of ICAL may be exploited
to enhance its physics reach \cite{Devi:2014yaa}, 
as shall be seen in this section.
This is implemented by performing an  analysis that employs binning 
in all these three quantities, so that no information is lost and all 
correlations are taken care of.

\subsection{The three-dimensional binning}
\label{3d-binning}

For event generation and inclusion of oscillation, we use the same 
procedure as described in Sec.~\ref{sec:emu_thmu}.
After incorporating the detector response for muons and hadrons
\cite{Chatterjee:2014vta,Devi:2013wxa}, 
the measured event distribution in terms of 
($E_\mu$, $\cos\theta_\mu$, $E'_{\rm had}$) is obtained.
A three-dimensional binning scheme using the measured quantities 
$E_\mu$, $\cos\theta_\mu$ and $E^{\prime}_{had}$ is employed for the
$\chi^2$ analysis.  In order to ensure significant statistics in each
bin and also to avoid large number of bins, we use a non-uniform binning
scheme for each polarity as shown in Table~\ref{table:3d-bin}. Thus for
each polarity, one has a total of ($10 \times 21 \times 4$) = 840 bins.

\begin{table}[htp] 
\centering 
\begin{tabular}{|c| c| c| c|} 
\hline 
Observable & Range & Bin width & Total bins \\ 
\hline
$E_{\mu}$ (GeV) & \makecell[c]{1 -- 4 \\ 4 -- 7 \\~~7 -- 11} & 
\makecell[c]{0.5 \\ 1 \\4} & 
$\left.\begin{tabular}{l}
6 \\ 3 \\ 1
\end{tabular}\right\}$  10\\
\hline
$\cos\theta_\mu$  & \makecell[c]{~~(-1) -- (-0.4) \\ 
(-0.4) -- 0~~~~~~\\0 -- 1} & \makecell[c]{0.05 \\ 0.1 \\0.2} & 
$\left.\begin{tabular}{l}
12 \\ 4 \\ 5
\end{tabular}\right\}$  21\\
\hline
$E^{\prime}_{had}$ (GeV)  & \makecell[c]{0 -- 2 \\ 2 -- 4\\ ~~4 -- 15} & \makecell[c]{1 \\ 2 \\11} & 
$\left.\begin{tabular}{l}
2 \\ 1 \\ 1
\end{tabular}\right\}$  4\\ \hline
\end{tabular}
\caption{The binning scheme in $E_\mu$, $\cos\theta_\mu$, and $E^{\prime}_{had}$ for each polarity}
\label{table:3d-bin}
\end{table}

For the statistical analysis, we define the Poissonian $\chi^2_-$ 
for the $\mu^-$ events as
\begin{eqnarray}
\chi^{2}_{-}
& = & {\min_{\xi_l}} \sum_{i=1}^{N_{E_{had}^{\prime}}} \sum_{j=1}^{N_{E_{\mu}}} 
\sum_{k=1}^{N_{\cos \theta_{\mu}}} \left[ 2(N_{ijk}^{\rm theory} -  N_{ijk}^{\rm data}) \right.
\nonumber \\
& & \phantom{ {\min_{\xi_l}} \sum_{i=1}^{N_{E_{had}^{\prime}}} \sum_{j=1}^{N_{E_{\mu}}} 
\sum_{k=1}^{N_{\cos \theta_{\mu}}} \left[ \right.}
\left. - 2 N_{ijk}^{\rm data} \: \ln \left( \frac{N_{ijk}^{\rm theory}}{N_{ijk}^{\rm data}} 
\right) \right]
+ \sum_{l=1}^{5} \xi_{l}^{2}\,,
\label{chisq}
\end{eqnarray}
where
\begin{equation}
N^{\rm theory}_{ijk} = N^{0~{\rm theory}}_{ijk}\bigg(1 + \sum_{l=1}^{5} \pi_{ijk}^{l} \xi_{l}\bigg)\,.
\label{n-theory-definition}
\end{equation}
Here $N_{ijk}^{\rm theory}$ and $N_{ijk}^{\rm
data}$ are the expected and observed number of $\mu^-$ events in a
given ( $E_{\mu}$, $\cos \theta_{\mu}$, $E_{\rm had}^{\prime}$)
bin, with $N_{E_{had}^{\prime}}$ = 4, $N_{E_{\mu}}$ = 10, and $N_{\cos\theta_{\mu}}$ = 21.
The systematic uncertainties have been included by using the pull method
\cite{Fogli:2002au,Huber:2002mx},
which uses the ``pull'' variables $\xi_l$ \cite{Ghosh:2012px,Thakore:2013xqa}.  

For $\mu^+$ events, $\chi^2_{+}$ is similarly defined.
The total  $\chi^2$ is obtained as
\begin{equation}
\chi^2_{\rm ~ICAL} =  \chi^{2}_{-} +  \chi^{2}_{+} + \chi^2_{\rm prior} \,,
\label{total-chisq}
\end{equation}
where
\begin{equation}
\chi^2_{\rm prior} \equiv \left(\frac{
\sin^2 2\theta_{13}- \sin^2 2\theta_{13}{\rm (true)}}{0.08 \times
\sin^2 2\theta_{13}{\rm (true)}} \right)^2 \; . 
\end{equation}
The 8\% prior on $\sin^2 2\theta_{13}$ corresponds to the current
accuracy in the measurement of this quantity.
No prior on $\theta_{23}$ or $\Delta m^2_{32}$ is used, since these 
parameters will be directly measured at the ICAL. 

While implementing the minimization procedure, $\chi^2_{\rm ~ICAL}$ 
is first minimized with respect to the pull variables $\xi_l$, 
and then marginalized over the $3\sigma$ allowed ranges of 
the oscillation parameters 
$\sin^2 \theta_{23}$, $\Delta m^2_{\rm eff}$ and $\sin^2 2\theta_{13}$,
wherever appropriate. 
We do not marginalize over $\delta_{\rm CP}, \Delta m^2_{21}$ and
$\theta_{12}$ since they have negligible effect on the relevant
oscillation probabilities at ICAL \cite{Blennow:2012gj,Gandhi:2007td}.
Also, $\delta_{\rm CP}=0$ throughout the analysis.

\subsection{Mass hierarchy sensitivity}
\label{3d-mh}

We quantify the statistical significance of the analysis to rule out the
wrong hierarchy by
\begin{equation}
\Delta \chi^2_{\rm ~ICAL-MH} = \chi^2_{\rm ~ICAL} (\rm{false~ MH}) - 
\chi^2_{\rm ~ICAL} (\rm{true ~MH}) .
\label{mh_chi2_def}
\end{equation}
Here, $\chisqical(\rm{true ~MH})$ and $\chisqical(\rm{false ~MH})$
are obtained by performing a fit to the ``observed'' data assuming
true and false mass hierarchy, respectively.

\begin{figure}[h]
\centering
\includegraphics[width=0.49\textwidth,height=0.35\textwidth]
{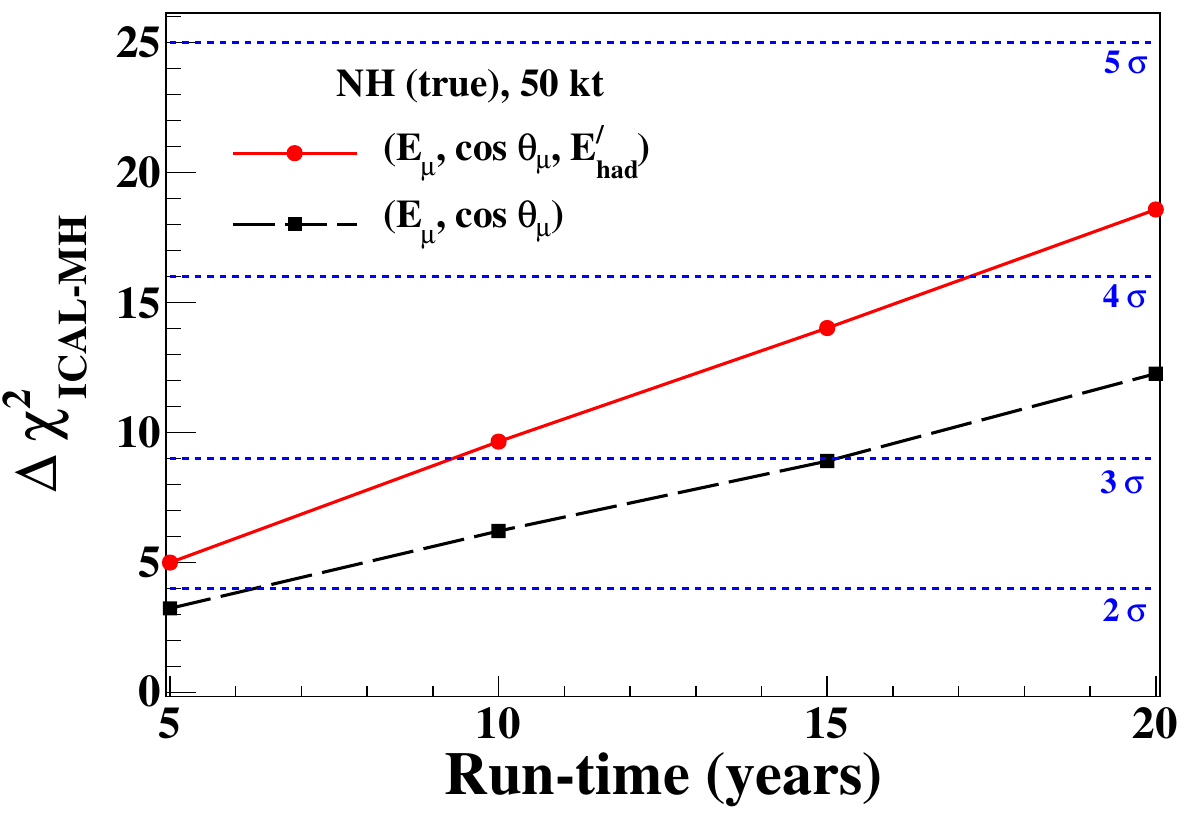}
\includegraphics[width=0.49\textwidth,height=0.35\textwidth]
{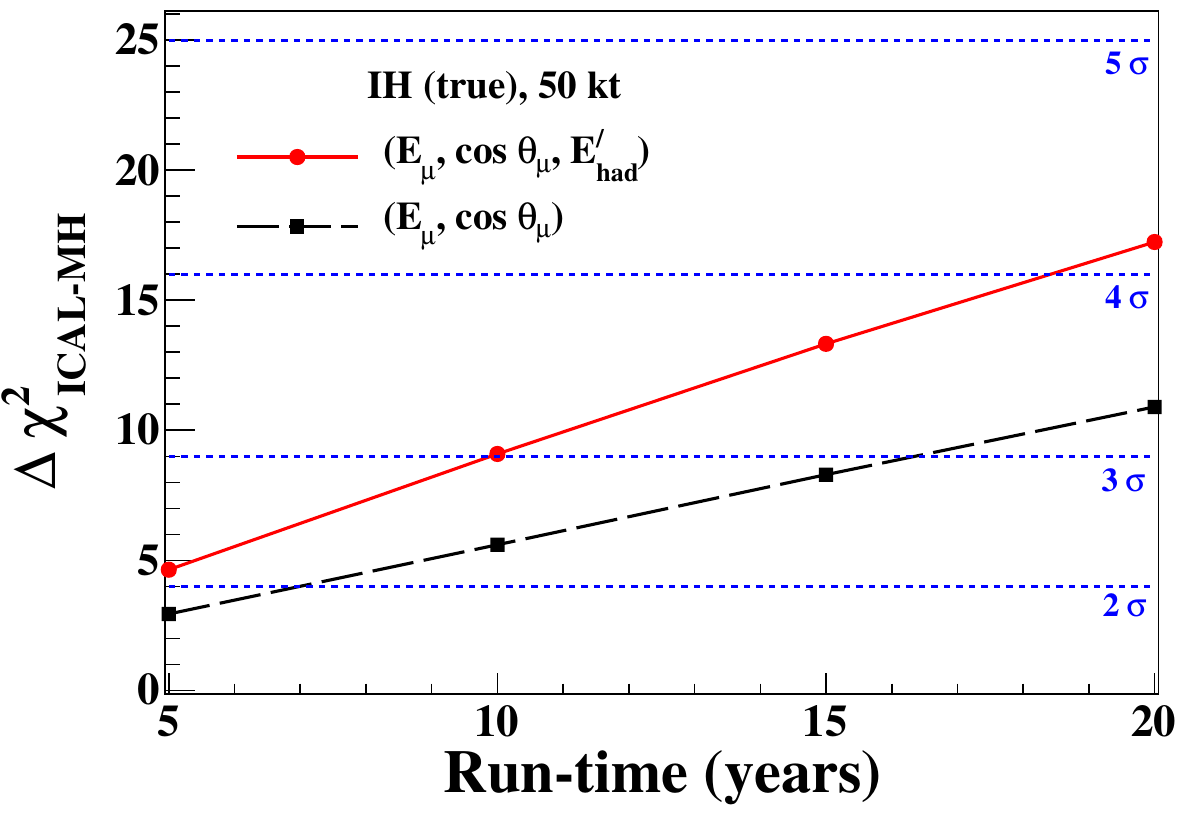}
\caption{$\chisqmh$ as a function of the exposure 
assuming NH (left panel) and IH (right panel) as true hierarchy.
The line labelled $(E_\mu, \cos\theta_\mu)$ denotes results without
including hadron information, while
the line labelled $(E_\mu, \cos\theta_\mu,E'_{\rm had})$ denotes 
improved results after including hadron energy information \cite{Devi:2014yaa}.
Here we have taken $\stch$(true) = 0.1 and $\sa$(true) = 0.5.}
\label{hierarchy}
\end{figure}

Figure~\ref{hierarchy} shows the mass hierarchy sensitivity of ICAL 
as a function of the run-time of the experiment. 
It is found that after including the hadron energy information, 
10 years of running of the 50 kt ICAL can rule out the wrong 
hierarchy with  $\chisqmh \approx 9.7$ (for true NH), and
$\chisqmh \approx 9.1$ (for true IH).
In other words, the wrong hierarchy can be ruled out to 
about 3$\sigma$ for either hierarchy. 
If the true values of $\theta_{23}$ and $\theta_{13}$ are varied
over their allowed $3\sigma$ range, the corresponding range for
$\chisqmh$ after 10 years is 7--12.
Compared to the results without using hadron information,
with the same binning scheme, the value of 
$\chisqmh$ increases by about 40\% when the correlated
hadron energy information is added.
This improvement is not merely due to using additional bins compared 
to the muon-only analysis, as can be checked by comparing the results
with those in Sec.~\ref{sec:emu_thmu}.

\subsection{Precision Measurement of Atmospheric Parameters}
\label{PM-results}

In order to quantify the precision in the measurements of a parameter 
$\lambda$ (here $\lambda$ may be $\sin^2\theta_{23}$ or $\Delta m^2_{32}$ 
or both), we use the quantity 
\begin{equation}
\chisqpm(\lambda) = 
\chisqical(\lambda) - \chi^2_0 \; ,
\end{equation}
where $\chi^2_0$ is the minimum value of $\chisqical$ in the
allowed parameter range.

\begin{figure}[h]
\centering
\includegraphics[width=0.49\textwidth,height=0.35\textwidth]
{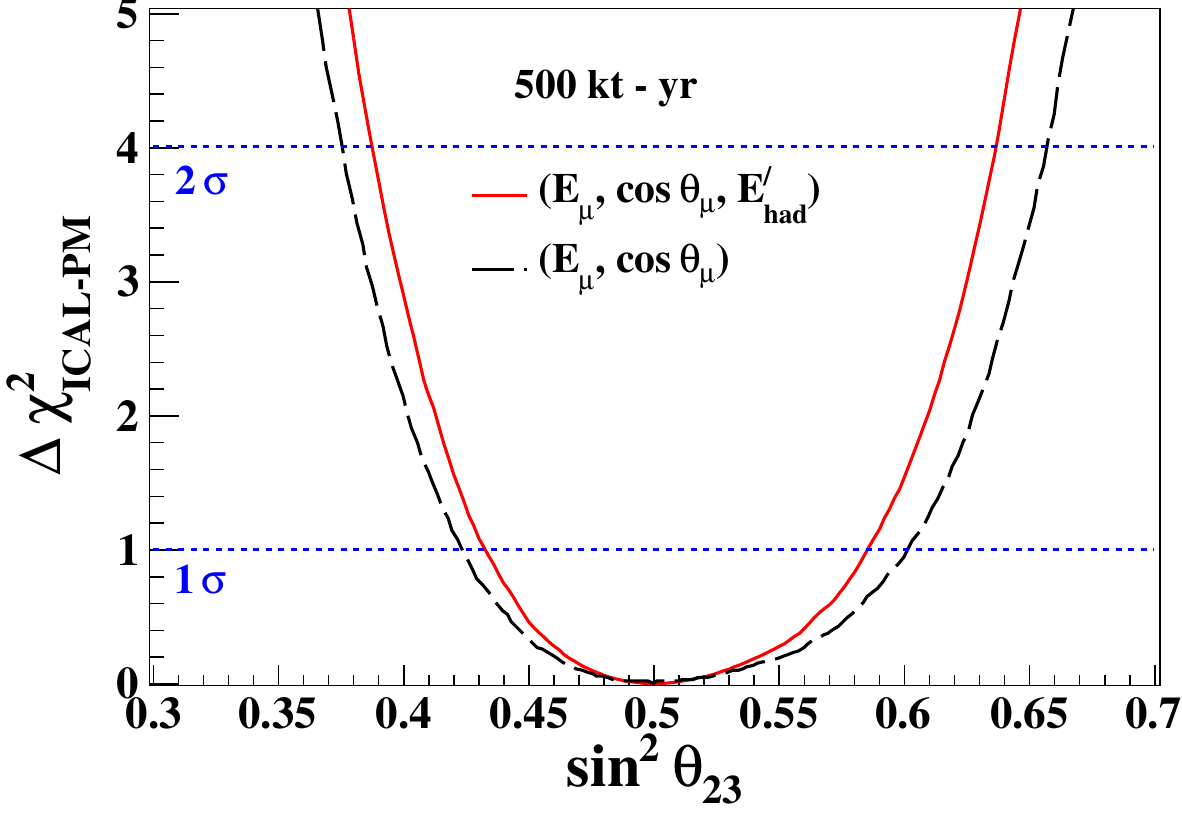}
\includegraphics[width=0.49\textwidth,height=0.35\textwidth]
{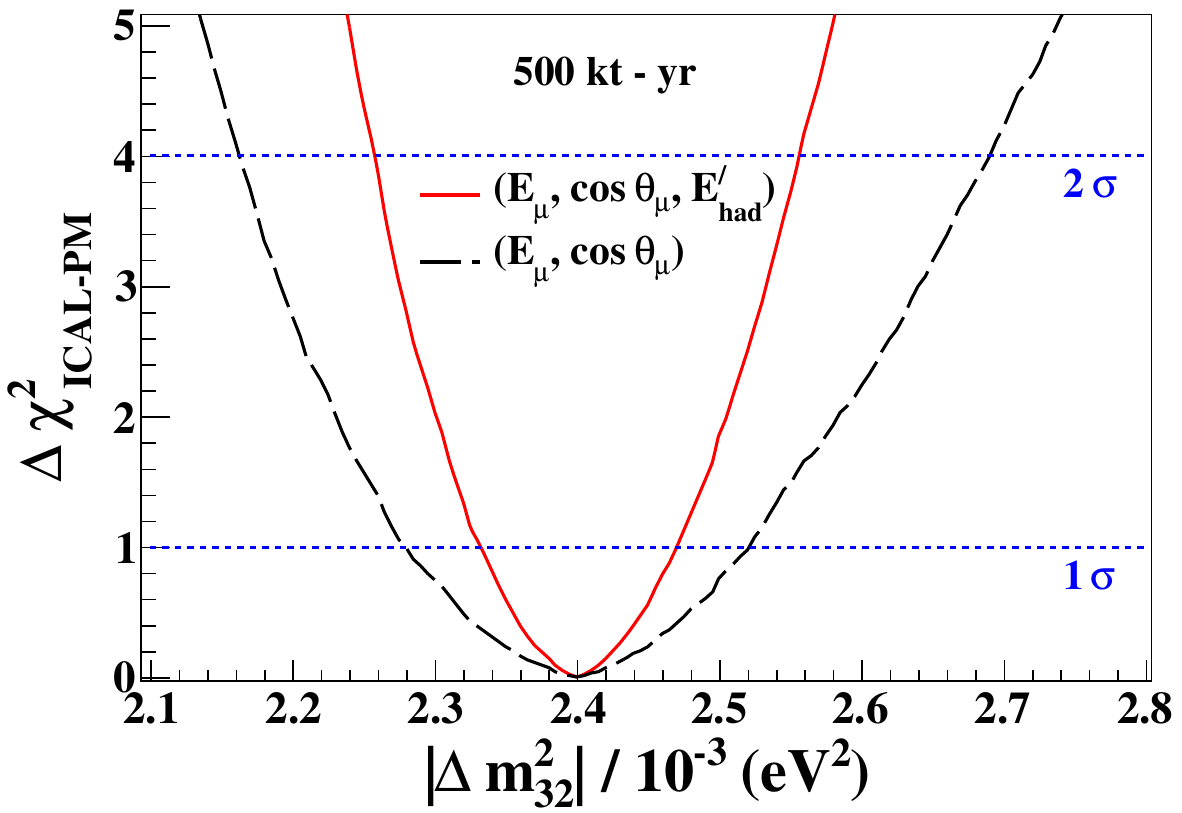}
\caption{The left panel shows  $\chisqpm(\sa)$ and 
the right panel shows $\chisqpm(|\Delta m^2_{32}|)$, assuming
NH as true hierarchy.
The lines labelled $(E_\mu, \cos\theta_\mu)$ denote results without
including hadron information, while
the lines labelled $(E_\mu, \cos\theta_\mu,E'_{\rm had})$ denote 
improved results after including hadron energy information \cite{Devi:2014yaa}.}
\label{pm-1d}
\end{figure}

The two panels of Fig.~\ref{pm-1d}, show the sensitivity of ICAL
to the two parameters $\sin^2\theta_{23}$ 
and $|\Delta m^2_{32}|$ separately, where the other 
parameter has been marginalized over, along with 
$\theta_{13}$ and the two possible mass hierarchies.
While the figure shows the results for NH as the true hierarchy,
the results with true IH are almost identical.
It may be observed from the figure that with the inclusion of hadron 
energy information, 500 kt-yr of ICAL exposure would be able to measure  
$\sin^2\theta_{23}$ to a $1\sigma$ precision of 12\% and 
$|\Delta m^2_{32}|$ to a $1\sigma$ precision of 2.9\%. 
This may be compared with the muon-only analysis with identical
$(E_\mu, \cos\theta_\mu)$  binning, which gives the precisions of 
13.7\% and 5.4\%, respectively.

\begin{figure}[h!]
\centering
\includegraphics[width=0.49\textwidth,height=0.35\textwidth]
{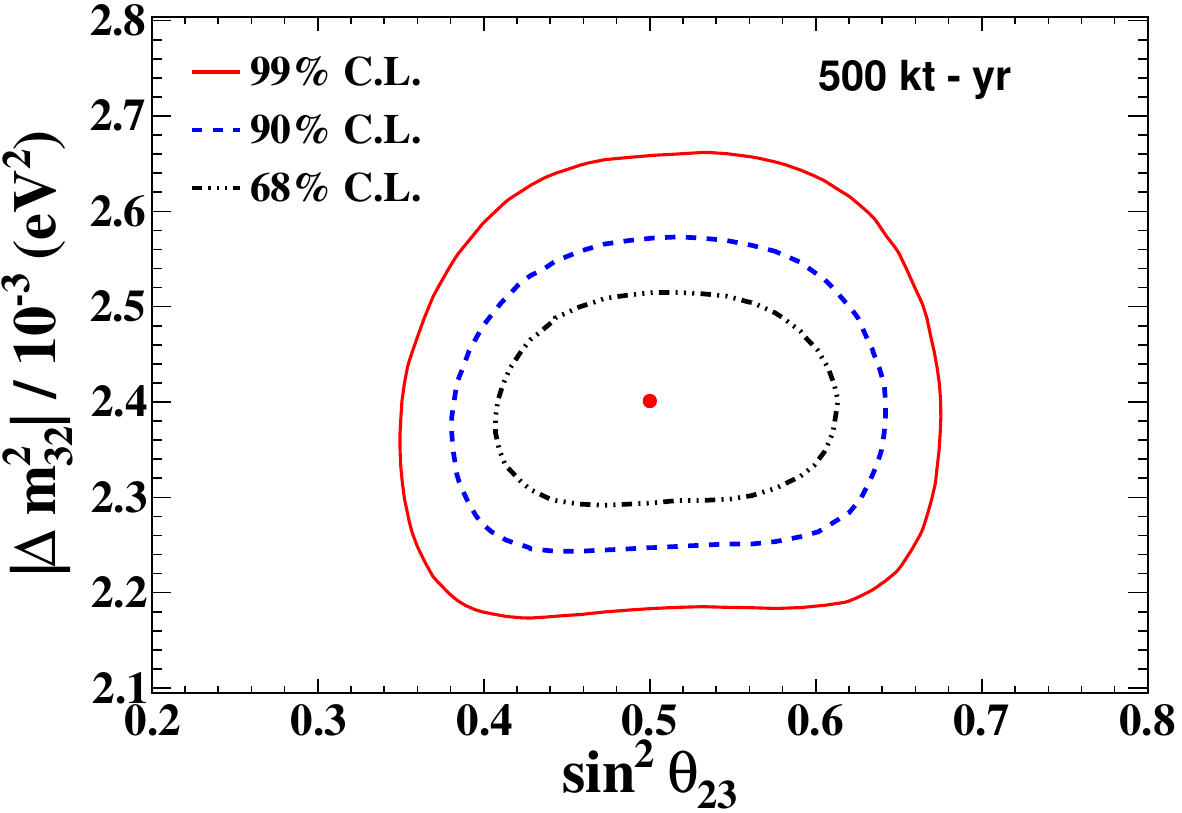}
\includegraphics[width=0.49\textwidth,height=0.35\textwidth]
{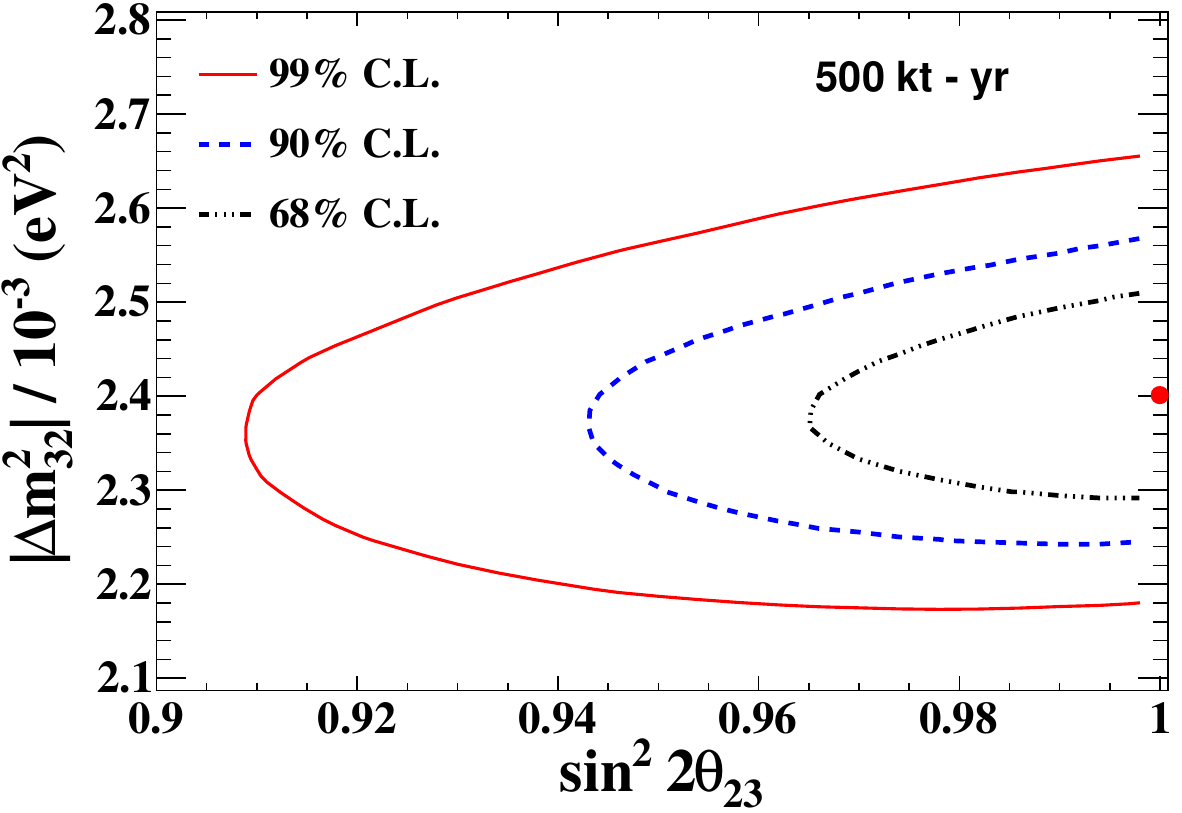}
\caption{$\chisqpm$ confidence level contours (2 dof) 
in the the $\sa$---$|\Delta m^2_{32}|$ plane (left panel) and in the 
$\sta$---$|\Delta m^2_{32}|$ plane (right panel),
using the hadron energy information, with NH as the true hierarchy.
The true choices of the parameters have been marked with a dot
\cite{Devi:2014yaa}.} 
\label{pm-ssqth-ssq2th-mmix}
\end{figure}

Figure~\ref{pm-ssqth-ssq2th-mmix} shows the $\chisqpm$ 
contours in the $\sa$---$|\Delta m^2_{32}|$ 
plane (left panel) and in the $\sta$---$|\Delta m^2_{32}|$ plane (right panel),
using the hadron energy information.
Here the true value of $\theta_{23}$ has been taken to be maximal,
so the contours in the left panel are almost symmetric in $\sa$. 

\begin{figure}[h!]
\centering
\includegraphics[width=0.75\textwidth]
{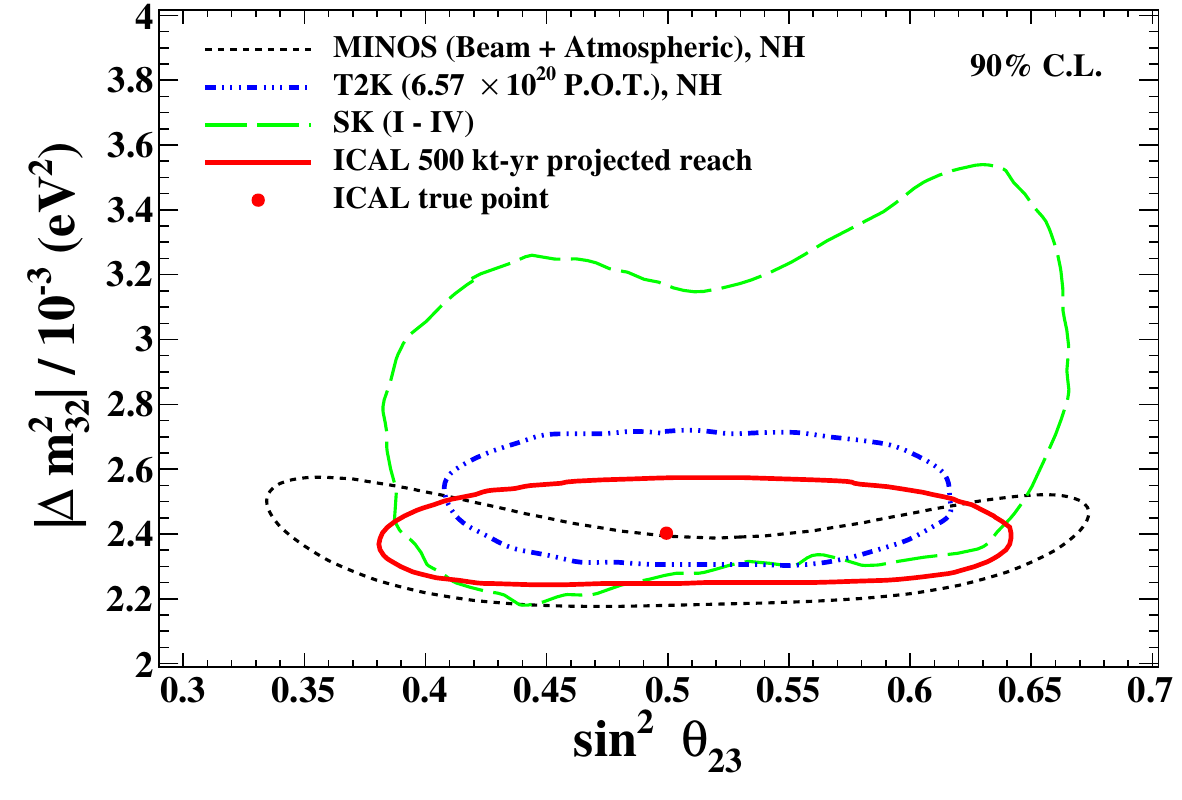}
\caption{90\% C.L. contours in the $\sa$---$|\Delta m^2_{32}|$ plane
(2 dof): the current limits from SuperKamiokande \cite{Himmel:2013jva},
MINOS \cite{Adamson:2014vgd} and T2K \cite{Abe:2014ugx}
have been shown along with the ICAL reach for the exposure of
500 kt-yr, assuming true NH. 
The true choices of the parameters for ICAL have been marked with a dot
\cite{Devi:2014yaa}.} 
\label{3d-pm-ssq2th-mmix}
\end{figure}

Figure~\ref{3d-pm-ssq2th-mmix} shows the comparison of the projected 
90\% C.L. precision reach of ICAL 
(500 kt-yr) in $\sa$--$|\Delta m^2_{32}|$ plane
with other experiments \cite{Himmel:2013jva,Adamson:2014vgd,Abe:2014ugx}.
Using hadron energy information, the ICAL will be able to achieve a
$\sa$ precision comparable to the current precision for Super-Kamiokande 
\cite{Himmel:2013jva} or T2K \cite{Abe:2014ugx}, and the $|\Delta m^2_{32}|$ 
precision comparable to the MINOS reach \cite{Adamson:2014vgd}. 
Of course, some of these experiments would have collected much more 
statistics by the time ICAL would have an exposure of 500 kt-yr, so
the ICAL will therefore not be competing with these experiments 
for the precision measurements of these mixing parameters. However
the ICAL measurements will serve as complementary information 
for the global fit of world neutrino data. 
Note that, as compared to the atmospheric neutrino analysis at
Super-Kamiokande, the ICAL precision on $|\Delta m^2_{32}|$ is far superior.
This is due to the better precision in the reconstruction
of  muon energy and direction at ICAL.

\begin{figure}[htp!]
\centering
\includegraphics[width=0.49\textwidth,height=0.35\textwidth]
{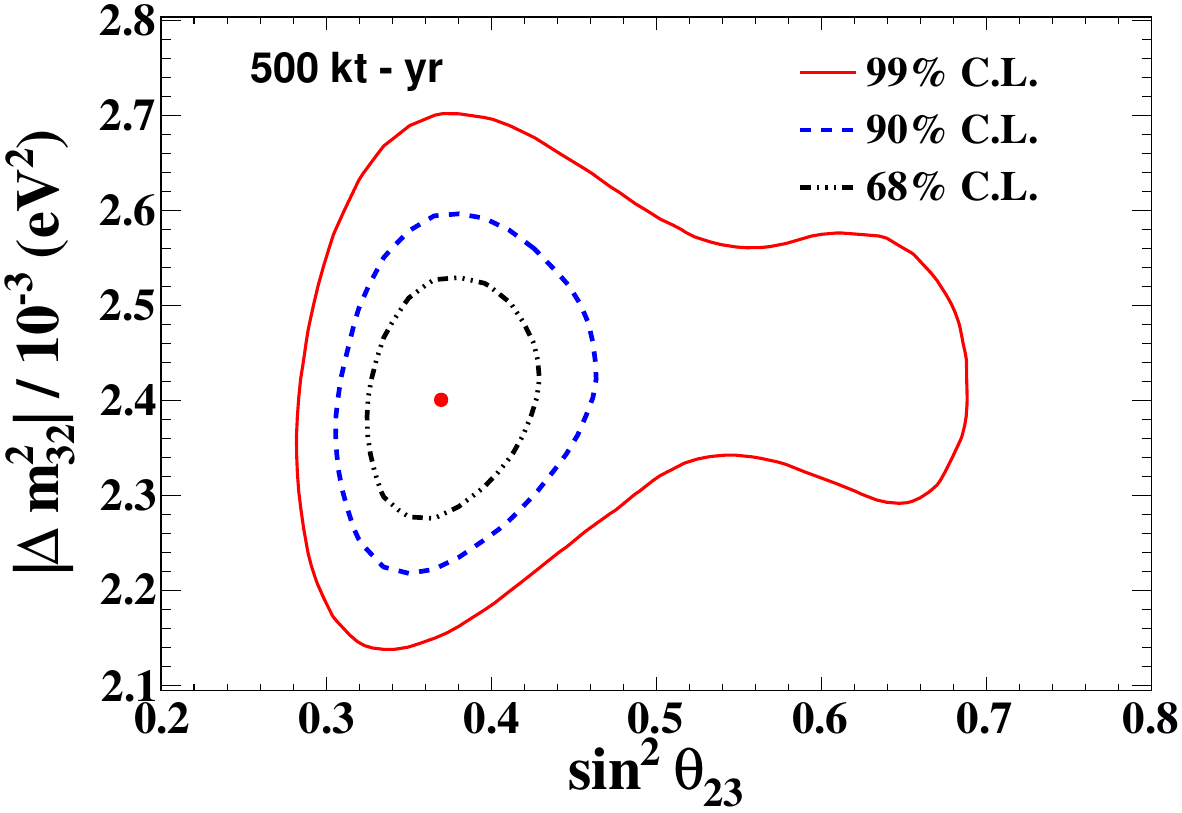}
\includegraphics[width=0.49\textwidth,height=0.35\textwidth]
{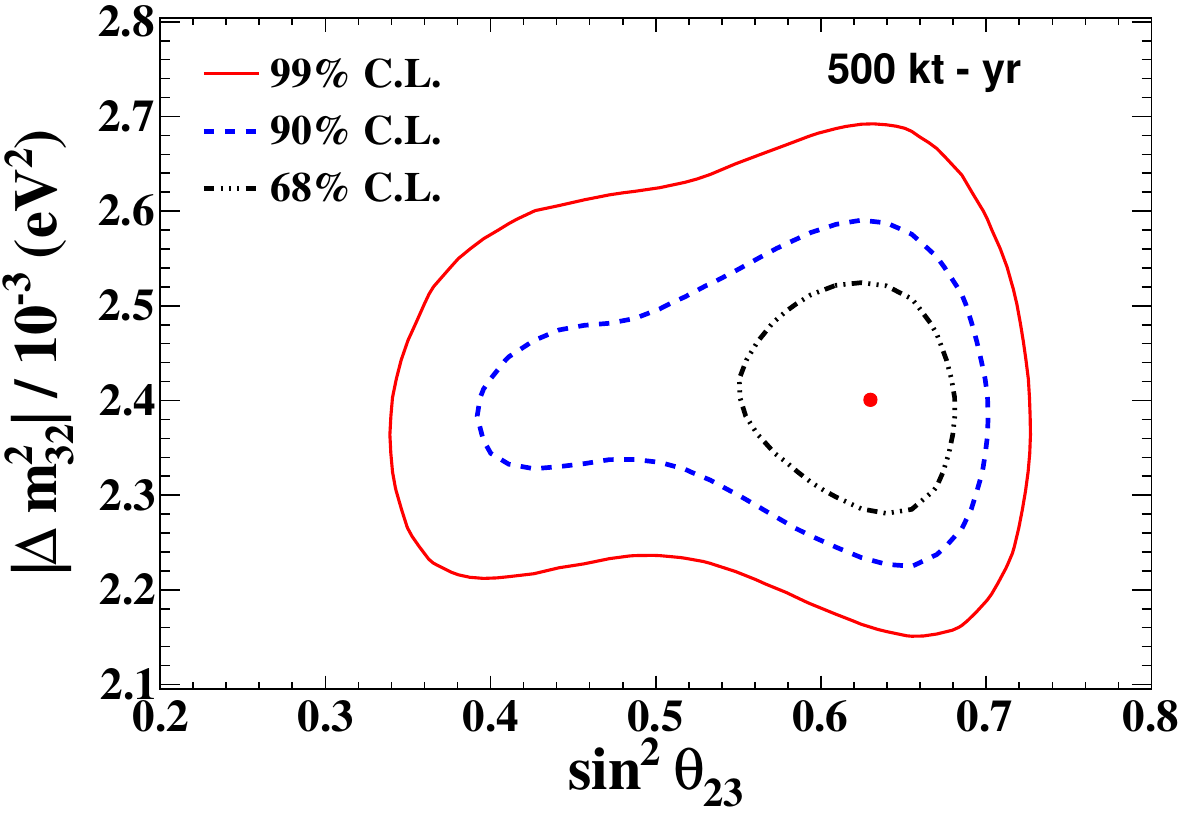}
\caption{$\chisqpm$ confidence level contours 
in the the $\sa$--$|\Delta m^2_{32}|$ plane, for $\sa$(true) = 0.37 (left panel)
and $\sa$(true)= 0.63 (right panel), 
using the hadron energy  information.
Here the true hierarchy is NH \cite{Devi:2014yaa}.}
\label{PM-nonmaximal}
\end{figure}

With a non-maximal true value of $\theta_{23}$, 
the bounds on $\sin^2 \theta_{23}$ range would be asymmetric about 0.5.
Figure~\ref{PM-nonmaximal} shows the sensitivity of ICAL for 
$\sta$ = 0.93 (i.e. $\sa$ = 0.37, 0.63). 
It may be observed that for $\theta_{23}$ in the lower octant, 
the maximal mixing can be ruled out with 99\% C.L. 
with 500 kt-yr of ICAL data. However, if $\theta_{23}$ is closer to the 
maximal mixing value, or in the higher octant, then the ICAL sensitivity 
to exclude maximal mixing would be much smaller. 

\subsection{Non-maximal $\theta_{23}$ and its octant}

\begin{figure}[h!]
\centering
\includegraphics[width=0.49\textwidth,height=0.35\textwidth]
{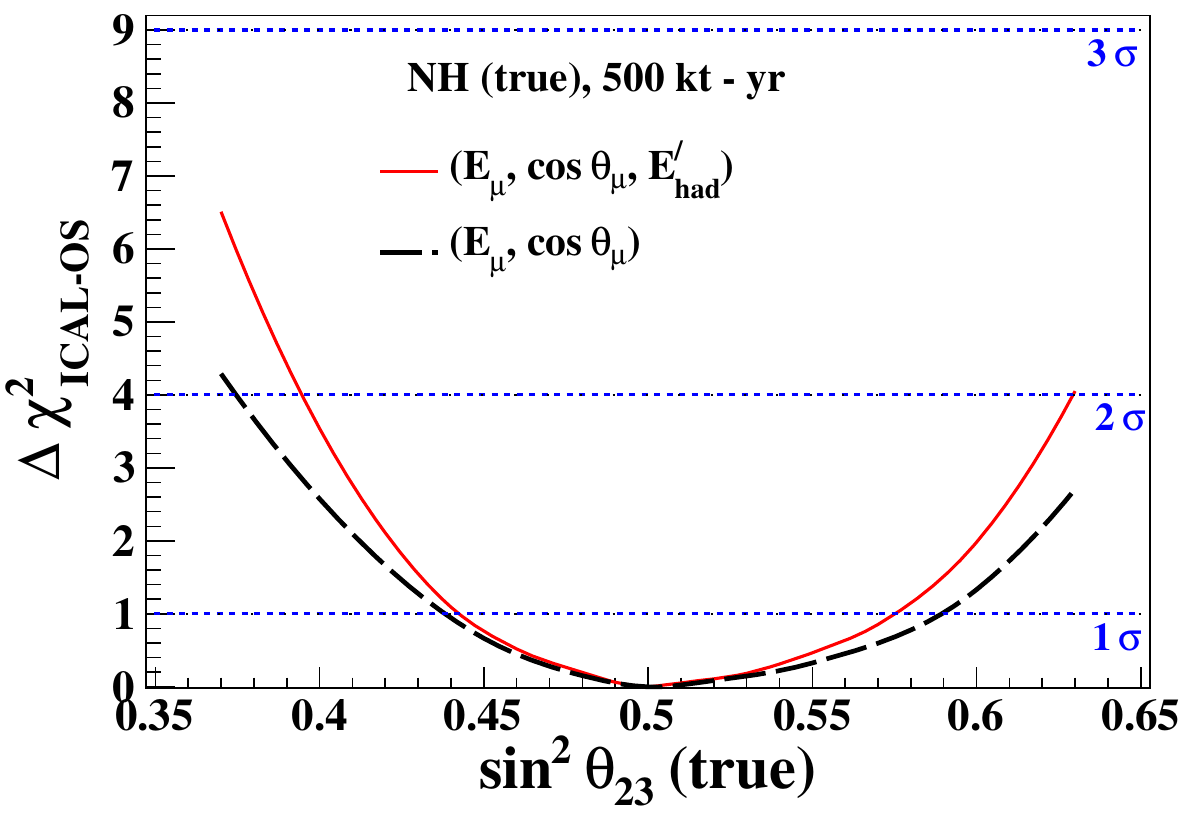}
\includegraphics[width=0.49\textwidth,height=0.35\textwidth]
{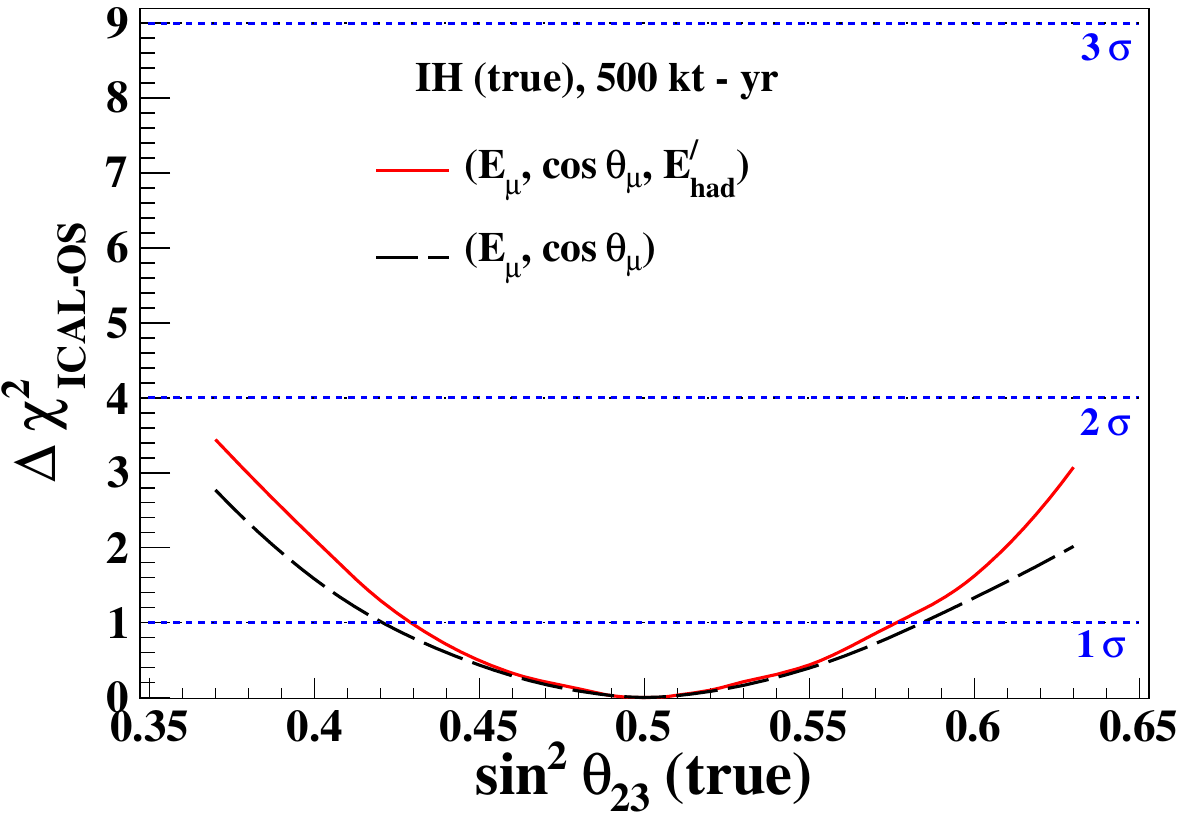}
\caption{$\chisqos$ 
for octant discovery potential as a function of true $\sa$.
The left panel (right panel) assumes NH (IH) as true hierarchy.
The line labelled $(E_\mu, \cos\theta_\mu)$ denotes results without
including hadron information, while
the line labelled $(E_\mu, \cos\theta_\mu,E'_{\rm had})$ denotes 
improved results after including hadron energy information.
ICAL exposure of 500 kt-yr is considered \cite{Devi:2014yaa}.}
\label{os-2d-vs-3d-th13-best-fit}
\end{figure}

In analogy with the mass hierarchy discovery sensitivity, 
the statistical significance of the analysis to rule out the wrong octant 
is quantified as
\begin{equation}
\chisqos = \chisqical(\rm{false~ octant}) - 
\chisqical (\rm{true ~octant}) .
\label{os_chi2_def}
\end{equation}
Here, $\chisqical(\rm{true\ octant})$ 
and $\chisqical(\rm{false\ octant})$
are obtained by performing a fit to the ``observed'' data assuming
the true octant and wrong octant, respectively.

Figure~\ref{os-2d-vs-3d-th13-best-fit} shows the sensitivity of ICAL 
to the identification of the $\theta_{23}$ octant,
with and without including the hadron energy information.
It may be observed that a $2\sigma$ identification of the octant
is possible with the 500 kt-yr INO data alone only when the true hierarchy 
is NH and the true octant is LO. In this case, without using the 
hadron energy information one can get a $2\sigma$ identification
only when $\sin^2 \theta_{23} (\rm{true)}<0.365$, which is almost close to 
the present
$3\sigma$ bound. With the addition of hadron energy information, this
task is possible as long as $\sin^2\theta_{23} (\rm{true)}<0.395$. 
If the true octant is HO or the true mass hierarchy is inverted,
then the descrimination of $\theta_{23}$ octant with the ICAL data alone 
becomes rather difficult. 
The reach is found not to be much sensitive to the exact value of 
$\theta_{13}$.
Clearly the octant discrimination becomes more and more difficult as
the true value of $\sa$ goes close to the maximal mixing. 
A combination of atmospheric as well as long baseline experiments is needed
to make this measurement \cite{Choubey:2013xqa}.

In conclusion, the inclusion of correlated hadronic information improves the
sensitivity of ICAL to mass hierarchy, precision mesurement of $|\Delta
m^2_{32}|$ and $\theta_{23}$, exclusion of maximal mixing, as well as
$\theta_{23}$ octant determination. This analysis appears to be 
the optimal one to extract information from ICAL data. However for
the potential of this method to be realized, a very good understanding
of the hadron response of the ICAL detector is crucial.

\section{ICAL physics potential: highlights}

In summary, the ICAL detector is extremely suitable for determining
the neutrino mass hierarchy, due to its capabilities of measurements of
muon and hadron energies, and identification of the muon charge. 
The cleanest and the simplest analysis of the ICAL data uses the
information on muons only. However the reach of the detector improves
tremendously if the information on hadron energy is also used in
addition. Simply adding this energy to the muon energy to reconstruct the
neutrino energy is, however, not enough, as it causes some dilution in
the accurately known muon information. The information on the
muon energy, muon direction, and hadron energy has to be kept separately
and used in the analysis. Such an analysis indicates that in 10 years,
a 50 kt ICAL can, by itself, distinguish between the normal and the
inverted hierarchy with a significance of more than $3\sigma$.

One important point to emphasize here is that this capability of ICAL
is independent of the actual value of $\delta_{\rm CP}$. Hence the results 
in this chapter are independent of $\delta_{\rm CP}$. Moreover, this
feature may be exploited by combining the ICAL information with that 
from the other CP-sensitive experiments, to improve the mass hierarchy
discrimination. Such synergies between ICAL and the other experiments 
will be explored in the next chapter.


\chapter{Synergy with Other Experiments}
\label{synergy}

\begin{flushright}
{\it
The whole is greater than the sum of its parts.\\
-- Aristotle
}
\end{flushright}


While ICAL has not yet started construction, long baseline experiments
like \nova \cite{Ayres:2002ws,Ayres:2004js,Ayres:2007tu} 
and T2K \cite{Abe:2011sj} have already been taking data, and are 
in principle sensitive to the mass hierarchy. Apart from them, major 
atmospheric neutrino detectors like HyperKamiokande (HK) \cite{Abe:2011ts},
Precision IceCube Next Generation Upgrade (PINGU) \cite{Aartsen:2014oha},
and Oscillation Research with Cosmic in the Abyss (ORCA) \cite{Katz:2014tta}
are being planned. The medium-baseline reactor oscillation experiments 
JUNO \cite{Li:2014qca} and RENO-50 \cite{Kim:2014rfa} also will aim to 
determine the mass hierarchy by performing a very precise, high statistics 
measurement of the neutrino energy spectrum. In this Section, we first 
comment on the individual sensitivities of these experiments, and later
explore the synergy between them.

While \nova and T2K get a large number of events due to intense
neutrino beams created at Fermilab and J-PARC respectively, the neutrino
baseline distances available to them are rather small (approximately 
800 km and 300 km, respectively), as compared to those for atmospheric 
neutrinos, which can be as long as 10000 km. As a result, the matter 
effects experienced by the neutrinos during their propagation, which
are crucial for the mass hierarchy identification, are small. In spite
of this, the large input flux allows \nova to be sensitive to mass
hierarchy, at least with favourable values of $\delta_{\rm CP}$.

HK is a planned water Cherenkov detector with a fiducial
volume of 500 kt, while PINGU is a megaton-size part of IceCube, 
where an increased density of the 
digital optical modules would bring down the threshold from 150 GeV to 
5 GeV so that atmospheric neutrinos may be detected. 
ORCA would be a deep sea neutrino-telescope in the Mediterranean Sea.
All these 
detectors can measure the energy and direction of muons as well as
electrons (PINGU can also detect hadron showers through their cascade events), 
however they do not have charge identification capabilities. 
The sensitivities of these atmospheric neutrino experiments will arise from 
their large sizes, which lead to a large number of events. 

The reactor experiments JUNO and RENO50, with baseline $\sim 50$ km, will
approach the mass hierarchy measurement by using the interference effects 
between the two oscillation frequencies \cite{Petcov:2001sy,Choubey:2003qx}.
Determination of mass hierarchy in such setups require a very good
knowledge of the shape uncertainty of the reactor neutrino fluxes and
the non-linearity of the detector response.
In a detailed analysis by the JUNO collaboration \cite{An:2015jdp},
it has been claimed that $\sim 3\sigma$ sensitivity to mass hierarchy is 
possible in 6 years including reasonable values of systematic uncertainties.

\begin{figure}[t]
\centerline{
\includegraphics[width=0.98\textwidth]{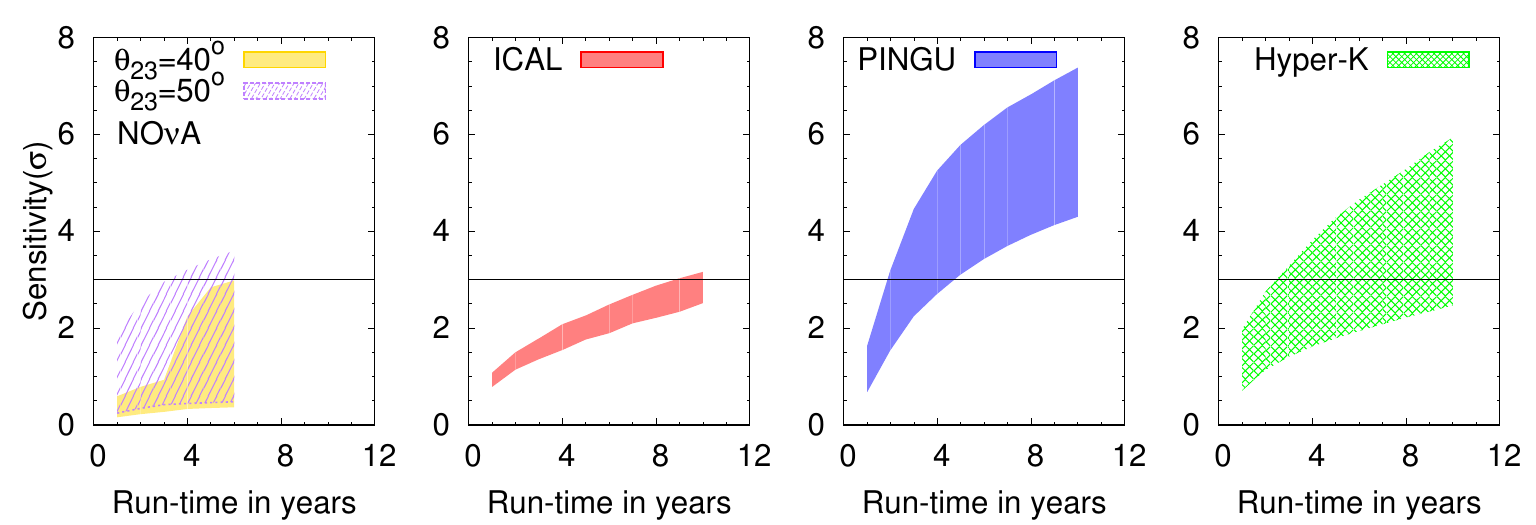}}
\caption{Projected hierarchy sensitivities of NO$\nu$A, ICAL, HK, 
and PINGU, if the true hierarchy is NH.
It is to be noted that that this figure is only indicative in nature; 
the actual sensitivities may get altered 
with changes in analysis techniques, systematics etc. 
While \nova has already started taking data, the other experiments 
are yet to start.}
\label{timeline}
\end{figure}

In Fig. \ref{timeline},  we compare the individual hierarchy sensitivities of 
\nova, HK and PINGU with the sensitivity of ICAL \cite{sg-ichep}. 
The figure is drawn following \cite{Blennow:2012gj}, wherein the median 
sensitivities of the experiments are shown when the true hierarchy is NH.
The \nova results, obtained by using GLoBES  \cite{Huber:2004ka,Huber:2007ji},
assume equal time runs for $\nu$ and $\bar\nu$ every year, with two
$\theta_{23}$ values, $40^\circ$ and $50^\circ$.
The projected sensitivities of ICAL from \cite{Devi:2014yaa},
HK from \cite{hkloi}, and PINGU  from \cite{Aartsen:2014oha}  
are used in generating this plot. 
Note that the sensitivity for ICAL shown here is higher than that
shown in \cite{Blennow:2012gj} due to the improvement coming from
the inclusion of hadronic information \cite{Devi:2014yaa}.

The width of the shaded/coloured regions for \nova arises from the variation of 
$\dcp$ in its full range, while for atmospheric neutrino experiments, which 
are insensitive to $\dcp$, the width of the shaded region is due mainly to 
the variation of $\theta_{23}$. (The $\theta_{23}$ ranges used are slightly
different for different experiments, depending on what they use for their 
analyses. The range used is $38^\circ$--$53^\circ$ For ICAL,
$40^\circ$--$50^\circ$ for HK, and $38.7^\circ$--$51.3^\circ$ for PINGU.) 
The lower end of the bands, indicating the worst sensitivity, 
corresponds to the lowest value of $\theta_{23}$.

The figure demonstrates that after 3 years of $\nu$ and
3 years of $\bar{\nu}$ run at \nova, which is the current plan, 
one may obtain a sensitivity anywhere between $0.5\sigma$ and
$3.5\sigma$, depending on the actual value of $\dcp$. 
HK can attain a $3\sigma$ sensitivity to mass hierarchy in approximately 
5 years for $\sin^2\theta_{23}=0.5$ \cite{Abe:2011ts}, while PINGU
can reach the same sensitivity in 3 years for this scenario, 
if the cascade events can be included in the analysis \cite{Aartsen:2014oha}. 
The hierarchy sensitivity of PINGU and HK depends strongly on 
$\theta_{23}$, growing very fast with increasing  $\theta_{23}$ because of 
the corresponding increase in the number of events. 
ORCA (not shown in the figure) can achieve  $3-5\sigma$ hierarchy sensitivity 
using the muon events for a 20 Mt-yearr exposure \cite{Katz:2014tta}.    
ICAL is expected to take about 9--10 years to reach the $3\sigma$
sensitivity by itself, however if true value of 
$\dcp$ is in the unfavourable region for \nova, then an early hint may 
be obtained from ICAL. Even later, for lower values of $\theta_{23}$ 
the hierarchy sensitivity of ICAL can be greater than that of HK. 

The relative importance of different experiments to the mass 
hierarchy determination thus depends crucially on what the value of
$\dcp$ is, and what the starting dates of the experiments are.
However note that ICAL is the only experiment among
these that has a magnetic field and the resulting charge identification 
capability that can distinguish between neutrinos and antineutrinos. 
Among the atmospheric neutrino experiments, it will be the only experiment 
that can perform a neutrino and antineutrino analysis independently.
Its importance in pinning down the mass hierarchy is thus expected
to be crucial.

So far in this Report, the expected data from ICAL
alone has been used. However the reach of INO will be enhanced due 
to the information available from earlier experiments. 
A consistent way of taking the impact of these experiments
into account is to include their data in a combined $\chi^2$ fit.
In this work we present the impact of the prior data from these 
experiments on the ICAL physics reach, as well as how the ICAL
data will help these experiments remove certain ambiguities from 
their analysis, and zero in on the actual neutrino mixing parameters.

\section{Combined mass hierarchy reach at $\delta_{\rm CP}=0$ 
for ICAL + T2K + \nova}
\label{dcp=0}

As we have seen in Chapter~\ref{analysis}, the 50 kt ICAL by itself can
identify the mass hierarchy with a significance of 
$\Delta \chi^2 \approx 9$ with 10 years of running. This reach is
independent of the actual value of the CP-violating phase $\delta_{\rm CP}$.
Currently running fixed baseline experiments like T2K and \nova\ 
would already have obtained some sensitivity to the mass hierarchy 
during their run. This sensitivity will depend on the actual
value of $\delta_{\rm CP}$, as we shall see in Sec.~\ref{sec:dcp}.  
In this section, we shall see the effect of using the information from
these experiments, in the case where the value of $\delta_{\rm CP}$ 
is taken to be zero. Note that the contribution of the current reactor 
experiments to the mass hierarchy measurements is negligible
\cite{Ghosh:2012px}.

\begin{figure}[t]
\centering
\includegraphics[width=0.49\textwidth,height=0.35\textwidth]
{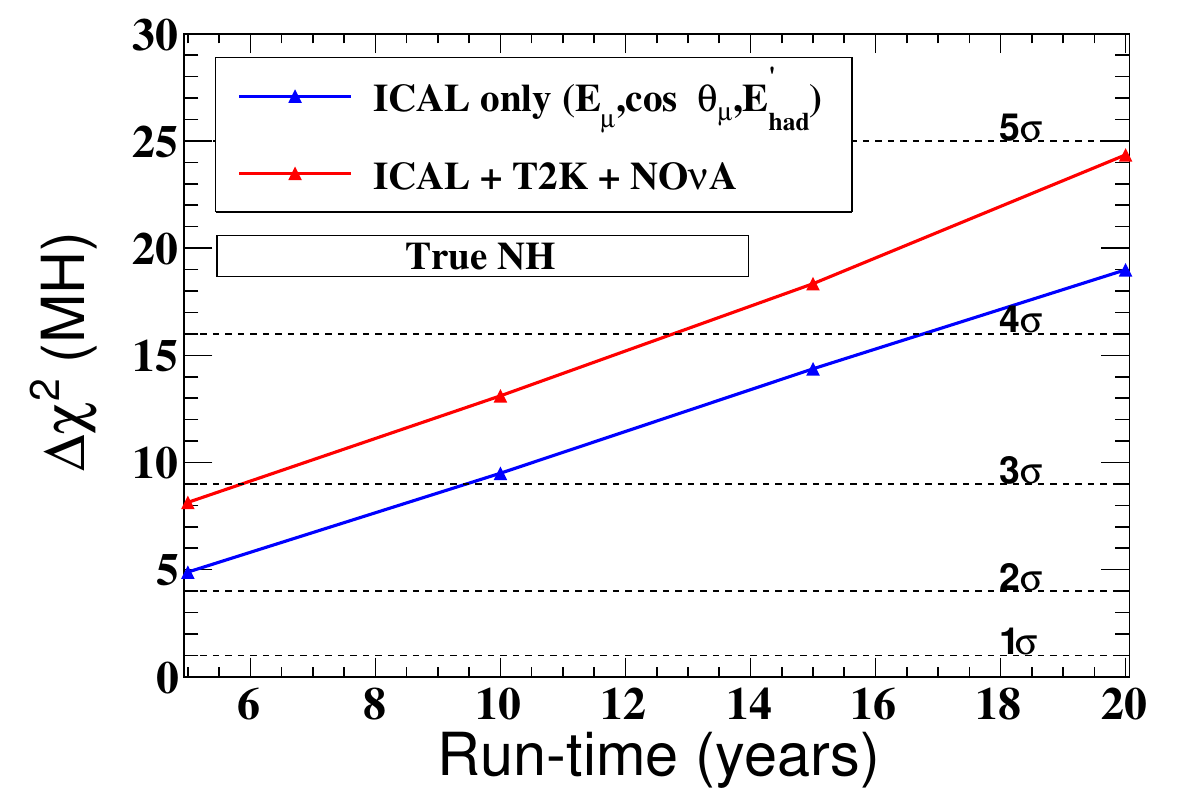}
\includegraphics[width=0.49\textwidth,height=0.35\textwidth]
{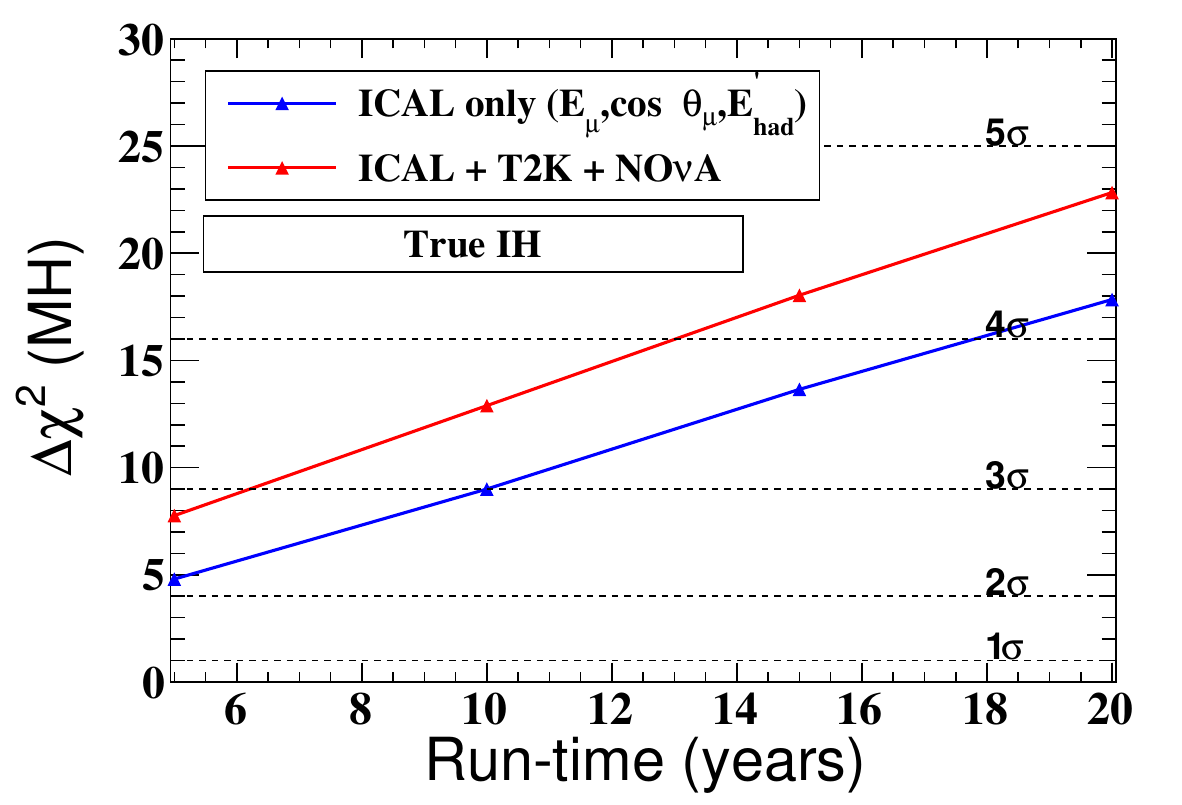}
\caption{Preliminary results on the $\Delta \chi^2$ for the wrong hierarchy 
obtained from a combined analysis of T2K (a luminosity of $8 \times 10^{21}$ 
pot in neutrino run), \nova\ (3 years neutrino and 3 years antineutrino run), 
and ICAL \cite{synergy-new}. 
The left (right) panel is for true normal (inverted) hierarchy.
We take $\stcht=0.1, \sat=0.5$ and $\delta_{\rm CP}=0$, 
and all parameters are allowed
to vary over their $3\sigma$ ranges as shown in Table \ref{tab:best-fit}.
}
\label{fig:hiersens}
\end{figure}

A preliminary estimate of the combined sensitivity to the neutrino mass 
hierarchy as a function of number of years of run of the ICAL atmospheric 
neutrino experiment is shown in Fig. \ref{fig:hiersens} \cite{synergy-new}.  
For each set of oscillation parameters, the joint $\chi^2$ from all experiments 
is given by
\be
\chi^2 = \chi^2_- + \chi^2_+ + \sum_i \chi^2_i + \chi^2_{\rm prior} ~,
\ee
where $\chi^2_-, \chi^2_+$ are as defined in 
Eqs.~(\ref{chisq}) and (\ref{n-theory-definition}), 
$\sum_i \chi^2_i$ is the contribution from the accelerator experiments 
($i$ runs over T2K and NO$\nu$A), and $\chi^2_{\rm prior}$ is the prior
on $\theta_{13}$ from the reactor experiments.
Here a prior of 4.5\% on $\stch$ has been taken, which matches the outcome 
of the recent global fit \cite{Gonzalez-Garcia:2014bfa}.
This joint $\chi^2$ is computed and marginalized over all oscillation 
parameters, to determine the minimized 
joint $\Delta \chi^2$ shown in the figure. Note that the $x$-axis in this 
figure shows the number of years of running of ICAL only.
We assume a complete T2K run (total luminosity $8 \times 10^{21}$ pot)
with neutrinos only, 3 years of
\nova\ run with neutrinos, and 3 years of \nova run with antineutrinos
have already been completed. We use the standard setup of these
experiments as in the GLoBES package \cite{Huber:2004ka,Huber:2007ji}.

The figure shows that the prior information from T2K and \nova\  implies
that the target of $\Delta \chi^2=9$ significance for the hierarchy 
identification may now be achieved within 6 years of the ICAL running,
as opposed to about 10 years with only the ICAL data.
Note that this is not just the effect of the $\Delta \chi^2$ provided
by the fixed baseline experiments. These experiments also yield
an improved precision in $\theta_{23}$, $|\Delta m^2_{\rm eff}|$ 
(through their disappearance channel), and in $\theta_{13}$ (through their 
appearance channel). As a result, the impact of marginalization over
these parameters in the ICAL analysis is greatly reduced. 
Indeed, for the 500 kt-yr of ICAL data, the contribution of 
$\Delta \chi^2_{\rm ICAL-MH}$ to the total $\Delta \chi^2$ in the case of true 
normal hierarchy increases from 9.5 (without the data from T2K and NO$\nu$A) 
to 10.4 (with the data from T2K and NO$\nu$A). Note that the contributions of 
T2K and NO$\nu$A themselves are expected to be 
$\Delta \chi^2 \approx 0.12$ and $2.6$, respectively.

The fact that the total $\Delta \chi^2$ in the ICAL, T2K and \nova~ 
experiments is greater than the sum of their individual $\Delta \chi^2$
values is the {\rm synergy} among these experiments. Though ICAL is
the dominant contributor to the hierarchy sensitivity in this case, it
clearly benefits tremendously from this synergy. Similary for other
physics issues like the precision measurements of mixing parameters,
the combination of data from long baseline experiments and ICAL will
help improve our overall understanding of these parameter values. Here
the long baseline experiments are expected to play a dominant role
since they are directly sensitive to $\theta_{23}$ and 
$|\Delta m^2_{\rm eff}|$ through their disappearance channel, while
ICAL is expected to play a complementary role.

So far we have discussed the case of $\dcp=0$. In the next section,
we shall explore the case of nonvanishing $\dcp$ in detail, since
it affects the mass hierarchy sensitivity of fixed baseline experiments.

\section{Ensuring mass hierarchy sensitivity for all $\dcp$}
\label{sec:dcp}

As seen in Sec.~\ref{insensitivity}, the insensitivity of ICAL to the
actual value of $\dcp$ comes from the fact that the muon neutrinos 
at ICAL come dominantly from the original unoscillated muon neutrinos,
while the muon neutrinos coming from the oscillated electron neutrinos
play a subdominant role. The survival probability $P_{\mu\mu}$ (See
Appendix~\ref{app:prob}) is therefore more relevant than the conversion
probability $P_{e\mu}$. In $P_{\mu\mu}$, the CP-violating phase $\dcp$
appears as a subdominant oscillating term, whose oscillations,
moreover, are averaged out due to the uncertainties in the energies
and directions of the incoming neutrinos \cite{Ghosh:2013yon}.

\begin{figure}[h!]
\centering
\includegraphics[width=0.49\textwidth,height=0.35\textwidth]
{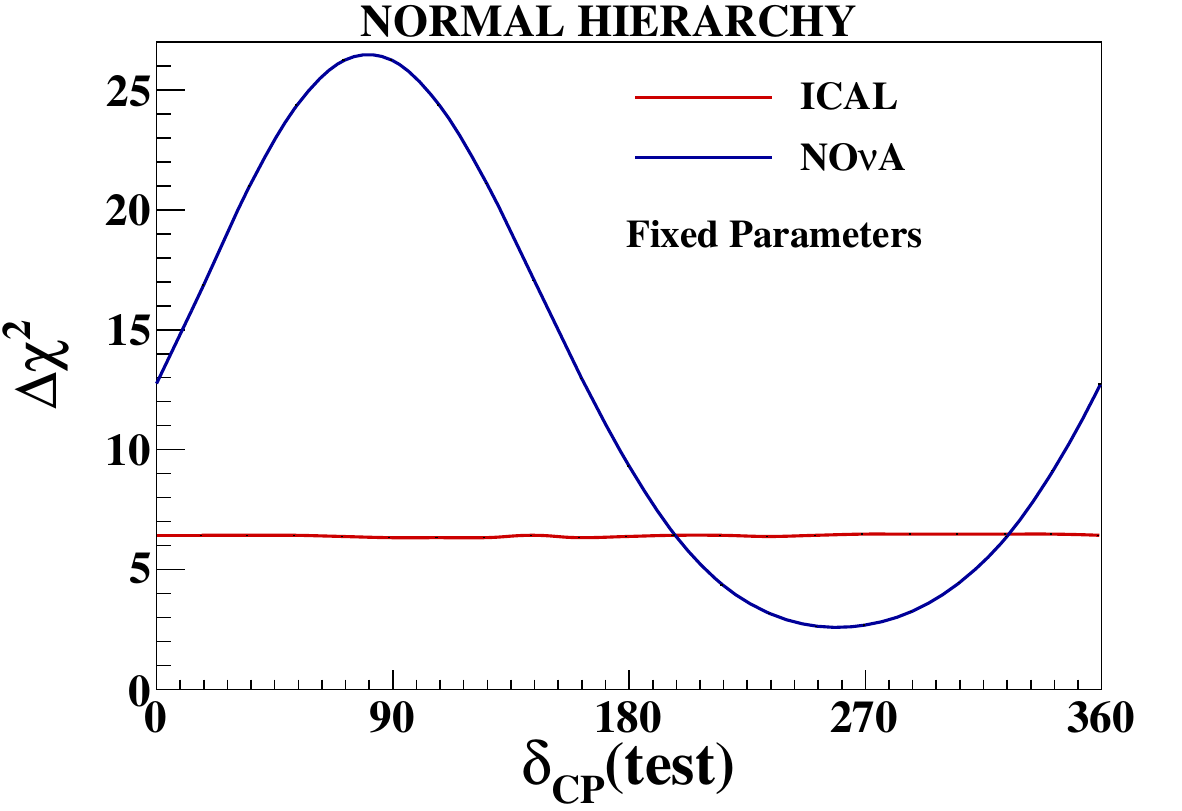}
\includegraphics[width=0.49\textwidth,height=0.35\textwidth]
{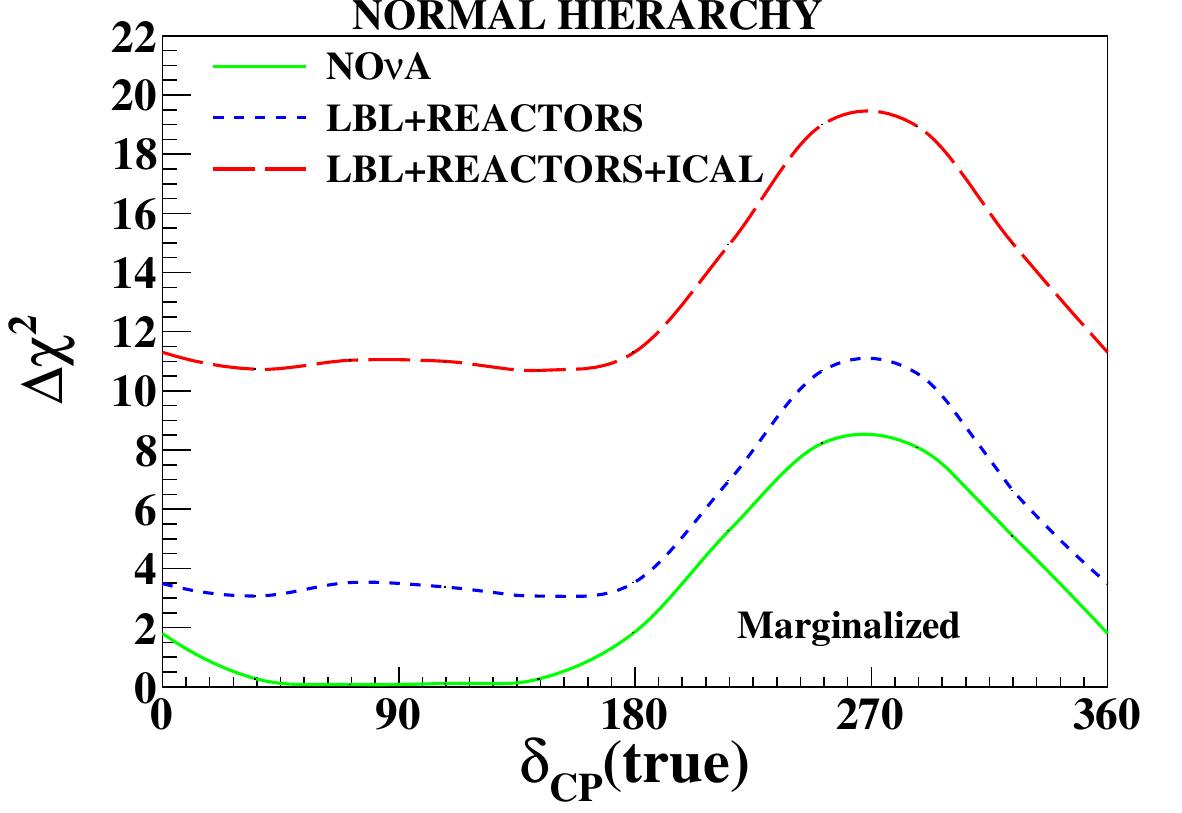}
\caption{The left panel shows the $\Delta \chi^2$ for mass hierarchy
sensitivity at \nova~ and ICAL, when the actual value of $\dcp$ is vanishing, 
and different test value of $\dcp$ (shown along the $x$-axis) are taken.
The right panel shows the dependence of mass hierarchy sensitivities of 
experiments (or their combinations) to the actual $\dcp$, when the test $\dcp$
is varied over all its range for minimizing $\Delta \chi^2$.
The full proposed runs of the long baseline and reactor experiments are
taken. The ICAL exposure is taken to be 500 kt-yr, and only the muon
energy and direction are used, ignoring the hadron information
\cite{Ghosh:2012px}.
}
\label{fig:hierlblcp}
\end{figure}

On the other hand, the fixed baseline experiments like T2K or \nova\
are sensitive to the conversion probability $P_{\mu e}$, whose dominant term 
depends on $\dcp$. Moreover, though this term is oscillating, these
experiments have a better knowledge of the neutrino energies, and the 
neutrino directions are very precisely known. As a result, the 
data at these experiments is highly sensitive to $\dcp$. 
In particular, while performing the fit to the data, the value of $\dcp$ 
can be adjusted, along with the value of $|\Delta m^2_{31}|$, 
to compensate for the wrong hierarchy in half
the $\dcp$ parameter space. As a result, even if the wrong hierarchy is
being fitted, a good fit may be obtained, albeit at a wrong value of
$\dcp$. Since the actual $\dcp$ value is unknown, this worsens the
hierarchy sensitivity of the experiment. This may be seen from 
the left panel of Fig.~\ref{fig:hierlblcp}, where the actual $\dcp$
is taken to be vanishing, however \nova~ gives a better fit at 
another value of $\dcp$, with a very reduced value of $\dcp$ as
compared to that at the true $\dcp$ value. In the absence of any
knowledge about the actual $\dcp$ value, this results in a reduced
performance of the detector.

The right panel of Fig. \ref{fig:hierlblcp} shows the $\Delta \chi^2$ 
for the mass hierarchy sensitivity as a function of $\delta_{\rm CP}$(true). 
The data are generated for normal hierarchy at each value of 
$\delta_{\rm CP}$(true) shown on the $x$-axis, and a fit is performed for 
inverted hierarchy by marginalizing over {\it all} oscillation parameters, 
including $\delta_{\rm CP}$. Clearly, the reach of NO$\nu$A alone for determining
the neutrino mass hierarchy is extremely sensitive to the actual
value of $\delta_{\rm CP}$. While the sensitivity is  $\Delta \chi^2 \approx 9$ 
for $\dcpt$ near $270^\circ$, it falls to almost zero for $\dcpt \simeq
[50^\circ - 150^\circ]$. 

When T2K and all reactor data are added, there is some improvement to the 
combined sensitivity. In particular, in the  
$\dcpt \simeq [50^\circ - 150^\circ]$ range where NO$\nu$A by itself 
gives no mass hierarchy sensitivity, the addition of T2K and 
reactor data takes $\Delta \chi^2$ to $\simeq 3.5$. The reason for this 
is the mismatch between the best-fits for different experiments.
For the same reason, even the reactor data that are
not sensitive to $\dcp$ but only to $|\ma|$ help, albeit marginally, 
in disfavouring the spurious best-fit minima for the wrong hierarchy, 
since they do not allow the fit value of $|\ma|$ to stray far from the 
actual one. A combined fit with all accelerator and reactor data thus give 
a best-fit at a point ($\dcp=198^\circ$, $\sa=0.48$ and $\stch=0.1$) where the
tension between these experiments gives a small hierarchy sensitivity
even in the disfavoured $\dcpt$ range ($[0^\circ - 180^\circ]$ for
normal hierarchy). 

Finally, addition of the ICAL data raises the $\Delta \chi^2$ by a constant 
amount for all values of $\dcpt$, and ensures the identification of 
hierarchy to more than $\Delta \chi^2 \approx 10$ for even those $\dcpt$
values for which the other experiments cannot rule out the wrong
hierarchy by themselves. In the best-case scenario, the hierarchy may be
identified to $\Delta \chi^2 \approx 20$. 

Note that the ICAL results in
the figure have been obtained by using only the information on muon energy
and momentum. If the hadron information is also added, the combined
$\Delta \chi^2$ of all the experiments is expected to be 12--22, depending on
whether the actual $\dcp$ value is favourable to the fixed baseline
experiments or not.

\section{Octant of $\theta_{23}$ from the ICAL, NO$\nu$A and T2K data}
\label{octant-ical-old-2d}

Combined analysis of all the neutrino oscillation data available disfavors
the maximal mixing solution for $\theta_{23}$ at $1.4\sigma$ confidence
level~\cite{NuFIT,GonzalezGarcia:2012sz} which is mostly driven by
the MINOS accelerator data in $\numu$ and $\anumu$ disappearance
modes~\cite{Adamson:2013whj}.  Now, if $\sin^22\theta_{23}$
turns out to be different from unity as suggested by the recent
oscillation data, this creates the problem of octant degeneracy of
$\theta_{23}$~\cite{Fogli:1996pv}. In a recent global fit work by
F. Capozzi {\it et al.}~\cite{Capozzi:2013csa}, the authors have found an
overall preference for the first or lower octant (LO) at 95\% confidence
level assuming normal hierarchy.  In case of inverted hierarchy, 
the higher octant (HO) seems to be preferred~\cite{NuFIT}.

\begin{figure}[htp]
\centering
\includegraphics[width=0.49\textwidth,height=0.35\textwidth]
{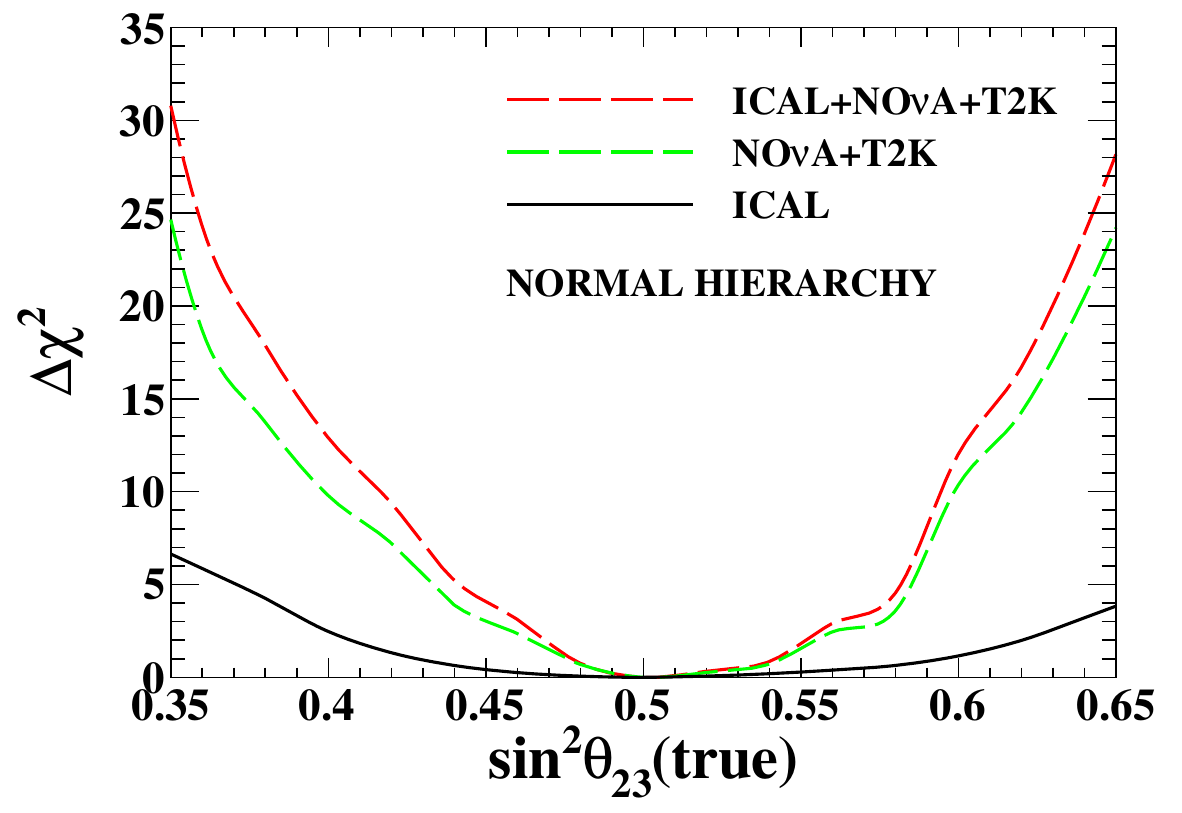}
\includegraphics[width=0.49\textwidth,height=0.35\textwidth]
{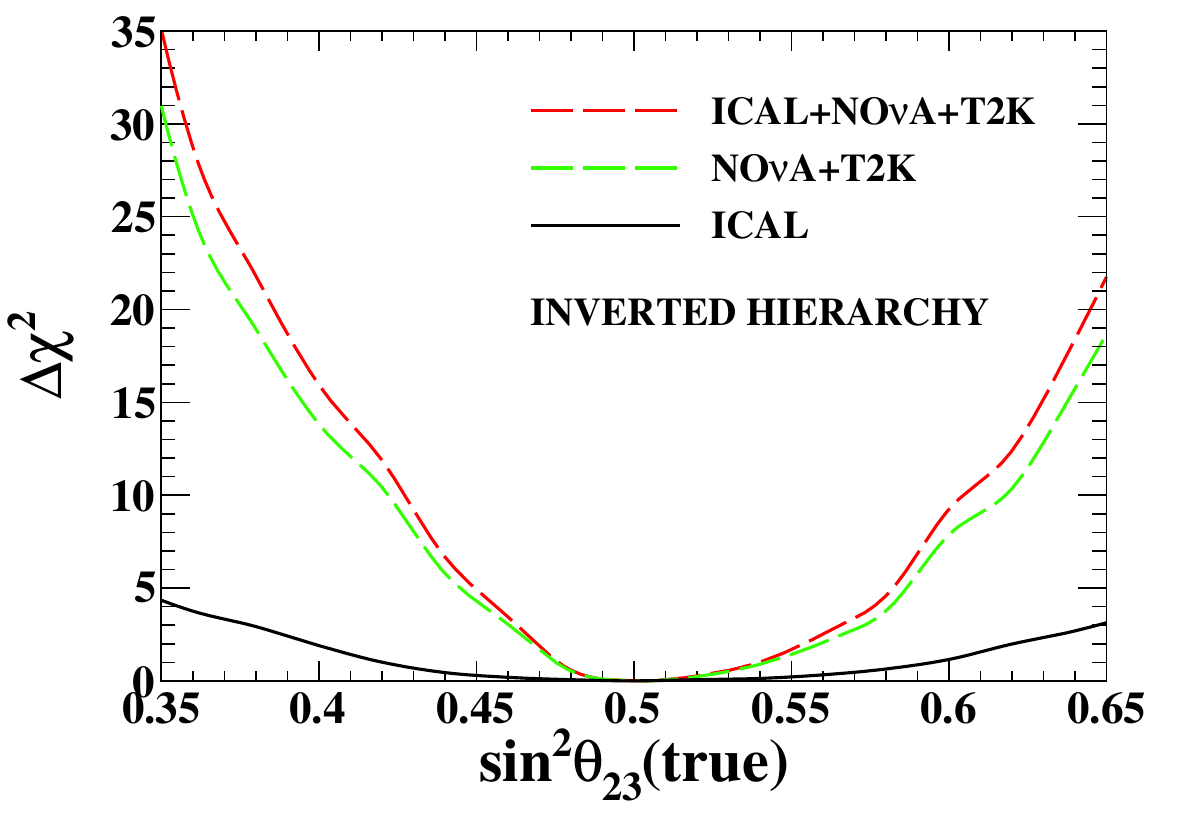}
\caption{$\Delta\chi^2_{\rm ~ICAL-OS}$ for octant discovery potential as
a function of true $\sin^2\theta_{23}$. The left (right) panel assumes
normal (inverted) hierarchy as true choice. In each panel, the black
solid line shows the performance of ICAL with an exposure of 500 kt-yr
using only the information on muon energy and muon direction.  The green
dotted line depicts the combined sensitivity of T2K 
(integrated luminosity of $4 \times 10^{21}$ pot in neutrinos and
$4 \times 10^{21}$ pot in antineutrinos) and NO$\nu$A (3 years of $\nu$ run +
3 years of $\bar\nu$ run). The red dashed graph presents the combined
results of ICAL, T2K, and NO$\nu$A \cite{octant-new}.
}
\label{octant-ical-old-2d-plus-lbl}
\end{figure}

In this section, we present the preliminary results \cite{octant-new} 
showing the discovery reach of the octant of $\theta_{23}$ with atmospheric 
neutrinos at ICAL in combination with projected 
T2K~\cite{Itow:2001ee,Abe:2011ks}
and NO$\nu$A~\cite{Ayres:2002ws,Ayres:2004js,Ayres:2007tu}
data.  We have performed the analysis for ICAL by using only
the information on muon energy and muon direction which has been
described in detail in Sec.~\ref{sec:emu_thmu}.  The Earth matter effect in
the $P_{\mu\mu}$ channel can be very useful to resolve the octant
ambiguity of $\theta_{23}$~\cite{Choubey:2005zy}. 

The potential of the experiments for excluding the wrong octant as 
a function of  true value of $\sin^2\theta_{23}$ is shown in 
Fig.~\ref{octant-ical-old-2d-plus-lbl}. 
For each given value of $\theta_{23}$ (true),
we marginalize over all the allowed values of $\theta_{23}$ in the
opposite octant, including the maximal mixing value. We take 5\% prior
on $\sin^22\theta_{13}$ with a true value of 0.1, and take $|\Delta
m^2_{\rm eff}|$(true) = 2.4 $\times 10^{-3}$ eV$^2$.  No priors have
been taken on atmospheric parameters. In the case of T2K and NO$\nu$A, 
we generate the data with $\delta_{\rm CP} = 0^{\circ}$, but while
performing the fit, we marginalize over the entire range of 
$\delta_{\rm CP}$ between 0 to 2$\pi$. For ICAL, we take 
$\delta_{\rm CP}= 0^{\circ}$ both in data and in the fit, 
as ICAL atmospheric analysis is not sensitive to $\delta_{\rm CP}$.

From Fig.~\ref{octant-ical-old-2d-plus-lbl}, we can see that 50 kt
ICAL in ten years can identify the correct octant at 2$\sigma$ if
$\sin^2\theta_{23}$ (true) $<$ 0.38 (0.35) only when the true hierarchy
is normal (inverted) and the true octant is LO.  The green dotted line
depicts the combined sensitivity of T2K (2.5 years of $\nu$ run + 2.5
years of $\bar\nu$ run) and NO$\nu$A (3 years of $\nu$ run + 3 years of
$\bar\nu$ run). The red dashed graph presents the combined results of
ICAL, T2K, and NO$\nu$A.

The projected data from T2K and NO$\nu$A will clearly play a crucial role 
in addressing the issue of $\theta_{23}$ octant. Adding the information
from these long-baseline experiments, we can improve the
octant discovery reach of ICAL significantly, suggesting the possible
synergy between the atmospheric and long-baseline data. 
While the contribution of ICAL itself is marginal, the combined
atmospheric and long-baseline data can establish the correct octant at
3$\sigma$ if $\sin^2\theta_{23}$ (true) $<$ 0.42 (0.43) assuming normal
(inverted) hierarchy and the lower octant is chosen by nature.



\section{ICAL data for improving CPV discovery potential of T2K and \nova} 
\label{synergy-cp}

The identification of mass hierarchy and measurement of CP violation 
are intrinsically interconnected at the fixed baseline experiments,
due to the leading term in the relevant conversion probability
$P_{\mu e}$ as given in Appendix~\ref{app:prob}.
Therefore in the absence of knowledge about hierarchy, it is possible
that the wrong hierarchy conspiring with an incorrect value of $\dcp$
may mimic the correct combination of hierarchy and $\dcp$.
As a result, in the unfavourable range of $\dcp$ as described in 
Sec.~\ref{sec:dcp}, i.e. $\dcp \in [0^\circ,180^\circ]$ if the actual hierarchy 
is normal NH and $\dcp \in [180^\circ,360^\circ]$ if the actual hierarchy is
inverted, the ability of  \nova\ and T2K 
to measure $\dcp$, in particular to discover nonzero $\dcp$ and hence
CP violation, is severely curtailed.
The hierarchy sensitivity of an atmospheric neutrino experiment like ICAL, 
which is independent of the actual $\dcp$ value, can restore the
ability of these fixed baseline experiments to discover CP violation
\cite{Ghosh:2013yon,Ghosh:2014dba}.

The ICAL analysis is performed using only the information on
muon energy and direction, neglecting hadronic information, and taking an 
exposure of 500 kt-yr. The input parameter ranges and marginalisation 
ranges as given in Table~\ref{tab:best-fit} are used, with the prior of 
5\% taken on $\theta_{13}$. 
The discovery potential for CPV is quantified by considering a variation of
$\dcp$  over the full range $[0^\circ,360^\circ]$ in the simulated data, 
and comparing it with $\dcp = 0^\circ$ or $180^\circ$
in the theory expectation. The definition of $\chi^2$ used is the same as
in Sec.~\ref{sec:emu_thmu}.

\begin{figure}[t]
\includegraphics[width =0.49\textwidth,height=0.35\textwidth]
{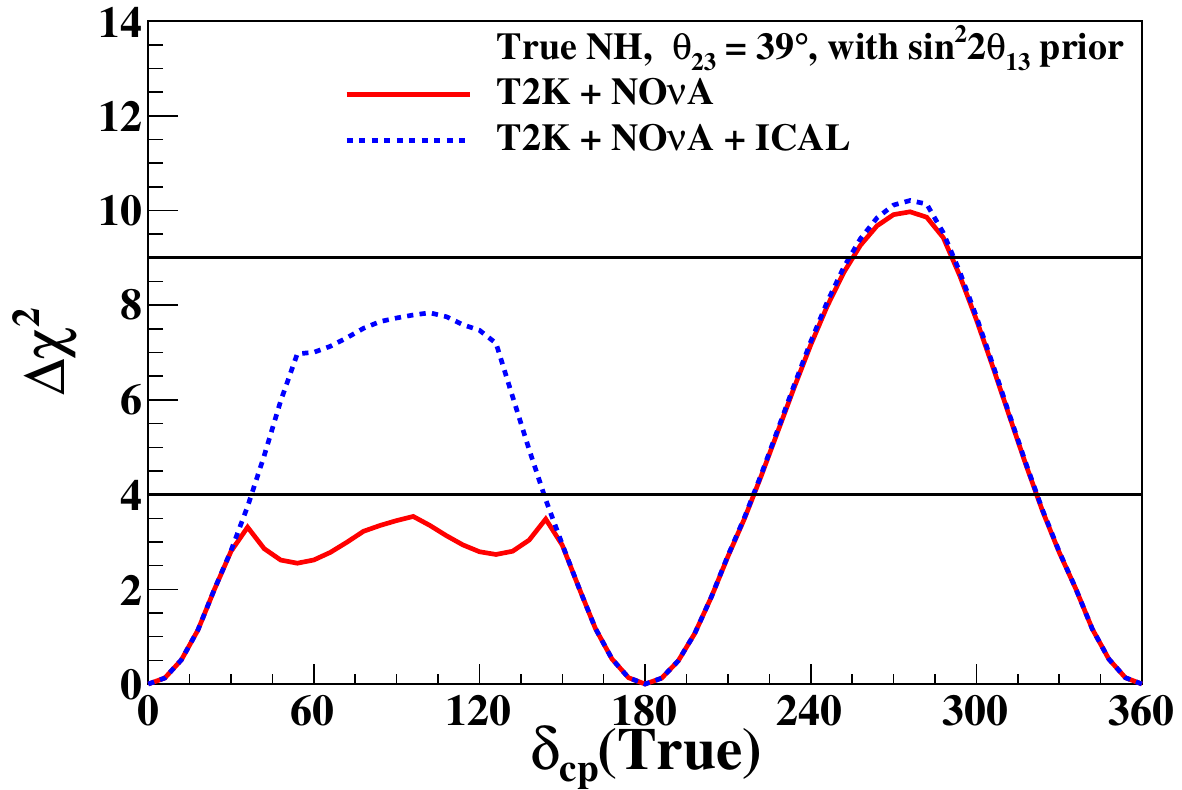}
\includegraphics[width =0.49\textwidth,height=0.35\textwidth]
{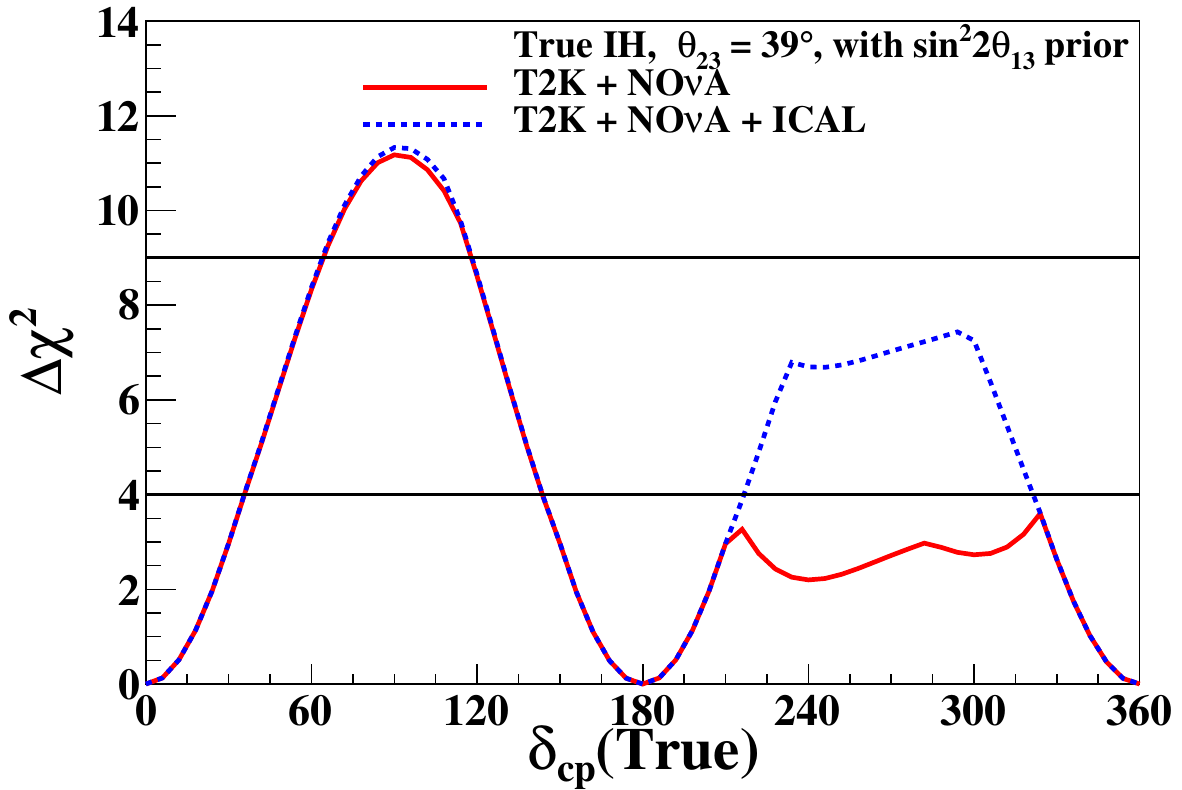}\\
\includegraphics[width =0.49\textwidth,height=0.35\textwidth]
{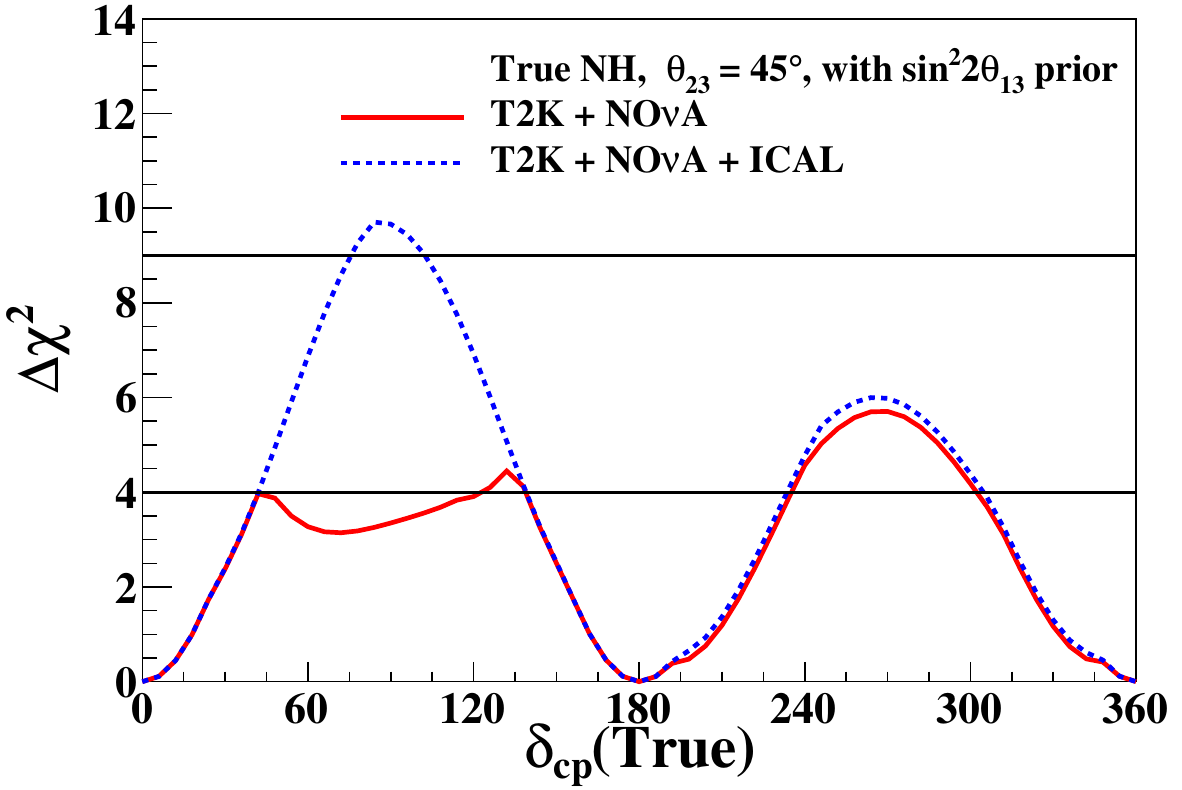}
\includegraphics[width =0.49\textwidth,height=0.35\textwidth]
{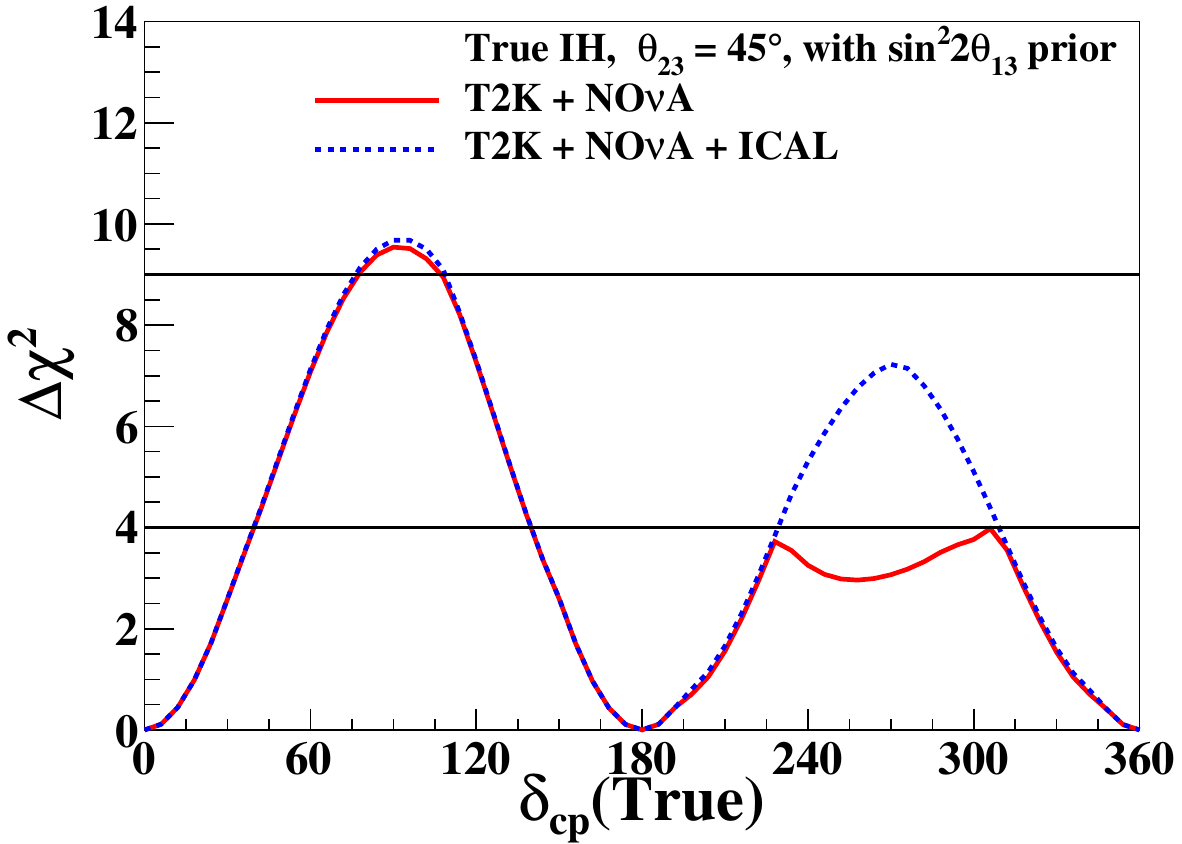}
\caption{CPV discovery vs true $\dcp$ for \nova+T2K and \nova+T2K+ICAL,
for $\theta_{23}$(true)$ = 39^{\circ}$ (upper row) and $45^\circ$ (lower row).
The left and right 
panels correspond to normal and inverted true hierarchies, respectively.
We use $\sin^2 2\theta_{13} = 0.1$, a prior of $\sigma_{\theta_{13}} = 0.005$,
and a 500 kt-yr exposure for the ICAL \cite{synergy-cp-new}. 
}
\label{cpv-discovery}
\end{figure}

In Fig.~\ref{cpv-discovery}, we plot the CPV discovery potential of 
T2K and \nova\ with and without information from ICAL. 
For T2K and \nova, the same specifications as in Sec.~\ref{dcp=0}
have been used. 
The figure shows that, as expected, the CP-violation sensitivity 
of the experiments is zero for true $\dcp=0$ and $\pi$, while it is close 
to maximum at the maximally CP violating values $\dcp= 90^\circ$ or
$\dcp=270^\circ$, depending on the hierarchy.  
The CPV discovery with \nova and T2K suffers a drop in one of the 
half-planes of $\dcp$, depending on the true hierarchy --- in the region 
$[0^\circ,180^\circ]$ if it is NH, and $[180^\circ,360^\circ]$ if it is IH.

The figure also shows that the additional information from ICAL would  
increase the sensitivities of these experiments in the unfavourable  
$\dcp$ half-plane.   
This corresponds to an almost two-fold increase in the range of $\dcp$ values 
for which CP violation can be discovered by the fixed-baseline experiments,
and it happens because the hierarchy sensitivity of ICAL excludes the
wrong-hierarchy minimum for the CPV discovery at $2\sigma$.
Thus it is quite possible that, though an atmospheric neutrino experiment 
like ICAL is not sensitive to the CP phase, the first signature of
CP violation may  well come by the addition of ICAL data to those of
these fixed baseline experiments, possibly even a few years after these
experiments have completed their runs \cite{Ghosh:2013yon,Ghosh:2014dba}.

The sensitivities of the fixed baseline experiments to CPV discovery,
and the relative improvement due to the addition of ICAL, have a clear
dependence on $\theta_{23}$, which may be discerned from 
Fig.~\ref{cpv-discovery}, where we compare the results with 
$\theta_{23}=39^\circ$ and $\theta_{23}=45^\circ$.
In the favourable $\dcp$ region, the CPV discovery potential worsens with
increasing $\theta_{23}$.  This is because the  $\dcp$-independent
leading term in Eq.~(\ref{P-emu}) increases with $\theta_{23}$, giving a
higher statistical error, while the CP-dependent term has only a weak
dependence on this parameter \cite{Huber:2009cw}.  In the unfavourable
region, on the other hand, the CPV discovery potential improves with increasing
$\theta_{23}$. This happens because here the minimum of $\chi^2$ from the
long baseline experiments comes with the wrong hierarchy, and the
atmospheric neutrino data is needed to bring it to the correct hierarchy.
The hierarchy identification capability of atmospheric neutrino data
increases for larger $\theta_{23}$, hence the improvement in CPV
discovery potential with higher $\theta_{23}$, as can be seen in 
Fig.~\ref{cpv-discovery}.
For $\theta_{23} \gtrsim  50^\circ$, the ICAL information is nearly 
superfluous, since the hierarchy sensitivity of the T2K and \nova\
combination itself is good enough to exclude the wrong hierarchy
CPV discovery minimum even for unfavourable $\dcp$ values.

In this section we have shown that for unfavourable values of $\dcp$,
atmospheric neutrino data from ICAL considerably improves the CPV
discovery potential of T2K and \nova, and could lead to a significant CPV
discovery using existing and upcoming facilities for a large fraction
($\gtrsim 50\%$) of $\dcp$ values.  Adding ICAL muon data to T2K and \nova\
results in an enhanced CPV discovery potential at $2\sigma$ for almost
twice the range of $\dcp$ values compared to the fixed baseline experiments
alone. For maximal CPV the significance of the signal can reach $3\sigma$ in
the unfavourable half-plane also. Indeed, if nature has chosen such
unfavourable combinations of parameters then the addition of ICAL
to T2K+\nova\  may give us the first signal of leptonic CP violation.

\chapter{Exploring New Physics at ICAL}
\label{exotic}

\begin{flushright}
{\it
The end of all our exploring will be
to arrive where we started \\
And know the place for the first time. \\
-- T.S. Eliot
}
\end{flushright}

In addition to determining the parameters describing neutrino masses
and mixing, the neutrino detection at ICAL may be used for probing
various sources of new physics that could affect neutrino oscillations.
Moreover, while ICAL is primarily designed for detecting muons and 
hadrons produced from neutrino interactions, it can also be sensitive
to more exotic particles like magnetic monopoles or dark matter particles
passing through the detector. 
In this chapter we briefly discuss a few such ideas.

\section{Probing Lorentz and CPT Violation} 

Invariance under the product of charge conjugation (C), parity (P)
and time reversal (T), {\it i.e} the CPT theorem \cite{W.Pauli,G.Grawert}, 
is an essential feature of quantum field 
theories that underlie particle physics. This is a consequence of
the invariance of the Lagrangian under proper Lorentz transformations.
However in a Standard Model Extended (SME) Lagrangian that does not
respect the Lorentz trasformation symmetry, the CPT violation (CPTV) may be
manifest \cite{Kostelecky:2003xn,Kostelecky:2003cr}, 
which may be measurable at the neutrino oscillation experiments.
Some bounds on the CPTV parameters have already been obtained, 
using the atmospheric neutrino data from SuperKamiokande \cite{Abe:2014wla}.
The ultrahigh energy neutrino data is expected to be especially
sensitive to CPTV; for instance see \cite{Bhattacharya:2010xj}.
If the effects of CPTV are not observed, one may obtain limits on CPT and 
Lorentz violating parameters. 

\subsection{CPTV effects at the probability level}

Lorentz violation may be introduced in the effective Lagrangian for a 
single fermion field as \cite{Colladay:1996iz} 
\begin{equation}
{\cal L} = i \bar{\psi} \partial_\mu \gamma^\mu \psi 
-m \bar{\psi} \psi 
- A_\mu \bar{\psi} \gamma^\mu \psi 
- B_\mu \bar{\psi} \gamma_5 \gamma^\mu \psi  \;,
\label{Lagran1}
\end{equation}
where $A_\mu$ and $B_\mu$ are constant 4-vectors. The terms containing 
$A_\mu$ and $B_\mu$ may be induced by new physics at higher energies,
for instance. Since $A_\mu$ and $B_\mu$ are invariant under boosts 
and rotations, they explicitly give rise to Lorentz violation,
which in turn leads to CPTV \cite{Greenberg:2002uu}. (CPT
violation may also occur if particle and antiparticle masses are
different. Such violation, however, also breaks the locality assumption
of quantum field theories\cite{Greenberg:2002uu}. This mode of CPTV is not
considered here.)
The effective CPTV contribution to the neutrino Lagrangian can then be
parametrized \cite{Coleman:1998ti} as
\begin{equation}
{\cal L}_\nu^{\rm CPTV} =  
\bar{\nu}_L^\alpha \, b^{\alpha \beta}_\mu \, 
\gamma^\mu \, \nu_L^{\beta} \; ,
\label{L-nu}
\end{equation} 
where $b_{\mu}^{\alpha \beta}$ are four Hermitian $3\times 3$ matrices
corresponding to the four Dirac indices $\mu$, while $\alpha, \beta$ are
flavor indices. The effective Hamiltonian in vacuum for
ultra-relativistic neutrinos with definite momentum $p$ then becomes
\begin{equation}
H \equiv \frac{M M^\dagger}{2 p} + b \; ,
\label{eff-H}
\end{equation}
where $M$ is the neutrino mass matrix in the CPT conserving limit
and $b$ is the CPT violating term.

For atmospheric neutrinos that may pass through appreciable amounts of
matter before reaching the detector, the effective Hamiltonian in the 
flavor basis, that takes the matter effects and CPTV effects into account,
may be written as
\begin{eqnarray}
H_{f} & = &  \frac{1}{2E}. {U}_{0}. D(0,\Delta{m}_{21}^{2},\Delta{m}_{31}^{2}).
U_{0}^{\dagger} \nonumber \\
& & + U_{b}. D_b(0,\delta{b}_{21},\delta{b}_{31}). U_{b}^{\dagger} 
+ D_m(V_{e},0,0) \; , 
\end{eqnarray}
where $U_{0}$ and $U_{b}$ are unitary matrices that diagonalize the
$M M^\dagger$ and $b$ matrices, respectively, while $D, D_m$ and $D_b$
are diagonal matrices with their elements as listed in brackets. 
Here $V_e$ = $\sqrt{2} G_{F} N_{e}$ with the electron number density $N_e$, 
and $\delta{b_{i1}} \equiv b_i - b_1$ for $i =2,3$, with $b_1$, $b_2$ and $b_3$ 
the eigenvalues of $b$. 

As in the standard convention, $U_{0}$ is parametrized by three mixing 
angles $(\theta_{12}, \theta_{23}, \theta_{13})$ and one phase $\delta_{\rm CP}$.
The matrix $U_{b}$ is prametrized by three mixing angles 
$(\thb_{12}, \thb_{23}, \thb_{13})$ and six phases. 
Thus, $H_f$ contains 6 mixing angles 
$(\theta_{12}, \theta_{23}, \theta_{13}, \thb_{12}, \thb_{23}, \thb_{13} )$ 
and seven phases.

The results will clearly depend on the mixing angles in the CPTV sector. 
In this section, we examine the effects of three different representative 
sets of mixing angles:
(1) small mixing:
($\thb_{12}=6\,^{\circ}, \thb_{23}=9\,^{\circ}, \thb_{13}=3\,^{\circ}$),
(2) large mixing: ($\thb_{12}=38\,^{\circ}, \thb_{23}=45\,^{\circ},
\thb_{13}=30\,^{\circ}$) and 
(3) Identical to the mixing angles in the PMNS matrix: 
($\thb_{12}=\theta_{12}, \thb_{23}=\theta_{23}, \thb_{13}=\theta_{13}$). 
For simplicity, all seven phases have been taken to be zero, and
the neutrino oscillation parameters as given in Table~\ref{tab:best-fit}
are used.
It is observed that in the probability expressions, $\delta{b}_{21}$
always appears with the smaller (by a factor of 30) mass squared
difference $\Delta m^2_{21}$. Thus its effects on oscillations are
subdominant, limiting the capability of atmospheric neutrinos to constrain
it, so that no useful constraints on $\delta{b}_{21}$ seem to be possible.

\begin{figure}[]
\centering
\subfloat[][Case 1, NH]{
 \includegraphics[width=0.49\textwidth,height=0.35\textwidth]
{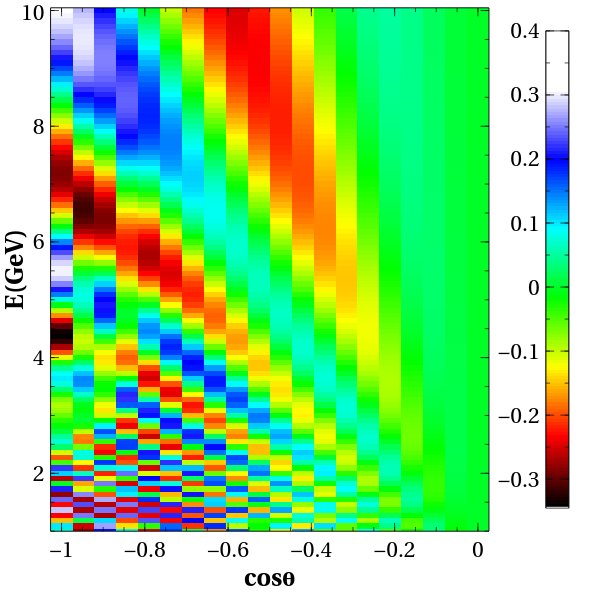}
  \label{fig:Oscgrm_Pmm_smallmix_NH}
}
\subfloat[][Case 1, IH]{
 \includegraphics[width=0.49\textwidth,height=0.35\textwidth]
{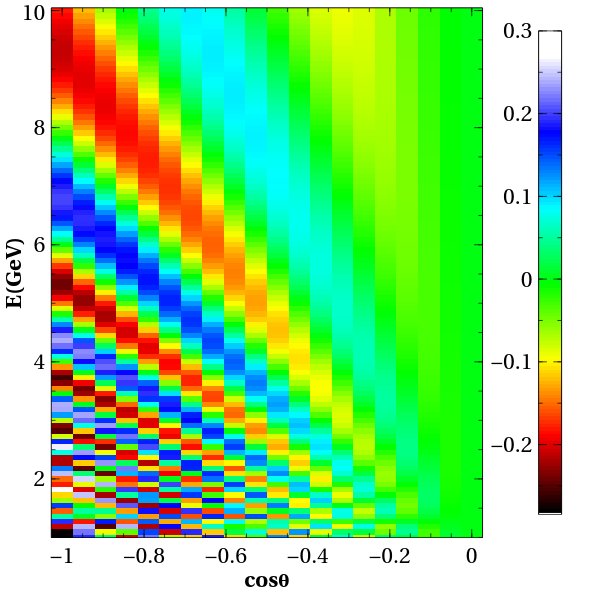}
  \label{fig:Oscgrm_Pmm_smallmix_IH}
}
\quad
\subfloat[][Case 2, NH]{
\includegraphics[width=0.49\textwidth,height=0.35\textwidth]
{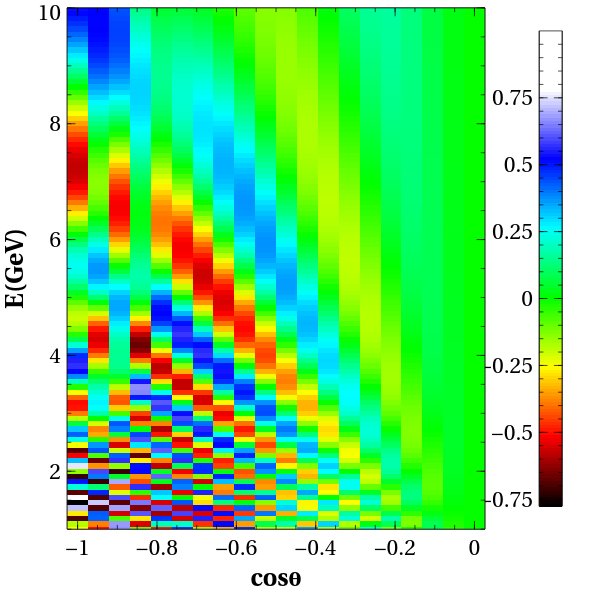}
  \label{fig:Oscgrm_Pmm_largmix_NH}
}
\subfloat[][Case 2, IH]{
\includegraphics[width=0.49\textwidth,height=0.35\textwidth]
{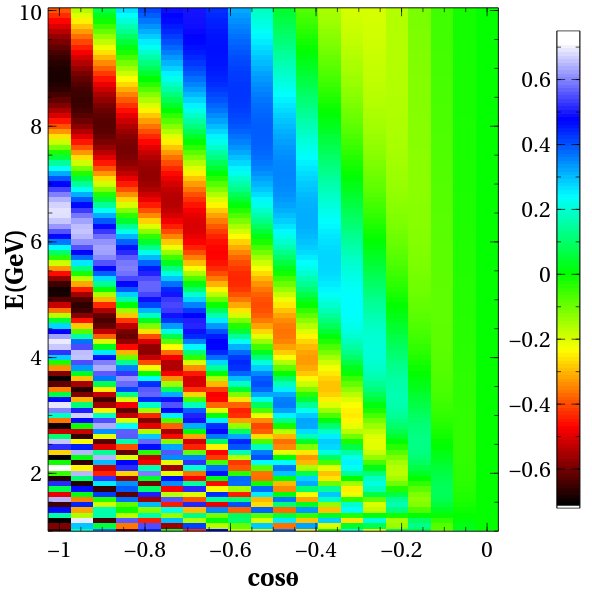}
  \label{fig:Oscgrm_Pmm_larmix_IH}
}
\quad
\subfloat[][Case 3, NH]{
\includegraphics[width=0.49\textwidth,height=0.35\textwidth]
{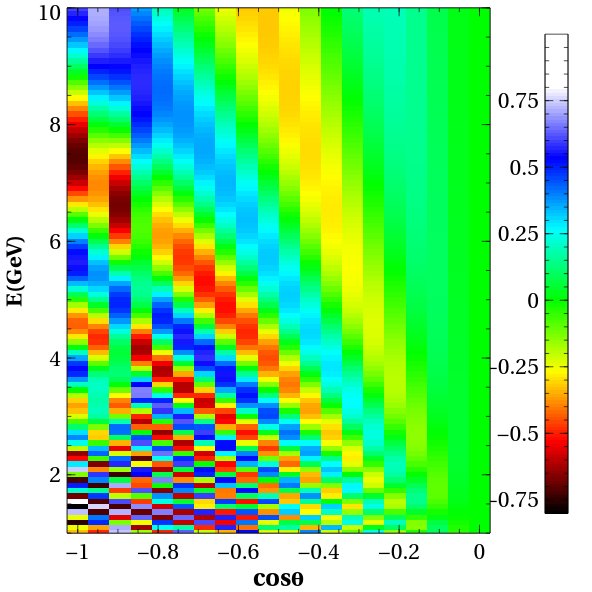}
 
  \label{fig:Osc_Pmm_numix_NH}
}
\subfloat[][Case 3, IH]{
\includegraphics[width=0.49\textwidth,height=0.35\textwidth]
{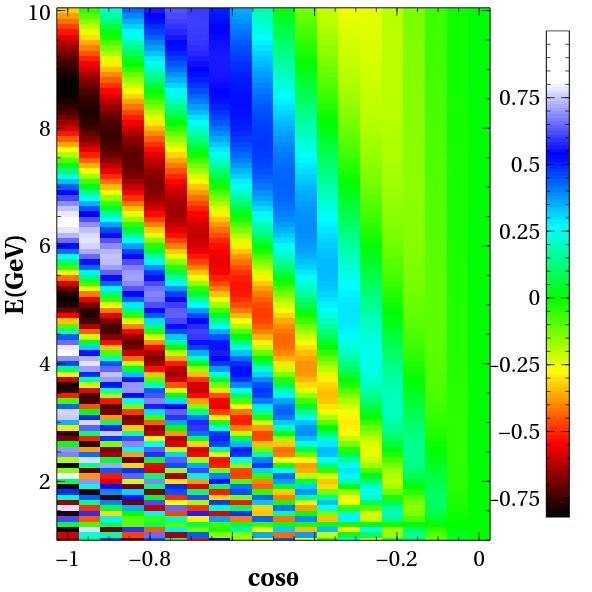}
\label{fig:Osc_Pmm_numix_IH}
}
\caption{
\label{fig:Osc_Pmm} 
The oscillograms of $\Delta{P} = (P_{\nu_{\mu}\nu_{\mu}}^{U_{b}\neq 0} - 
P_{\nu_{\mu}\nu_{\mu}}^{U_{b}= 0}$) for 3 different mixing cases as described
in the text. 
The left and right panels are for Normal and Inverted hierarchy, respectively.
The value of $\delta{b}_{31}=3\times 10^{-23}$ GeV has been taken for
illustration \cite{Chatterjee:2014oda}. }
\end{figure}

For ilustration, the oscillograms for the difference of the survival 
probability of $\nu_\mu$
with and without CPTV for $\delta{b}_{31}=3\times 10^{-23}$ GeV are shown in 
Fig.~\ref{fig:Osc_Pmm}. Several general features may be observed.
First, the CPTV effects are larger at larger baselines for all energies.
This is as expected from the results of 
the two-flavour analysis \cite{Datta:2003dg},  which showed that  
the survival probability difference in vaccuum is proportional 
to $\sin(\frac{\Delta{m}^{2} L}{2E})\sin(\delta{b} L)$. 
Second, as is well-known, matter effects are large and resonant
for neutrinos with NH, and for antineutrinos with IH. Thus in both
these cases, they mask the (smaller) effect of CPTV stemming from $U_{b}$. 
Hence for neutrino events, the CPTV sensitivity is significantly higher
if the hierarchy is inverted as opposed to normal, and the converse is
true for antineutrino events. 
Finally, the CPTV effects are largest for the cases 2 and 3, 
as compared to case 1. This is due to the fact that mixing in case 1
is very small compared to other two. The origin of the difference for the 
cases 2 and 3 is likely due to the fact that CPT violating effects are 
smaller when $\theta_{b13}$ is large \cite{Chatterjee:2014oda}. 

\subsection{Simulation procedure and results}

The oscillation probabilities are calculated with the true values of 
oscillation parameters corresponding to Table~\ref{tab:best-fit} and 
assuming no CPTV. The re-weighting algorithm \cite{Ghosh:2012px} has been used  
to generate oscillated events. The energy resolutions, efficiencies,
and charge misidentification errors are taken into account.
Oscillated muon events are binned as 
a function of muon energy (10 bins) and muon zenith angle (40 bins). 
These binned data are folded with detector efficiencies and resolution 
functions. These data (labelled as $N^{ex}$) are then fitted with another 
set of data (labelled as $N^{th}$), where CPTV is allowed. 
The statistical significance of the difference between these two sets of
data is calculated, using the pull method to include systematic errors
as in the analyses in Chapter~\ref{analysis}.
The $\chi^2$ is calculated separately for the $\mu^+$ and $\mu^-$ 
events and then added, to exploit the charge identification capability
of ICAL. Marginalization has been carried out with respect to the
neutrino oscillation parameters  $\Delta{m}_{31}^{2}$, $\theta_{23}$, 
$\theta_{13}$, which are varied over their $3\sigma$ allowed ranges. 
The CP violating phase $\delta_{\rm CP}$ is varied over its whole range, 
while $\delta{b_{21}}$ is marginalized over the range 0 to
$5 \times 10^{-23}$ GeV.

\begin{figure}[htp]
\centering
\subfloat[]{
\includegraphics[width=0.49\textwidth,height=0.35\textwidth]
{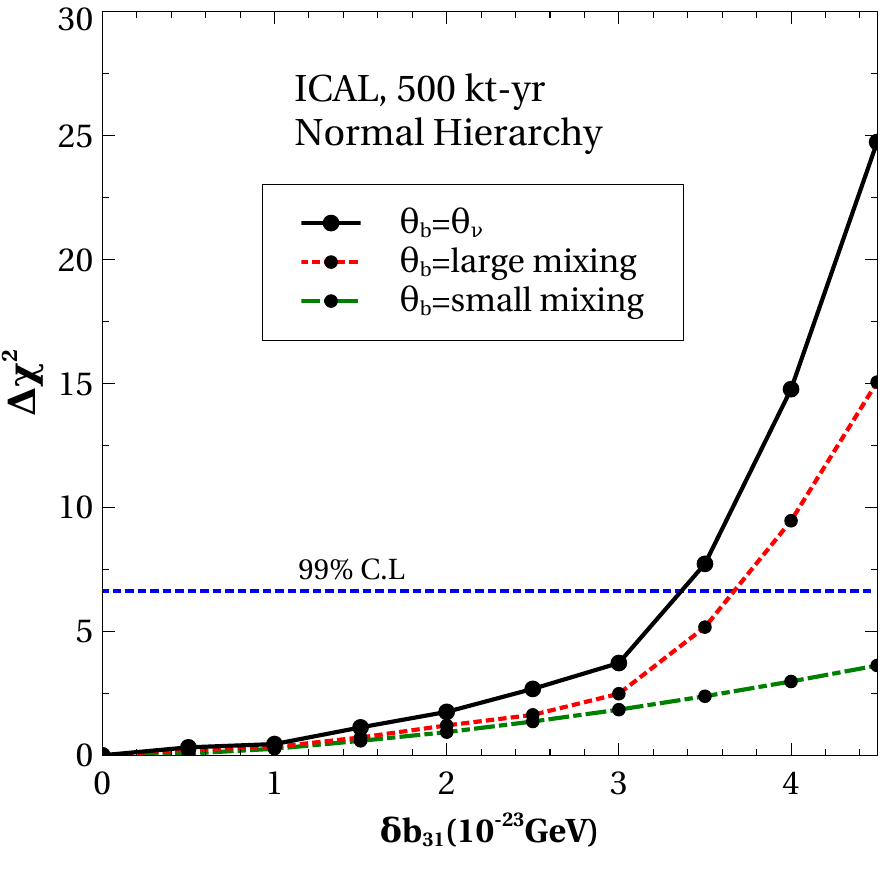}
\label{fig:3mix_compar_NH_b21margin}
}
\subfloat[]{
\includegraphics[width=0.49\textwidth,height=0.35\textwidth]
{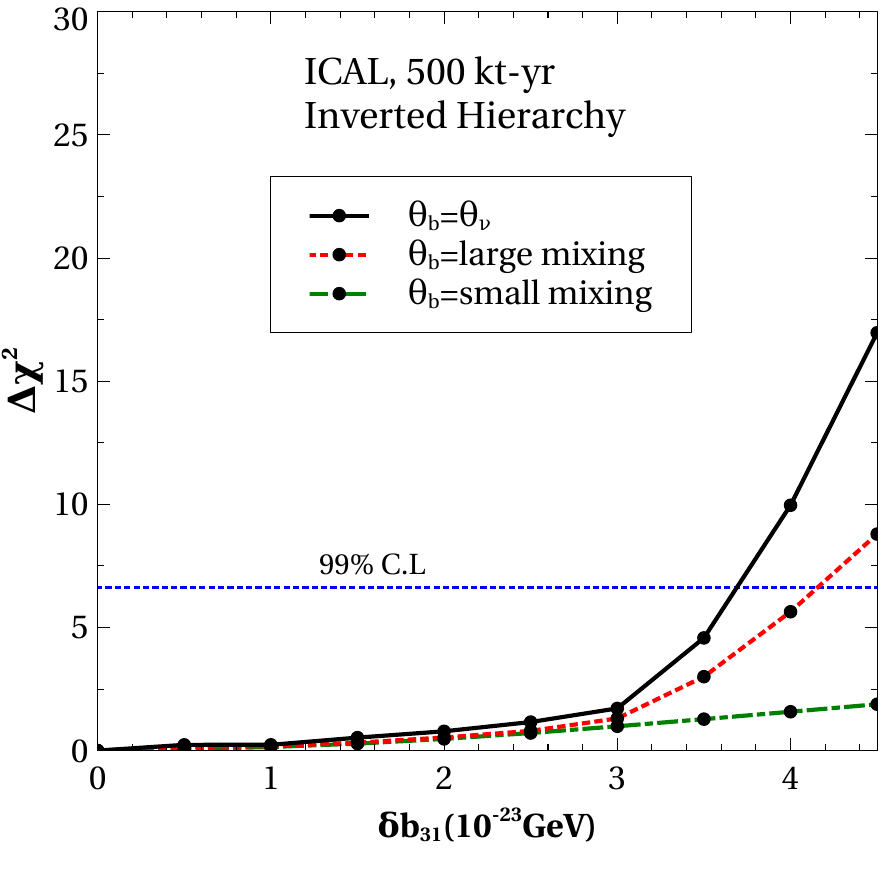}
\label{fig:3mix_compar_IH_b21margin}
}
\caption{
The bounds on $\delta{b}_{31}$ for different cases for the mixing angles
$\theta_{b}$. 
The results are marginalized over $\theta_{23}$, $\theta_{13}$,
$\delta_{\rm CP}$, $\Delta{m}_{31}^{2}$ and $\delta{b_{21}}$. Left and right
panel are for Normal and Inverted hierarchy respectively 
\cite{Chatterjee:2014oda}.}
\label{NH_Plot} 
\end{figure}

Figure~\ref{NH_Plot} illustrates the results of the analysis. 
It is observed that the best bounds on $\delta b_{31}$ are obtained
for both hierarchies in case 3, where mixing in the CPTV sector 
is the same as the neutrino mixing. 
Good bounds are also obtainable if the mixing in the CPTV sector
is large, as in case 2. 
While the analysis for all mixing angles of $U_b$ has not been carried 
out, these results indicate that, as long as the mixing angles are
not too small, limits on $\delta b_{31}$ of the order of
$4\times 10^{-23}$ GeV would be possible at 99\% C.L.. 
It should be noted that the above analysis assumes that the
mass hierarchy is known. Indeed, if the marginalization over
hierarchy is carried out, the results are considerably weaker.

Note that this study pertains to the type of CPTV that may be parametrized 
by Eq.~\ref{Lagran1}, which stems from explicit Lorentz violation, and 
the analysis is restricted to the muon detection channel. 
The CPTV that gives rise to differing masses for particles and antiparticles
has not been considered. Given the smaller statistics and flux uncertainties 
that typify atmospheric data, it would be difficult to obtain good 
sensitivities to this type of CPTV, which breaks the locality assumption 
of quantum field theories. Finally, we note that in order to obtain good 
sensitivity to CPTV, knowing the hierarchy will be an important asset.
This will anyway be the focus of many other analyses at ICAL. 


\section{Search for Magnetic Monopoles}

The possible existence of magnetic monopoles (MM) is predicted by 
unification theories. These monopoles can have their magnetic charge 
$g$ quantized via $e g = n \hbar c/2$, where $n$ can take positive
or negative integer values \cite{Dirac:1931hr,Saha:1936hr}.   
MM solutions of the classical equations of motion for spontaneously broken
non-abelian gauge theories \cite{Hooft:1974hr,Polyakov:1974hr} 
lead to a lower bound on 
the mass of the monopole: $M_{\rm MM} \geq M_X/G$, where $M_X$ is the mass of the 
carrier of the unified interaction, and G is the unified coupling constant.
The MMs are therefore expected to have large masses.
They are expected to be created during the big-bang, and being heavy, their
relative speed with respect to the Earth should be of the same order
of magnitude as the speed of the galaxy, i.e. $\sim 10^{-3} c$.
The MMs that enter the Earth lose their energy gradually by electromagnetic
interactions with the surrounding medium \cite{Derkaoui:1998hr}. If their mass
is small, so will be their kinetic energy and they will be stopped in
the Earth matter. The heavier MM's can however penetrate large distances,
and can reach deep inside the Earth, like in the ICAL cavern.

Monopoles would be accelerated in the magnetic field of the galaxy, and 
hence energy would be drained from the galactic magnetic field. 
Since the rate of this energy loss should be small over the time
scale of regeneration of the galactic magnetic field, an upper bound on 
the MM flux can be obtained \cite{Parker:1970hr}.
This bound on the MM flux, called the Parker limit, is about 
$10^{-15}$ cm$^{-2}$ s$^{-1}$ sr$^{-1}$ for $M_{\rm MM} \leq 10^{17}$ GeV. 
For higher masses, the bound increases linearly.
The detectors used for detecting MM are mainly based either on the principle
of induction or excitation / ionization. The induction method, 
where the induced magnetic 
field is measured using a Superconducting QUantum Interferometer Device 
(SQUID) \cite {Cabrera:1982hr,Bermon:1990hr}, has yielded the upper 
bound on the MM flux 
to be $\sim 3.8 \times 10^{-13}$ cm$^{-2}$ s$^{-1}$ sr$^{-1}$. 
The ionization method has been used in
Cherenkov detectors \cite{Kajita:1985hr,Aynutdinov:2008hr}, as well as 
in scintillators and gaseous detectors 
\cite{Thron:1992hr,Ambrosio:2002hr,Balestra:2008hr}.
The bounds from these experiments are much tighter -- for example, 
the current best bound for $10^{10}$ GeV $< M_{\rm MM} < 10^{16}$ GeV is 
$\sim 2.8 \times 10^{-16}$ cm$^{-2}$ s$^{-1}$ sr$^{-1}$, 
from the MACRO experiment \cite{Ambrosio:2002hr}.
MMs have also been looked for in accelerator-based experiments 
\cite{Abulencia:2006hr,Kalbfleisch:2000hr,Aad:2006hr},
 
When MMs pass through ICAL, the ionization produced in the RPC detectors 
can be detected as a pulse which carries the information that there has 
been a hit, and the time of the hit.
Since the momentum of the MMs would be large, they will transfer only
a small fraction of it to the detector, and will travel through the
detector in a straight line, almost unaffected. 
The large surface area of ICAL, combined with the large number of layers
the MMs would be able to pass through, makes the MM detection possible.
The track of the MM in the ICAL and 
the characteristic sequence of trigger times of consecutive layers of RPCs 
will help in identifying the MM against the background. 
Here we focus on MMs with masses $10^7$--$10^{17}$ GeV and 
speeds $10^{-3}c$--$0.7c$ \cite{Dash:2015qha}. 
(The results are presented in terms of
$\beta \equiv v/c$ of the particle.)
In this parameter space, if the MM flux is near the Parker limit, a few
events per year may be expected at ICAL. On the other hand,
the non-observation of such events
would allow ICAL to put strong bounds on the incoming MM flux.

\subsection{Monopole simulation for ICAL}

The ICAL detector response for the monopole events is simulated using GEANT4,
wherein the ICAL detector geometry is defined.
A rock mass of density $2.89$ g/cc of height $1.3$ km from the top 
surface of the detector is considered in addition to the ICAL itself.
Particles are incident on the surface of the rock, and pass through it
before being detected in ICAL. An isotropic flux of downward-going MM's
is taken into account by smearing it over the zenith angle (cos$\theta$) 
from $0$ to ${\pi}/{2}$, and uniformly over the $2\pi$ range of the 
azimuthal angle $\phi$.

The stopping power in the rock of the earth and the iron of the ICAL are 
taken care of using \cite{Ahlen:1978hr,Ahlen:1982hr,Derkaoui:1998gf}
\begin{equation}
-\frac{dE}{dX}=\frac{4\pi {N_{e}}^{2}g^{2}}{m_{e}c^{2}} \Bigg[ \ln(\frac{2m_{e}c^{2}\beta^{2}\gamma^{2}}{I_{m}})-\frac{1}{2}-\frac{\delta_{m}}{2}
-B(\mid g \mid)+\frac{K(\mid g \mid)}{2} \Bigg] \; .
\end{equation}
Particles also lose energy in the active region of the detector which has 
a gas mixture [R134A (95.15$\%$), Iso-Butane (4.15$\%$) and $SF_6$ (0.34$\%$)],
and register hits, whose position and time information is recorded. 
This allows the reconstruction of the velocity of the particle.

\begin{figure}[h]
\includegraphics[width=0.49\textwidth,height=0.35\textwidth]
{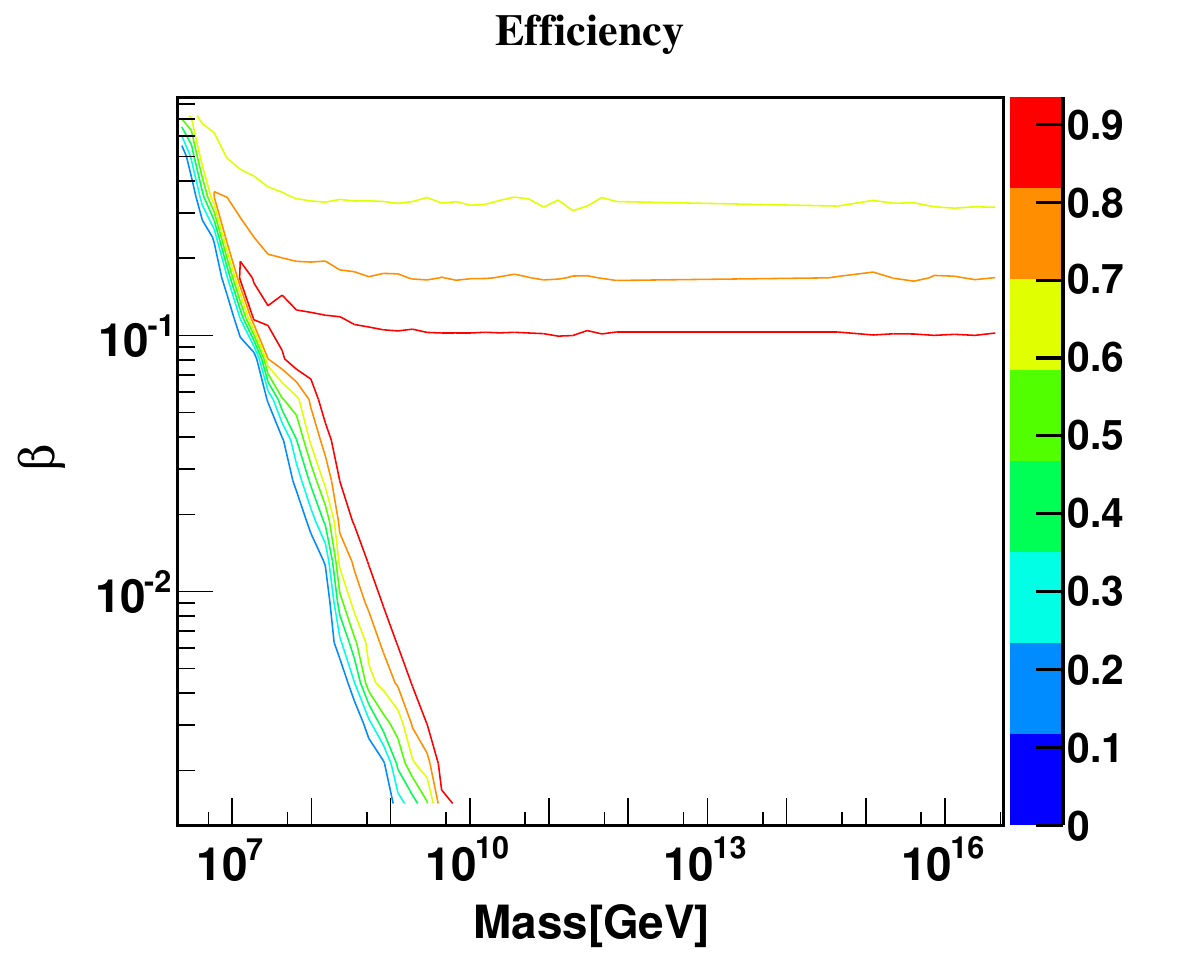}
\includegraphics[width=0.49\textwidth,height=0.35\textwidth]
{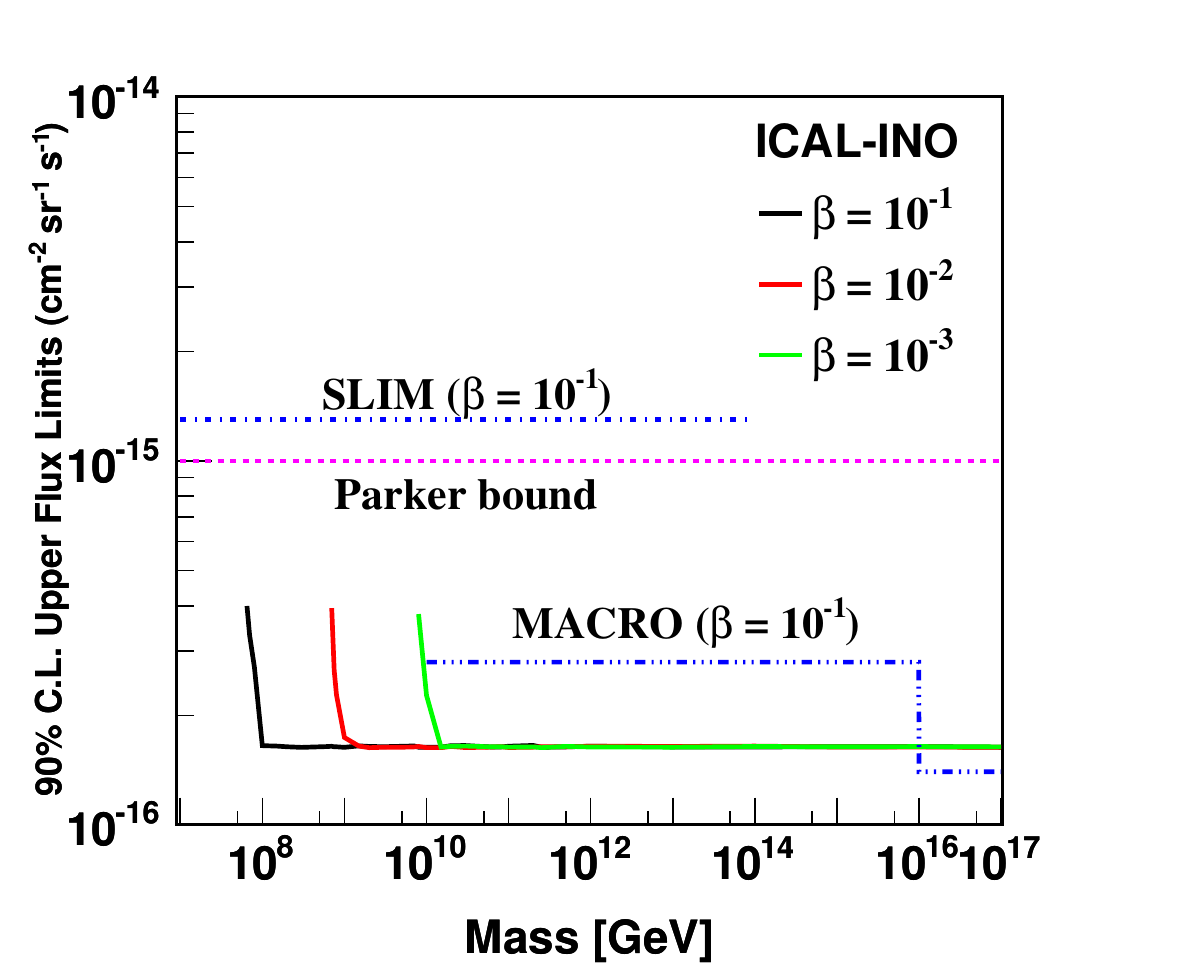}
\caption{\label{efficiency-fluxlim} 
{The left panel shows the efficiency
of the detector for MM events, with cuts on the maximum value of
$\beta$ and the minumum number of layers to be traversed by the particle.
The right panel indicates the 90\% upper bounds on the MM flux
expected from 10 year running of the 50 kt ICAL, if the number of 
candidate events is zero \cite{Dash:2015qha}.
The upper bounds obtained from the MACRO \cite{Ambrosio:2002hr} 
and SLIM \cite{Balestra:2008hr} have also been shown.}}
\end{figure}

We use the time-of-flight method for identifying the MM using ICAL. 
For relativistic monopoles, the high energy muons will constitute the main 
background. We avoid this background by focussing on the events with
$\beta < 0.8$. In the smaller velocity region, 
the background will be due to chance coincidences. This can also be 
minimised by choosing only those events which cross a certain minimum 
number of layers: if the noise rate per pickup strip is ${\cal R}$
and the speed of the particle is $\beta c$, the probability of getting
random hits in $n$ consecutive layers in the relevant time window is
$n R^n (n d/\beta c)^{n-1}$, where $d$ is the distance between two layers.
Thus for $R \sim 200$ Hz and $\beta =0.1$, for example, each additional
layer would decrease the probability of chance coincidence by 
$\sim 10^{-5}$. We require the particle to cross at least 10 RPC layers, 
which would suppress both the high-$\beta$ and low-$\beta$ backgrounds 
to negligible levels. 
For each mass and $\beta$, we use a sample of 10,000 events to 
estimate the detector efficiency. The result is shown in the left panel of
Fig.~\ref{efficiency-fluxlim}. It may be observed that the efficiency is 
almost 90\% for $M_{\rm MM} > 10^{12}$ GeV and $10^{-3} < \beta < 0.1$. 

\subsection{Reach for limits on the monopole flux}

The event rate expected at ICAL may be estimated by taking the area 
$A$ of the top surface as the effective area, and the solid angle in
which the MM would cross the cut of minimum number of layers as
the effective solid angle $\Omega$. If the MM flux is $f$ and the detector 
efficiency is $\epsilon$, the expected number of events $N_{\rm ex}$ 
after a running time $T$ of the detector would be 
\begin{equation}
N_{\rm ex}=f({\rm cm}^{-2}~{\rm sr}^{-1}~{\rm s}^{-1})~A(cm^2)~
\Omega({\rm sr})~T({\rm s})~\epsilon \; .
\end{equation}

If the total observed number of events $N_{\rm obs}$ is not significantly
greater than the expected number of background events $N_{\rm BG}$, then an
upper bound on the MM flux may be obtained. In 
the right panel of Fig.~\ref{efficiency-fluxlim},
we present the upper bound that will be obtained at the 50 kt ICAL 
in 10 years, in the scenario $N_{\rm obs} = N_{\rm BG} =0$ 
for different $\beta$ values.
From the figure, it may be observed that an upper bound of 
$\sim 2 \times 10^{-16}$ cm$^{-2}$ s$^{-1}$ sr$^{-1}$ (90\% C.L.)
should be possible with an exposure of 500 kt-yr of the ICAL detector.
This is fairly competitive with the current strongest bound
coming from MACRO \cite{Ambrosio:2002hr}.
Indeed a direct comparison for $\beta = 0.1$ shows that for 
$M_{\rm MM} \lesssim 10^{16}$ GeV, the ICAL reach is clearly better.
Monopoles with higher masses can penetrate through the Earth, and
the additional up-going events accessible to the detector cause the
increase in the sensitivity of MACRO as seen in the figure.
Our analysis presented here is restricted to the flux from the upper 
hemispehere, and the upward-going events at high monopole masses 
have not been taken into account.
With additional detectors on the walls and ceiling of the cavern,
the sensitivity of ICAL to MMs can be increased by about a factor of 2 
\cite{Dash:2015qha}.

\chapter{Concluding Remarks}

This report presents the first systematic study of the
physics capabilities and potential of the ICAL detector. This includes
the response of ICAL to the muons and hadrons produced in the 
$\nu_\mu$ interactions in the detector, and an understanding of 
the physics reach of the experiment for the identification of neutrino 
mass hierarchy and precision measurements of the atmospheric mixing parameters.

The ICAL detector geometry, and its elements such as iron plates, RPCs, 
the magnetic field, etc. are coded in the GEANT4 simulations framework.
At this stage the atmospheric neutrino fluxes used are those at the
Kamioka site, however the fluxes at the INO site
will soon be incorporated. The atmospheric muon neutrino-induced
events are generated using the NUANCE neutrino generator.

The ICAL detector is sensitive to muons in the GeV range. The muon
momenta are reconstructed using a Kalman filter algorithm that uses
the bending of muons in the magnetic field. It enables the reconstruction 
of muons with an efficiency of more than 80\%, and
the measurement of its momentum with a precision of $\sim$ 20\% (10\%)
at 1 GeV (10 GeV). The muon charge is identified correctly with an 
efficiency of more than 98\% for $E_\mu > 4$ GeV, while the zenith
angle of the muon at the point of production can be reconstructed to
within a degree.

The reconstruction of multi-GeV hadrons is a unique feature of the ICAL.
Using calibration against the number of hits in the detector, the energy
of a hadron shower can be reconstructed to $\sim$ 85\% (35\%) at 1 GeV
(15 GeV). The addition of the hadron energy information 
enhances the reach of ICAL much above that from the muon information alone. 
The optimal analysis method to obtain the best sensitivity seems to be the 
analysis that uses these three quantities in each event, without trying
to reconstruct the neutrino energy or direction itself.

The simulations studies have included various systematic uncertainties from
neutrino fluxes and cross sections. However some detector systematics,
like the background contributions from mis-identified neutral current 
events, have not yet been included in the analyses. 
The studies show that with 10 years run of the 50 kt ICAL, 
the mass hierarchy may be identified with a significance of $\chi^2 \geq 9$, 
i.e. more than $3\sigma$. At the same time, the values of the atmospheric 
mixing parameters $|\Delta m^2_{32}|$ ($\sin^2 \theta_{23}$) may be determined to 
a precision of 3\% (12\%). The identification of the octant of $\theta_{23}$
with the ICAL alone is limited to a $2\sigma$ significance even for
favourable ranges of the parameter. 

Detection of the atmospheric neutrinos with ICAL is rather insensitive to 
the value of $\delta_{\rm CP}$. This means that its reach for the mass hierarchy
identification does not depend on the actual value of $\delta_{\rm CP}$, 
unlike the ongoing fixed baseline experiments.  
Since it provides a data set that is free of the 
ambiguities due to the unknown $\delta_{\rm CP}$, the addition of the
ICAL data to the long baseline data of the ongoing experiments would
enhance the $\delta_{\rm CP}$ reach of these experiments. This synergy of
the ICAL with other concurrent experiments should be exploited. 

Given the multipurpose nature of ICAL, it can also be used for 
exploring exotic physics scenarios like CPT violation, magnetic monopoles, 
dark matter, etc., and more such avenues of using the ICAL data are sure 
to be found in years to come.
The simulation studies presented in this report make a strong physics case
for the ICAL detector. The codes and algorithms used in the analyses 
will be improved and fine-tuned to make them more realistic and efficient. 
This is expected to be a continually ongoing effort.

In parallel with the simulations described in this Report, efforts have 
gone on to finalise the design and structure of the ICAL detector modules, 
including the magnet, the RPCs, and their associated electronics. 
A prototype 1:8 scale model is planned to be built in Madurai, India, 
where most of the design will be validated. 
In the meanwhile, many more exciting and novel ideas are in the pipeline, 
and would be presented in future versions of this report.

The Government of India has recently (Dec 2014) given the final approval 
for the establishment of INO, giving a big boost to the project.
The INO collaboration currently has more than 20 participating institutes
from India, and welcomes the participation of high energy physicists
from all over the world.

\chapter*{Acknowledgements}
\addcontentsline{toc}{chapter}{Acknowledgements}

The simulation and physics analysis of the ICAL detector would not have been
possible without those who have set up the simulation framework for INO on 
multiple platforms; we would like to especially thank Pavan Kumar and
P. Nagaraj, as well as the computing resource personnel in all the 
institutions where this work was carried out. We are grateful to Dave Casper 
for making the NUANCE neutrino generator software freely available, and 
answering a long list of questions on its use in the initial stages.
We also thank Sunanda Banerjee for sharing his expertise on GEANT.  
We would also like to express our gratitude towards M. Honda, who has 
provided us the detailed theoretical predictions of atmospheric neutrino 
fluxes, at the SuperKamiokande site as well as at the INO site.

Such a large endaevour is not possible without direct or indirect support
from many of our colleagues. While it is not possible to mention the names
of everyone here, we would like to thank Debajyoti Choudhury, Anindya Datta, 
Anjan Joshipura, Namit Mahajan, Subhendra Mohanty, and many of our 
international colleagues who have worked on phenomenological aspects of 
atmospheric neutrino oscillations, and whose ideas have been developed 
into some of the physics analysis issues discussed in this report.
Many students and postdocs have contributed to the development and testing
of the simulation framework in the initial stages, and to the early results.
We would like to make a special mention of Abhijit Bandyopadhyay, 
Subhendu Chakrabarty, G. K. Padmashree, Arnab Pal, Subhendu Rakshit, 
Vikram Rentala and Abhijit Samanta, who have worked on simulations
related to atmospheric neutrino oscillations.

While this report deals with the simulations of the ICAL detector, 
these simulations are based on the detector development that has
been taking place in parallel, and the feedback of our hardware colleagues
has been crucial in the fine-tuning of these simulations.
We would therefore like to thank D. N. Badodkar, M. S. Bhatia, 
N. S. Dalal, S. K. Thakur, S. P. Prabhakar and Suresh Kumar 
for the development and simulation of the magnet; 
Anita Behere, Nitin Chandrachoodan, S. Dhuldhaj, Satyajit Jena, S. R. Joshi, 
N. Krishnapurna, S. M. Lahamge, P. K. Mukhopadhyay, P. Nagaraj, B. K. Nagesh,
S. Padmini, Shobha K. Rao, Anil Prabhakar, Mandar Saraf, 
R. S. Shastrakar, K. M. Sudheer, M. Sukhwani and S. S. Upadhya for
electronics development; 
Sarika Bhide, Manas Bhuyan, Suresh Kalmani, L. V. Reddy, R. R. Shinde and
Piyush Verma for the detector development. We would also like
to thank Manoj Kumar Jaiswal, N. Raghuram and Vaibhav Pratap Singh,
students who have completed their projects related to ICAL hardware.

We would like to thank K. S. Gothe, Kripa Mahata, Vandana Nanal,
Sreerup Raychaudhuri and Aradhana Shrivastava for their invaluable help 
in running the INO Graduate Training Program (INO GTP).
We gratefully acknowledge the guidance from C. V. K. Baba (deceased),
Pijushpani Bhattacharjee, Ramanath Cowsik, 
J. N. Goswami, S. K. Ghosh, H. S. Mani, V. S.  Narasimham, 
Sudhakar Panda, Raj Pillay, Amit Roy, D. P. Roy, Probir Roy, Utpal Sarkar, 
C. P. Singh and Bikash Sinha from time to time.

We thank the Homi Bhabha National Institute (HBNI) for accepting
the INO GTP in its ambit.
We are grateful to TIFR, which has been the host institute for INO,
for making its resources available for INO, including for the GTP. 
We would also like to acknowledge BARC for all the support is has provided 
over the years, and for acting as the host Constituent Institution 
of HBNI for the INO GTP.

Finally, we are grateful 
to the Department of Atomic Energy and the Department of Science and
Technology, Government of India, for funding the INO project.
We would also like to acknowledge the funding provided by the 
University Grants Commission for many of
the students working on INO and ICAL, without which this report would 
not have been possible.

\appendix
\chapter{Neutrino Fluxes at the Kamioka and INO sites}
\label{honda-ino-flux}

In this section, we briefly discuss the preliminary atmospheric 
neutrino flux calculations corresponding to the INO site, and compare
them with those for the Kamioka site. This work is based on 
Refs.~\cite{Honda:2011nf} and \cite{Honda:2006qj}. 
The  primary cosmic ray flux model 
based on AMS~\cite{Alcaraz:2000vp} and BESS~\cite{Sanuki:2000wh,Haino:2004nq} 
data has been used, 
and the hadronic interactions have been implemented with  
DPMJET-III~\cite{Roesler:2000he} above 32 GeV, and JAM below 32 GeV. 
For the propagation of cosmic rays in the atmosphere, the model 
NRLMSISE-00~\cite{NRLMSISE-00}, which takes into account the 
temperatures and densities of the atmosphere's components and takes care 
of the position dependence and the time variation in a year, is used. 
The geomagnetic field model used is IGRF2010 \cite{igrf}. The 
atmospheric neutrino flux is obtained using a 3-dimensional scheme below 
32 GeV, and a 1-dimensional scheme above that. The flux calculated in 
both the schemes agree with each other at 32 GeV \cite{Honda:2011nf}.

\begin{figure}
\centering
\includegraphics[width=0.49\textwidth,height=0.35\textwidth]
{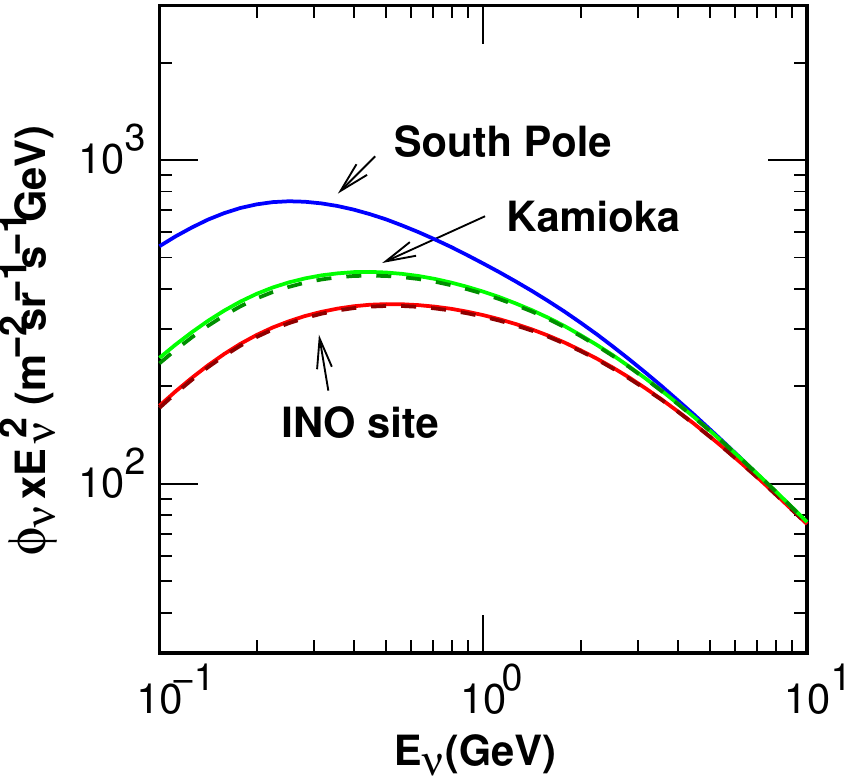}
\caption{Atmospheric neutrino flux averaged over all directions
and summed over $\nu_e~+~\bar\nu_e~+~\nu_\mu~+~\bar\nu_\mu~$,
as a function of neutrino energy for Super-Kamiokande, INO and South Pole sites. This is for the all the energy range of the calculation. 
The fluxes calculated with US-standard76 are also plotted 
in dashed line for the Kamioka and the INO sites.
}
\label{fig:alldir}
\end{figure}

Figure~\ref{fig:alldir} shows the calculated atmospheric neutrino flux 
averaged over one year by folding over all directions and summing over 
all types of neutrinos ($\nu_e,\bar\nu_e,\nu_\mu,\bar\nu_\mu$) for the 
Super-Kamiokande, INO and South Pole sites. These results, obtained 
with the NRLMSISE-00 model, agree well with those obtained from the
US-standard76 \cite{Sanuki:2006yd} model. However the fluxes with the
NRLMSISE-00, as used here, have an advantage in the study of seasonal 
variations, which could be appreciable at the INO site.

\begin{figure}[h!]
\includegraphics[width=0.49\textwidth,height=0.45\textwidth]
{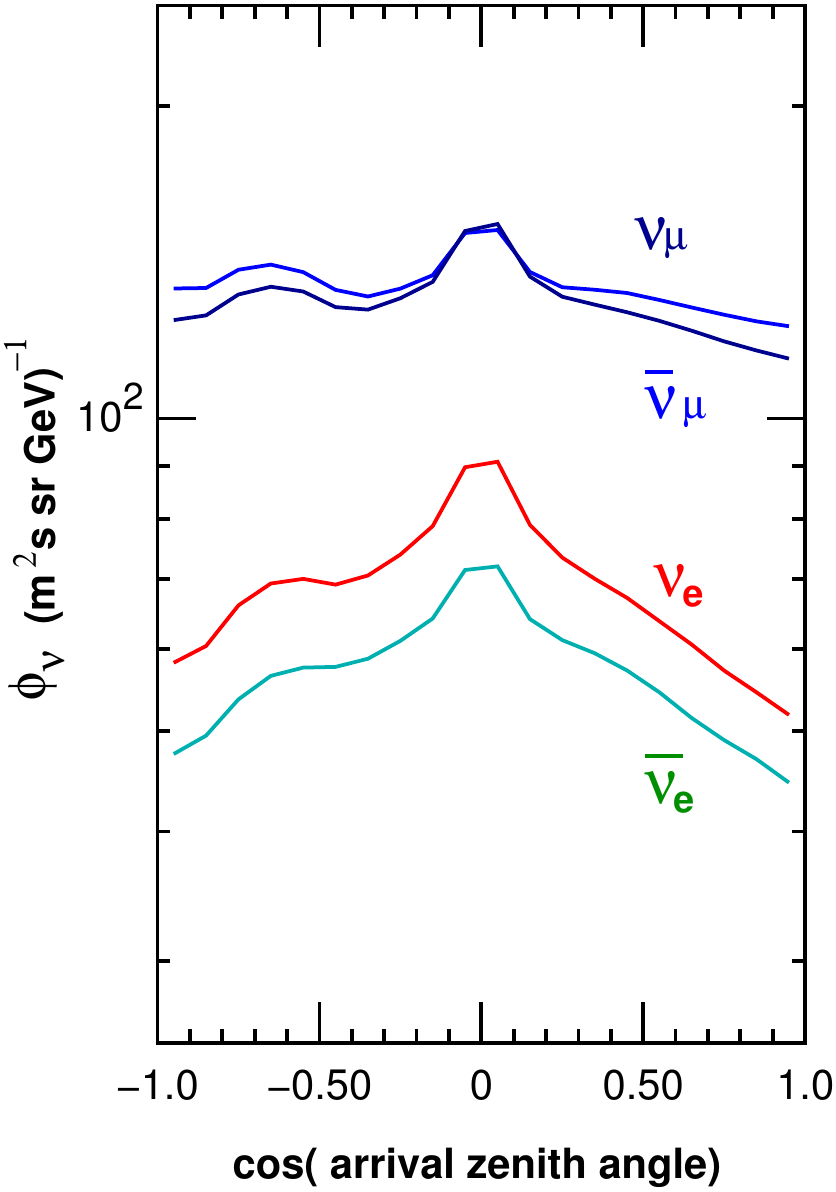}
\includegraphics[width=0.49\textwidth,height=0.45\textwidth]
{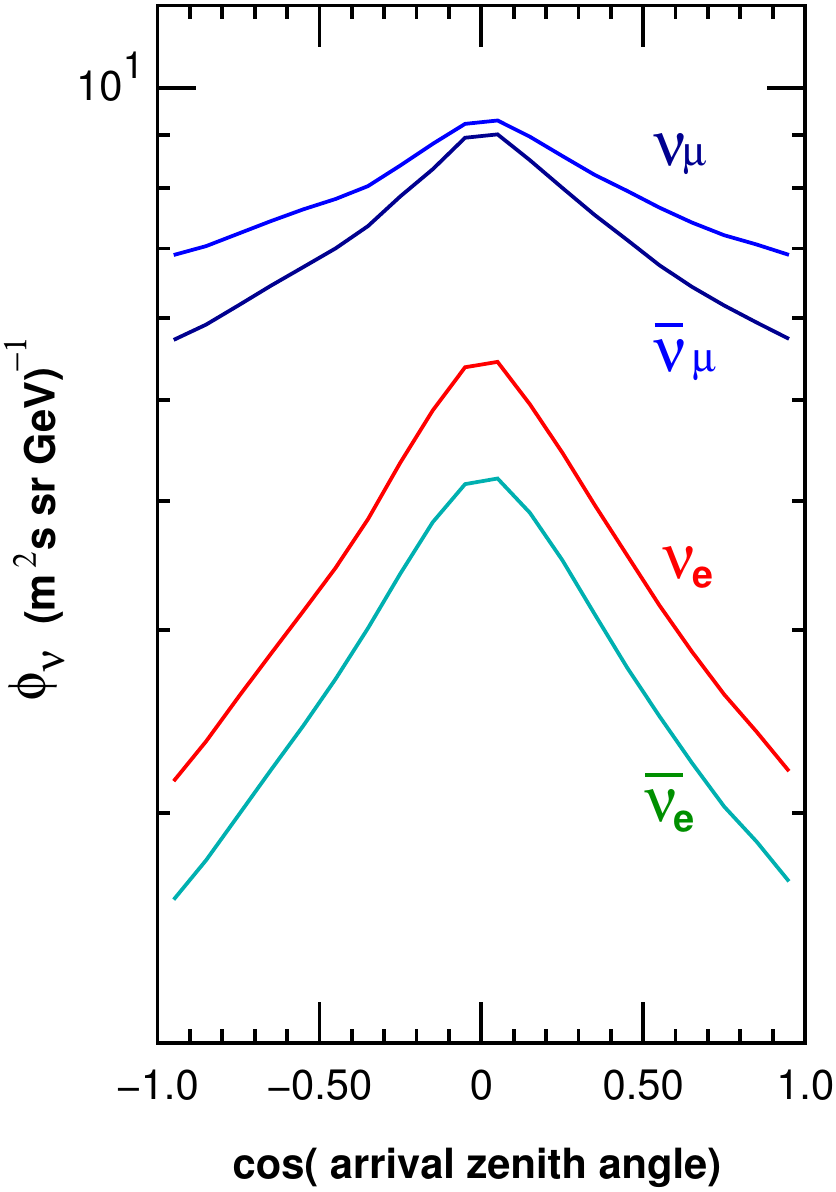}
\includegraphics[width=0.49\textwidth,height=0.45\textwidth]
{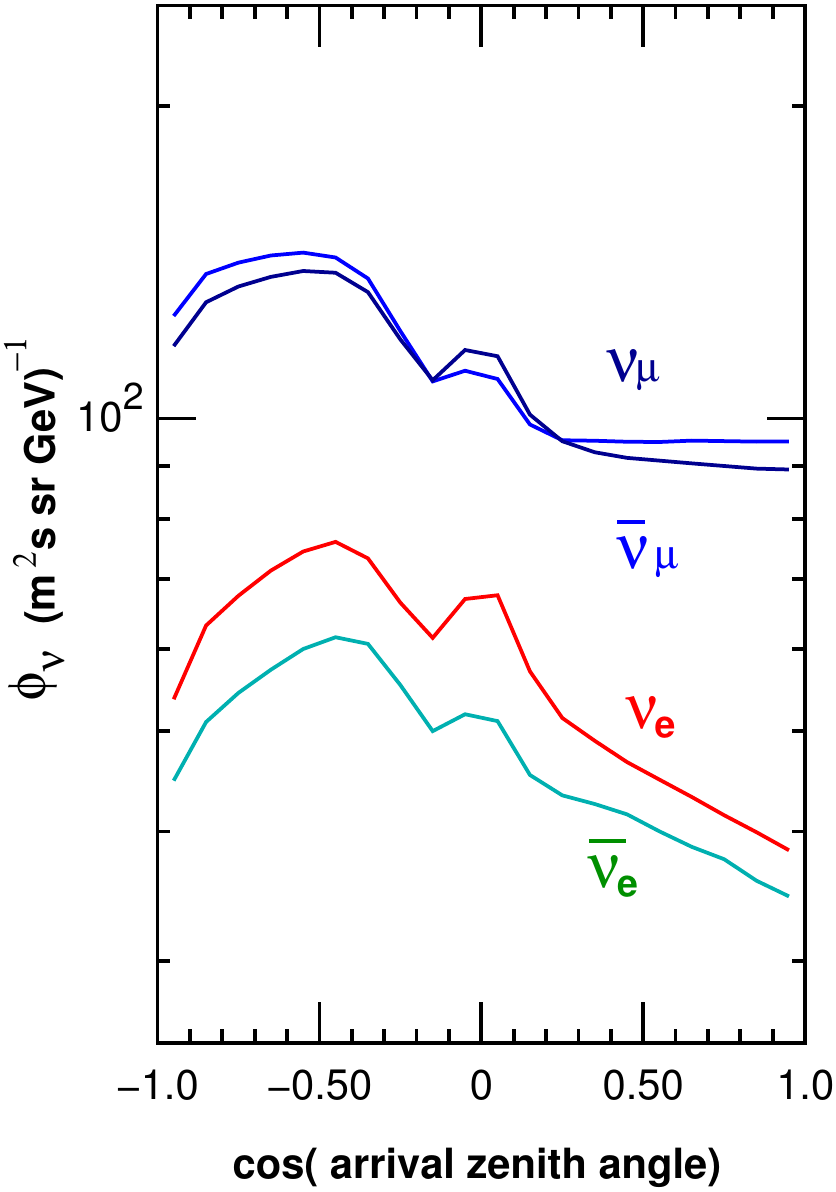}
\includegraphics[width=0.49\textwidth,height=0.45\textwidth]
{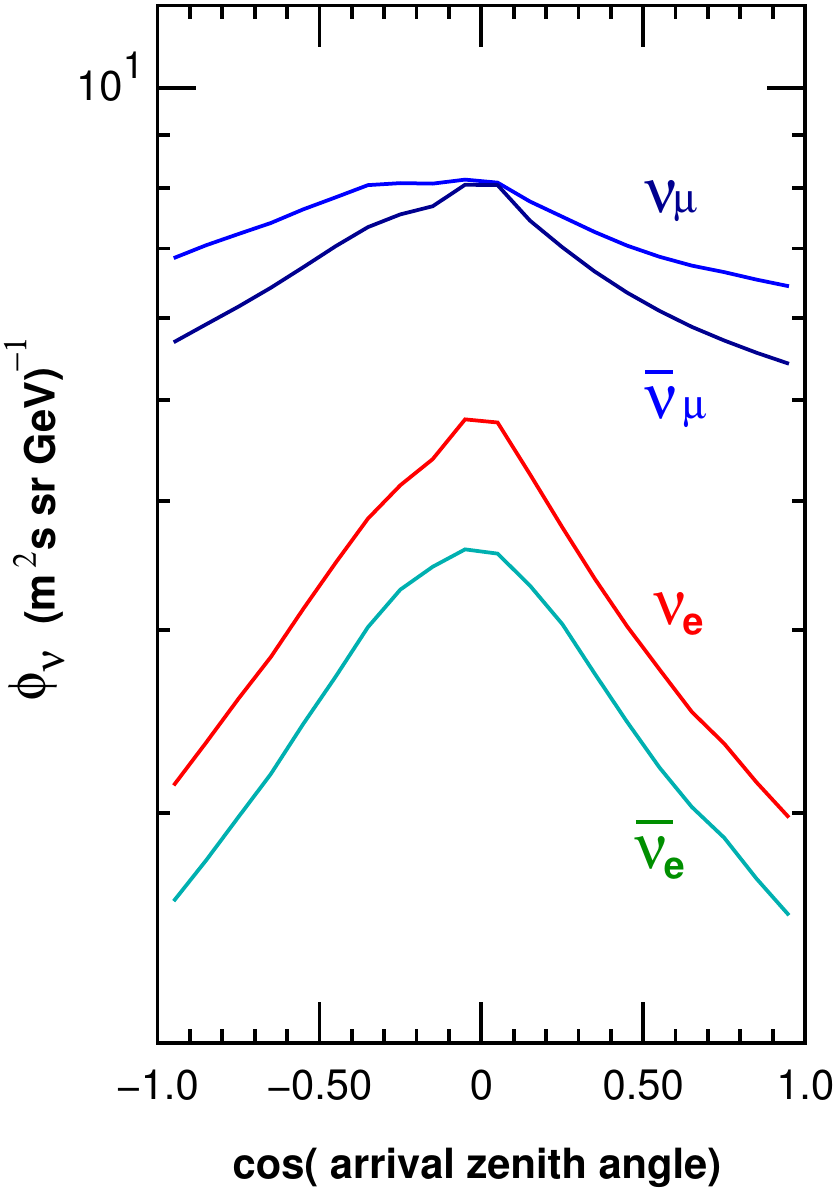}
\caption{The zenith angle dependence of atmospheric neutrino flux at 
E=1.0~GeV(Left) and E=3.2~GeV(Right),
averaged over all azimuthal angles calculated for the Super-Kamiokande (top) and the INO (bottom) sites. 
Here $\theta$ is the arrival direction of the neutrino, with $\cos\theta=1$
for vertically downward going neutrinos, and $\cos\theta=-1$ for
vertically upward going neutrinos.}
\label{fig:sk-ino-zenith}
\end{figure}

It is observed that the total flux at INO is slightly smaller than that 
at Kamioka at low energies ($E \lesssim 3$ GeV), but the difference
becomes small with the increase in neutrino energy. It may be noted that 
this is true
only for the angle-integrated fluxes. Figure~\ref{fig:sk-ino-zenith} shows
the zenith angle dependence (integrating over all azimuthal angles) 
of fluxes at Kamioka and at the INO site at two values of energy. 
It is found that at 1 GeV, there are large up-down asymmetries in the
fluxes at the INO site; the upward-going flux is larger than the 
downard-going one. 
These asymmetries decrease with the increase in neutrino 
energy and almost disappear at 10 GeV.

\begin{figure}[t]
\includegraphics[width=0.49\textwidth,height=0.45\textwidth]
{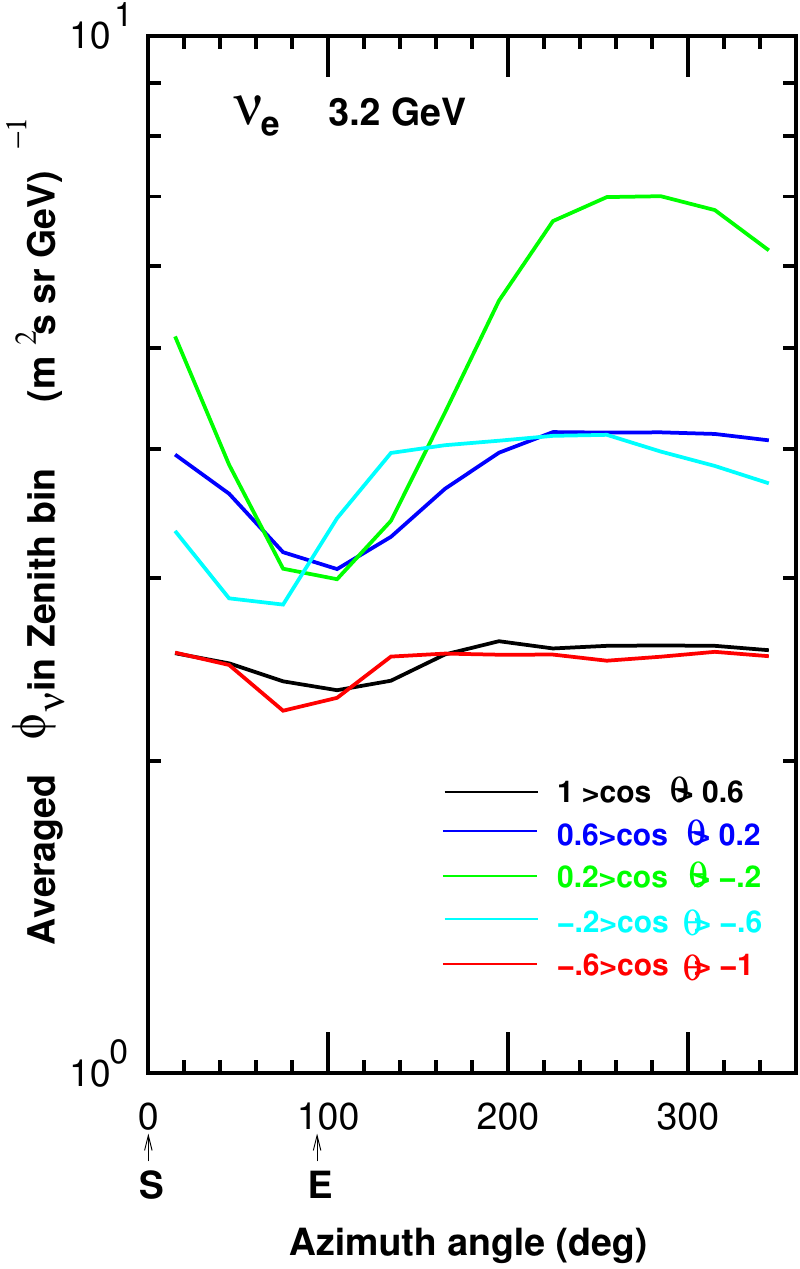}
\includegraphics[width=0.49\textwidth,height=0.45\textwidth]
{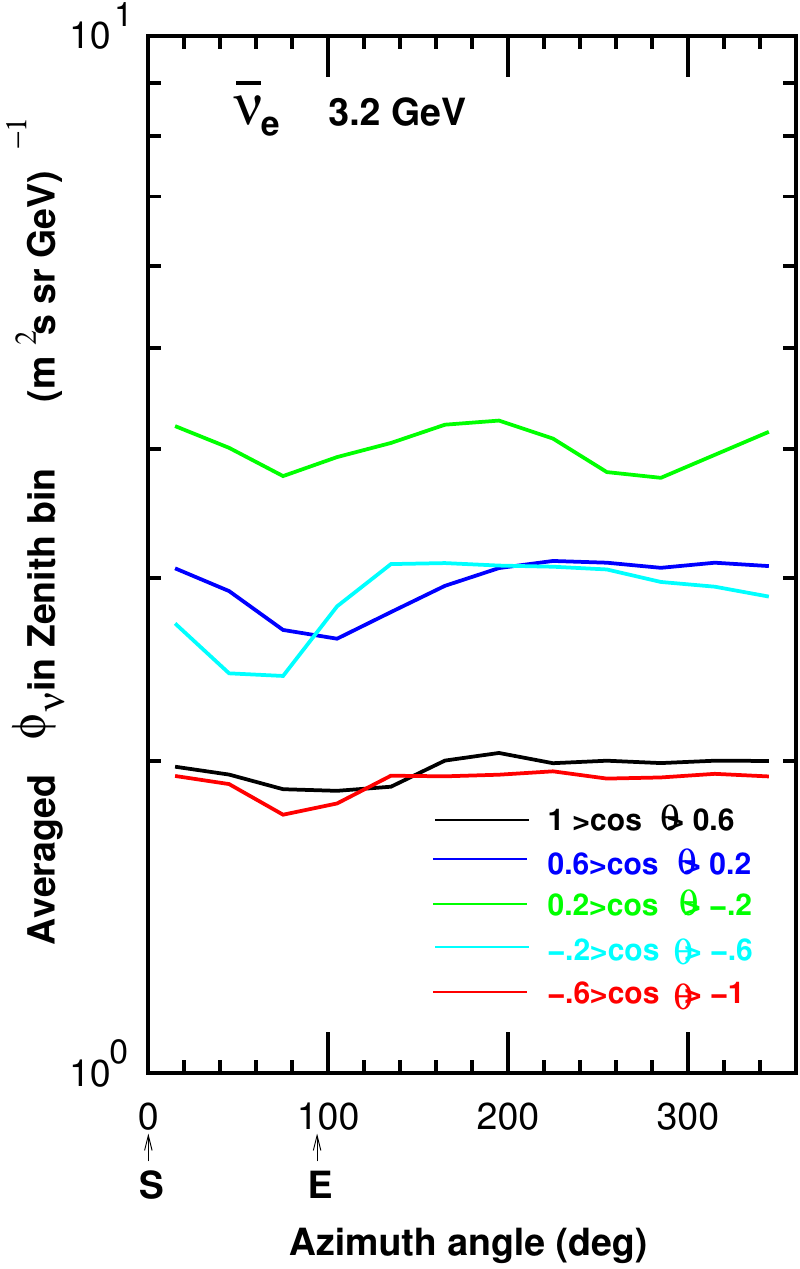}\\
\includegraphics[width=0.49\textwidth,height=0.45\textwidth]
{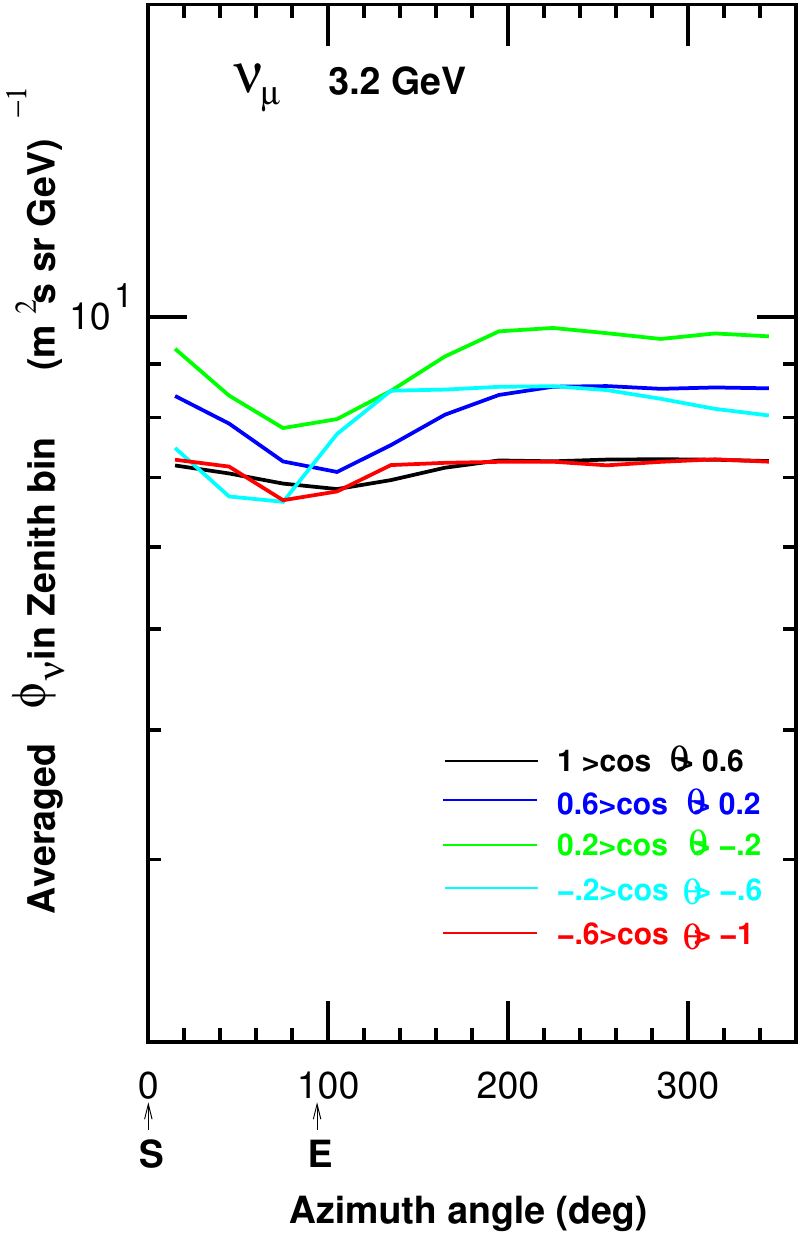}
\includegraphics[width=0.49\textwidth,height=0.45\textwidth]
{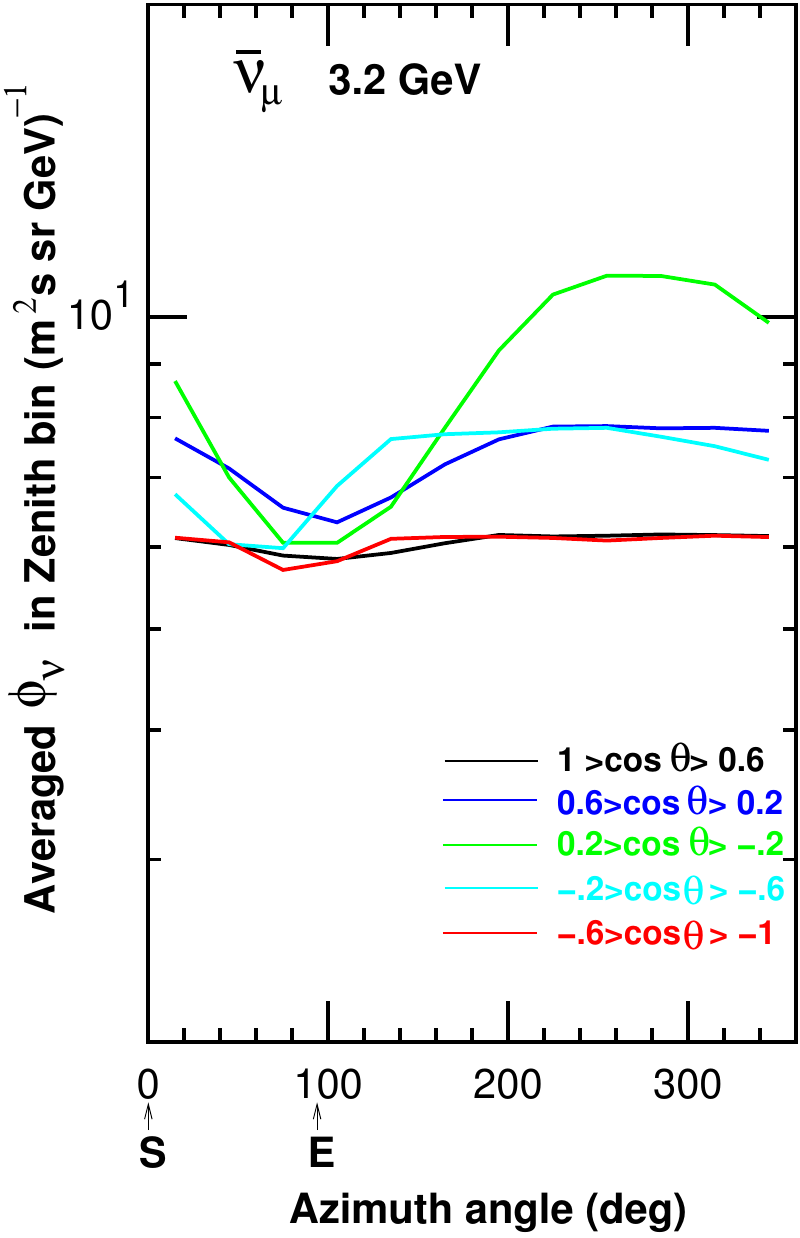}
\caption{The azimuthal angle dependence of atmospheric neutrino flux, 
averaged over zenith angle bins calculated for the 
Super-Kamiokande site for $\nu_e$ at E=3.2~GeV.}
\label{fig:nue0.3-k-GeV}
\end{figure}
\begin{figure}[h]
\includegraphics[width=0.49\textwidth,height=0.45\textwidth]
{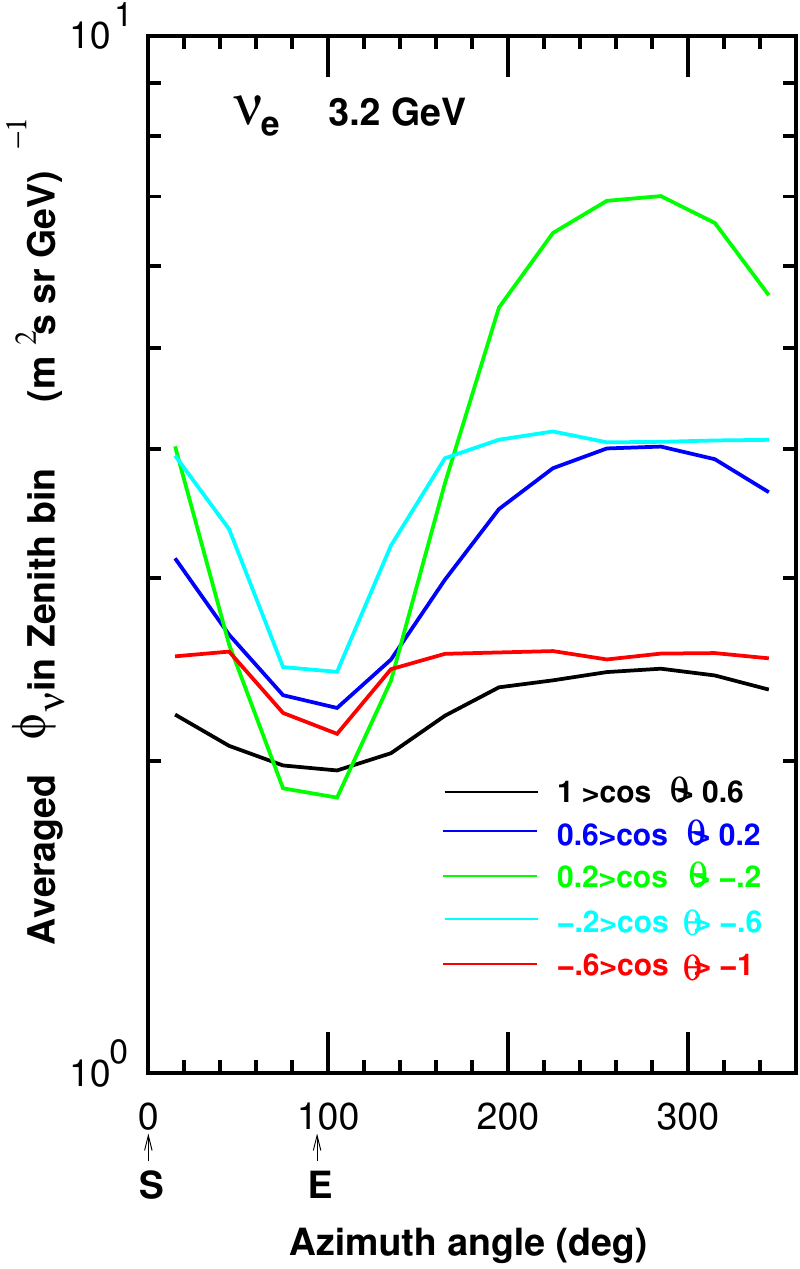}
\includegraphics[width=0.49\textwidth,height=0.45\textwidth]
{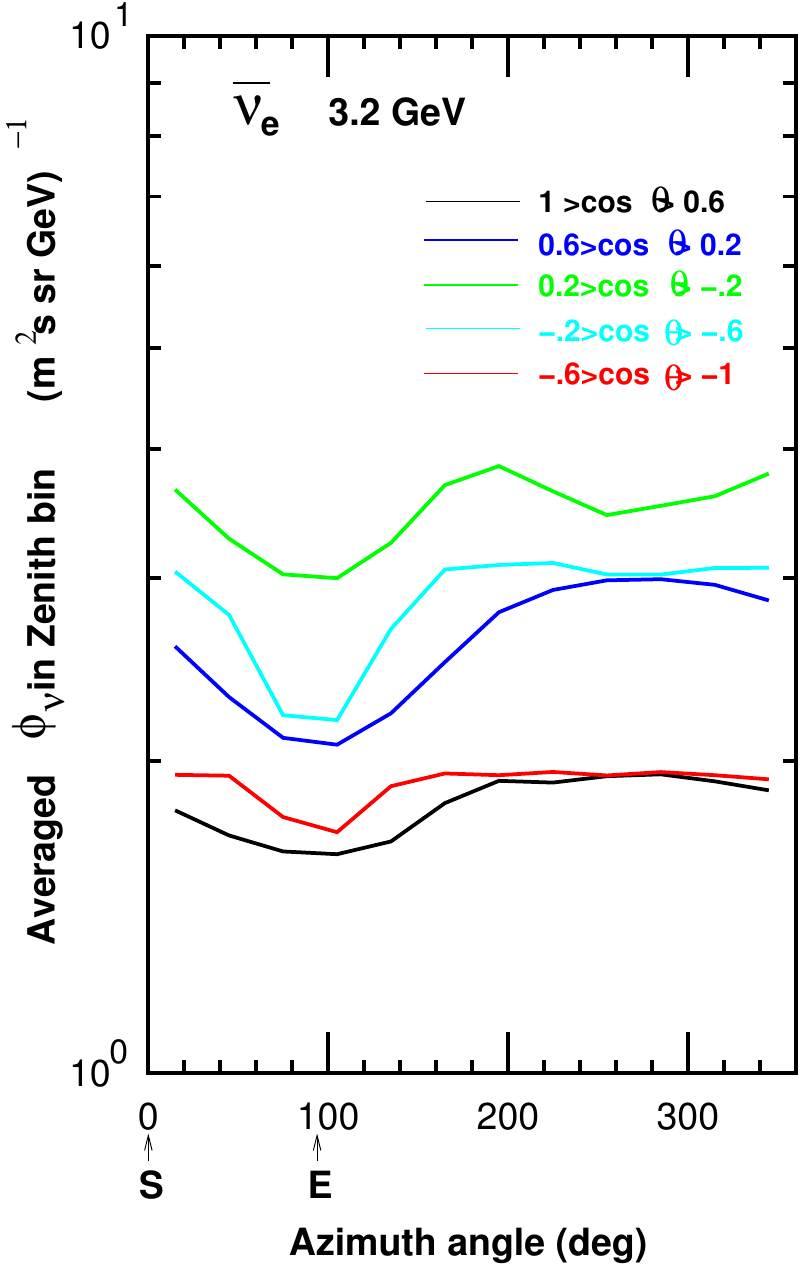}\\
\includegraphics[width=0.49\textwidth,height=0.45\textwidth]
{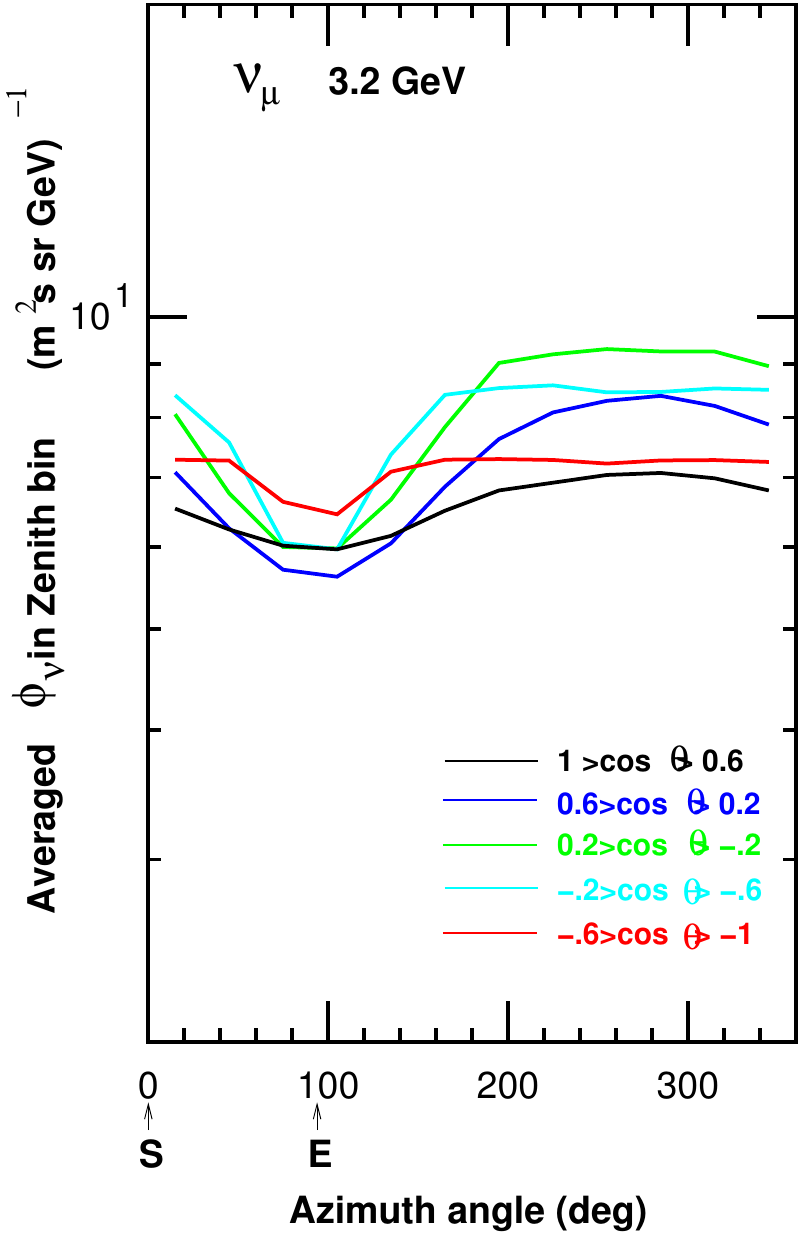}
\includegraphics[width=0.49\textwidth,height=0.45\textwidth]
{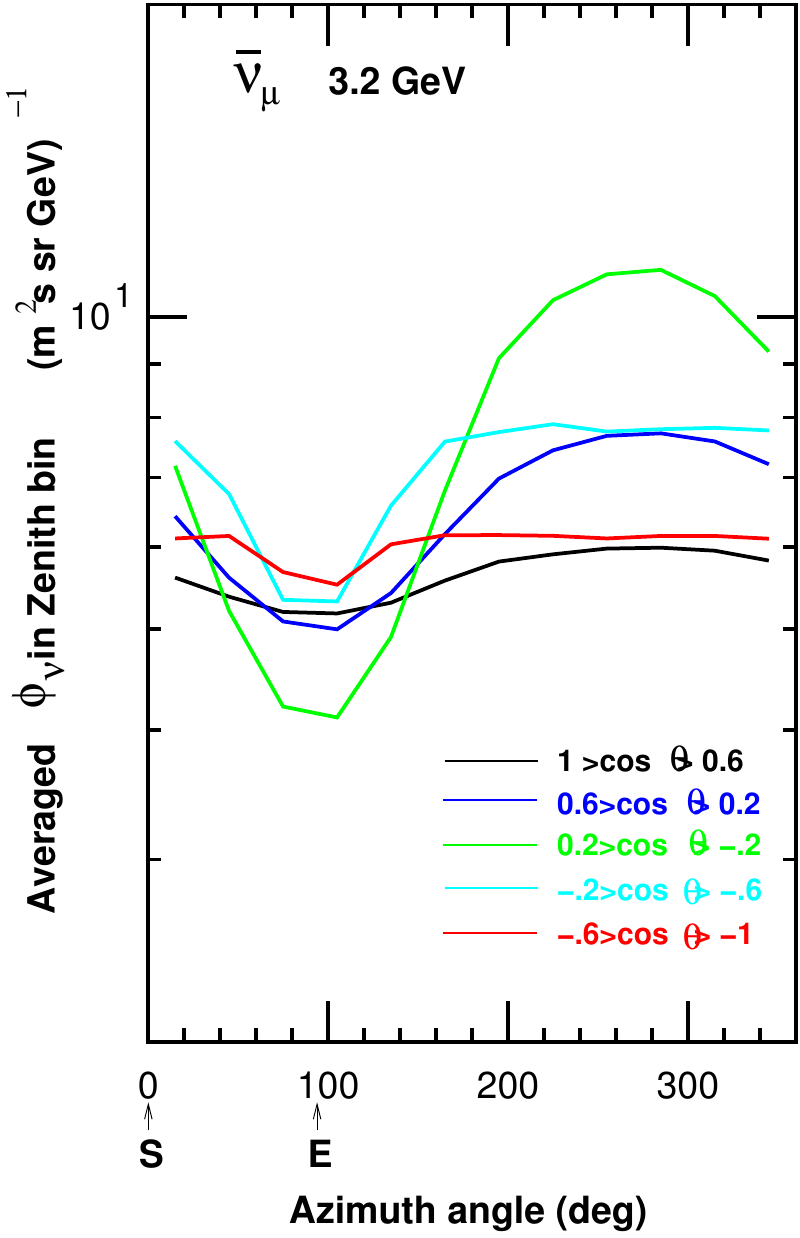}
\caption{The azimuthal angle dependence of atmospheric neutrino flux, 
averaged over zenith angle bins calculated for the 
INO site for $\nu_e$ at E=3.2~GeV.}
\label{fig:nue0.3GeV}
\end{figure}

The results for the atmospheric neutrino fluxes as functions of the 
azimuthal angle $\phi$, in five zenith angle bins, are shown in 
Fig.~\ref{fig:nue0.3-k-GeV} for Kamioka and in Fig.~\ref{fig:nue0.3GeV} 
for the INO site. These results are presented for $\nu_e$, $\bar\nu_e$, 
$\nu_\mu$ and $\bar\nu_\mu$, for the (anti)neutrino energy of 3.2 GeV. 
It is observed that even for such a high energy, the variation of the 
atmospheric neutrino flux has a complex structure. This is a result of
the rigidity cutoff and muon bending in the geomagnetic field. 
At the INO site, the horizontal component of the geomagnetic field is
$\sim$ 40 $\mu T$, larger than that at Kamioka, where it is $\sim$ 
30 $\mu T$, so the azimuthal angle dependence is also more complex.
This complex azimuthal angle dependence continues even above 10 GeV
for the near horizontal directions. 

The atmospheric electron neutrino flux also shows a rapid variation in 
the azimuth angles, but the 
statistical errors in the simulations of production of atmospheric
neutrinos are still large, and more statistics is needed for a 
better understanding. Since the INO site is close to the equator, 
the seasonal variation 
is also expected, the calculation for which is also in progress. These 
updates are expected to be reported after the accumulation of sufficient 
statistics.

\begin{figure}[h!]
\includegraphics[width=0.49\textwidth,height=0.45\textwidth]
{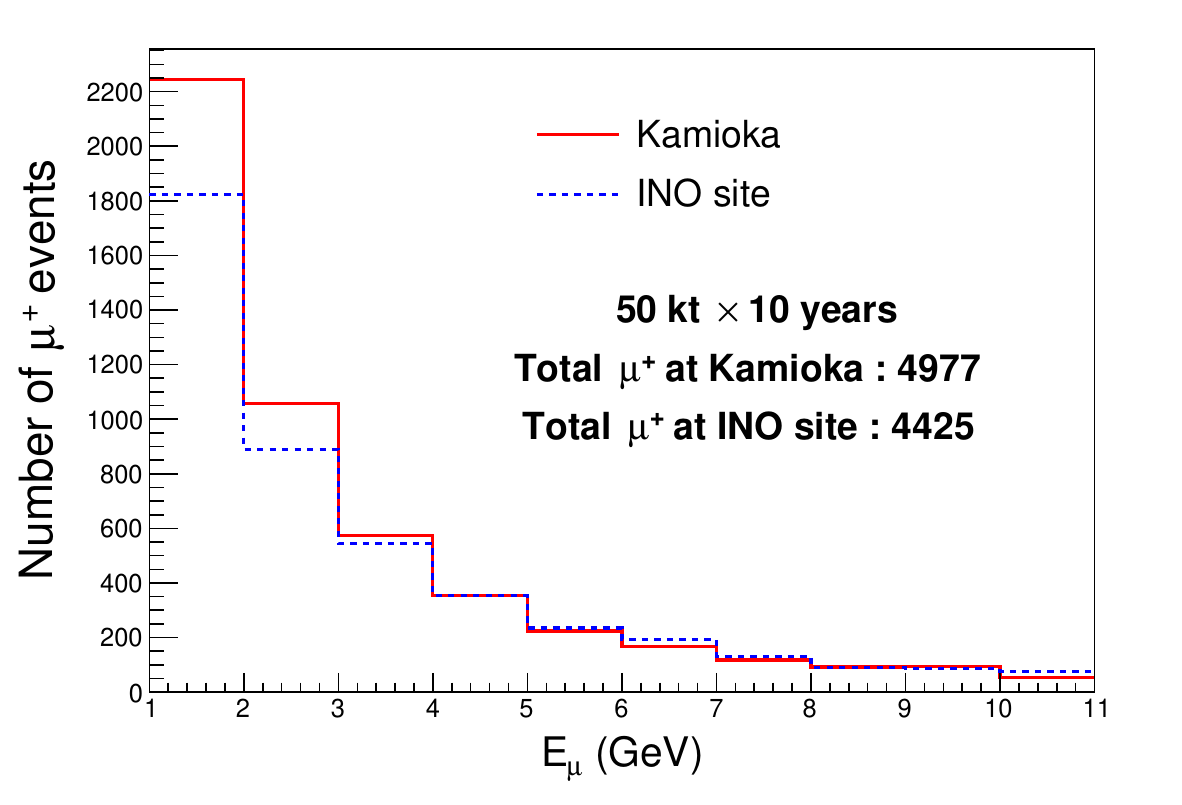}
\includegraphics[width=0.49\textwidth,height=0.45\textwidth]
{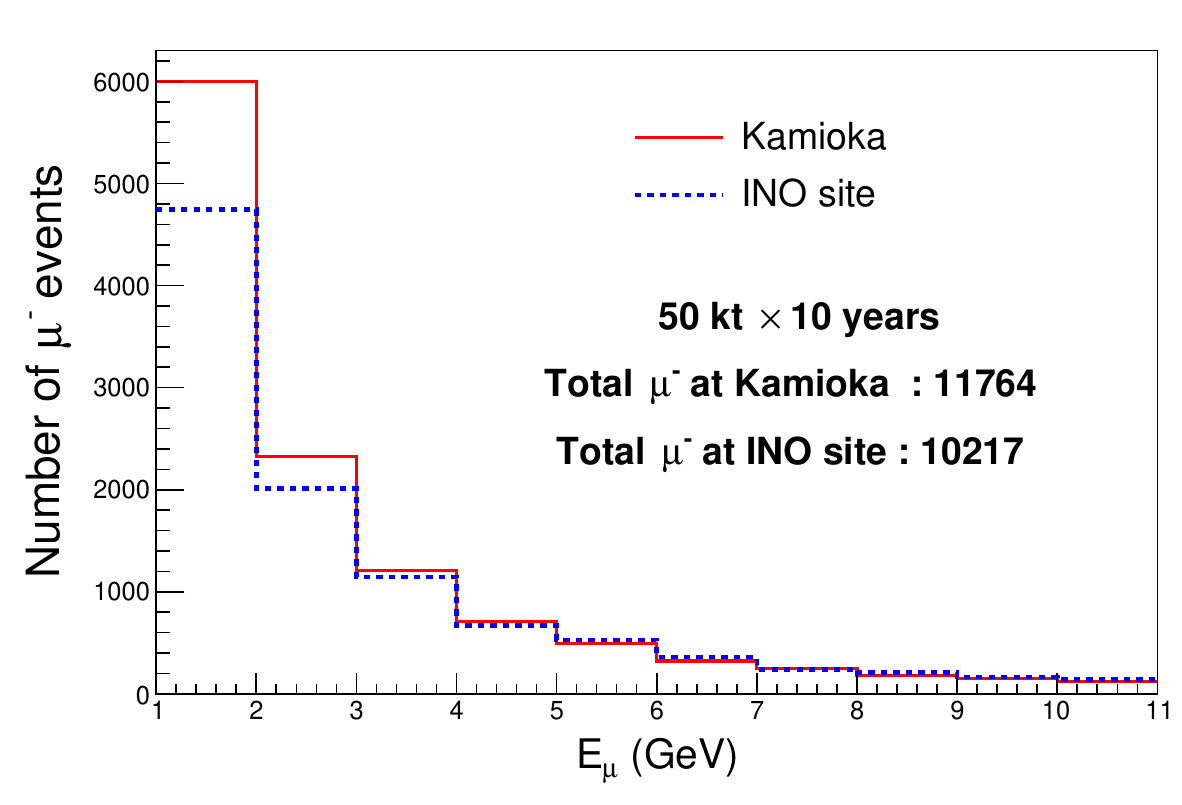}
\caption{Comparisons of energy distributions of $\mu^-$ and $\mu^+$ events
in 500 kt-yr of ICAL, with no oscillations, 100\% efficiency for
detection and charge identification of muons, and extremely accurate
energy measurement. }
\label{compare-sk-theni-events}
\end{figure}

Detailed physics reanalyses with the fluxes at the INO site need to be 
carried out in order to determine the final physics potential of ICAL.
However, the effect of the change of flux may be estimated by comparing 
the number of events calculated using the fluxes at Kamioka and at the 
INO site. The comparison of the number of $\mu^-$ and $\mu^+$ events
at these two sites, as a function of the muon energy, is shown in
Fig.~\ref{compare-sk-theni-events}. For the sake of this sample
comparison, we have considered charged-current muon events in the 
energy range 1--11 GeV, with no oscillations, 100\% efficiency for 
detection and charge identification of muons, and extremely accurate 
energy measurement.

The total number of muon events will be less with the fluxes at the INO site.
As a result, the performance may be expected to be slightly worse than
that calculated from the Kamioka fluxes. The extent to which the 
performance will be affected will depend on the quantity of interest,
though. For example, the accuracy in the measurement of $\sin^2 2\theta_{23}$ 
typically is controlled by the total number of events. Since the total
number of events with the fluxes at the INO site are about 14\% smaller,
we expect to take about 14\% additional exposure to obtain the same
level of accuracy as described in this report. On the other hand, as
has been pointed out in Fig.~5 of \cite{Devi:2014yaa}, the hierarchy sensitivity
comes mainly from the events with muon energy greater than 4 GeV. 
As Fig.~\ref{compare-sk-theni-events} indicates, the numbers of such 
muon events calculated using the two fluxes are nearly the same within
statistical uncertainties. The results for the mass hierarchy determination
are thus expexted to be unaffected.



\chapter{Neutrino oscillation probabilities in matter} 
\label{app:prob} 

Neutrino oscillation probabilities in matter are obtained 
by solving the propagation equation, which may be written in the 
flavour basis as
\begin{equation} 
i\frac{d|\nu_\alpha(x)\rangle}{dx} = H \mid{\nu_\alpha(x)}\rangle
\end{equation} 
where $|\nu_\alpha(x)\rangle = (\nu_e(x), \nu_\mu(x),\nu_\tau(x))^T$. 
$H$ is the effective Hamiltonian, given as
\begin{equation} 
H = \frac{1}{2E} ~U ~ {\rm diag}\left(0,\Delta m^2_{21}, \Delta m^2_{31}\right)~ 
U^\dagger + {\rm diag}\left(V(x),0,0 \right) \; .
\label{Hmat} 
\end{equation}  
Here 
$E$ is the energy of the neutrino and $\Delta m^2_{ij} = m_i^2 - m_j^2$ 
is the mass-squared difference between the neutrino mass eigenstates. The 
PMNS mixing matrix $U$ relates the neutrino flavour eigenstates and mass 
eigenstates.
$V(x)$ is the matter potential arising due to the charged-current 
interaction of $\nu_e$ with electrons and is given 
in terms of the electron density $n_e$ by 
$V(x) = \sqrt{2} G_F n_e(x)$. 
For antineutrinos,  $U \rightarrow U^{*}$ 
and $V \rightarrow -V$.  
In general, for an arbitrary density profile one needs to solve the 
above equation numerically to obtain the probabilities. 
However, simplified analytic expressions can be obtained assuming 
constant matter density. 
In such cases one can diagonalize the above Hamiltonian 
to obtain 
\begin{equation} 
H = \frac{1}{2E}~ U^{\rm m}~ {\rm diag} \left( (m_1^{\rm m})^2, 
(m_2^{\rm m})^2, (m_3^{\rm m})^2 \right)~U^{{\rm m}\dagger} 
\label{Hmatdiag} 
\end{equation} 
where $m_i^{\rm m}$ and $U^{\rm m}$ denote the mass eigenvalues and mixing matrix 
in matter respectively. 
For a neutrino travelling a distance $L$, the flavour conversion 
probability in  matter of constant  density 
has an analogous expression as in the case of vacuum 
[See Eq.~(\ref{eq:pab})], and can be expressed as 
\begin{eqnarray} 
P_{\alpha\beta} (L) =& \delta_{\alpha\beta}& - 4 \sum_{j>i} {\rm Re}
\left(U_{\alpha i}^{\rm m} U_{\beta i}^{{\rm m}^\star} U_{\alpha j}^{{\rm m}^\star}
U_{\beta j}^{\rm m}\right) \sin^2 \Delta_{ij}^m  
\nonumber \\
&&+ 2 \sum_{j>i} {\rm Im} \left(U_{\alpha i}^{\rm m} U_{\beta i}^{{\rm m}^\star}
U_{\alpha j}^{{\rm m}^\star} U_{\beta j}^{\rm m}\right) \sin (2 \Delta_{ij}^m) \; , 
\label{pmat} 
\end{eqnarray} 
where the quantity $\Delta_{ij}^m$ in presence of matter is defined as  
\[ \Delta_{ij}^m =  \frac{ 1.27 ~ (\Delta m_{ij}^2)^{\rm m} (\mathrm{eV}^2) ~L (\mathrm{km}) }
{ ~E (\mathrm{GeV}) } \; , \]
with $(\Delta m_{ij}^2)^{\rm m} = ({m_i^{\rm m}})^2 - ({m_j^{\rm m}})^2$ 
the difference between
the squares of the mass eigenvalues $m_i^{\rm m}$ and $m_j^{\rm m}$ in matter. 

To obtain tractable expressions, further assumptions need to be made. 
Many approximate analytic expressions for probabilities exist in the
literature. However, different assumptions that lead to different
approximate forms have different regimes for validity.
To understand the results presented in this Report, the probability expressions 
obtained under the following two approximations are mostly relevant: 
\begin{itemize} 
\item the one mass scale dominance (OMSD) approximation  
which assumes $\Delta m^2_{21}=0$, and 
\item the double expansion in terms of small parameters 
$\alpha = \Delta m^2_{21}/\Delta m^2_{31}$ and $\sin\theta_{13}$
\cite{Freund:2001pn,Akhmedov:2004ny}. 
\end{itemize} 
The condition on the neutrino energy and baseline for the validity of both 
approximations can be expressed as $\Delta m^2_{21} L /E << 1$. 
This translates  to $L/E << 10^4$ km/GeV for typical 
values of the solar mass-squared difference $\Delta m^2_{21}$, and 
hence to L$\leq 10^{4}$ km for  neutrinos of energy ${\cal{O}}({\rm GeV})$. 
Thus these approximations are valid for most of the energy and path-length 
ranges considered here. The OMSD approximation is exact in $\theta_{13}$ 
and works better near the resonance region. Below we give the probabilities 
relevant for the study presented in this report in both OMSD and double 
expansion approximations, and discuss in which $L$ and $E$ regimes 
these are appropriate. Note that we give the expressions only for
neutrino propagation through a constant matter density. This approximation
is not applicable for neutrinos passing through the Earth's core. However it
is enough for an analytic understanding of our arguments. All our
numerical calculations take the variation of earth's density into 
account through the Preliminary Reference Earth Model (PREM).

\section{One Mass Scale Dominance approximation} 

In this approximation, the Hamiltonian in Eq.~(\ref{Hmat}) 
can be exactly diagonalized analytically.
Below we give the expressions for the muon neutrino survival 
probability 
$P(\nu_\mu \to \nu_\mu) \equiv P_{\mu \mu}$ 
and conversion probability of electron neutrinos to muon neutrinos 
$P(\nu_e \to \nu_\mu) \equiv P_{e \mu}$, which are relevant for the 
atmospheric neutrinos at ICAL since the detector is sensitive to 
muon flavour.  In the OMSD approximation, these can be expressed as  
\begin{eqnarray} 
P_{\mu \mu} 
&=& 
1 - \cos^2 \theta_{13}^{\rm m} \ \sin^2 2 \theta_{23}
\sin^2\left[1.27 (\Delta m^2_{31} + A + (\Delta m_{31}^2)^{\rm m} ) L/2E \right]
\nonumber \\
&-& 
\sin^2 \theta_{13}^{\rm m} \ \sin^2 2 \theta_{23}  \ 
\sin^2\left[1.27 (\Delta m^2_{31} + A - (\Delta m_{31}^2)^{\rm m}) L/2E \right]
\nonumber \\
&-& 
\sin^4 \theta_{23} \ \sin^2 2\theta_{13}^{\rm m} \ 
\sin^2 \left[ 1.27 \Delta m^2_{31} L/ E \right] \; ,
\label{eq:pmumumat1}
\end{eqnarray} 
and 
\be
P_{e \mu}  =
\sin^2 \theta_{23} \ \sin^2 2 \theta^{\rm m}_{13} \ 
\sin^2 \left[ 1.27 (\Delta m_{31}^2)^{\rm m} L/E \right] \; .
\label{eq:pemumat}
\ee
In the above expressions, 
$(\Delta m_{31}^2)^{\rm m}$ and $\sin 2 \theta_{13}^{\rm m}$,
the mass-squared difference
and mixing angle in matter, respectively, are given by
\bea
(\Delta m_{31}^2)^{\rm m} &=& 
\sqrt{(\Delta m^2_{31} \cos 2 \theta_{13} - A)^2 +
(\Delta m^2_{31} \sin 2 \theta_{13})^2} \; ,
\nn \\ 
\sin 2 \theta_{13}^{\rm m} &=& 
\sin 2 \theta_{13}
\frac{\Delta m^2_{31}}{(\Delta m_{31}^2)^{\rm m}} \; ,
\label{eq:dm31}
\eea
where
$$A ({\rm eV}^2) = 2EV= 2\sqrt{2} G_F n_e E = 0.76 \times 10^{-4}
~\rho({\rm g/cc}) ~E ({\rm GeV})  \; .$$
When $A = \Delta m^2_{31} \cos 2 \theta_{13}$, we see a resonance. The 
resonance energy is given by 
\be 
E_{\rm res} = \frac{\Delta m^2_{31} \cos 2\theta_{13}} 
{2 \sqrt{2} G_F n_e} 
\label{eq:eres}
\ee
In Tab.~\ref{Tab:eres}, we give the average resonance energies for neutrinos
travelling a given distance $L$ through the Earth, for baselines ranging
from 1000--10000 km.

\begin{table}[t]
\begin{center}
\begin{tabular}{| c | c | c | }
\hline
\hspace{0.5cm}{ {L ({\rm km})}}\hspace{0.5cm} & 
\hspace{0.5cm} ${{\rho_{\rm avg} ({\rm g/cc})}}$ \hspace{0.5cm} 
& \hspace{0.5cm}$
{E_{\rm res}}$ {({\rm GeV})}\hspace{0.5cm}
\\
\hline
 1000 &  3.00  &  9.9 \\
 3000 &  3.32  &  9.4 \\
 5000  &   3.59 &  8.7 \\
 7000   & 4.15 & 7.5 \\
 10000  & 4.76 &  6.6 \\
\hline
\end{tabular}
\caption{Values of ${E_\mathrm{res}}$ 
at various baselines using the line-averaged PREM \cite{prem} density
${\mathrm{\rho_{avg}}}$.
We have used $\Delta m^2_{31}=2.5 \times 10^{-3}$ $\mathrm{eV}^2$
and $\sin^2 2\theta_{13}=0.1$.
}
\label{Tab:eres}
\end{center}
\end{table}

It is seen from the table that the resonance energy is 
in the range 6--10 GeV for path lengths in the range 1000--10000 km.
These ranges are relevant for atmospheric neutrinos passing through earth
and hence provide an excellent avenue to probe resonant earth  matter 
effects.  The importance of this can be understood by noting that 
the resonance  condition depends  on the sign of $\Delta m^2_{31}$. 
For  $\Delta m^2_{31} > 0$ there is a matter enhancement in $\theta_{13}^{\rm m}$
for neutrinos, and a matter suppression in $\theta_{13}^{\rm m}$ for antineutrinos 
(since $A \to -A$). The situation is reversed for $\Delta m^2_{31} < 0$.  
Thus matter effects can differentiate between the two hierarchies 
and detectors with charge sensitivity (like ICAL) are very suitable 
for probing this. 

For atmospheric neutrinos in ICAL, the most relevant probability is   
$P_{\mu\mu}$. The significance of the $P_{e \mu}$ channel is  
less than that of $P_{\mu\mu}$ for two reasons: the number of electron
neutrinos produced in the atmosphere is smaller, and more importantly,
the probability of their conversion to muon neutrinos is also 
usually smaller than $P_{\mu \mu}$,
so that their contribution to the total number of events in ICAL is small.
It is not completely negligible though, since the value of $\theta_{13}$
is moderately large. 

Note that $P_{e \mu}$ does not attain its maximum value 
at $E = E_{\rm res}$ even though $\sin 2 \theta_{13}^{\rm m}$ 
achieves its maximum 
value of unity at this energy, because the mass-squared difference 
$(\Delta m_{31}^2)^{\rm m}$ hits a minimum \cite{Banuls:2001zn}.  
The values of $(\Delta m_{31}^2)^{\rm m} \sin 2 \theta_{13}^{\rm m}$ 
and $P_{\mu e}$ remain small
for path-lengths of $L \lesssim 1000$ km. If $L$ is chosen
suitably large so as to satisfy $(1.27 \Delta m^2_{31} \sin 2
\theta_{13} L/E) \geq \pi/4$, then $P_{e \mu}$ can
reach values $\geq 0.25$ for $\sin^2 2\theta_{23} = 1$. 
One needs $L \gtrsim 6000$ Km to satisfy the above condition. 
For such baselines and in the energy range 6-8 GeV, the resonant earth matter
effects lead to $P_{e \mu}$ in matter being significantly greater 
than its vacuum value \cite{Gandhi:2004md}. 

\begin{figure}[t]
\centering
 \includegraphics[width=0.49\textwidth,height=0.35\textwidth]
 {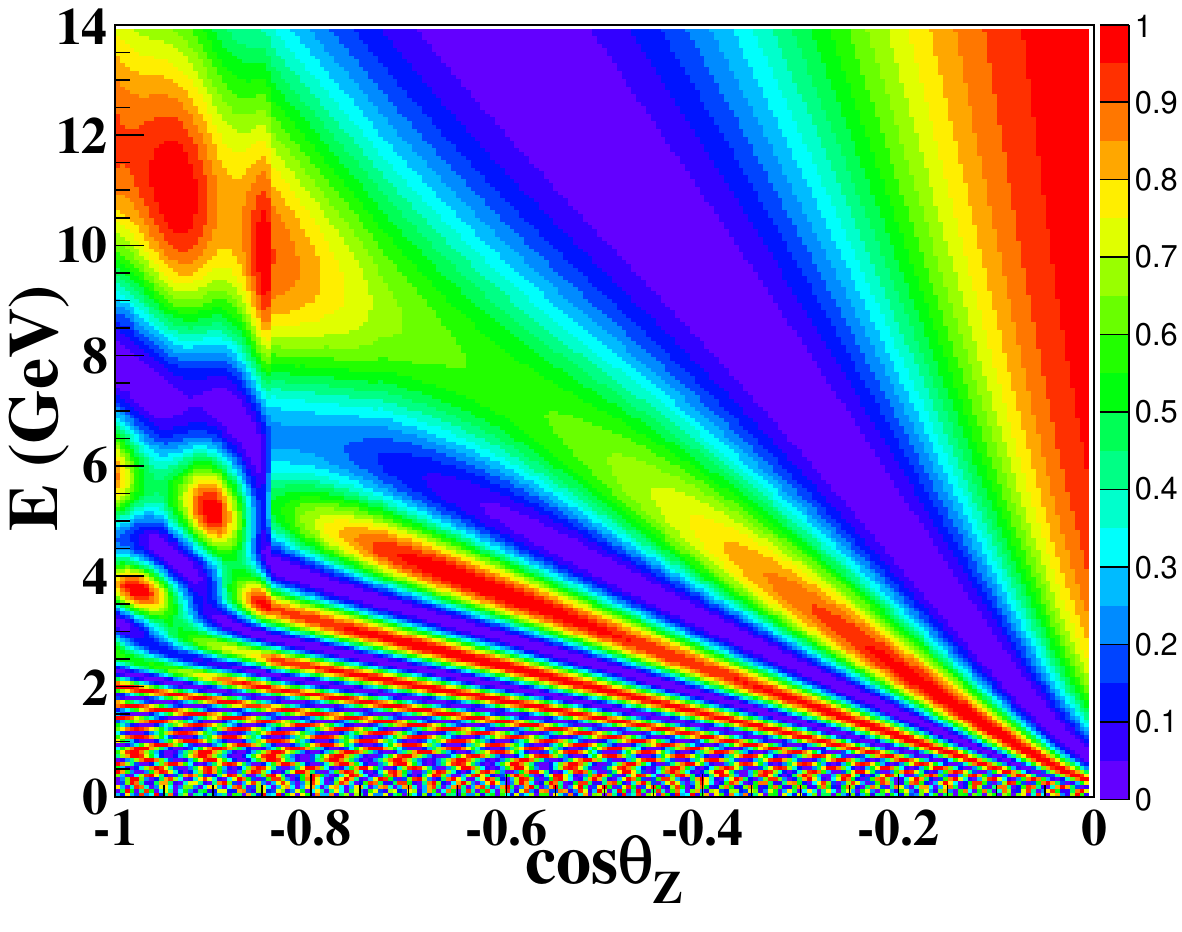}
 \includegraphics[width=0.49\textwidth,height=0.35\textwidth]
 {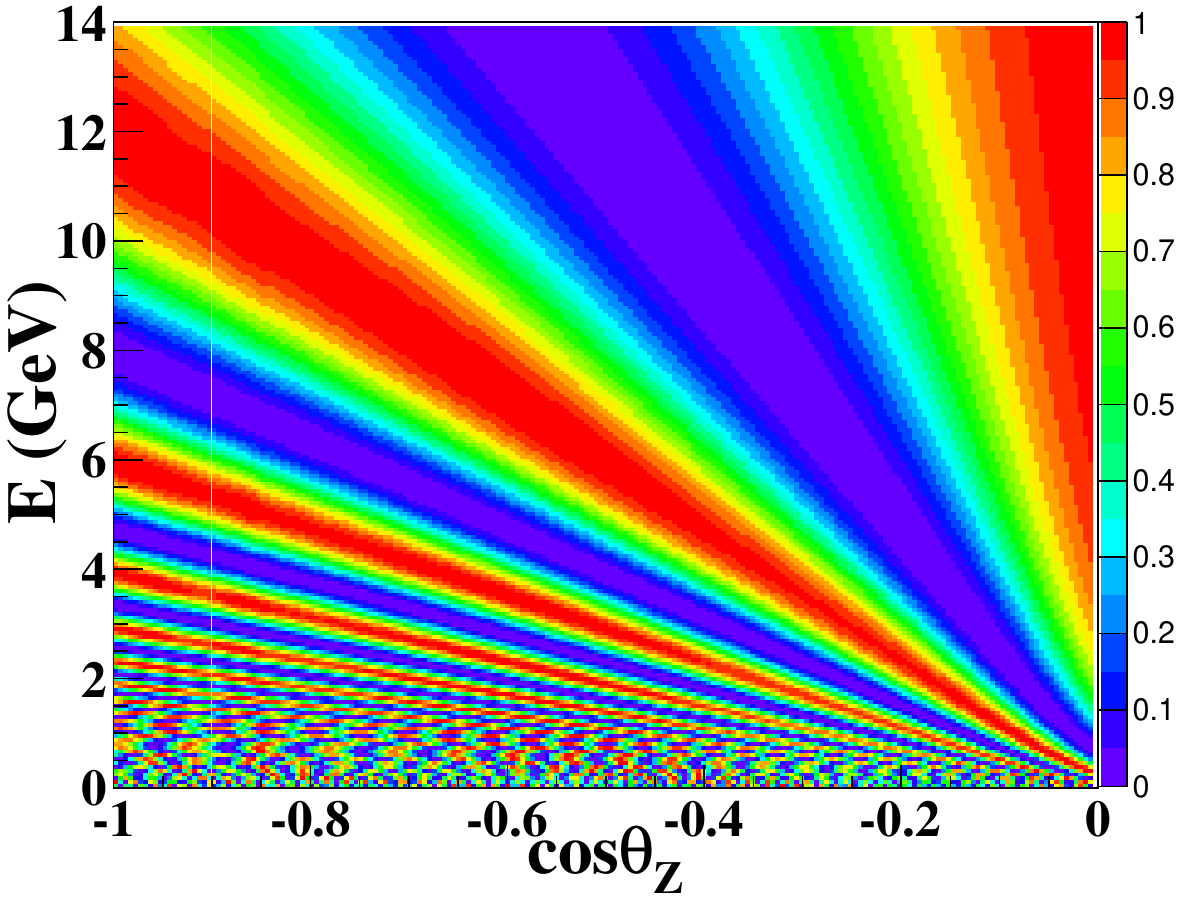}
\caption{The oscillograms for the muon neutrino (left panel) and
antineutrino (right panel) 
survival probabilities during their passage through earth in 
E-$\cos\theta_{z}$ plane. The oscillation parameters used are 
$ \theta_{23} = 45^\circ$, $\delta_{\rm CP} = 0$, 
$ \Delta m^2_{31} = +2.45 \times 10^{-3} \textrm{eV}^2$ (NH) and
$ \sin^2 2\theta_{13} = 0.1$.
}
\label{fig:pmumu}
\end{figure}

The muon neutrino survival probability is a more complicated 
function and can  show both fall and rise above the vacuum value 
for longer baselines ($\sim$ 10000 km). Thus the energy and angular
smearing effects are more for $P_{\mu \mu}$. The maximum hierarchy sensitivity 
is achieved  in this channel when the resonance occurs close to a 
vacuum peak or dip, thus maximizing the matter effects. 
This is because when there is resonant matter effect for one hierarchy,
the probability for the other hierarchy closely follows the vacuum value. 
Figure~\ref{fig:pmumu} shows oscillograms for muon neutrino and antineutrino 
survival probabilities in the case of normal hierarchy, 
in the plane of neutrino energy and cosine of the zenith angle $\theta_z$. 
The plot in the left panel shows the resonant effect in the muon neutrino 
probabilities in the region  between $\cos\theta_{z}$ between -0.6 to -0.8
and energy in the range 6--8 GeV. This feature is not present in the right 
panel since for the normal hierarchy, the muon antineutrinos do not encounter 
any resonance effect. The plot also shows the enhanced oscillation features 
due to the effects of the Earth's core
($\cos\theta_{z}$ between -0.8 and -1.0) for neutrinos. 
For the inverted hierarchy, the muon antineutrino survival probability 
will show resonance effects, whereas the neutrino probabilities will not. 
The ICAL detector  being charge sensitive can differentiate between neutrino
and antineutrino effects and hence between the two hierarchies. 

Equations~(\ref{eq:pmumumat1}) and (\ref{eq:pemumat}) can also help
us understand the octant sensitivity of atmospheric neutrinos  
arising due to resonant matter effects. The leading order term in 
$P_{e\mu}$ in vacuum depends on $\sin^2 \theta_{23} \sin^2 2\theta_{13}$.
Although this term is sensitive to the octant of $\theta_{23}$, the 
uncertainty in the value of $\theta_{13}$ may give rise to octant degeneracies. 
In matter, the $\sin^2 2\theta_{13}^{\rm m}$ term gets amplified near resonance,
and  the combination $\sin^2 \theta_{23} \sin^2 2 \theta_{13}^{\rm{m}}$ 
breaks the degeneracy of the octant with $\theta_{13}$.
Also, the strong octant-sensitive nature of the term $\sin^4 \theta_{23} 
\sin^2 2 \theta_{13}^{\rm{m}}$ near resonance can overcome the degeneracy 
due to the $\sin^2 2\theta_{23}$-dependent terms.
Unfortunately, the muon events in ICAL get contribution from both 
$P_{\mu \mu}$ channel and $P_{e \mu}$ channel, and the matter effect in these 
two channels act in opposite directions for most of the baselines.
This causes a worsening in the octant sensitivity of muon events at 
atmospheric neutrino experiments. 

The OMSD probabilities are  in the limit $\Delta m^2_{21} =0$ and have no 
dependence on the CP phase. These  expressions match well with the 
numerical probabilities obtained by solving the propagation equation
in the resonance region, i.e. for the baseline range 6000--10000 km. 
Accelerator-based experiments like T2K and NO$\nu$A have shorter 
baselines and lower matter effects, and lie far from resonance. 
For these experiments, the dominant terms in $P_{\mu\mu}$ 
are insensitive to the hierarchy and octant. Consequently, 
the relative change in probability 
due to the hierarchy/octant-sensitive sub-dominant terms is small. Therefore 
the $P_{\mu\mu}$ oscillation channel does not 
contribute much to the hierarchy and octant sensitivity of T2K and NO$\nu$A. 
However, these experiments 
get their sensitivity primarily from the $\nu_\mu \to \nu_e$ conversion 
probability. For this case, the double expansion up to second order in 
$\alpha$ and $\sin\theta_{13}$ works better.

\section{Double expansion in $\alpha$ and $\sin\theta_{13}$} 

In accelerator experiments, high energy pions or kaons decay 
to give muons and muon neutrinos/antineutrinos. 
One can study the muon neutrino conversion probability 
$P(\nu_\mu \to \nu_e) \equiv P_{\mu e}$ 
in these with a  detector sensitive to electron flavour.  
In order to study the effect due to $\Delta m^2_{21}$ and the 
CP phase $\delta_{CP}$ it is convenient to write down 
the probabilities as an expansion in  terms of the two small parameters,
$\alpha = \Delta m^2_{21}/\Delta m^2_{31}$ and $\sin\theta_{13}$, to second order 
(i.e. terms up to $\alpha^2, \sin^2\theta_{13}$
and $\alpha \sin\theta_{13}$ are kept) \cite{Freund:2001pn,Akhmedov:2004ny}.

\begin{eqnarray} 
P_{\mu e }&=
& \sin^2 2\theta_{13} \sin^2\theta_{23} 
\frac{\sin^2{[(1 -\hat A)\Delta]}}{(1-\hat A)^2}
\nonumber \\
&&
+ \alpha \sin{2\theta_{13}}  \sin{2\theta_{12}} \sin{2\theta_{23}}
\cos{(\Delta + \delta_{\rm CP})}
\frac{\sin(\hat A \Delta)}{\hat A} \frac{\sin{[(1-\hat A)\Delta]}}{(1-\hat A)}
\nonumber \\
&&
+ \alpha^2 \sin^2\theta_{12} \cos^2\theta_{23}
\frac{\sin^2 (\hat A \Delta)}{\hat A^2}
+ {\cal{O}}(\alpha^3,\alpha^2 s_{13},\alpha s_{13}^2,s_{13}^3) ~.
\label{P-emu}
\end{eqnarray} 

The notations used in writing 
the probability expressions are: 
$\Delta \equiv \Delta m^2_{31}L/4E$, 
$s_{ij} (c_{ij}) \equiv \sin{\theta_{ij}}(\cos{\theta_{ij}})$, 
$\hat{A} = 2\sqrt{2} G_F n_e E / \Delta m^2_{31}$. 
For neutrinos, the signs of $\hat{A}$ and $\Delta$ are
positive (negative) for NH (IH). 
The sign of $\hat{A}$ as well as $\delta_{\rm CP}$ reverse for antineutrinos. 
This  probability is sensitive to all the three current unknowns in 
neutrino physics --- hierarchy, octant of $\theta_{23}$ as well 
as $\delta_{\rm CP}$ --- and is often hailed as the golden channel. 
However the dependences are interrelated and 
extraction of each of these unknowns depends on the 
knowledge of the others. Specially the complete lack of knowledge 
of $\delta_{\rm CP}$ gives rise to the hierarchy-$\delta_{\rm CP}$ degeneracy   
as well as the octant-$\delta_{\rm CP}$ degeneracy in these experiments,
through the second term in Eq.~(\ref{P-emu}). 
 
The above expressions reduce to the vacuum expressions for 
shorter baselines for which $A \rightarrow 0$.  
For such cases there is no hierarchy sensitivity. 
The hierarchy sensitivity increases with increasing baseline and is maximum 
in the resonance region.   
The resonance energy at shorter baselines  is $>$ 10 GeV  
and therefore these experiments cannot probe resonant earth matter 
effects.

\section{Probability for Reactor neutrinos } 
\label{sec:nue-numu}

A crucial input in the analysis presented in this Report
is the value  of $\theta_{13}$, measured by the reactor neutrino experiments. 
The probability relevant for reactor neutrinos is 
the survival probability for electron antineutrinos 
$P(\overline{\nu}_e \to \overline{\nu}_e) \equiv P_{\overline{e} \overline{e}}$. 
Since reactor neutrinos have very low energy (order of MeV) and they travel 
very short distances (order of km), they experience negligible matter effects. 
The exact formula for the survival probability in vacuum is given by 
\cite{Choubey:2003yp,Gandhi:2004md}
\bea
P_{\overline{e} \overline{e}} & =  & 1 -
c_{13}^4 \sin^2 2\theta_{12} \sin^2 \Delta_{21}
 - \sin^2 2\theta_{13} \sin^2 \Delta_{31} \nonumber \\
& & +\sin^2 2\theta_{13} s_{12}^2 \left[\sin^2 \Delta_{31}
 - \sin^2 \Delta_{32} \right] \; . 
\eea
Since this is independent of $\delta_{\rm CP}$ and matter effects, 
the probability is the same for neutrinos and antineutrinos (assuming CPT 
conservation, i.e. the same mass and mixing parameters describing neutrinos 
and antineutrinos). 



\chapter{The Vavilov distribution function} 
\label{sec:appA} 
 
The Vavilov probability distribution function is found to be the suitable 
one to represent the hit distributions of hadrons of a given energy in
the ICAL, as has been observed from Fig.~\ref{histocomp}.
The Vavilov probability density function in the standard form is defined by 
\cite{vavilov}
\begin{equation} 
P(x; \kappa, \beta^2) =  
  \frac{1}{2 \pi i}\int_{c-i\infty}^{c+i\infty} \phi(s) e^{x s} ds \; , 
\label{app1} 
\end{equation} 
where  
\begin{equation} 
\phi(s) = e^{C} e^{\psi(s)},~~~C = \kappa (1+\beta^2 \gamma )  \; ,
\label{app2} 
\end{equation} 
and   
\begin{equation} 
\psi(s)= s \ln \kappa + (s+\beta^2 \kappa) 
\cdot \left [ \int \limits_{0}^{1} 
\frac{1 - e^{-st/\kappa}}{t} \,d t~-~ \gamma \right ] 
- \kappa \, e^{-s/\kappa} \; ,
\label{app3} 
\end{equation} 
where $\gamma = 0.577\dots$ is the Euler's constant. 

The parameters mean and variance ($\sigma^2$) of the distribution in 
Eq. (\ref{app1}) are given by  
\begin{equation} 
{\hbox{mean}}= \gamma -1 - \ln\kappa-\beta^2;~~~  
\sigma^2 = \frac{2-\beta^2}{2\kappa}.  
\end{equation} 
For $\kappa\le 0.05$, the Vavilov distribution may be approximated by the
Landau distribution, while for $\kappa\ge10$, it may be approximated by
the Gaussian approximation, with the corresponding mean and variance. 
 
We have used the Vavilov distribution function $P(x; \kappa, \beta^2)$ 
defined above, which is also built into ROOT, as the basic distribution 
for the fit. However the hadron hit distribution itself is fitted to the 
modified distribution 
$\left({\rm P}_4/{\rm P}_3 \right)\;
P( ({\rm x}- {\rm P}_2)/{\rm P}_3; 
~{\rm P}_0,~{\rm P}_1)$,  
to account for the x-scaling (${\rm P}_3$), normalization ${\rm P}_4$ and the 
shift of the peak to a non-zero value, ${\rm P}_2$.
Clearly ${\rm P}_0=\kappa$ and ${\rm P}_1 = \beta^2$. 
The modified mean and variance are then 
\begin{equation} 
\mbox{Mean}_{\rm Vavilov} = (\gamma-1-\ln {\rm P}_0 - {\rm P}_1)
{\rm P}_3+ {\rm P}_2\; , \quad 
\sigma^2_{\rm Vavilov}= \frac{(2- {\rm P}_1)}{2 {\rm P}_0} {\rm P}_3^2 \; . 
\end{equation}  
These are the quantities used while presenting the energy response of
hadrons in the ICAL detector.

\chapter{Hadron energy resolution as a function of plate thickness}
\label{app:thickness}

A potentially crucial factor in the determination of hadron energy and 
direction is the thickness of absorber material, namely iron plate 
thickness in ICAL. In all the simulation studies reported here, we have 
assumed that the thickness of the iron plate is 5.6 cm, which is the 
default value. While not much variation in this thickness is possible 
due to constraints imposed by total mass, physical size, location of the 
support structure and other parameters like the cost factor etc, we look 
at possible variation of this thickness in view of optimising the hadron 
energy resolution \cite{Mohan:2014qua}. 
The hadron energy resolution is a crucial limiting 
factor in reconstructing the neutrino energy in atmospheric neutrino 
interactions in the ICAL detector. This information is also helpful 
since ICAL is modular in form and future modules may come in for further 
improvements using such analyses.

Naively, this can be achieved by simply changing the angle of the 
propagation of the particle in the simulation, since the effective thickness 
is $(t/\cos\theta)$. 
In the case of muons this itself may be sufficient to study the effect 
of plate thickness. However, in an actual detector, the detector 
geometry --- including support structure, orientation as well as the 
arrangement of detector elements --- imposes additional nontrivial dependence 
on thickness. Therefore we study hadron energy resolution with the 
present arrangement of ICAL by varying the plate thickness, while other 
parameters are fixed. The analysis was done by propagating pions in the 
simulated ICAL detector at various fixed energies, averaged over all 
directions in each case.

\begin{figure}[htp]
\centering
\includegraphics[width=0.59\textwidth,height=0.45\textwidth]
{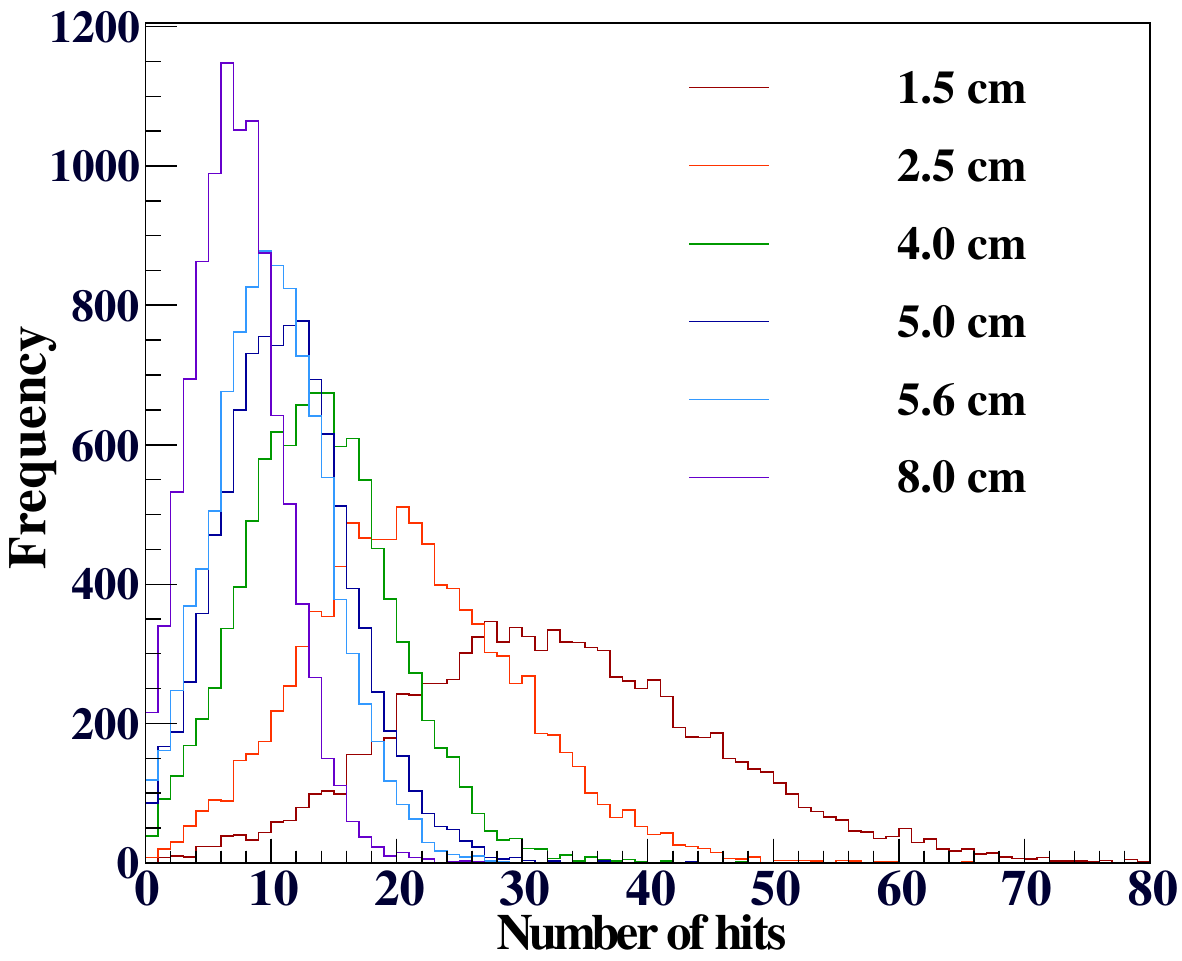}
\caption{Hit distribution of 5 GeV pions propagated through sample
iron plate thicknesses \cite{Mohan:2014qua}.}
\label{apxt1}
\end{figure}

The hit distribution patterns for 5 GeV pions propagated through sample 
plate thicknesses in the central region are shown in  Fig.~\ref{apxt1}. 
The methodology is already discussed in
Chapter~\ref{response} and we will not repeat it here.
For comparing the resolutions with different thicknesses we use the mean
and rms width ($\sigma$) of the hit distributions as functions of energy.

\begin{figure}[htp]
\centering
\includegraphics[width=0.49\textwidth,height=0.35\textwidth]
{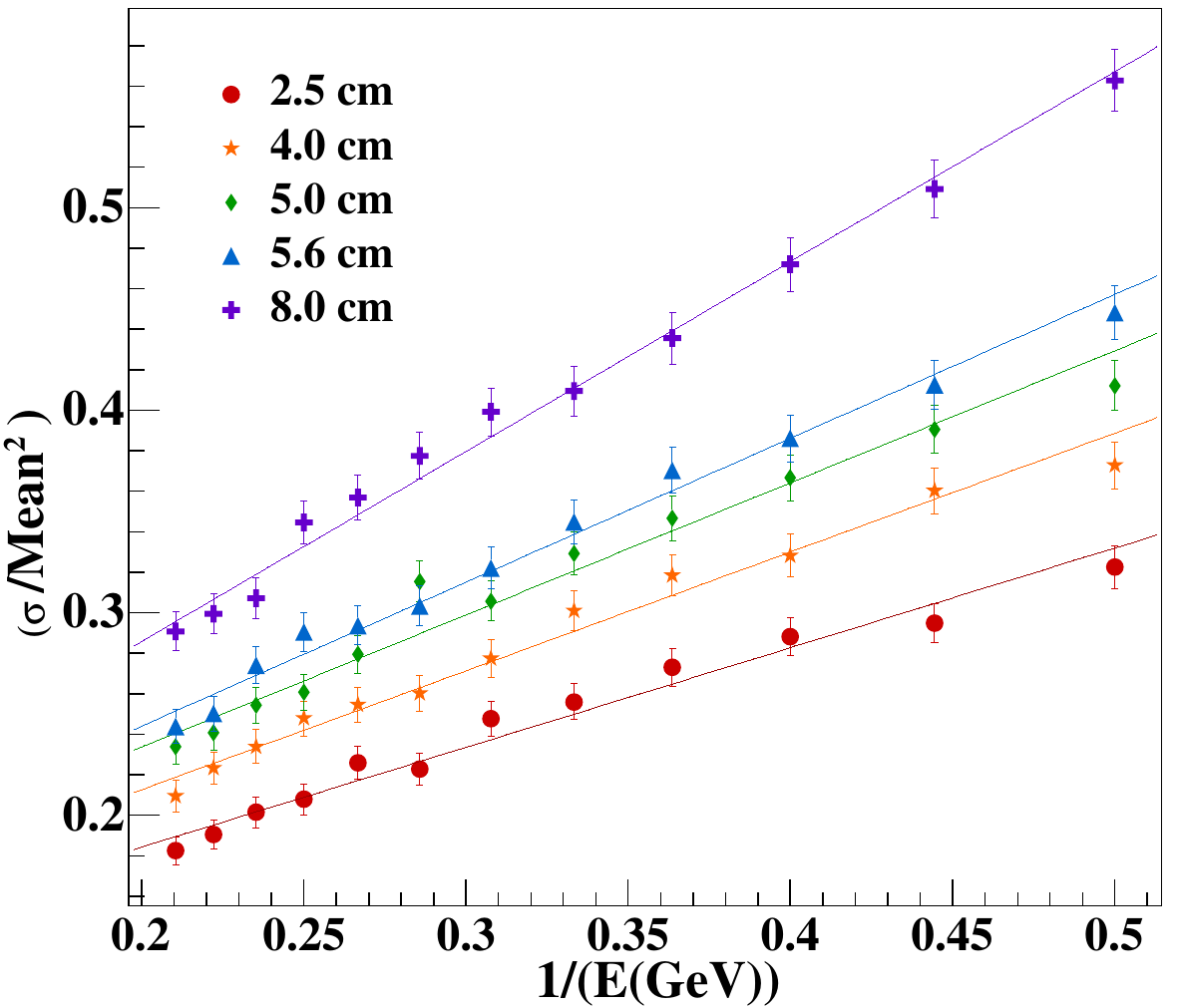}
\includegraphics[width=0.49\textwidth,height=0.35\textwidth]
{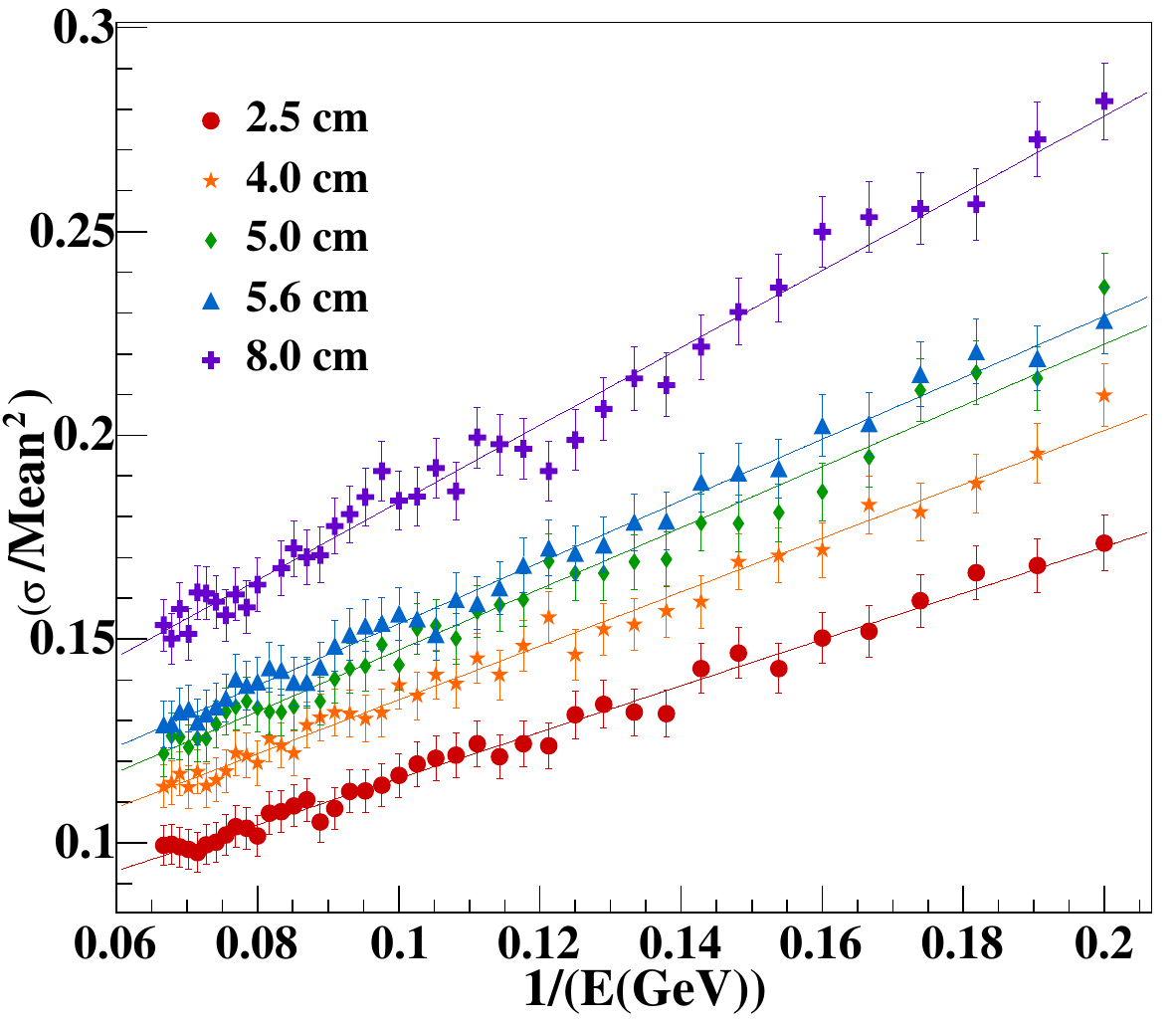}
\caption{Plots of $(\sigma/mean)^2$ as a function of $1/E$. The data
as wells as fits to Eq.\ref{apxeq1} are shown in the energy range
2-4.75 GeV (left) and 5-15 GeV (right). The thickness is varied from 2.5cm
to 8cm \cite{Mohan:2014qua}.}
\label{apxt2}
\end{figure}

The hadron energy resolution is parametrised as 
\begin{equation}
\left(\frac{\sigma}{E}\right)^2 =\frac{a^2}{E}+b^2,
\label{apxeq1}
\end{equation}
where $a$ is the stochastic coefficient and $b$ is a constant, both of 
which depend on the thickness. We divide the relevant 
energy range 2-15 GeV into two sub-ranges, below 5 GeV and above 5 GeV. 
Below 5 GeV, the quasi-elastic, resonance and deep inelastic processes  
contribute to the production of hadrons in neutrino interactions in 
comparable proportions, while above 5 GeV the hadron production is 
dominated by the deep inelastic scattering. 
The results for the energy resolution as a function of plate thickness
are shown in Fig.~\ref{apxt2}.
Note that here we show the square of the resolution instead
of the resolution itself.

\begin{figure}[htp]
\centering
\includegraphics[width=0.59\textwidth,height=0.45\textwidth]
{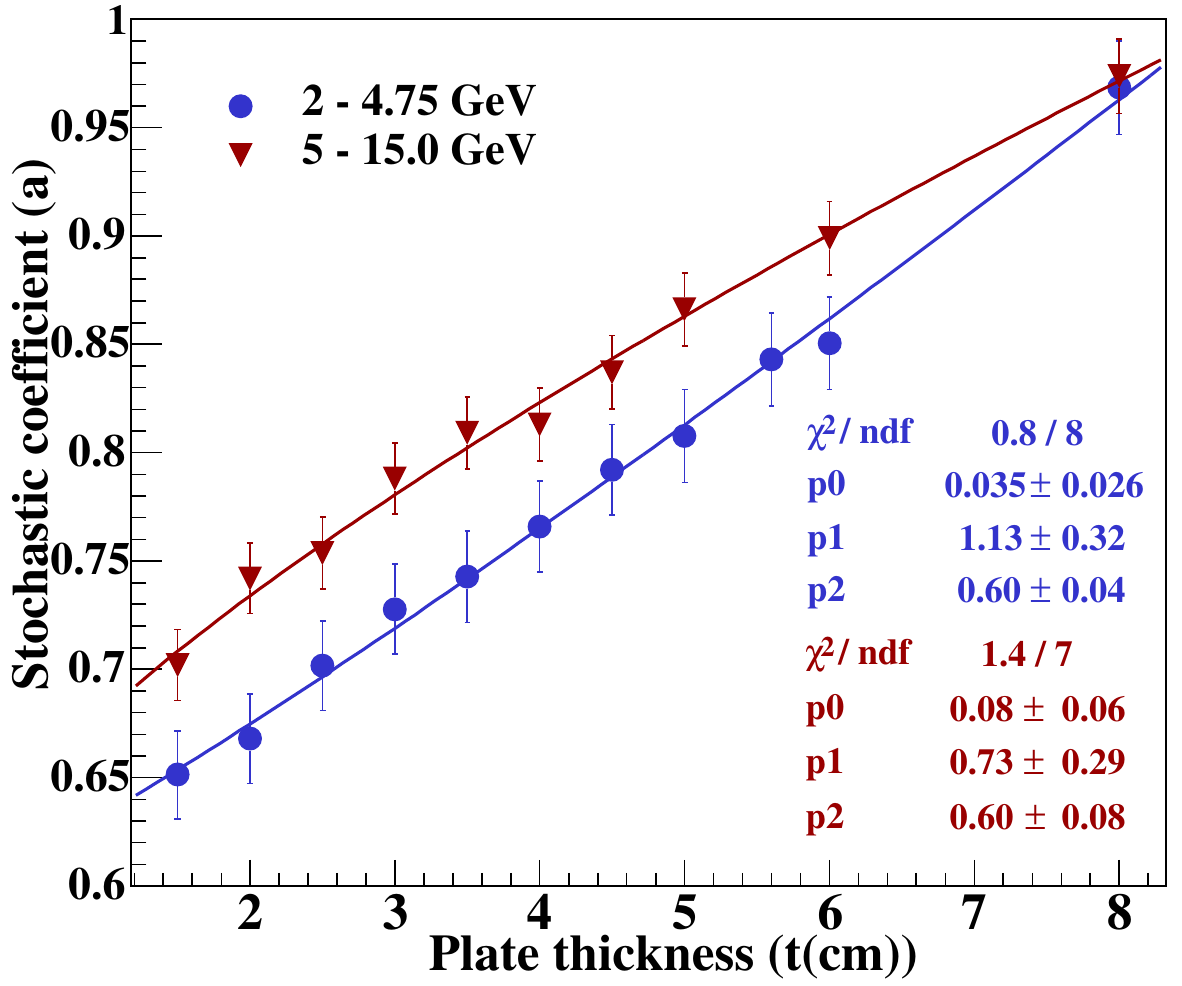}
\caption{The stochastic coefficient $a$ obtained from the fit to 
Eq.\ref{apxeq1} is shown in the two energy ranges as function of plate
thickness \cite{Mohan:2014qua}.}
\label{apxt3}
\end{figure}

The stochastic coefficient $a$ as a function of thickness is 
obtained from a fit to the hadron energy resolution, and is shown in 
Fig.~\ref{apxt3} as a function of plate thickness for the two
energy ranges as in Fig.~\ref{apxt2}.

The analysis in the two energy ranges shows that the thickness dependence
is stronger than $\sqrt{t}$ which is observed in hadron calorimeters at
high energies (tens of GeV) \cite{dang}. 
In fact at the energies of relevance to us,
the thickness dependence is not uniform but dependent on the energy. This
is borne out by two independent analyses: in the first one we obtain
the thickness dependence of the stochastic coefficient $a$ and in the second
analysis we directly parametrise the energy resolution as a function of
thickness at each energy. Typically instead of $t^{0.5}$, we find the
power varying from about 0.65--0.98 depending on the energy. 

Finally we compare the ICAL simulations with varying thickness with 
MONOLITH and MINOS and their test beam runs. This is a useful comparison
since no test beam runs with ICAL prototype have been done up to present.
The data from the above detectors can however be used for 
the validation of the ICAL simulations results.

\begin{figure}[htp]
\centering
\includegraphics[width=0.59\textwidth,height=0.45\textwidth]
{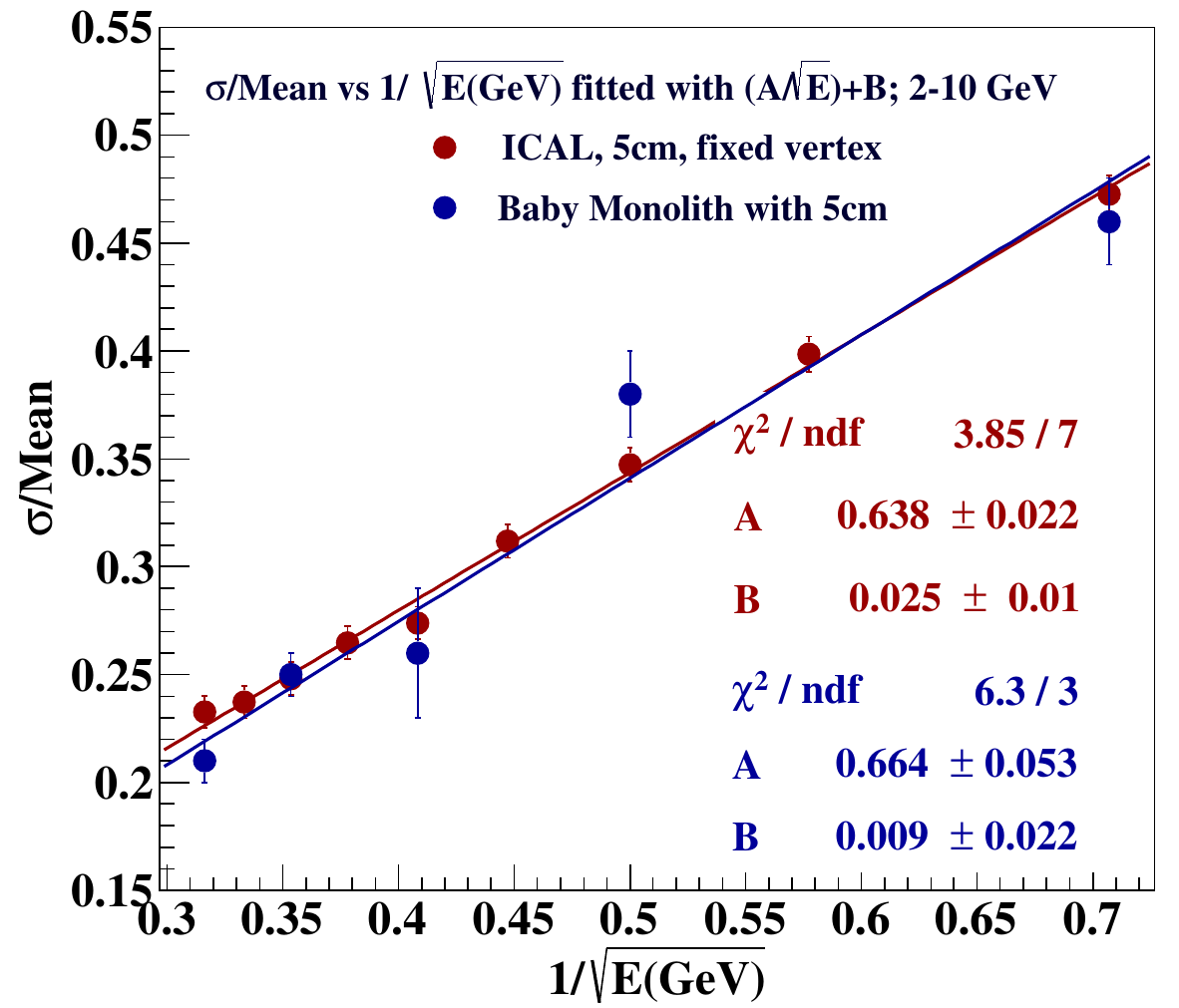}
\caption{The energy response of ICAL detector with 5cm thick iron plates
with single pions in the energy range 2-10 GeV, propagated from a
fixed vertex in the vertical direction \cite{Mohan:2014qua}
compared with the data 
from MONOLITH test beam run \cite{Ambrosio:2000ry,Bari:2003bt}.}
\label{apxt5}
\end{figure}
 
The test beam results for the Baby MONOLITH (BM) detector at CERN
with 5 cm thick iron plates \cite{Ambrosio:2000ry,Bari:2003bt}
have been obtained with the beam energy is in the range 2-10 GeV. 
In order to provide a comparison, we have simulated the ICAL detector 
response with 5cm iron plates, for single pions of energy 2-10 GeV 
incident normally on the detector at a fixed vertex. 
Also, in order to be consistent with
the BM parametrization, the energy resolution $\sigma_E/E$
is fitted to the function $A/\sqrt{E} + B$. A comparison of the ICAL
simulated results with the BM beam results, along with the
respective fits, is shown in the Fig.~\ref{apxt5}. 

For BM, an energy resolution of $\sigma_E/E = 68\%/\sqrt{E} + 2\%$ was 
reported \cite{Ambrosio:2000ry}, however no errors on the parameters
$A$ and $B$ were specified. Our fit to the same BM data gives
$A_{\rm BM} = (66 \pm 5)\%$ and $B_{\rm BM} = (1\pm 2)\%$, which
also gives an estimation of errors on these parameters. 
The fit for the ICAL resolution gives the parameter values
$A_{\rm ICAL} = (64\pm 2)\%$ and $B_{\rm ICAL} = (2 \pm 1)\%$.
The consistency of our simulated results with the beam results 
of BM testifies to the correctness of our approach.  

In the test beam run of MINOS with Aluminium Proportional Tubes (APT)
active detectors and 1.5 inch (4 cm) steel plates, a hadron energy resolution
of $71\%/\sqrt{E}\pm 6\%$ was reported in the range 2.5-30 GeV \cite{minotest}. 
ICAL simulation with 4cm iron plates in the same energy range gave
$61\%/\sqrt{E}\pm 14\%$. The results are compatible within
errors, since the two detector geometries are very different. 

Obviously the final choice of the plate thickness depends not only
on the behaviour of hadrons but also on the energy range of interest to
the physics goals of the experiment. There are also issues of cost, 
sensitivity to muons and even possibly electrons. The thickness
dependence study summarized here provides one such input to the final design.


\end{document}